%% file: main.tex
\documentclass[conference]{IEEEtran}
\IEEEoverridecommandlockouts
\usepackage{cite}
\usepackage{amsmath,amssymb,amsfonts}
\usepackage[section]{placeins}
\usepackage{algorithmic}
\usepackage{multicol}
\usepackage{graphicx}
\usepackage{textcomp}
\usepackage[dvipsnames]{xcolor}
\usepackage{pgfplots}
\pgfplotsset{compat=1.17}
\usepackage{subfiles}
\usepackage{booktabs}
\usepackage{textcomp}
\usepackage{hyperref}[prefix=tikz/]
\hypersetup{
    colorlinks=true,
    linkcolor=black,
    urlcolor=blue,
    citecolor=black
}

\def\BibTeX{{\rm B\kern-.05em{\sc i\kern-.025em b}\kern-.08em
    T\kern-.1667em\lower.7ex\hbox{E}\kern-.125emX}}
\begin{document}
\providecommand{\keywords}[1]
{
  \small	
  \textbf{\textit{Keywords---}} #1
}
\title{CrowdLogo: crowd simulation in NetLogo\\}

\author{\IEEEauthorblockN{Davide Foini}
\IEEEauthorblockA{\textit{Departamento de Engenharia Informática} \\
\textit{Faculdade de Engenharia da Universidade do Porto}\\
up202202244@fe.up.pt}
\and
\IEEEauthorblockN{Magdalena Rzyska}
\IEEEauthorblockA{\textit{Departamento de Engenharia Informática} \\
\textit{Faculdade de Engenharia da Universidade do Porto}\\
up202203499@fe.up.pt}
\and
\IEEEauthorblockN{Katharina Baschmakov}
\IEEEauthorblockA{\textit{Departamento de Engenharia Informática} \\
\textit{Faculdade de Engenharia da Universidade do Porto}\\
up202209100@fe.up.pt}
\and
\IEEEauthorblockN{Sergio Murino}
\IEEEauthorblockA{\textit{Departamento de Engenharia Informática} \\
\textit{Faculdade de Engenharia da Universidade do Porto}\\
up202203049@fe.up.pt}
}

\maketitle

\begin{abstract}
Planning the evacuation of people from crowded places, such as squares, stadiums, or indoor arenas during emergency scenarios is a fundamental task that authorities must deal with. This article summarizes the work of the authors to simulate an emergency scenario in a square using NetLogo, a multi-agent programmable modeling environment. The emergency scenario is based on a real event, which took place in Piazza San Carlo, Turin, on the 3rd of June 2017. The authors have developed a model and conducted various experiments, the results of which are presented, discussed and analyzed. The article concludes by offering suggestions for further research and summarizing the key takeaways.
\end{abstract}

%\begin{IEEEkeywords}
%\end{IEEEkeywords}

\section{Introduction}
\subsection{Context and motivation}
Emergency situations where a crowd is involved are many and they happen in different areas of the world and in different situations, ranging from sports events \cite{b6} to concerts or celebrations \cite{b7}. In that kind of setting, if adequate safety precautions and procedures are overlooked or totally ignored, the evacuation can result in a stampede, causing injuries and victims. \par
Our efforts have been focused on modeling the event that happened in Piazza San Carlo, Turin, on the 3rd of June 2017, when after firecrackers exploded in the middle of the crowd, the stampede that broke out caused three victims and more than one thousand injuries, and the investigations carried out proved that safety norms were not observed \cite{b5}. In particular, the biggest neglects were that iron barriers were placed at the entrances of the square to screen the people entering, but when the evacuation started they acted like a trap blocking the way out and nurturing obstructions. Another fundamental safety norm that was not enforced was the ban on glass bottles: a lot of street vendors were allowed to sell glass beer bottles, and when the crowd started escaping all those bottles started to break, creating more panic, easing slipping and increasing the risk of cutting with shards of glass and being overrun by others.\par
\begin{figure}[ht]
\centerline{\includegraphics[width=0.7\columnwidth, height=60mm]{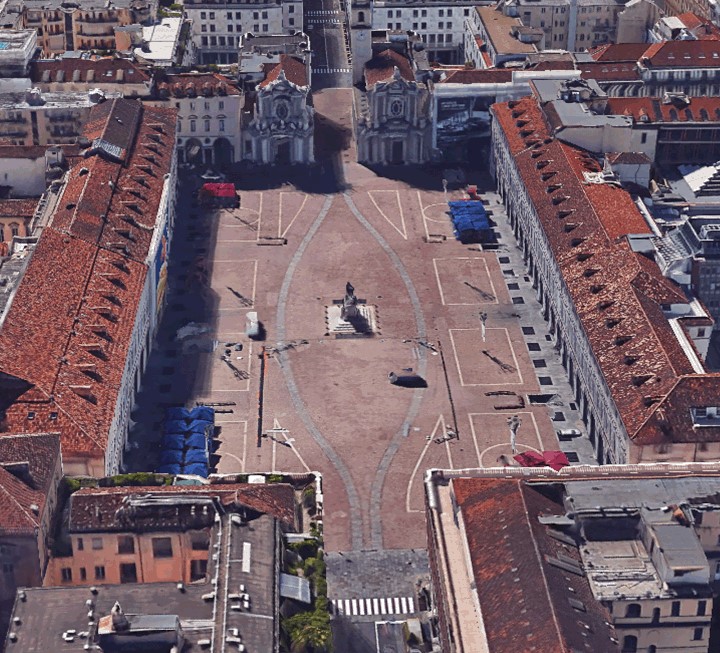}}
\caption{\footnotesize Piazza San Carlo seen from the northern side \cite{b8}}
\label{SanCarloPic}
\end{figure}

In Fig. \ref{SanCarloPic} a bird-eye view of the square is displayed. The northern-side view shows the street (northern gate) that was closed during the evacuation in 2017. However, the simulation model enables to use of this exit for evacuation as well as the other five of them.

\subsection{Goals and expected contributions} \label{Goals and expected contributions}
The main goals of this study have tackled two different simulation approaches: the first was to create a model accurate enough that represented crowd behavior (\emph{descriptive approach}), and the second goal was to verify what could have changed if more attention would have been given to safety measures, like a smartphone application that points the user to the nearest exit (\emph{speculative approach}).\par
We hope our work, still considering its limited extension,  will be a useful tool to analyze situations that can be considered dangerous and that it will be used as a source for further research and developments in this area.\par
The full code of the model is available at \cite{b10}.

\subsection{Structure of the manuscript}
The paper is structured as follows. In Section \ref{LitRev} we analyze the works of different authors on the same topic as the one we focused on, in Section \ref{MetAppr} we first give a formalization of the problem, then we explain our model, both in terms of approach and in details, including also the metrics and the KPIs used. The following Section \ref{Res} will show the results obtained experimenting with the model. The discussion ends by summing up the discussion and with proposals for future works.

\section{Literature Review}\label{LitRev}
A comprehensive overview of evacuation models is presented in  \cite{b11}, here the authors suggest that agent-based models may be the most suitable for the developing of what-if scenarios, among a large list of different approaches, such as cellular automata models or lattice gas models. Moreover, they present a categorization of the evacuation models, these are a classical model, a hybridized model, and a generic model. Each of the latter models is further split into subcategories, hence giving a really comprehensive, but fragmented list of possibilities. Following this scheme, our model could mostly be classified as: classical model \textrightarrow microscopic model \textrightarrow information of individual movement \textrightarrow agent-based.\par
In \cite{b1} the authors developed in NetLogo an evacuation scenario, but their focus was on the architecture of a closed public space, and only an adult population was considered. The experiments carried out varied the number of exits and their width and observed better performances when the Portuguese Fire Code requirements were respected.\par
Concerning the “Review of Pedestrian and Evacuation Simulations” by G. Keith Still, \cite{b12} the past situations have an impact on creating new simulation models – as it was after the Word Trade Centre attack on 9th of September 2001. The simulations aimed to analyze the problems that occurred during the evacuation to prevent the same mistakes in the future and adjust the safety rules in the buildings. The main goal of the simulations described in the paper was to reach the lowest time of evacuation (due to the fire in the building and the high probability of building collapse).  When it comes to open spaces the evacuation on San Carlo square happened in 2017 and the safety procedures were not scrupulously obeyed, which led to the death of 3 people and thousands injured. By the simulation model described in this report, we would like to analyze the main problems that can occur during mass events. Since the main concern in such events is panic, which may cause the growth of injured participants and victims, the goal is not to obtain the time of the evacuation minimized, but to keep the lowest possible number of victims at the lowest possible time of the evacuation.\par
In \cite{b13} another approach for an agent-based model of evacuations inside buildings is modeled and investigated. In particular, this study is trying to include psychological factors into its model, such as group decision-making, leader-follower, and consensus. The results indicate that evacuating individually is faster than evacuating in groups and evacuation time increases together with the size of the group. Inspired by that we are also considering implementing psychological factors, in our case we are choosing the awareness fraction and panic fraction.\par
Lastly, in the research of \cite{b14}, evacuation models and results can be implemented into cyber-physical systems (CPS) with the main aim to support decision-making processes in evacuation scenarios.\\

\section{Methodological approach}\label{MetAppr}
\subsection{Problem formalization}
The aim of the project is to show the behavior of pedestrians during an emergency situation - their decision-making process and its influence on others. The simulation will be expanded by the possibility of adding new gates and take into account if a certain percentage of people is aware of the best evacuation path, for example with a mobile application installed on their smartphones. The analysis will show how the evacuation time and damage during emergency situation change in the different organizational patterns.
\subsection{Modelling approach}
\begin{figure}[htbp]
    \centerline{\includegraphics[width=\columnwidth, height=60mm]{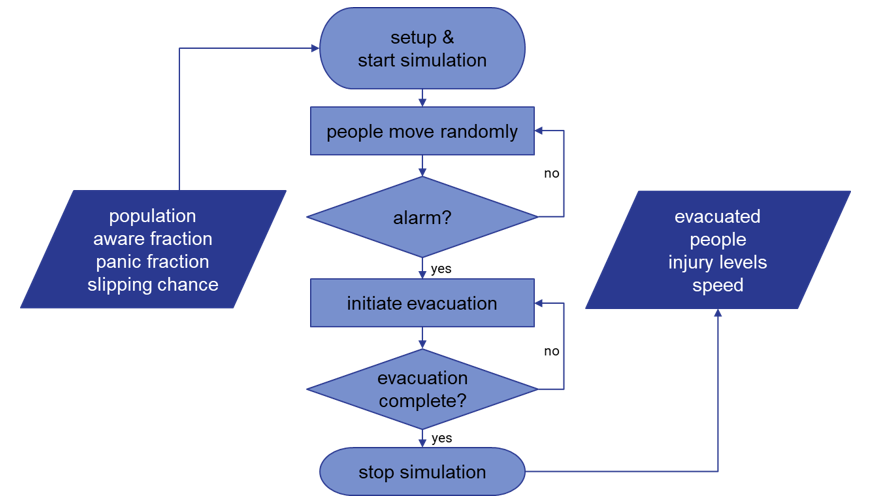}}
    \caption{Flow chart of the simulation process.}
    \label{flowchart}
\end{figure}

The simulation starts with the setup of the map of Piazza San Carlo and a certain number of pedestrians spawning within. When the alarm starts, indicating the beginning of the evacuation, their role is to find the closest exit which may be accessed in the shortest possible time. Pedestrians have to adjust their velocity according to other pedestrians and be able to change direction. They own attributes like velocity, position, direction, the time they need to leave the area to be evacuated, and index of the health state. Velocity and health state depends on the interactions between pedestrians, more specifically on the density of people per patch. The health state is categorized into seven levels of injury, which are taken from the Abbreviated Injury Scale \cite{b4}. Health status, evacuation time, and speed are displayed in plots whilst the simulation is running. Furthermore, we introduce two variables in the model: the aware fraction and the panic fraction. The aware fraction indicates the fraction of pedestrians who are aware of the coordinates of the exits so aware pedestrians move then directly towards these. Panic fraction defines the number of pedestrians who are experiencing panic, hence they move more randomly. The simulation ends when all people leave the square. Figure \ref{flowchart} describes the model as a flow chart.
\subsection{Detailed description}
In this subsection, we are going to describe deeper how the procedure we implemented works and what they are willing to render about a real situation.\par
The first aim has been to reproduce a fair abstraction of the place where the event took place, Piazza San Carlo. We decided to create a bijection between a patch color and its type, namely, we associated one color with each of the following types: gate, wall, outside of the square, inside of the square, and obstacles (e.g. the statue in the center of the square).\par
Then the start\_simulation procedure was created, which aim is indeed to initialize the environment, spanning a number of people equal to the global variable population, and make them move randomly in the square before the evacuation starts. After an alarm is activated they will perform some actions based on some combination of their self variable, for example escaping, rational, or panic.\par
The evacuation starts once the alarm button is pushed, with this procedure we set the escaping status of every turtle to true, and make the turtles face their destination, which is a random gate if aware was (randomly) set to false or the nearest gate if aware was set to true. The scope of this aware binary variable is to reproduce a scenario (hence it refers to a speculative approach) in which people may receive a message, through an application specially created for the event, that computes for every user their nearest gate at every moment; this could be very useful since it may help to minimize the total evacuation time.\par
Once all the people have their status set to escaping they will follow either the procedure move\_person or the procedure follow\_crowd, based on the value of their panic binary variable. With this binary variable we wanted to reproduce how panic influences people's actions, the main assumption is that if panic is present people will tend to randomly follow the crowd instead of take a rational decision, like checking their smartphones and heading themselves to the nearest gate. The global variable panic\_fraction allows us to choose how much we want panic to influence the model, moreover once panic is present every person which is actually panicking will have a different amount of panic, sampled randomly from a uniform distribution, which will then be the parameter of a Bernoulli distribution governing the variable rational (if rationality is not present people will follow the crowd). The procedure move\_person is the most detailed of our model, and it aims to govern how people should move in the square once the evacuation started and if their rational status is set to true; the high amounts of details in it come from the fact that we had to handle with a lot of different combinations of patches' color, number of people in every patch.\par
At every iteration of our simulation, namely every tick, we also perform an update of people's health status, this is done through the procedure updating people's status and is based on \cite{b4}. After the health status is updated also the speed is, with the procedure update\_speed, which takes into account the health status of every person together with her/his gender and age.\par
Finally, we have the, previously mentioned, follow\_crowd procedure, which either makes the turtle follow the most crowded patch neighbor or makes him/her exit the evacuation if a patch gate is one of its neighbors. This is based on the assumption that even if a person is panicking, once he or she has a gate really nearby, he will go through the exit.

\subsection{Metrics and KPIs}
To evaluate the simulation results following metrics were taken into consideration:
\begin{itemize}
 \item \textbf{gates throughput (evacuation speed)}: number of people exiting per second;
 \item \textbf{evacuation time}: time between the beginning of the life-threatening situation and the end of the evacuation. The evacuation is finished when all people have left the area; 
 \item \textbf{average speed}: average velocity of people in the simulation;
 \item \textbf{injury level}: number of people with the specified health status based on the Injury Severity Score \cite{b4}: healthy, minor, moderate, serious, severe, critical and fatal.\par

\end{itemize}
Furthermore, a subsequent assessment considered the following Key Performance Indicators: 
\begin{itemize}
 \item \textbf{number of victims}: number of people with injury level specified as fatal;
 \item \textbf{number of injured}: number of people with minor, moderate, serious, severe, critical injury level.  

\end{itemize}

\section{Results and discussion}\label{Res}
\subfile{sections/results_and_discussion}

\section{Conclusion}\label{Con}
\subfile{sections/conclusion}

\subfile{sections/bibliography}
\end{document}

%% file: sections/results_and_discussion.tex
In this section, we are going to explain the different scenarios we have simulated, which operational policies we used, the experiments we performed, and the results obtained.
\subsection{Scenarios}
As already introduced in Section B, we simulated two different scenarios: the \emph{descriptive scenario}, and the \emph{speculative scenario}.
The \emph{descriptive scenario} aims at reproducing the real event as realistically as possible.
The \emph{speculative scenario} has been developed as a tool to assess what could have happened if the safety norms would have been observed, meaning reducing the number of people in the square, the absence of glass bottles and the more accessible gates, and the availability of the mobile application to know the nearest exit.\par

\subsection{Operation policies}
The operational policies employed differ in the two scenarios. In the \emph{descriptive scenario}, just 50\% of the population is aware of the nearest exit, while in the \emph{speculative scenario} this rate is increased in each experiment. We also analyzed the impact of glass bottles: if their presence is flagged as true people have the chance of slipping and therefore not moving. The diffusion of a smartphone application indicating the nearest exit is also considered. Another operational policy is mediated via the panic fraction, which influences how people behave.  The final operational policy used is the accessibility of gates, meaning that the maximum number of people on the patch is the same as a "regular" one.

\subsection{Experiments}
All of the experiments performed were executed using Netlogo \emph{behavior space}, more in detail one experiment per scenario, for a total of six experiments.\par
The first experiment is referred to the descriptive scenario, so with the aim of reproducing the real event, with a focus on the number of victims. This experiment will also represent the baseline to measure the results of the later experiments, the full parameters setup is available in Table \ref{param_tab}.
\begin{table}[htbp]
    \centering
    \caption{Baseline parameters setup}
    \label{param_tab}
\begin{tabular}{@{} *5l @{}}    \toprule
\emph{Parameter} & \emph{Value} &&&  \\\midrule
population & 30000\\
aware\_fraction & 50\\
panic\_fraction & 0\\ 
glass\_bottles & on\\
real\_exits & on\\
female\_fraction & 50\\
adult\_fraction & 80\\
elderly\_fraction & 10\\
children\_fraction & 10\\
injury\_weight & 0.1365\\
speed\_enabled & on\\
scale & 2\\
slipping\_chance & 1\\
people\_dim & 0.75\\
wall-thickness & 0\\\bottomrule
 \hline
\end{tabular}
\vspace{1mm}
   
\end{table}

From the second to the last experiments the impact of different situations has been analyzed, based on the different operational policies pointed out in the previous section.
In Table \ref{experiments_tab} all of the experiments are described in terms of the number of runs and composition of such runs.
\begin{table}[htbp]
    \centering
    \caption{Experiments description}
    \label{experiments_tab}
\resizebox{\columnwidth}{!}{%
\begin{tabular}{@{} *5l @{} *5l @{}}    \toprule
\emph{Experiment} & \emph{Runs} & \emph{Description} &&&  \\\midrule
Descriptive & 10 & ten runs to verify the accuracy of the simulation\\
Number of people & 5 & starting from 30,000 people to 20,000, with a step of 2,500\\
Glass bottles & 6 & three runs with glass bottles and three runs without\\ 
Mobile application & 6 & starting from 50\%  to 100\%, with a step of 10\%\\
Panic fraction & 6 & starting from 0\%  to 50\%, with a step of 10\%\\
Accessible exits & 6 & three runs without accessible exits and three with accessible exits\\\bottomrule
 \hline
\end{tabular}
}
\vspace{1mm}
    
\end{table}

\subsection{Results}
In this section, we are going to show and discuss the results obtained from running the experiments, starting from the descriptive experiments and proceeding afterward to the speculative ones. \par
The first result is the duration of the evacuation, which is shown in Table \ref{evac_times_tab}, where the average value is reported for the runs of the experiment. The experiment that impacted the most on the duration of the evacuation was the panic fraction experiment, where increasing the panic fraction from 0\% to 10\% was almost enough to double the time duration of the evacuation. Another interesting feedback was that increasing the diffusion of the mobile application from 90\% to 100\% nearly halved the evacuation time. \par
%table for evacuation duration
\begin{table}[htbp]
    \centering
    \caption{Duration of the evacuation in different experiments}
    \label{evac_times_tab}
\begin{tabular}{@{} *5l @{}}    \toprule
\emph{Experiment} & \emph{Evac. Duration [s]} &&&  \\\midrule
Descriptive & 351  \\ 
No glass bottles & 345\\ 
Accessible exits & 353\\
Mobile app & \\
\begin{tabular}{r}
60\%
\end{tabular} & 351\\
\begin{tabular}{r}
70\% 
\end{tabular} & 351\\
\begin{tabular}{r}
80\%
\end{tabular} & 347\\
\begin{tabular}{r}
90\%
\end{tabular} & 338\\
\begin{tabular}{r}
100\%
\end{tabular} & 171\\
Number of people & \\
\begin{tabular}{r}
27500
\end{tabular} & 348\\
\begin{tabular}{r}
25000
\end{tabular} & 355\\
\begin{tabular}{r}
22500
\end{tabular} & 348\\
\begin{tabular}{r}
20000
\end{tabular} & 346\\
Panic fraction & \\
\begin{tabular}{r}
10\%
\end{tabular} & 666\\
\begin{tabular}{r}
20\%
\end{tabular} & 649\\
\begin{tabular}{r}
30\%
\end{tabular} & 671\\
\begin{tabular}{r}
40\%
\end{tabular} & 682\\
\begin{tabular}{r}
50\%
\end{tabular} & 639\\\bottomrule
 \hline
\end{tabular}
\vspace{1mm}
    
\end{table}

\textbf{Descriptive Scenario}\par
\begin{figure}[htbp]
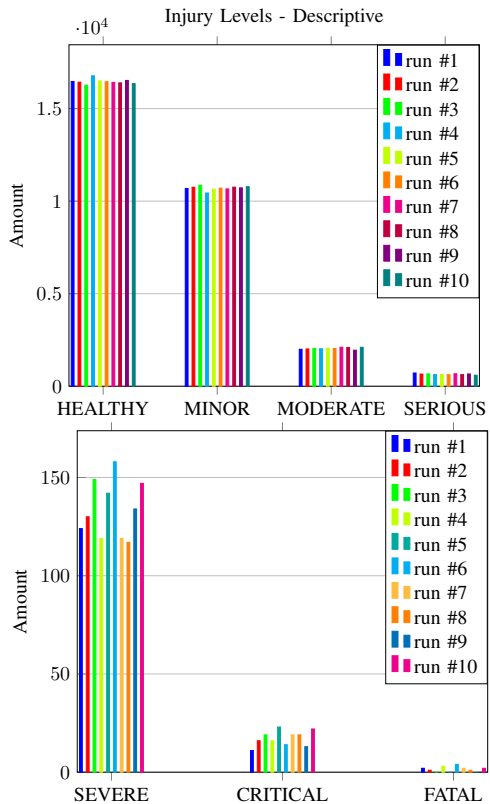

\centering
\subfile{../plots/descriptive/injury_levels}
\caption{Injury levels per run in the descriptive experiment}
\label{injury_levels_descriptive}
\end{figure}

The focus of this experiment is on the number of victims resulting from the evacuation. In Fig. \ref{injury_levels_descriptive} is possible to note in the last column of the second histogram how in all the runs the number of victims is between zero and five, due to the difference in generating the initial position of people, which is close to the real number of three victims. It is also worth noting that all the other levels of injury are on the same level independently of the run.

\begin{figure}[htbp]
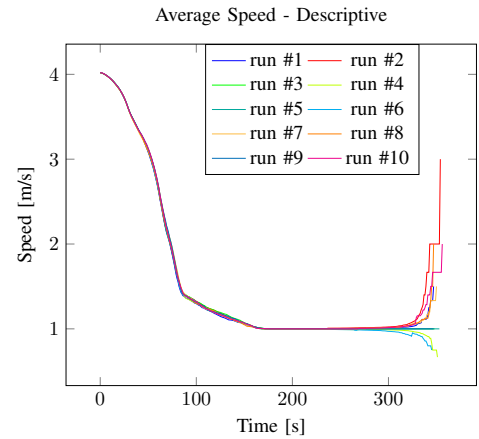

    \centering
    \subfile{../plots/descriptive/avg_speed}
    \caption{Average speed per run in the descriptive scenario}
    \label{avg_speed_descriptive}
\end{figure}

\begin{figure}[htbp]
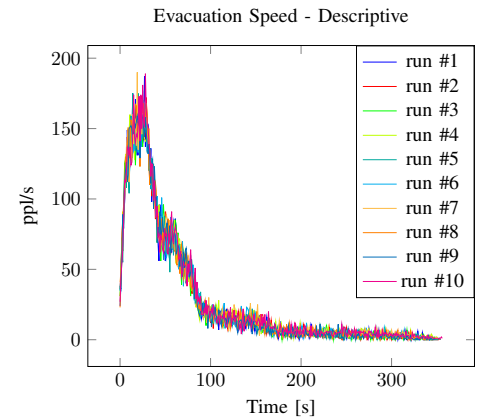

    \centering
    \subfile{../plots/descriptive/evac_speed}
    \caption{Evacuation speed per run in the descriptive scenario. }
    \label{evac_speed_descriptive}
\end{figure}

With regards to the average speed and the evacuation speed, reported in Figures \ref{avg_speed_descriptive} and \ref{evac_speed_descriptive}, it should be noted that the values and their evolution over time are homogeneous, except for the last part for the average speed, that is due to some people that were stuck in the crowd and find a path when most of the others were already evacuated.

The last result obtained was the evolution of the evacuation through the simulation, displayed in Fig. \ref{evac_time_num_ppl}, which have been divided in two plots for clarity. It is possible to see how run \#10 and run \#2 have a lower evacuation time and the others have an higher one. 
\begin{figure}[htbp]
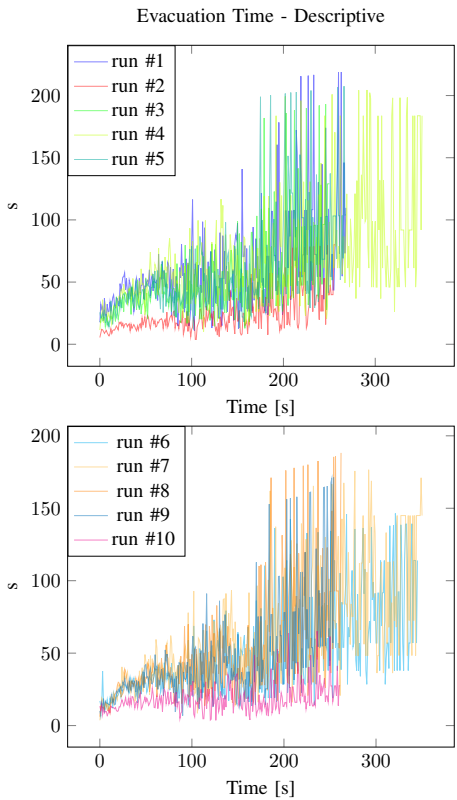

    \centering
    \subfile{../plots/descriptive/evac_time}
    \caption{Evacuation time per run in the descriptive scenario}
    \label{evac_time_descriptive}
\end{figure}

\textbf{Speculative Scenario}\par
\textbf{\emph{Number of People}}\par
\begin{figure}[htbp]
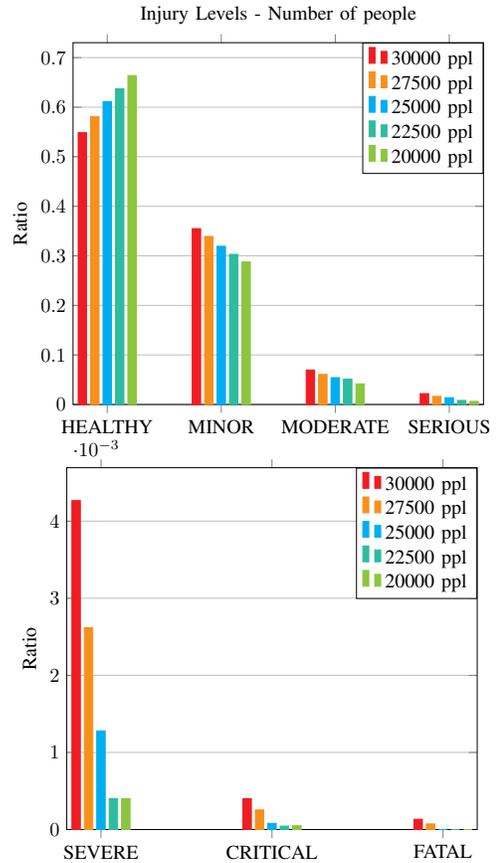

    \centering
    \subfile{../plots/num_ppl/injury_levels}
    \caption{Injury levels per run in the experiment based on the number of people}
    \label{injury_levels_num_ppl}
\end{figure}
\begin{figure}[htbp]
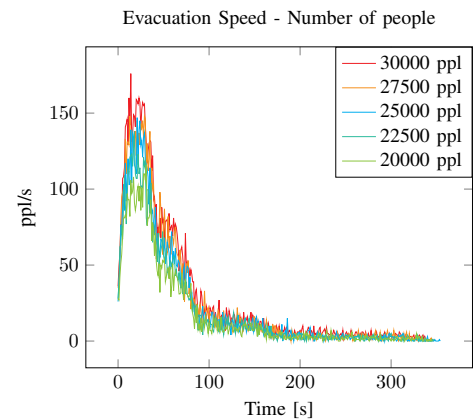

    \centering
    \subfile{../plots/num_ppl/evac_speed}
    \caption{Evacuation speed per run in the experiment based on the number of people}
    \label{evac_speed_num_ppl}
\end{figure}

With regards to the experiment based on the number of people, reported in Fig. \ref{injury_levels_num_ppl}, it is worth noting that decreasing the number of people gradually decreases the number of the first level of injury, but for the last three categories the ratio is almost halved every run and that with five thousand fewer people there are no victims.
What emerged from the experiments about the evacuation speed (Fig. \ref{evac_speed_num_ppl}) and the evacuation time (Fig. \ref{evac_time_num_ppl}) is that all the runs follow the same trend, but with slightly higher values based on the number of people, meaning that that the values on the y axis decrease starting from the first run o the last one.
\begin{figure}[htbp]
    \centering
    \subfile{../plots/num_ppl/evac_time}
    \caption{Evacuation time per run in the experiment based on the number of people}
    \label{evac_time_num_ppl}
\end{figure}

\begin{figure}[htbp]
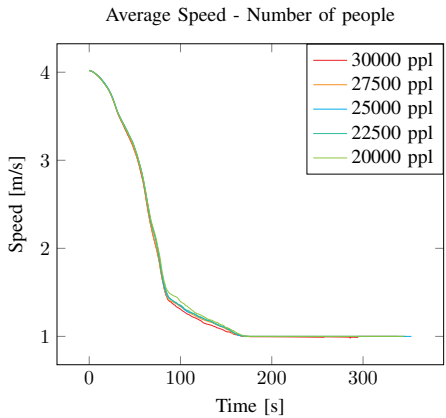

    \centering
    \subfile{../plots/num_ppl/avg_speed}
    \caption{Average speed per run in the experiment based on the number of people}
    \label{avg_speed_num_ppl}
\end{figure}

The experiment about the average speed of the simulation showed that the values and their evolution do not change based on the number of people (Fig. \ref{avg_speed_num_ppl}).

\vspace{0.5cm}
\textbf{\emph{Mobile Application}}\par
\begin{figure}[htbp]
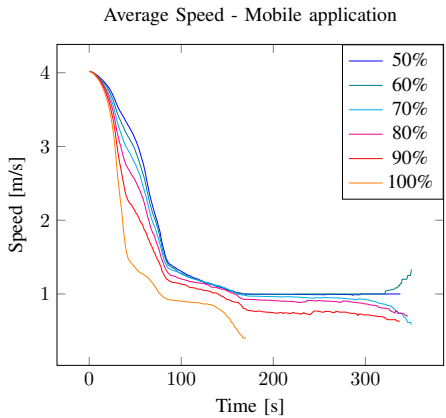

    \centering
    \subfile{../plots/aware_fraction/avg_speed}
    \caption{Average speed per run in the experiment based on the diffusion of the mobile application}
    \label{avg_speed_aware_fraction}
\end{figure}
\begin{figure}[htbp]
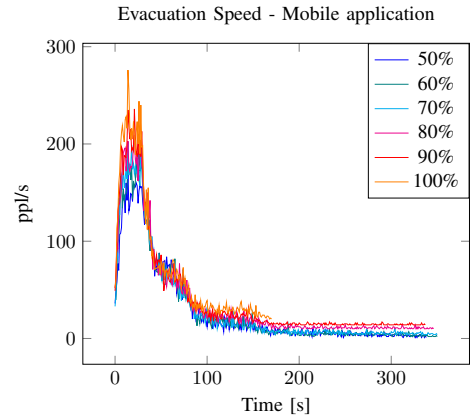

    \centering
    \subfile{../plots/aware_fraction/evac_speed}
    \caption{Evacuation speed per run in the experiment based on the diffusion of the mobile application}
    \label{evac_speed_aware_fraction}
\end{figure}
\begin{figure}[htbp]
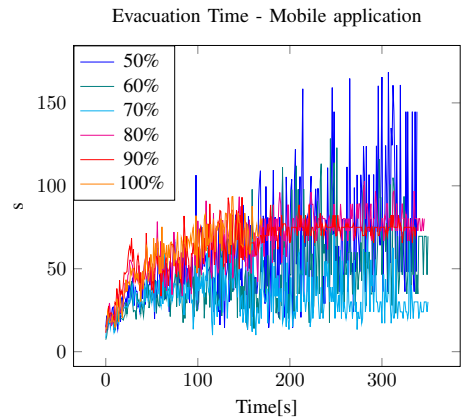

    \centering
    \subfile{../plots/aware_fraction/evac_time}
    \caption{Evacuation time per run in the experiment based on the diffusion of the mobile application}
    \label{evac_time_aware_fraction}
\end{figure}
\begin{figure}[htbp]
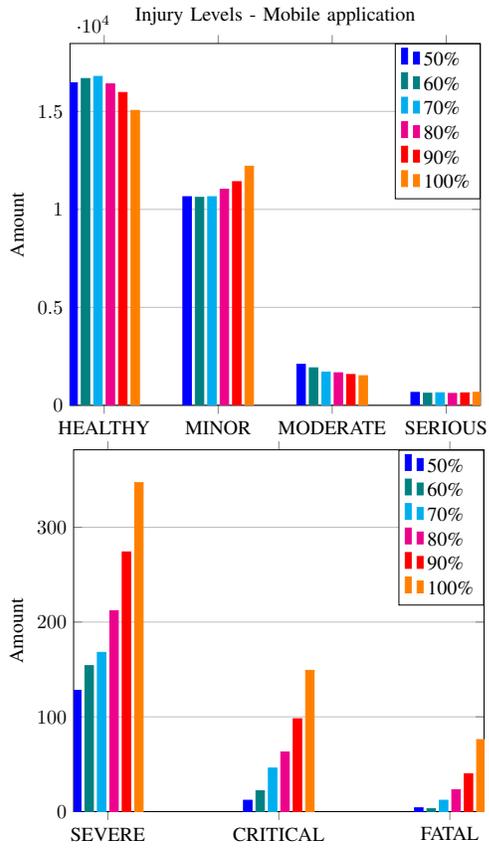

    \centering
    \subfile{../plots/aware_fraction/injury_levels}
    \caption{Injury levels per run in the experiment based on the diffusion of the mobile application}
    \label{injury_levels_aware_fraction}
\end{figure}

This experiment is the one that has given the most interesting results. In Fig. \ref{avg_speed_aware_fraction} it is possible to notice how the rapidity with which the average speed decreases with higher levels of diffusion of the smartphone application, while in Fig. \ref{evac_speed_aware_fraction} the evacuation speed has higher peaks with higher degrees of diffusion. 
With regard to the evacuation time, plotted in Fig. \ref{evac_time_aware_fraction}, the best results have been obtained with a diffusion ratio of 70\% and the more stable ones for the higher ratios of 90\% and 100\%. It is also possible to note already in this graph that the maximum ratio of diffusion obtained the best evacuation time. This result is also shown in Table \ref{evac_times_tab}. 
The downside of having a higher percentage of people knowing the nearest exit is displayed in Fig. \ref{injury_levels_aware_fraction}. With the exception of moderate and serious injuries, all the other levels registered a worse value when increasing the diffusion ratio. This is most probably caused by the fact that a higher density is more easily obtained when a lot of people reach their exit in a short time, creating a blockage in the proximity of the gates.
To conclude, a better evacuation time is obtained but in exchange, more injuries and victims are observed.

\vspace{0.5cm}
\textbf{\emph{Accessible Exits}}\par
\begin{figure}[htbp]
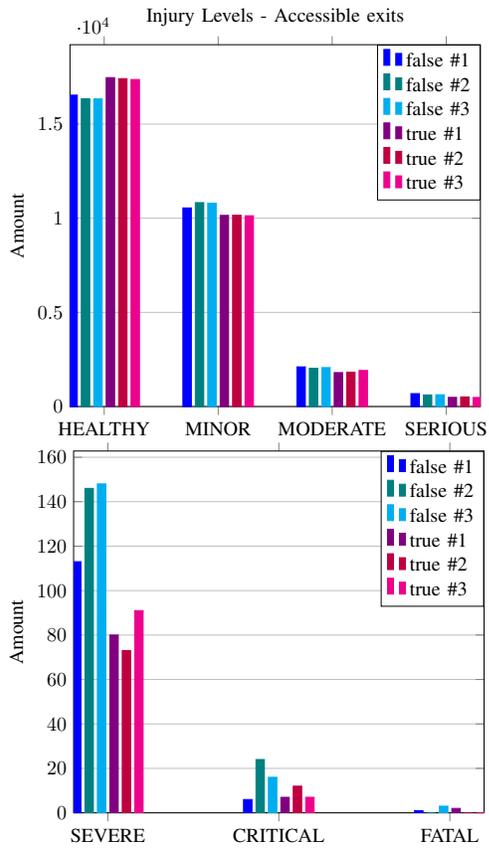

    \centering
    \subfile{../plots/accessible_exits/injury_levels}
    \caption{Injury levels per run in the experiment based on the accessibility of exits}
    \label{injury_levels_accessible_exits}
\end{figure}

Enabling the accessibility of exits has given the expected results in terms of evacuation time and in the levels of injury, even if with fewer improvements with respect to other experiments. In Fig. \ref{evac_time_accessible_exits} is possible t note how the runs with an easier way to exit the square have a lower evacuation time and in \ref{injury_levels_accessible_exits} is shown how the injury levels have lower values when exits are accessible.
\begin{figure}[htbp]
    \centering
    \subfile{../plots/accessible_exits/evac_time}
    \caption{Evacuation time per run in the experiment based on the accessibility of exits}
    \label{evac_time_accessible_exits}
\end{figure}
\begin{figure}[htbp]
    \centering
    \subfile{../plots/accessible_exits/avg_speed}
    \caption{Average speed per run in the experiment based on the accessibility of exits}
    \label{avg_speed_accessible_exits}
\end{figure}
 With what concerns the average speed and the evacuation speed, plotted in Figures \ref{avg_speed_accessible_exits} and \ref{evac_speed_accessible_exits}, the evolution of the values does not change substantially in the runs.
\begin{figure}[htbp]
    \centering
    \subfile{../plots/accessible_exits/evac_speed}
    \caption{Evacuation speed per run in the experiment based on the accessibility of exits}
    \label{evac_speed_accessible_exits}
\end{figure}

\vspace{0.5cm}
\textbf{\emph{Glass Bottles}}\par
\begin{figure}[htbp]
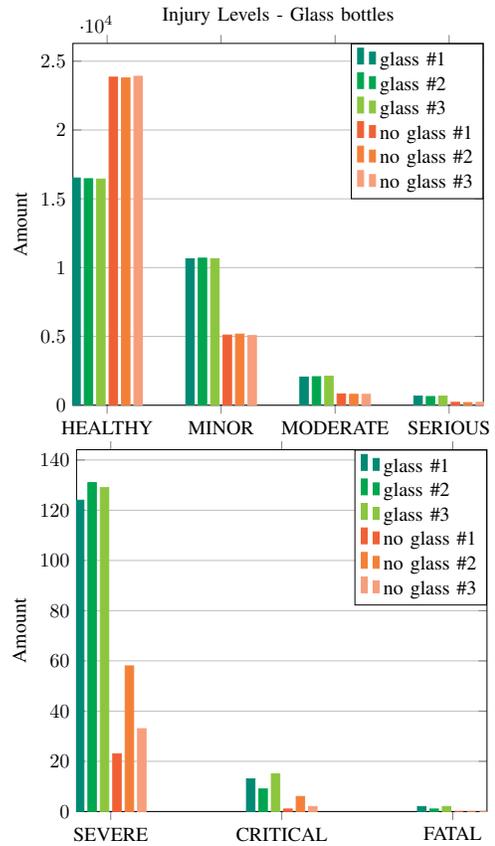

    \centering
    \subfile{../plots/glass_bottles/injury_levels}
    \caption{Injury levels per run in the experiment based on the presence of glass bottles}
    \label{injury_levels_glass_bottles}
\end{figure}
\begin{figure}[htbp]
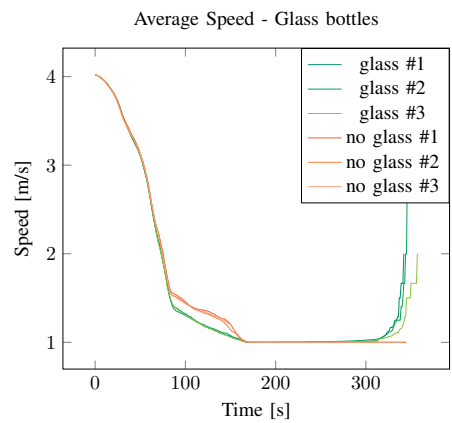

    \centering
    \subfile{../plots/glass_bottles/avg_speed}
    \caption{Average speed per run in the experiment based on the presence of glass bottles}
    \label{avg_speed_glass_bottles}
\end{figure}

Removing the glass bottles from the simulation has given the best results in the number of injured people. In Fig. \ref{injury_levels_glass_bottles} is displayed how the number of healthy people increased by a ratio of almost 60\%, while the low levels of injury are halved and the higher levels are more than halved.
Slipping left some people with a higher residual speed in the simulation, how it is possible to note in Fig. \ref{avg_speed_glass_bottles}, while the absence of glass bottles did not cause an improvement in the evacuation time (Fig. \ref{evac_time_glass_bottles}) and in the average speed (Fig. \ref{evac_speed_glass_bottles}).
\begin{figure}[htbp]
    \centering
    \subfile{../plots/glass_bottles/evac_speed}
    \caption{Evacuation speed per run in the experiment based on the presence of glass bottles}
    \label{evac_speed_glass_bottles}
\end{figure}
\begin{figure}[htbp]
    \centering
    \subfile{../plots/glass_bottles/evac_time}
    \caption{Evacuation time per run in the experiment based on the presence of glass bottles}
    \label{evac_time_glass_bottles}
\end{figure}

\vspace{0.5cm}
\textbf{\emph{Panic Fraction}}\par
The panic fraction had an impact on both the average speed and the evacuation speed. It is possible to see how in Figures \ref{avg_speed_panic_fraction} and \ref{evac_speed_panic_fraction} after 300 seconds into the simulation the average speed increases again while the evacuation speed starts to become almost constant. That effect is due to the number of people that keep panicking and that after some time they exit that state and resume the evacuation.  
\begin{figure}[htbp]
    \centering
    \subfile{../plots/panic_fraction/avg_speed}
    \caption{Average speed per run in the experiment based on the panic fraction}
    \label{avg_speed_panic_fraction}
\end{figure}
\begin{figure}[hbtp]
        \centering
        \subfile{../plots/panic_fraction/evac_speed}
        \caption{Evacuation speed per run in the experiment based on the panic fraction}
        \label{evac_speed_panic_fraction}
\end{figure}
With regard to the evacuation time, it is possible to note how the higher percentages of panic cause a more stable evacuation time with respect to the lower ones.
\begin{figure}
        \centering
        \subfile{../plots/panic_fraction/evac_time}
        \caption{Evacuation time per run in the experiment based on the panic fraction}
        \label{evac_time_panic_fraction}
\end{figure}
As expected the injury levels increase their values with higher panic fractions with the exception of the moderate injuries and serious injuries that are not observed. Moreover, three segments are identifiable, where the levels of injuries increasing not proportionally but every 20\%, with the exception of the victims that are more than doubled from 40\% to 50\%.
\begin{figure}[hbtp]
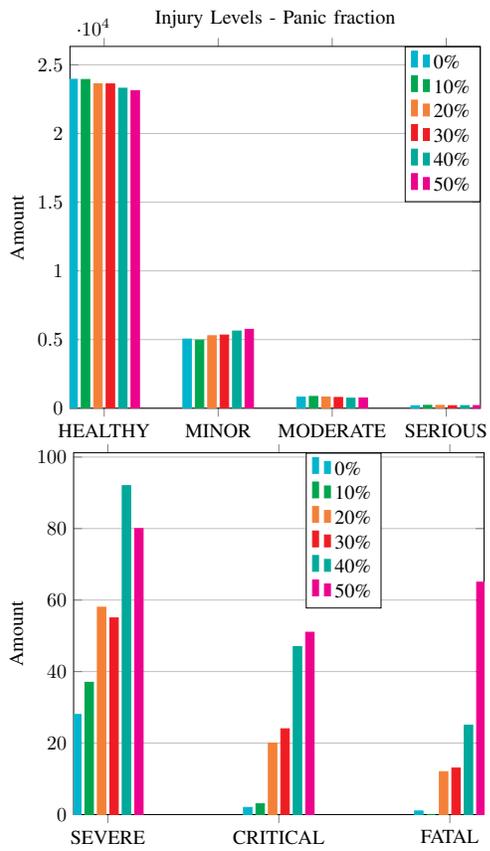

    \centering
    \subfile{../plots/panic_fraction/injury_levels}
    \caption{Injury levels per run in the experiment based on the panic fraction}
    \label{injury_levels_panic_fraction}
\end{figure}

%% file: plots/descriptive/injury_levels.tex
\begin{tikzpicture}[scale=0.75]
\begin{axis}[
width=\columnwidth,
title={Injury Levels - Descriptive},
ybar, % style of the histogram: clustered columns
%width=\clumnwidth, % width of the plot
ylabel={Amount}, % label for the y-axis
symbolic x coords={HEALTHY,MINOR,MODERATE,SERIOUS}, % labels for the x-ticks
xtick=data, % position the x-ticks at the data points
ymin=0, % minimum value of the y-axis
legend style={at={(1,1)}, anchor=north east}, % position of the legend
legend cell align=left, % alignment of the legend cells
ymajorgrids=true, % display major grids
bar width=0.05cm % width of the bars
]
\addplot[color=blue, fill] coordinates{(HEALTHY,16466)(MINOR,10685)(MODERATE,1996)(SERIOUS,716)
}; % plot of the first histogram
\addplot[color=red, fill] coordinates{(HEALTHY,16427)(MINOR,10752)(MODERATE,2015)(SERIOUS,659)
}; % plot of the second histogram
\addplot[color=green, fill] coordinates{(HEALTHY,16264)(MINOR,10857)(MODERATE,2043)(SERIOUS,668)
}; % plot of the third histogram
\addplot[color=cyan, fill] coordinates{(HEALTHY,16771)(MINOR,10440)(MODERATE,2025)(SERIOUS,626)
}; % plot of the fourth histogram
\addplot[color=lime, fill] coordinates{(HEALTHY,16492)(MINOR,10649)(MODERATE,2058)(SERIOUS,636)
}; % plot of the fifth histogram
\addplot[color=orange, fill] coordinates{(HEALTHY,16453)(MINOR,10703)(MODERATE,2043)(SERIOUS,625)
}; % plot of the first histogram
\addplot[color=magenta, fill] coordinates{(HEALTHY,16416)(MINOR,10661)(MODERATE,2109)(SERIOUS,674)
}; % plot of the second histogram
\addplot[color=purple, fill, fill] coordinates{(HEALTHY,16390)(MINOR,10755)(MODERATE,2091)(SERIOUS,627)
}; % plot of the third histogram
\addplot[color=violet, fill] coordinates{(HEALTHY,16522)(MINOR,10718)(MODERATE,1949)(SERIOUS,664)
}; % plot of the fourth histogram
\addplot[color=teal, fill] coordinates{(HEALTHY,16347)(MINOR,10781)(MODERATE,2105)(SERIOUS,596)
}; % plot of the fifth histogram
\legend{run \#1,  run \#2, run \#3, run \#4, run \#5, run \#6,  run \#7, run \#8, run \#9, run \#10} % legend of the plot
\end{axis}
\end{tikzpicture}

\begin{tikzpicture}[scale=0.75]
\begin{axis}[
width=\columnwidth,
ybar, % style of the histogram: clustered columns
%width=\clumnwidth, % width of the plot
symbolic x coords={SEVERE,CRITICAL,FATAL}, % labels for the x-ticks
ylabel={Amount}, % label for the y-axis
xtick=data, % position the x-ticks at the data points
ymin=0, % minimum value of the y-axis
legend style={at={(1,1)}, anchor=north east}, % position of the legend
legend cell align=left, % alignment of the legend cells
ymajorgrids=true, % display major grids
bar width=0.05cm % width of the bars
]
\addplot[color=blue, fill] coordinates{(SEVERE,124)(CRITICAL,11)(FATAL,2)
}; % plot of the first histogram
\addplot[color=red, fill] coordinates{(SEVERE,130)(CRITICAL,16)(FATAL,1)
}; % plot of the second histogram
\addplot[color=green, fill] coordinates{(SEVERE,149)(CRITICAL,19)(FATAL,0)
}; % plot of the third histogram
\addplot[color=lime, fill] coordinates{(SEVERE,119)(CRITICAL,16)(FATAL,3)
}; % plot of the fourth histogram
\addplot[color=Emerald, fill] coordinates{(SEVERE,142)(CRITICAL,23)(FATAL,0)
}; % plot of the fifth histogram
\addplot[color=cyan, fill] coordinates{(SEVERE,158)(CRITICAL,14)(FATAL,4)
}; % plot of the first histogram
\addplot[color=Dandelion, fill] coordinates{(SEVERE,119)(CRITICAL,19)(FATAL,2)
}; % plot of the second histogram
\addplot[color=orange, fill] coordinates{(SEVERE,117)(CRITICAL,19)(FATAL,1)
}; % plot of the third histogram
\addplot[color=NavyBlue, fill] coordinates{(SEVERE,134)(CRITICAL,13)(FATAL,0)
}; % plot of the fourth histogram
\addplot[color=Magenta, fill] coordinates{(SEVERE,147)(CRITICAL,22)(FATAL,2)
}; % plot of the fifth histogram
\legend{run \#1,  run \#2, run \#3, run \#4, run \#5, run \#6,  run \#7, run \#8, run \#9, run \#10} % legend of the plot
\end{axis}
\end{tikzpicture}

%% file: plots/descriptive/evac_speed.tex
\begin{tikzpicture}[scale=0.75]

% Set the axis labels and title
\begin{axis}[    xlabel={Time [s]},    ylabel={ppl/s},    title={Evacuation Speed  - Descriptive},   legend style={at={(1,1)}, anchor=north east},]

\addplot[color=blue] coordinates{(0,24)(1,40)(2,72)(3,73)(4,94)(5,97)(6,112)(7,116)(8,149)(9,132)(10,137)(11,132)(12,144)(13,138)(14,175)(15,158)(16,137)(17,171)(18,138)(19,131)(20,158)(21,131)(22,167)(23,137)(24,147)(25,145)(26,142)(27,187)(28,137)(29,167)(30,154)(31,153)(32,148)(33,115)(34,136)(35,98)(36,124)(37,112)(38,113)(39,100)(40,100)(41,83)(42,68)(43,56)(44,96)(45,88)(46,96)(47,70)(48,77)(49,64)(50,56)(51,72)(52,74)(53,74)(54,67)(55,83)(56,81)(57,62)(58,80)(59,84)(60,73)(61,86)(62,79)(63,69)(64,67)(65,63)(66,57)(67,54)(68,49)(69,54)(70,52)(71,47)(72,53)(73,40)(74,59)(75,58)(76,53)(77,53)(78,44)(79,42)(80,44)(81,37)(82,33)(83,30)(84,31)(85,28)(86,27)(87,20)(88,27)(89,23)(90,15)(91,19)(92,16)(93,22)(94,24)(95,19)(96,19)(97,14)(98,21)(99,15)(100,13)(101,16)(102,15)(103,19)(104,16)(105,22)(106,18)(107,13)(108,15)(109,7)(110,17)(111,20)(112,19)(113,9)(114,20)(115,16)(116,15)(117,14)(118,22)(119,13)(120,18)(121,14)(122,16)(123,19)(124,19)(125,11)(126,19)(127,18)(128,11)(129,10)(130,13)(131,7)(132,10)(133,14)(134,12)(135,12)(136,14)(137,13)(138,10)(139,8)(140,18)(141,12)(142,10)(143,11)(144,14)(145,6)(146,7)(147,11)(148,11)(149,6)(150,13)(151,12)(152,10)(153,13)(154,17)(155,17)(156,12)(157,8)(158,6)(159,12)(160,4)(161,12)(162,16)(163,5)(164,11)(165,8)(166,9)(167,8)(168,8)(169,6)(170,4)(171,8)(172,9)(173,6)(174,5)(175,1)(176,5)(177,4)(178,3)(179,7)(180,4)(181,7)(182,3)(183,7)(184,9)(185,8)(186,7)(187,8)(188,7)(189,3)(190,9)(191,3)(192,7)(193,5)(194,2)(195,7)(196,9)(197,8)(198,0)(199,2)(200,5)(201,6)(202,2)(203,6)(204,5)(205,4)(206,8)(207,6)(208,5)(209,1)(210,3)(211,11)(212,9)(213,4)(214,6)(215,4)(216,4)(217,2)(218,4)(219,5)(220,7)(221,6)(222,5)(223,6)(224,2)(225,4)(226,2)(227,3)(228,3)(229,5)(230,2)(231,2)(232,2)(233,5)(234,5)(235,7)(236,3)(237,4)(238,2)(239,4)(240,4)(241,2)(242,4)(243,3)(244,3)(245,3)(246,2)(247,5)(248,4)(249,6)(250,1)(251,4)(252,2)(253,2)(254,3)(255,4)(256,2)(257,3)(258,5)(259,3)(260,3)(261,2)(262,2)(263,2)(264,2)(265,6)(266,3)(267,3)(268,3)(269,4)(270,4)(271,5)(272,2)(273,3)(274,1)(275,8)(276,4)(277,3)(278,5)(279,3)(280,3)(281,6)(282,5)(283,5)(284,2)(285,5)(286,4)(287,3)(288,3)(289,2)(290,3)(291,4)(292,4)(293,3)(294,2)(295,5)(296,5)(297,3)(298,1)(299,2)(300,3)(301,3)(302,2)(303,1)(304,1)(305,1)(306,5)(307,0)(308,3)(309,4)(310,2)(311,2)(312,3)(313,3)(314,1)(315,6)(316,1)(317,0)(318,2)(319,2)(320,2)(321,4)(322,0)(323,3)(324,2)(325,3)(326,3)(327,3)(328,2)(329,6)(330,2)(331,0)(332,1)(333,1)(334,5)(335,0)(336,0)(337,0)(338,1)(339,1)(340,2)(341,2)(342,0)(343,1)(344,0)(345,1)(346,1)(347,0)
};
\addplot[color=red] coordinates{(0,36)(1,51)(2,71)(3,87)(4,89)(5,100)(6,130)(7,118)(8,144)(9,131)(10,145)(11,141)(12,158)(13,140)(14,142)(15,147)(16,156)(17,144)(18,166)(19,144)(20,137)(21,149)(22,153)(23,155)(24,173)(25,142)(26,169)(27,142)(28,148)(29,172)(30,153)(31,141)(32,121)(33,125)(34,125)(35,102)(36,110)(37,100)(38,106)(39,86)(40,86)(41,86)(42,76)(43,74)(44,77)(45,73)(46,67)(47,78)(48,78)(49,80)(50,91)(51,72)(52,74)(53,80)(54,70)(55,60)(56,61)(57,75)(58,74)(59,77)(60,77)(61,65)(62,57)(63,70)(64,69)(65,68)(66,60)(67,51)(68,58)(69,61)(70,47)(71,57)(72,61)(73,52)(74,55)(75,46)(76,48)(77,44)(78,49)(79,33)(80,33)(81,40)(82,39)(83,46)(84,26)(85,27)(86,29)(87,19)(88,16)(89,29)(90,24)(91,18)(92,19)(93,17)(94,25)(95,19)(96,21)(97,15)(98,16)(99,24)(100,12)(101,21)(102,17)(103,8)(104,19)(105,19)(106,21)(107,18)(108,12)(109,13)(110,18)(111,17)(112,18)(113,13)(114,14)(115,16)(116,19)(117,19)(118,17)(119,8)(120,12)(121,17)(122,15)(123,10)(124,11)(125,12)(126,13)(127,11)(128,10)(129,13)(130,12)(131,20)(132,16)(133,15)(134,23)(135,9)(136,16)(137,17)(138,13)(139,15)(140,10)(141,13)(142,12)(143,13)(144,13)(145,10)(146,14)(147,9)(148,17)(149,14)(150,12)(151,16)(152,11)(153,8)(154,10)(155,9)(156,10)(157,18)(158,9)(159,10)(160,9)(161,8)(162,4)(163,9)(164,6)(165,15)(166,7)(167,14)(168,9)(169,8)(170,9)(171,10)(172,8)(173,6)(174,7)(175,9)(176,4)(177,7)(178,6)(179,6)(180,2)(181,6)(182,6)(183,3)(184,4)(185,7)(186,5)(187,2)(188,3)(189,1)(190,3)(191,5)(192,8)(193,5)(194,10)(195,4)(196,2)(197,1)(198,4)(199,5)(200,3)(201,4)(202,7)(203,11)(204,8)(205,6)(206,4)(207,3)(208,3)(209,4)(210,4)(211,7)(212,5)(213,3)(214,6)(215,7)(216,2)(217,2)(218,2)(219,6)(220,3)(221,5)(222,6)(223,5)(224,2)(225,5)(226,2)(227,4)(228,4)(229,4)(230,3)(231,9)(232,4)(233,5)(234,2)(235,6)(236,4)(237,4)(238,2)(239,2)(240,4)(241,2)(242,1)(243,1)(244,1)(245,0)(246,0)(247,5)(248,3)(249,6)(250,9)(251,4)(252,2)(253,1)(254,3)(255,1)(256,2)(257,5)(258,4)(259,5)(260,1)(261,4)(262,2)(263,4)(264,6)(265,6)(266,0)(267,2)(268,4)(269,5)(270,2)(271,2)(272,3)(273,0)(274,1)(275,2)(276,3)(277,4)(278,3)(279,2)(280,5)(281,2)(282,3)(283,5)(284,3)(285,4)(286,3)(287,1)(288,2)(289,1)(290,1)(291,1)(292,5)(293,5)(294,3)(295,0)(296,2)(297,4)(298,2)(299,4)(300,0)(301,2)(302,4)(303,2)(304,3)(305,2)(306,3)(307,3)(308,3)(309,5)(310,5)(311,1)(312,2)(313,1)(314,6)(315,3)(316,1)(317,2)(318,0)(319,5)(320,3)(321,2)(322,2)(323,2)(324,2)(325,2)(326,1)(327,2)(328,2)(329,1)(330,3)(331,2)(332,1)(333,2)(334,1)(335,1)(336,2)(337,0)(338,1)(339,1)(340,0)(341,0)(342,1)(343,0)(344,0)(345,0)(346,0)(347,0)(348,0)(349,0)(350,0)(351,0)(352,0)(353,1)(354,0)
};
\addplot[color=green] coordinates{(0,31)(1,50)(2,70)(3,63)(4,95)(5,125)(6,122)(7,124)(8,123)(9,132)(10,121)(11,138)(12,145)(13,129)(14,143)(15,144)(16,140)(17,124)(18,146)(19,167)(20,155)(21,137)(22,169)(23,169)(24,160)(25,137)(26,167)(27,155)(28,167)(29,150)(30,158)(31,136)(32,127)(33,126)(34,120)(35,119)(36,124)(37,110)(38,101)(39,99)(40,97)(41,69)(42,91)(43,81)(44,83)(45,56)(46,76)(47,89)(48,96)(49,79)(50,87)(51,60)(52,69)(53,78)(54,72)(55,68)(56,78)(57,82)(58,72)(59,75)(60,85)(61,85)(62,81)(63,71)(64,76)(65,60)(66,68)(67,50)(68,39)(69,60)(70,58)(71,60)(72,45)(73,57)(74,54)(75,54)(76,48)(77,44)(78,31)(79,57)(80,40)(81,33)(82,34)(83,38)(84,33)(85,28)(86,22)(87,24)(88,19)(89,20)(90,28)(91,23)(92,19)(93,23)(94,18)(95,10)(96,21)(97,13)(98,23)(99,14)(100,21)(101,23)(102,13)(103,12)(104,22)(105,18)(106,15)(107,12)(108,15)(109,17)(110,18)(111,17)(112,13)(113,17)(114,10)(115,14)(116,21)(117,19)(118,18)(119,11)(120,18)(121,18)(122,15)(123,12)(124,13)(125,12)(126,13)(127,13)(128,17)(129,15)(130,17)(131,14)(132,12)(133,12)(134,12)(135,16)(136,7)(137,16)(138,14)(139,12)(140,17)(141,15)(142,14)(143,14)(144,16)(145,15)(146,12)(147,9)(148,13)(149,19)(150,9)(151,10)(152,6)(153,16)(154,13)(155,12)(156,13)(157,5)(158,14)(159,9)(160,5)(161,9)(162,10)(163,7)(164,7)(165,6)(166,10)(167,4)(168,2)(169,3)(170,9)(171,8)(172,7)(173,7)(174,3)(175,3)(176,4)(177,7)(178,6)(179,0)(180,5)(181,4)(182,6)(183,2)(184,4)(185,2)(186,5)(187,3)(188,8)(189,2)(190,6)(191,3)(192,4)(193,4)(194,4)(195,5)(196,3)(197,8)(198,4)(199,5)(200,8)(201,5)(202,3)(203,4)(204,4)(205,7)(206,5)(207,4)(208,3)(209,4)(210,3)(211,4)(212,4)(213,6)(214,6)(215,3)(216,4)(217,3)(218,6)(219,3)(220,4)(221,4)(222,3)(223,6)(224,3)(225,6)(226,5)(227,7)(228,10)(229,7)(230,3)(231,4)(232,2)(233,1)(234,6)(235,6)(236,0)(237,5)(238,4)(239,6)(240,2)(241,3)(242,3)(243,6)(244,7)(245,4)(246,3)(247,1)(248,4)(249,4)(250,2)(251,5)(252,2)(253,1)(254,2)(255,4)(256,0)(257,3)(258,4)(259,4)(260,2)(261,4)(262,4)(263,5)(264,4)(265,0)(266,0)(267,4)(268,4)(269,4)(270,3)(271,2)(272,3)(273,2)(274,4)(275,2)(276,4)(277,0)(278,4)(279,1)(280,6)(281,5)(282,4)(283,4)(284,1)(285,3)(286,4)(287,2)(288,3)(289,2)(290,3)(291,1)(292,0)(293,7)(294,2)(295,1)(296,3)(297,3)(298,2)(299,5)(300,1)(301,0)(302,1)(303,3)(304,0)(305,3)(306,1)(307,4)(308,2)(309,2)(310,2)(311,2)(312,2)(313,0)(314,4)(315,1)(316,2)(317,7)(318,1)(319,3)(320,3)(321,0)(322,1)(323,2)(324,3)(325,2)(326,1)(327,2)(328,1)(329,0)(330,1)(331,1)(332,0)(333,1)(334,1)(335,4)(336,3)(337,0)(338,0)(339,1)(340,1)(341,2)(342,1)(343,0)(344,0)(345,0)(346,1)(347,2)(348,1)(349,0)
};
\addplot[color=lime] coordinates{(0,31)(1,42)(2,59)(3,91)(4,87)(5,107)(6,132)(7,124)(8,125)(9,129)(10,149)(11,160)(12,128)(13,127)(14,149)(15,136)(16,128)(17,153)(18,157)(19,137)(20,138)(21,137)(22,150)(23,160)(24,153)(25,148)(26,163)(27,143)(28,141)(29,142)(30,141)(31,142)(32,129)(33,128)(34,123)(35,111)(36,118)(37,115)(38,116)(39,95)(40,95)(41,87)(42,78)(43,96)(44,71)(45,72)(46,67)(47,86)(48,66)(49,64)(50,86)(51,76)(52,82)(53,66)(54,78)(55,73)(56,80)(57,69)(58,68)(59,71)(60,85)(61,70)(62,54)(63,69)(64,66)(65,72)(66,71)(67,47)(68,63)(69,57)(70,64)(71,59)(72,43)(73,48)(74,57)(75,41)(76,51)(77,40)(78,44)(79,35)(80,44)(81,34)(82,38)(83,33)(84,23)(85,31)(86,23)(87,17)(88,14)(89,28)(90,27)(91,18)(92,19)(93,22)(94,23)(95,18)(96,15)(97,22)(98,17)(99,18)(100,23)(101,24)(102,14)(103,23)(104,18)(105,18)(106,25)(107,22)(108,21)(109,28)(110,14)(111,17)(112,17)(113,18)(114,13)(115,14)(116,17)(117,13)(118,12)(119,14)(120,13)(121,17)(122,17)(123,15)(124,22)(125,15)(126,12)(127,14)(128,21)(129,18)(130,7)(131,15)(132,12)(133,15)(134,18)(135,15)(136,11)(137,9)(138,17)(139,11)(140,12)(141,12)(142,9)(143,21)(144,11)(145,14)(146,10)(147,14)(148,21)(149,13)(150,14)(151,11)(152,10)(153,14)(154,16)(155,14)(156,4)(157,14)(158,7)(159,11)(160,8)(161,10)(162,12)(163,6)(164,8)(165,9)(166,5)(167,8)(168,5)(169,3)(170,7)(171,8)(172,13)(173,7)(174,7)(175,10)(176,6)(177,10)(178,12)(179,5)(180,6)(181,5)(182,6)(183,6)(184,5)(185,8)(186,5)(187,4)(188,7)(189,4)(190,5)(191,6)(192,1)(193,6)(194,7)(195,7)(196,3)(197,5)(198,5)(199,12)(200,7)(201,4)(202,3)(203,6)(204,4)(205,7)(206,5)(207,5)(208,5)(209,5)(210,7)(211,8)(212,5)(213,3)(214,2)(215,3)(216,6)(217,4)(218,6)(219,2)(220,4)(221,7)(222,3)(223,6)(224,8)(225,2)(226,6)(227,7)(228,2)(229,3)(230,9)(231,5)(232,6)(233,4)(234,4)(235,5)(236,7)(237,3)(238,5)(239,5)(240,4)(241,7)(242,8)(243,2)(244,3)(245,1)(246,7)(247,9)(248,6)(249,7)(250,4)(251,4)(252,2)(253,3)(254,4)(255,3)(256,6)(257,5)(258,2)(259,3)(260,5)(261,3)(262,1)(263,3)(264,7)(265,7)(266,7)(267,8)(268,7)(269,5)(270,3)(271,5)(272,4)(273,6)(274,7)(275,8)(276,3)(277,5)(278,5)(279,5)(280,3)(281,3)(282,4)(283,4)(284,2)(285,6)(286,6)(287,1)(288,6)(289,4)(290,6)(291,4)(292,5)(293,3)(294,3)(295,2)(296,4)(297,4)(298,4)(299,5)(300,4)(301,4)(302,4)(303,1)(304,2)(305,3)(306,5)(307,1)(308,3)(309,3)(310,2)(311,3)(312,4)(313,5)(314,3)(315,4)(316,4)(317,3)(318,1)(319,2)(320,4)(321,7)(322,2)(323,3)(324,4)(325,3)(326,4)(327,2)(328,2)(329,2)(330,2)(331,2)(332,3)(333,4)(334,1)(335,2)(336,3)(337,2)(338,2)(339,2)(340,2)(341,1)(342,3)(343,1)(344,4)(345,3)(346,1)(347,1)(348,1)(349,1)(350,2)(351,1)
};  
\addplot[color=Emerald] coordinates{(0,35)(1,44)(2,62)(3,55)(4,100)(5,115)(6,122)(7,108)(8,138)(9,116)(10,104)(11,135)(12,159)(13,141)(14,175)(15,153)(16,143)(17,134)(18,148)(19,161)(20,175)(21,148)(22,137)(23,164)(24,161)(25,143)(26,170)(27,156)(28,149)(29,153)(30,133)(31,139)(32,120)(33,136)(34,107)(35,129)(36,114)(37,125)(38,103)(39,99)(40,88)(41,85)(42,83)(43,82)(44,70)(45,82)(46,71)(47,80)(48,57)(49,88)(50,67)(51,72)(52,66)(53,61)(54,69)(55,48)(56,71)(57,71)(58,80)(59,60)(60,77)(61,70)(62,69)(63,71)(64,60)(65,61)(66,52)(67,53)(68,55)(69,58)(70,62)(71,47)(72,55)(73,42)(74,44)(75,45)(76,49)(77,40)(78,54)(79,48)(80,40)(81,41)(82,35)(83,26)(84,27)(85,35)(86,36)(87,21)(88,24)(89,34)(90,20)(91,20)(92,20)(93,26)(94,21)(95,18)(96,25)(97,18)(98,25)(99,21)(100,17)(101,16)(102,16)(103,15)(104,15)(105,20)(106,14)(107,18)(108,18)(109,20)(110,17)(111,16)(112,21)(113,19)(114,17)(115,13)(116,14)(117,16)(118,17)(119,10)(120,12)(121,13)(122,12)(123,13)(124,10)(125,20)(126,13)(127,12)(128,13)(129,13)(130,13)(131,11)(132,14)(133,11)(134,14)(135,15)(136,14)(137,14)(138,13)(139,14)(140,16)(141,14)(142,19)(143,11)(144,12)(145,10)(146,22)(147,15)(148,12)(149,15)(150,19)(151,16)(152,15)(153,9)(154,13)(155,7)(156,9)(157,11)(158,9)(159,11)(160,6)(161,5)(162,9)(163,6)(164,7)(165,13)(166,7)(167,5)(168,8)(169,9)(170,4)(171,9)(172,5)(173,5)(174,9)(175,5)(176,4)(177,7)(178,7)(179,5)(180,2)(181,3)(182,8)(183,8)(184,5)(185,1)(186,6)(187,9)(188,4)(189,6)(190,5)(191,8)(192,2)(193,4)(194,5)(195,3)(196,6)(197,3)(198,1)(199,6)(200,7)(201,4)(202,2)(203,2)(204,4)(205,3)(206,7)(207,5)(208,6)(209,6)(210,2)(211,5)(212,3)(213,4)(214,5)(215,3)(216,6)(217,7)(218,6)(219,2)(220,4)(221,2)(222,2)(223,5)(224,5)(225,8)(226,4)(227,5)(228,1)(229,4)(230,3)(231,1)(232,4)(233,7)(234,2)(235,1)(236,7)(237,5)(238,2)(239,5)(240,5)(241,6)(242,3)(243,1)(244,2)(245,3)(246,5)(247,1)(248,0)(249,4)(250,5)(251,4)(252,3)(253,4)(254,4)(255,4)(256,7)(257,7)(258,5)(259,4)(260,0)(261,5)(262,5)(263,7)(264,3)(265,4)(266,4)(267,6)(268,2)(269,1)(270,7)(271,3)(272,4)(273,3)(274,5)(275,2)(276,6)(277,5)(278,4)(279,2)(280,3)(281,3)(282,3)(283,2)(284,0)(285,3)(286,3)(287,1)(288,0)(289,1)(290,2)(291,4)(292,2)(293,5)(294,5)(295,3)(296,3)(297,3)(298,3)(299,1)(300,2)(301,4)(302,1)(303,4)(304,2)(305,4)(306,3)(307,4)(308,2)(309,3)(310,1)(311,2)(312,2)(313,4)(314,2)(315,2)(316,1)(317,2)(318,4)(319,4)(320,3)(321,0)(322,4)(323,2)(324,3)(325,3)(326,0)(327,4)(328,1)(329,4)(330,1)(331,2)(332,3)(333,1)(334,2)(335,0)(336,0)(337,1)(338,2)(339,0)(340,0)(341,0)(342,0)(343,0)(344,1)(345,0)(346,0)(347,0)(348,0)(349,0)(350,0)(351,0)(352,1)(353,1)(354,0)
}; 
\addplot[color=cyan] coordinates{(0,35)(1,48)(2,65)(3,66)(4,83)(5,108)(6,115)(7,115)(8,126)(9,139)(10,115)(11,125)(12,154)(13,157)(14,151)(15,134)(16,136)(17,144)(18,154)(19,139)(20,133)(21,144)(22,165)(23,137)(24,155)(25,177)(26,160)(27,169)(28,158)(29,164)(30,134)(31,146)(32,131)(33,118)(34,111)(35,108)(36,104)(37,118)(38,123)(39,92)(40,88)(41,102)(42,82)(43,81)(44,84)(45,80)(46,101)(47,83)(48,75)(49,79)(50,70)(51,58)(52,62)(53,79)(54,65)(55,73)(56,87)(57,81)(58,64)(59,79)(60,76)(61,70)(62,76)(63,65)(64,62)(65,50)(66,66)(67,63)(68,45)(69,70)(70,38)(71,55)(72,56)(73,49)(74,56)(75,45)(76,47)(77,48)(78,60)(79,38)(80,53)(81,53)(82,28)(83,41)(84,34)(85,31)(86,34)(87,22)(88,33)(89,16)(90,18)(91,21)(92,27)(93,20)(94,26)(95,18)(96,20)(97,22)(98,18)(99,22)(100,15)(101,19)(102,25)(103,14)(104,18)(105,13)(106,18)(107,20)(108,20)(109,15)(110,21)(111,11)(112,17)(113,18)(114,19)(115,26)(116,8)(117,15)(118,11)(119,15)(120,8)(121,13)(122,17)(123,17)(124,11)(125,11)(126,18)(127,10)(128,9)(129,13)(130,10)(131,13)(132,16)(133,18)(134,24)(135,10)(136,14)(137,12)(138,13)(139,17)(140,23)(141,22)(142,20)(143,22)(144,14)(145,17)(146,17)(147,8)(148,13)(149,19)(150,15)(151,17)(152,8)(153,13)(154,5)(155,11)(156,7)(157,11)(158,14)(159,9)(160,10)(161,8)(162,7)(163,16)(164,8)(165,7)(166,5)(167,8)(168,8)(169,8)(170,3)(171,10)(172,6)(173,5)(174,4)(175,5)(176,5)(177,13)(178,7)(179,5)(180,8)(181,9)(182,8)(183,6)(184,8)(185,9)(186,6)(187,4)(188,8)(189,4)(190,3)(191,4)(192,7)(193,7)(194,8)(195,5)(196,4)(197,3)(198,5)(199,4)(200,4)(201,10)(202,4)(203,1)(204,5)(205,3)(206,12)(207,5)(208,4)(209,6)(210,4)(211,4)(212,7)(213,5)(214,6)(215,7)(216,7)(217,7)(218,4)(219,5)(220,5)(221,11)(222,9)(223,4)(224,4)(225,2)(226,9)(227,3)(228,3)(229,3)(230,4)(231,3)(232,6)(233,8)(234,7)(235,4)(236,2)(237,2)(238,4)(239,5)(240,4)(241,7)(242,4)(243,3)(244,5)(245,10)(246,5)(247,7)(248,3)(249,4)(250,6)(251,2)(252,7)(253,6)(254,6)(255,1)(256,5)(257,6)(258,3)(259,2)(260,6)(261,7)(262,7)(263,5)(264,5)(265,5)(266,5)(267,6)(268,4)(269,4)(270,4)(271,7)(272,8)(273,5)(274,4)(275,5)(276,2)(277,2)(278,6)(279,3)(280,3)(281,3)(282,2)(283,8)(284,3)(285,1)(286,2)(287,3)(288,4)(289,3)(290,4)(291,2)(292,9)(293,6)(294,2)(295,2)(296,5)(297,6)(298,6)(299,3)(300,2)(301,2)(302,3)(303,3)(304,3)(305,1)(306,4)(307,3)(308,3)(309,1)(310,4)(311,6)(312,2)(313,2)(314,7)(315,2)(316,2)(317,5)(318,1)(319,2)(320,3)(321,3)(322,2)(323,3)(324,3)(325,3)(326,3)(327,2)(328,1)(329,2)(330,3)(331,1)(332,2)(333,2)(334,1)(335,3)(336,2)(337,2)(338,1)(339,2)(340,2)(341,1)(342,1)(343,1)(344,2)(345,1)
};  
\addplot[color=Dandelion] coordinates{(0,41)(1,49)(2,65)(3,86)(4,83)(5,99)(6,114)(7,137)(8,142)(9,151)(10,112)(11,151)(12,129)(13,136)(14,160)(15,139)(16,154)(17,156)(18,138)(19,190)(20,135)(21,160)(22,134)(23,163)(24,146)(25,165)(26,160)(27,151)(28,150)(29,164)(30,151)(31,135)(32,150)(33,122)(34,120)(35,120)(36,122)(37,113)(38,93)(39,106)(40,96)(41,83)(42,90)(43,71)(44,67)(45,81)(46,75)(47,76)(48,74)(49,76)(50,82)(51,89)(52,86)(53,83)(54,66)(55,85)(56,73)(57,65)(58,77)(59,77)(60,72)(61,78)(62,76)(63,75)(64,66)(65,58)(66,62)(67,63)(68,61)(69,60)(70,57)(71,50)(72,55)(73,42)(74,46)(75,60)(76,57)(77,39)(78,51)(79,38)(80,40)(81,44)(82,32)(83,37)(84,32)(85,23)(86,23)(87,24)(88,25)(89,22)(90,20)(91,20)(92,15)(93,17)(94,20)(95,20)(96,12)(97,18)(98,30)(99,13)(100,17)(101,15)(102,17)(103,14)(104,20)(105,16)(106,17)(107,21)(108,16)(109,17)(110,18)(111,14)(112,17)(113,22)(114,15)(115,17)(116,14)(117,20)(118,19)(119,14)(120,6)(121,14)(122,16)(123,14)(124,18)(125,13)(126,25)(127,19)(128,15)(129,14)(130,15)(131,14)(132,15)(133,9)(134,19)(135,17)(136,13)(137,9)(138,14)(139,15)(140,15)(141,15)(142,11)(143,18)(144,26)(145,15)(146,14)(147,14)(148,10)(149,9)(150,13)(151,17)(152,25)(153,8)(154,3)(155,12)(156,10)(157,17)(158,8)(159,15)(160,10)(161,12)(162,13)(163,11)(164,11)(165,11)(166,6)(167,10)(168,5)(169,8)(170,7)(171,7)(172,9)(173,5)(174,7)(175,7)(176,14)(177,10)(178,7)(179,7)(180,2)(181,9)(182,7)(183,5)(184,5)(185,2)(186,4)(187,3)(188,4)(189,7)(190,5)(191,5)(192,5)(193,7)(194,6)(195,3)(196,7)(197,5)(198,7)(199,7)(200,7)(201,3)(202,4)(203,7)(204,7)(205,5)(206,5)(207,9)(208,4)(209,9)(210,4)(211,3)(212,4)(213,3)(214,7)(215,4)(216,10)(217,8)(218,7)(219,5)(220,4)(221,4)(222,5)(223,5)(224,4)(225,3)(226,7)(227,3)(228,7)(229,2)(230,5)(231,3)(232,2)(233,4)(234,5)(235,6)(236,7)(237,3)(238,4)(239,5)(240,6)(241,3)(242,4)(243,5)(244,5)(245,4)(246,2)(247,4)(248,2)(249,7)(250,4)(251,4)(252,2)(253,6)(254,2)(255,2)(256,5)(257,4)(258,3)(259,4)(260,7)(261,7)(262,3)(263,5)(264,4)(265,2)(266,7)(267,4)(268,4)(269,7)(270,6)(271,8)(272,6)(273,4)(274,7)(275,7)(276,4)(277,3)(278,3)(279,2)(280,3)(281,3)(282,3)(283,8)(284,3)(285,3)(286,1)(287,6)(288,4)(289,7)(290,5)(291,7)(292,3)(293,3)(294,1)(295,2)(296,7)(297,6)(298,1)(299,4)(300,7)(301,4)(302,3)(303,3)(304,6)(305,5)(306,3)(307,2)(308,4)(309,5)(310,2)(311,4)(312,6)(313,3)(314,5)(315,3)(316,3)(317,4)(318,4)(319,3)(320,2)(321,6)(322,5)(323,1)(324,2)(325,3)(326,2)(327,1)(328,1)(329,1)(330,1)(331,1)(332,1)(333,3)(334,3)(335,1)(336,2)(337,1)(338,2)(339,3)(340,1)(341,2)(342,1)(343,3)(344,1)(345,1)(346,1)(347,1)(348,1)(349,2)(350,1)
}; 
\addplot[color=orange] coordinates{(0,23)(1,47)(2,72)(3,88)(4,96)(5,109)(6,130)(7,112)(8,136)(9,150)(10,137)(11,129)(12,156)(13,154)(14,147)(15,154)(16,152)(17,167)(18,166)(19,148)(20,158)(21,160)(22,123)(23,159)(24,171)(25,161)(26,168)(27,152)(28,159)(29,168)(30,160)(31,140)(32,120)(33,110)(34,123)(35,114)(36,118)(37,94)(38,109)(39,112)(40,106)(41,85)(42,88)(43,75)(44,68)(45,74)(46,86)(47,69)(48,76)(49,70)(50,78)(51,80)(52,73)(53,86)(54,68)(55,81)(56,73)(57,65)(58,81)(59,68)(60,63)(61,63)(62,55)(63,53)(64,66)(65,59)(66,47)(67,65)(68,68)(69,52)(70,43)(71,47)(72,55)(73,53)(74,57)(75,47)(76,50)(77,56)(78,38)(79,36)(80,42)(81,37)(82,39)(83,40)(84,39)(85,24)(86,25)(87,22)(88,24)(89,19)(90,23)(91,20)(92,19)(93,24)(94,28)(95,20)(96,23)(97,15)(98,16)(99,18)(100,21)(101,20)(102,18)(103,14)(104,17)(105,17)(106,22)(107,21)(108,20)(109,15)(110,20)(111,17)(112,17)(113,9)(114,11)(115,13)(116,18)(117,19)(118,15)(119,20)(120,13)(121,12)(122,8)(123,17)(124,12)(125,21)(126,14)(127,13)(128,16)(129,13)(130,7)(131,12)(132,16)(133,22)(134,10)(135,12)(136,13)(137,18)(138,12)(139,16)(140,11)(141,15)(142,13)(143,17)(144,10)(145,19)(146,16)(147,8)(148,11)(149,10)(150,10)(151,20)(152,10)(153,10)(154,11)(155,18)(156,12)(157,11)(158,8)(159,9)(160,4)(161,11)(162,12)(163,13)(164,11)(165,7)(166,6)(167,5)(168,4)(169,4)(170,6)(171,4)(172,5)(173,3)(174,1)(175,6)(176,3)(177,9)(178,10)(179,8)(180,3)(181,9)(182,7)(183,4)(184,11)(185,4)(186,7)(187,12)(188,6)(189,6)(190,3)(191,4)(192,8)(193,3)(194,6)(195,5)(196,6)(197,6)(198,2)(199,3)(200,4)(201,3)(202,2)(203,4)(204,4)(205,2)(206,7)(207,5)(208,3)(209,6)(210,2)(211,5)(212,5)(213,9)(214,4)(215,3)(216,8)(217,5)(218,2)(219,6)(220,4)(221,10)(222,7)(223,4)(224,5)(225,4)(226,2)(227,2)(228,4)(229,4)(230,5)(231,4)(232,2)(233,3)(234,4)(235,1)(236,5)(237,6)(238,6)(239,7)(240,3)(241,5)(242,3)(243,4)(244,5)(245,4)(246,1)(247,5)(248,2)(249,4)(250,4)(251,5)(252,3)(253,4)(254,5)(255,6)(256,3)(257,4)(258,2)(259,4)(260,6)(261,3)(262,3)(263,1)(264,4)(265,3)(266,4)(267,3)(268,3)(269,3)(270,6)(271,6)(272,2)(273,4)(274,3)(275,3)(276,2)(277,4)(278,1)(279,2)(280,0)(281,4)(282,3)(283,5)(284,1)(285,2)(286,2)(287,3)(288,0)(289,3)(290,1)(291,3)(292,4)(293,2)(294,7)(295,3)(296,2)(297,2)(298,3)(299,1)(300,3)(301,3)(302,3)(303,2)(304,0)(305,2)(306,5)(307,7)(308,1)(309,2)(310,3)(311,2)(312,1)(313,2)(314,2)(315,2)(316,1)(317,2)(318,2)(319,1)(320,0)(321,7)(322,0)(323,1)(324,1)(325,2)(326,0)(327,0)(328,1)(329,2)(330,1)(331,2)(332,1)(333,1)(334,1)(335,0)(336,0)(337,1)(338,0)(339,0)(340,0)(341,2)(342,2)(343,1)(344,1)(345,0)(346,1)(347,0)
}; 
\addplot[color=NavyBlue] coordinates{(0,38)(1,36)(2,61)(3,89)(4,79)(5,109)(6,120)(7,126)(8,112)(9,117)(10,150)(11,151)(12,152)(13,150)(14,146)(15,147)(16,154)(17,151)(18,155)(19,159)(20,154)(21,158)(22,129)(23,155)(24,160)(25,157)(26,150)(27,144)(28,158)(29,147)(30,139)(31,132)(32,139)(33,126)(34,111)(35,125)(36,128)(37,93)(38,123)(39,99)(40,101)(41,69)(42,80)(43,74)(44,91)(45,74)(46,77)(47,87)(48,77)(49,76)(50,72)(51,80)(52,83)(53,71)(54,69)(55,81)(56,62)(57,72)(58,77)(59,68)(60,77)(61,74)(62,61)(63,63)(64,67)(65,65)(66,52)(67,61)(68,60)(69,45)(70,62)(71,51)(72,63)(73,65)(74,55)(75,48)(76,56)(77,45)(78,48)(79,57)(80,48)(81,42)(82,29)(83,48)(84,35)(85,26)(86,22)(87,19)(88,17)(89,20)(90,20)(91,26)(92,17)(93,26)(94,21)(95,22)(96,19)(97,8)(98,17)(99,20)(100,20)(101,9)(102,14)(103,21)(104,18)(105,23)(106,17)(107,16)(108,12)(109,21)(110,19)(111,15)(112,10)(113,16)(114,18)(115,18)(116,14)(117,15)(118,11)(119,17)(120,13)(121,17)(122,18)(123,11)(124,12)(125,14)(126,15)(127,17)(128,17)(129,8)(130,5)(131,12)(132,13)(133,14)(134,14)(135,6)(136,10)(137,17)(138,11)(139,11)(140,13)(141,16)(142,8)(143,15)(144,11)(145,14)(146,4)(147,13)(148,11)(149,9)(150,10)(151,4)(152,7)(153,7)(154,9)(155,11)(156,8)(157,9)(158,8)(159,14)(160,10)(161,9)(162,6)(163,14)(164,8)(165,4)(166,8)(167,3)(168,8)(169,5)(170,6)(171,11)(172,9)(173,5)(174,2)(175,5)(176,5)(177,4)(178,4)(179,4)(180,8)(181,6)(182,6)(183,4)(184,5)(185,3)(186,10)(187,5)(188,6)(189,2)(190,4)(191,8)(192,6)(193,4)(194,6)(195,6)(196,6)(197,2)(198,1)(199,3)(200,2)(201,6)(202,6)(203,4)(204,3)(205,8)(206,4)(207,4)(208,3)(209,5)(210,7)(211,3)(212,3)(213,3)(214,5)(215,3)(216,2)(217,10)(218,2)(219,5)(220,4)(221,11)(222,5)(223,3)(224,4)(225,4)(226,4)(227,2)(228,5)(229,4)(230,3)(231,6)(232,1)(233,4)(234,4)(235,7)(236,6)(237,5)(238,5)(239,2)(240,2)(241,11)(242,3)(243,5)(244,5)(245,1)(246,2)(247,0)(248,3)(249,7)(250,3)(251,4)(252,8)(253,4)(254,4)(255,4)(256,2)(257,4)(258,1)(259,2)(260,4)(261,1)(262,4)(263,3)(264,6)(265,1)(266,2)(267,2)(268,3)(269,1)(270,3)(271,5)(272,4)(273,2)(274,3)(275,3)(276,2)(277,2)(278,1)(279,1)(280,4)(281,4)(282,3)(283,5)(284,3)(285,3)(286,7)(287,1)(288,3)(289,1)(290,5)(291,3)(292,3)(293,6)(294,3)(295,2)(296,2)(297,1)(298,2)(299,3)(300,1)(301,3)(302,4)(303,2)(304,0)(305,1)(306,1)(307,3)(308,2)(309,1)(310,3)(311,4)(312,4)(313,4)(314,0)(315,1)(316,1)(317,3)(318,2)(319,1)(320,1)(321,3)(322,1)(323,1)(324,0)(325,0)(326,0)(327,1)(328,0)(329,1)(330,0)(331,1)(332,1)(333,2)(334,1)(335,3)(336,2)(337,1)(338,2)(339,0)(340,1)(341,3)(342,0)(343,1)(344,1)(345,0)(346,0)(347,1)(348,0)
};  
\addplot[color=magenta] coordinates{(0,27)(1,45)(2,54)(3,64)(4,86)(5,113)(6,112)(7,119)(8,137)(9,135)(10,131)(11,137)(12,124)(13,126)(14,147)(15,174)(16,144)(17,140)(18,146)(19,144)(20,152)(21,173)(22,141)(23,174)(24,150)(25,181)(26,155)(27,151)(28,189)(29,163)(30,153)(31,151)(32,139)(33,134)(34,116)(35,125)(36,124)(37,99)(38,113)(39,96)(40,87)(41,95)(42,73)(43,75)(44,62)(45,74)(46,72)(47,64)(48,62)(49,79)(50,74)(51,75)(52,62)(53,61)(54,75)(55,71)(56,83)(57,91)(58,71)(59,69)(60,76)(61,80)(62,77)(63,74)(64,65)(65,57)(66,59)(67,63)(68,45)(69,48)(70,55)(71,53)(72,53)(73,51)(74,38)(75,54)(76,48)(77,54)(78,64)(79,51)(80,38)(81,36)(82,39)(83,51)(84,29)(85,41)(86,27)(87,30)(88,26)(89,31)(90,18)(91,18)(92,18)(93,22)(94,19)(95,28)(96,18)(97,24)(98,24)(99,16)(100,18)(101,22)(102,19)(103,25)(104,18)(105,15)(106,14)(107,12)(108,16)(109,17)(110,12)(111,17)(112,16)(113,16)(114,15)(115,20)(116,13)(117,9)(118,9)(119,15)(120,20)(121,17)(122,23)(123,11)(124,15)(125,15)(126,21)(127,13)(128,16)(129,16)(130,19)(131,13)(132,8)(133,11)(134,4)(135,14)(136,22)(137,6)(138,12)(139,13)(140,14)(141,16)(142,17)(143,13)(144,15)(145,9)(146,12)(147,14)(148,18)(149,13)(150,13)(151,11)(152,13)(153,15)(154,13)(155,7)(156,18)(157,10)(158,16)(159,12)(160,18)(161,8)(162,9)(163,10)(164,20)(165,10)(166,9)(167,8)(168,9)(169,8)(170,9)(171,11)(172,8)(173,7)(174,6)(175,3)(176,7)(177,11)(178,7)(179,5)(180,3)(181,7)(182,6)(183,8)(184,5)(185,8)(186,4)(187,6)(188,5)(189,11)(190,5)(191,5)(192,4)(193,12)(194,11)(195,2)(196,8)(197,6)(198,4)(199,9)(200,4)(201,4)(202,7)(203,4)(204,10)(205,9)(206,7)(207,6)(208,6)(209,6)(210,3)(211,10)(212,3)(213,5)(214,6)(215,7)(216,4)(217,3)(218,6)(219,6)(220,3)(221,2)(222,6)(223,5)(224,5)(225,3)(226,5)(227,5)(228,6)(229,6)(230,5)(231,5)(232,6)(233,6)(234,3)(235,3)(236,2)(237,5)(238,4)(239,4)(240,5)(241,5)(242,5)(243,6)(244,5)(245,11)(246,5)(247,9)(248,9)(249,4)(250,4)(251,2)(252,5)(253,7)(254,4)(255,3)(256,7)(257,5)(258,5)(259,8)(260,2)(261,7)(262,5)(263,5)(264,6)(265,4)(266,6)(267,5)(268,5)(269,6)(270,2)(271,4)(272,3)(273,4)(274,1)(275,6)(276,5)(277,6)(278,5)(279,3)(280,3)(281,5)(282,9)(283,3)(284,2)(285,7)(286,2)(287,4)(288,3)(289,3)(290,7)(291,3)(292,5)(293,4)(294,5)(295,4)(296,3)(297,4)(298,5)(299,3)(300,4)(301,2)(302,3)(303,4)(304,3)(305,5)(306,3)(307,4)(308,3)(309,2)(310,3)(311,2)(312,4)(313,3)(314,2)(315,2)(316,6)(317,3)(318,3)(319,3)(320,1)(321,1)(322,1)(323,5)(324,2)(325,4)(326,3)(327,2)(328,2)(329,3)(330,1)(331,1)(332,1)(333,2)(334,2)(335,3)(336,1)(337,2)(338,2)(339,1)(340,3)(341,1)(342,1)(343,2)(344,2)(345,2)(346,1)(347,1)(348,1)(349,1)(350,1)(351,1)(352,1)(353,1)(354,1)(355,2)(356,1)
};
\legend{run \#1,  run \#2, run \#3, run \#4, run \#5, run \#6,  run \#7, run \#8, run \#9, run \#10} % legend of the plot

\end{axis}
\end{tikzpicture}

%% file: plots/num_ppl/injury_levels.tex
\begin{tikzpicture}[scale=0.75]
\begin{axis}[
title={Injury Levels - Number of people},
width=\columnwidth,
ybar, % style of the histogram: clustered columns
height=8cm, % height of the plot
ylabel={Ratio}, % label for the y-axis
symbolic x coords={HEALTHY,MINOR,MODERATE,SERIOUS}, % labels for the x-ticks
xtick=data, % position the x-ticks at the data points
ymin=0, % minimum value of the y-axis
legend style={at={(1,1)}, anchor=north east}, % position of the legend
legend cell align=left, % alignment of the legend cells
ymajorgrids=true, % display major grids
bar width=0.15cm % width of the bars
]
\addplot[color=Red, fill] coordinates{(HEALTHY,0.5486333333333333)(MINOR,0.3549333333333333)(MODERATE,0.06966666666666667)(SERIOUS,0.021966666666666666)
}; % plot of the first histogram
\addplot[color=BurntOrange, fill] coordinates{(HEALTHY,0.5807636363636364)(MINOR,0.3391636363636364)(MODERATE,0.06072727272727273)(SERIOUS,0.0164)
}; % plot of the second histogram
\addplot[color=Cyan, fill] coordinates{(HEALTHY,0.61132)(MINOR,0.31936)(MODERATE,0.0542)(SERIOUS,0.01376)
}; % plot of the third histogram
\addplot[color=SeaGreen, fill] coordinates{(HEALTHY,0.6370222222222223)(MINOR,0.30302222222222225)(MODERATE,0.05128888888888889)(SERIOUS,0.008222222222222223)
}; % plot of the fourth histogram
\addplot[color=LimeGreen, fill] coordinates{(HEALTHY,0.6638)(MINOR,0.28775)(MODERATE,0.04165)(SERIOUS,0.00635)
}; % plot of the fifth histogram
\legend{30000 ppl,  27500 ppl, 25000 ppl, 22500 ppl, 20000 ppl} % legend of the plot
\end{axis}
\end{tikzpicture}

\begin{tikzpicture}[scale=0.75]
\begin{axis}[
ybar, % style of the histogram: clustered columns
width=\columnwidth,
height=8cm, % height of the plot
ylabel={Ratio}, % label for the y-axis
symbolic x coords={SEVERE,CRITICAL,FATAL}, % labels for the x-ticks
xtick=data, % position the x-ticks at the data points
ymin=0, % minimum value of the y-axis
legend style={at={(1,1)}, anchor=north east}, % position of the legend
legend cell align=left, % alignment of the legend cells
ymajorgrids=true, % display major grids
bar width=0.15cm % width of the bars
]
\addplot[color=Red, fill] coordinates{(SEVERE,0.004266666666666667)(CRITICAL,0.0004)(FATAL,0.00013333333333333334)
}; % plot of the first histogram
\addplot[color=BurntOrange, fill] coordinates{(SEVERE,0.002618181818181818)(CRITICAL,0.00025454545454545456)(FATAL,7.272727272727273e-05)
}; % plot of the second histogram
\addplot[color=Cyan, fill] coordinates{(SEVERE,0.00128)(CRITICAL,8e-05)(FATAL,0.0)
}; % plot of the third histogram
\addplot[color=SeaGreen, fill] coordinates{(SEVERE,0.0004)(CRITICAL,4.4444444444444447e-05)(FATAL,0.0)
}; % plot of the fourth histogram
\addplot[color=LimeGreen, fill] coordinates{(SEVERE,0.0004)(CRITICAL,5e-05)(FATAL,0.0)
}; % plot of the fifth histogram
\legend{30000 ppl,  27500 ppl, 25000 ppl, 22500 ppl, 20000 ppl} % legend of the plot
\end{axis}
\end{tikzpicture}

%% file: plots/num_ppl/evac_speed.tex
\begin{tikzpicture}[scale=0.75]

% Set the axis labels and title
\begin{axis}[    xlabel={Time [s]},    ylabel={ppl/s},    title={Evacuation Speed - Number of people},   legend style={at={(1,1)}, anchor=north east},]

\addplot[color=Red] coordinates{(0,36)(1,49)(2,58)(3,77)(4,70)(5,107)(6,108)(7,114)(8,141)(9,143)(10,146)(11,129)(12,160)(13,142)(14,176)(15,131)(16,136)(17,149)(18,148)(19,141)(20,159)(21,158)(22,156)(23,160)(24,157)(25,139)(26,152)(27,157)(28,155)(29,156)(30,141)(31,130)(32,117)(33,125)(34,130)(35,132)(36,131)(37,112)(38,122)(39,93)(40,95)(41,85)(42,90)(43,64)(44,75)(45,71)(46,81)(47,68)(48,88)(49,78)(50,80)(51,65)(52,79)(53,79)(54,80)(55,68)(56,84)(57,73)(58,75)(59,76)(60,69)(61,81)(62,73)(63,77)(64,75)(65,70)(66,63)(67,46)(68,52)(69,60)(70,62)(71,56)(72,47)(73,40)(74,71)(75,49)(76,52)(77,49)(78,48)(79,43)(80,47)(81,44)(82,39)(83,39)(84,20)(85,32)(86,34)(87,26)(88,17)(89,16)(90,20)(91,33)(92,30)(93,21)(94,17)(95,21)(96,17)(97,23)(98,20)(99,18)(100,12)(101,22)(102,20)(103,18)(104,22)(105,23)(106,10)(107,19)(108,14)(109,12)(110,19)(111,27)(112,13)(113,16)(114,17)(115,15)(116,13)(117,13)(118,7)(119,22)(120,17)(121,12)(122,20)(123,17)(124,9)(125,15)(126,14)(127,18)(128,17)(129,10)(130,15)(131,14)(132,13)(133,12)(134,16)(135,10)(136,14)(137,9)(138,19)(139,13)(140,13)(141,14)(142,16)(143,8)(144,11)(145,13)(146,19)(147,13)(148,17)(149,15)(150,9)(151,15)(152,9)(153,12)(154,11)(155,8)(156,13)(157,13)(158,10)(159,10)(160,15)(161,7)(162,9)(163,7)(164,9)(165,12)(166,9)(167,8)(168,6)(169,3)(170,10)(171,8)(172,4)(173,5)(174,10)(175,8)(176,6)(177,10)(178,6)(179,7)(180,5)(181,7)(182,6)(183,9)(184,6)(185,9)(186,8)(187,5)(188,7)(189,6)(190,6)(191,4)(192,3)(193,4)(194,7)(195,5)(196,3)(197,6)(198,6)(199,6)(200,7)(201,5)(202,8)(203,10)(204,4)(205,3)(206,4)(207,6)(208,7)(209,9)(210,8)(211,3)(212,7)(213,6)(214,2)(215,10)(216,4)(217,5)(218,4)(219,6)(220,4)(221,8)(222,9)(223,3)(224,5)(225,6)(226,7)(227,8)(228,5)(229,9)(230,5)(231,5)(232,4)(233,4)(234,6)(235,4)(236,5)(237,7)(238,5)(239,4)(240,6)(241,6)(242,2)(243,4)(244,4)(245,5)(246,2)(247,7)(248,1)(249,3)(250,5)(251,6)(252,4)(253,5)(254,4)(255,3)(256,7)(257,4)(258,4)(259,3)(260,4)(261,6)(262,6)(263,4)(264,3)(265,3)(266,5)(267,3)(268,6)(269,6)(270,5)(271,3)(272,6)(273,3)(274,2)(275,5)(276,3)(277,3)(278,7)(279,2)(280,4)(281,3)(282,4)(283,4)(284,6)(285,2)(286,3)(287,2)(288,4)(289,2)(290,4)(291,3)(292,4)(293,6)(294,4)(295,2)(296,2)(297,2)(298,5)(299,3)(300,3)(301,4)(302,2)(303,2)(304,4)(305,5)(306,4)(307,4)(308,3)(309,6)(310,5)(311,4)(312,2)(313,1)(314,3)(315,3)(316,4)(317,2)(318,3)(319,3)(320,2)(321,3)(322,3)(323,2)(324,3)(325,4)(326,1)(327,3)(328,1)(329,3)(330,3)(331,4)(332,3)(333,1)(334,3)(335,4)(336,1)(337,3)(338,1)
}; % plot of the first histogram
\addplot[color=BurntOrange] coordinates{(0,35)(1,38)(2,52)(3,80)(4,95)(5,98)(6,93)(7,91)(8,133)(9,110)(10,131)(11,131)(12,129)(13,148)(14,125)(15,138)(16,136)(17,127)(18,133)(19,138)(20,122)(21,145)(22,128)(23,137)(24,129)(25,143)(26,140)(27,145)(28,134)(29,148)(30,146)(31,110)(32,130)(33,112)(34,120)(35,138)(36,85)(37,114)(38,100)(39,103)(40,82)(41,79)(42,54)(43,60)(44,68)(45,71)(46,98)(47,87)(48,76)(49,57)(50,73)(51,87)(52,69)(53,60)(54,58)(55,82)(56,61)(57,69)(58,57)(59,69)(60,68)(61,64)(62,71)(63,75)(64,59)(65,57)(66,56)(67,52)(68,60)(69,54)(70,40)(71,51)(72,40)(73,26)(74,42)(75,51)(76,45)(77,53)(78,31)(79,37)(80,21)(81,27)(82,23)(83,28)(84,27)(85,20)(86,24)(87,23)(88,21)(89,17)(90,24)(91,21)(92,12)(93,18)(94,19)(95,33)(96,19)(97,15)(98,20)(99,22)(100,13)(101,18)(102,17)(103,18)(104,20)(105,13)(106,14)(107,10)(108,15)(109,9)(110,20)(111,18)(112,13)(113,12)(114,9)(115,12)(116,20)(117,14)(118,16)(119,12)(120,15)(121,13)(122,11)(123,10)(124,11)(125,9)(126,6)(127,11)(128,10)(129,10)(130,11)(131,22)(132,13)(133,17)(134,13)(135,15)(136,12)(137,6)(138,8)(139,13)(140,11)(141,7)(142,11)(143,7)(144,9)(145,11)(146,9)(147,8)(148,12)(149,16)(150,16)(151,10)(152,8)(153,13)(154,6)(155,9)(156,7)(157,10)(158,6)(159,6)(160,9)(161,6)(162,9)(163,14)(164,9)(165,7)(166,6)(167,5)(168,6)(169,3)(170,11)(171,3)(172,10)(173,10)(174,6)(175,1)(176,9)(177,4)(178,7)(179,7)(180,3)(181,5)(182,4)(183,2)(184,5)(185,2)(186,6)(187,2)(188,2)(189,4)(190,0)(191,7)(192,3)(193,3)(194,5)(195,5)(196,5)(197,1)(198,3)(199,4)(200,6)(201,6)(202,2)(203,2)(204,3)(205,2)(206,3)(207,3)(208,5)(209,2)(210,4)(211,3)(212,3)(213,3)(214,3)(215,3)(216,4)(217,3)(218,1)(219,6)(220,5)(221,3)(222,2)(223,5)(224,5)(225,1)(226,1)(227,3)(228,0)(229,2)(230,3)(231,2)(232,3)(233,3)(234,3)(235,4)(236,4)(237,4)(238,3)(239,2)(240,1)(241,6)(242,4)(243,3)(244,1)(245,3)(246,2)(247,3)(248,1)(249,2)(250,4)(251,2)(252,3)(253,5)(254,4)(255,5)(256,4)(257,2)(258,0)(259,2)(260,4)(261,4)(262,3)(263,3)(264,7)(265,3)(266,1)(267,2)(268,5)(269,4)(270,4)(271,2)(272,5)(273,3)(274,2)(275,3)(276,2)(277,1)(278,3)(279,1)(280,4)(281,3)(282,5)(283,7)(284,2)(285,1)(286,2)(287,3)(288,6)(289,0)(290,0)(291,2)(292,4)(293,1)(294,5)(295,4)(296,7)(297,4)(298,3)(299,1)(300,1)(301,4)(302,2)(303,7)(304,3)(305,4)(306,2)(307,1)(308,0)(309,4)(310,4)(311,1)(312,1)(313,2)(314,5)(315,1)(316,4)(317,6)(318,0)(319,0)(320,5)(321,1)(322,1)(323,3)(324,2)(325,1)(326,1)(327,3)(328,1)(329,3)(330,0)(331,2)(332,1)(333,2)(334,2)(335,0)(336,1)(337,0)(338,0)(339,1)(340,0)(341,1)(342,1)(343,1)(344,2)(345,0)(346,1)(347,0)
}; % plot of the second histogram
\addplot[color=Cyan] coordinates{(0,26)(1,40)(2,50)(3,76)(4,75)(5,87)(6,103)(7,86)(8,117)(9,92)(10,109)(11,120)(12,103)(13,103)(14,139)(15,104)(16,137)(17,134)(18,116)(19,114)(20,110)(21,147)(22,116)(23,139)(24,145)(25,129)(26,132)(27,121)(28,134)(29,137)(30,119)(31,116)(32,113)(33,108)(34,90)(35,114)(36,101)(37,84)(38,86)(39,77)(40,70)(41,88)(42,68)(43,63)(44,68)(45,60)(46,55)(47,67)(48,66)(49,64)(50,59)(51,68)(52,51)(53,56)(54,65)(55,62)(56,58)(57,47)(58,66)(59,72)(60,50)(61,59)(62,51)(63,61)(64,50)(65,49)(66,53)(67,52)(68,49)(69,36)(70,40)(71,31)(72,47)(73,50)(74,31)(75,45)(76,49)(77,48)(78,35)(79,42)(80,28)(81,32)(82,25)(83,24)(84,18)(85,22)(86,14)(87,16)(88,17)(89,17)(90,8)(91,15)(92,17)(93,23)(94,13)(95,12)(96,19)(97,11)(98,13)(99,14)(100,11)(101,9)(102,14)(103,19)(104,19)(105,9)(106,14)(107,17)(108,10)(109,9)(110,14)(111,14)(112,14)(113,11)(114,8)(115,19)(116,10)(117,14)(118,17)(119,14)(120,10)(121,12)(122,11)(123,11)(124,4)(125,11)(126,14)(127,12)(128,7)(129,10)(130,12)(131,8)(132,17)(133,6)(134,7)(135,11)(136,12)(137,8)(138,9)(139,14)(140,15)(141,6)(142,9)(143,10)(144,8)(145,8)(146,9)(147,13)(148,15)(149,6)(150,9)(151,14)(152,16)(153,10)(154,13)(155,7)(156,8)(157,8)(158,8)(159,15)(160,4)(161,6)(162,8)(163,4)(164,5)(165,4)(166,10)(167,4)(168,2)(169,6)(170,4)(171,1)(172,3)(173,3)(174,5)(175,4)(176,4)(177,4)(178,3)(179,3)(180,11)(181,0)(182,5)(183,5)(184,7)(185,4)(186,15)(187,3)(188,3)(189,5)(190,7)(191,3)(192,4)(193,0)(194,4)(195,2)(196,3)(197,5)(198,5)(199,6)(200,4)(201,5)(202,1)(203,2)(204,1)(205,2)(206,4)(207,1)(208,4)(209,7)(210,7)(211,9)(212,4)(213,3)(214,5)(215,0)(216,0)(217,2)(218,4)(219,6)(220,2)(221,4)(222,5)(223,3)(224,3)(225,4)(226,0)(227,3)(228,4)(229,4)(230,1)(231,1)(232,1)(233,2)(234,1)(235,3)(236,2)(237,3)(238,1)(239,3)(240,6)(241,3)(242,2)(243,1)(244,1)(245,3)(246,4)(247,5)(248,2)(249,2)(250,4)(251,2)(252,3)(253,4)(254,3)(255,1)(256,2)(257,2)(258,2)(259,3)(260,2)(261,3)(262,3)(263,3)(264,4)(265,2)(266,2)(267,2)(268,2)(269,2)(270,1)(271,3)(272,2)(273,1)(274,4)(275,2)(276,2)(277,1)(278,3)(279,2)(280,1)(281,4)(282,1)(283,1)(284,2)(285,2)(286,1)(287,2)(288,1)(289,0)(290,3)(291,2)(292,4)(293,0)(294,4)(295,2)(296,2)(297,4)(298,5)(299,3)(300,3)(301,3)(302,3)(303,2)(304,1)(305,1)(306,0)(307,1)(308,4)(309,3)(310,3)(311,4)(312,3)(313,5)(314,4)(315,1)(316,1)(317,1)(318,0)(319,1)(320,0)(321,1)(322,1)(323,1)(324,0)(325,0)(326,1)(327,1)(328,1)(329,1)(330,0)(331,2)(332,0)(333,0)(334,2)(335,1)(336,0)(337,2)(338,0)(339,1)(340,3)(341,0)(342,0)(343,0)(344,0)(345,1)(346,0)(347,0)(348,1)(349,0)(350,0)(351,0)(352,0)(353,1)(354,0)
}; % plot of the third histogram
\addplot[color=SeaGreen] coordinates{(0,28)(1,39)(2,62)(3,63)(4,76)(5,67)(6,90)(7,109)(8,90)(9,77)(10,93)(11,113)(12,111)(13,121)(14,122)(15,123)(16,115)(17,126)(18,138)(19,110)(20,107)(21,107)(22,123)(23,110)(24,114)(25,101)(26,102)(27,101)(28,121)(29,115)(30,119)(31,81)(32,76)(33,100)(34,96)(35,84)(36,104)(37,80)(38,65)(39,77)(40,71)(41,69)(42,58)(43,52)(44,71)(45,50)(46,56)(47,59)(48,63)(49,72)(50,56)(51,53)(52,56)(53,61)(54,45)(55,48)(56,44)(57,53)(58,52)(59,49)(60,54)(61,56)(62,55)(63,43)(64,47)(65,42)(66,40)(67,45)(68,38)(69,35)(70,37)(71,33)(72,39)(73,38)(74,32)(75,39)(76,31)(77,29)(78,27)(79,28)(80,27)(81,28)(82,19)(83,25)(84,23)(85,17)(86,21)(87,12)(88,7)(89,17)(90,17)(91,17)(92,14)(93,4)(94,12)(95,11)(96,12)(97,13)(98,14)(99,13)(100,10)(101,12)(102,11)(103,11)(104,7)(105,6)(106,12)(107,15)(108,15)(109,13)(110,8)(111,6)(112,14)(113,7)(114,16)(115,10)(116,4)(117,8)(118,12)(119,14)(120,11)(121,16)(122,10)(123,12)(124,13)(125,14)(126,15)(127,19)(128,14)(129,5)(130,9)(131,16)(132,13)(133,10)(134,9)(135,12)(136,14)(137,11)(138,8)(139,15)(140,8)(141,8)(142,11)(143,12)(144,15)(145,7)(146,7)(147,10)(148,7)(149,8)(150,7)(151,12)(152,8)(153,9)(154,5)(155,10)(156,10)(157,8)(158,8)(159,6)(160,8)(161,7)(162,10)(163,2)(164,10)(165,6)(166,5)(167,3)(168,10)(169,3)(170,4)(171,5)(172,5)(173,2)(174,2)(175,0)(176,3)(177,5)(178,5)(179,2)(180,3)(181,1)(182,4)(183,5)(184,5)(185,2)(186,2)(187,3)(188,1)(189,2)(190,1)(191,3)(192,4)(193,5)(194,4)(195,4)(196,2)(197,5)(198,4)(199,4)(200,1)(201,3)(202,5)(203,4)(204,2)(205,0)(206,3)(207,0)(208,4)(209,3)(210,3)(211,1)(212,3)(213,3)(214,4)(215,2)(216,1)(217,3)(218,3)(219,2)(220,3)(221,5)(222,3)(223,2)(224,2)(225,2)(226,3)(227,1)(228,4)(229,3)(230,2)(231,0)(232,3)(233,2)(234,1)(235,4)(236,4)(237,2)(238,3)(239,2)(240,1)(241,1)(242,4)(243,2)(244,1)(245,3)(246,3)(247,4)(248,3)(249,2)(250,3)(251,4)(252,3)(253,3)(254,3)(255,3)(256,3)(257,3)(258,2)(259,3)(260,0)(261,0)(262,2)(263,3)(264,4)(265,1)(266,4)(267,2)(268,1)(269,4)(270,0)(271,0)(272,2)(273,2)(274,1)(275,1)(276,1)(277,1)(278,2)(279,2)(280,4)(281,2)(282,3)(283,1)(284,3)(285,4)(286,0)(287,2)(288,4)(289,4)(290,1)(291,2)(292,0)(293,1)(294,2)(295,2)(296,1)(297,0)(298,3)(299,4)(300,5)(301,1)(302,1)(303,1)(304,3)(305,2)(306,2)(307,1)(308,1)(309,1)(310,0)(311,0)(312,0)(313,3)(314,0)(315,2)(316,2)(317,2)(318,4)(319,3)(320,2)(321,1)(322,1)(323,2)(324,1)(325,0)(326,0)(327,2)(328,2)(329,4)(330,2)(331,0)(332,0)(333,1)(334,0)(335,0)(336,0)(337,0)(338,0)(339,1)(340,0)(341,0)(342,0)(343,0)(344,0)(345,0)(346,1)(347,0)
}; % plot of the fourth histogram
\addplot[color=LimeGreen] coordinates{(0,28)(1,27)(2,42)(3,51)(4,57)(5,69)(6,81)(7,81)(8,83)(9,81)(10,97)(11,93)(12,93)(13,82)(14,106)(15,97)(16,99)(17,108)(18,91)(19,86)(20,90)(21,99)(22,102)(23,92)(24,102)(25,86)(26,101)(27,88)(28,110)(29,110)(30,106)(31,116)(32,76)(33,81)(34,81)(35,84)(36,93)(37,66)(38,78)(39,72)(40,56)(41,71)(42,52)(43,46)(44,43)(45,51)(46,32)(47,50)(48,48)(49,53)(50,52)(51,54)(52,39)(53,42)(54,47)(55,48)(56,41)(57,49)(58,29)(59,37)(60,51)(61,33)(62,43)(63,46)(64,43)(65,42)(66,39)(67,31)(68,31)(69,32)(70,45)(71,35)(72,27)(73,35)(74,41)(75,31)(76,31)(77,29)(78,29)(79,32)(80,38)(81,21)(82,18)(83,15)(84,8)(85,22)(86,12)(87,24)(88,8)(89,11)(90,7)(91,15)(92,6)(93,12)(94,14)(95,9)(96,16)(97,13)(98,10)(99,5)(100,12)(101,7)(102,7)(103,5)(104,11)(105,17)(106,9)(107,15)(108,12)(109,7)(110,13)(111,7)(112,6)(113,16)(114,11)(115,6)(116,6)(117,7)(118,9)(119,8)(120,10)(121,5)(122,7)(123,10)(124,11)(125,12)(126,7)(127,6)(128,9)(129,9)(130,7)(131,8)(132,6)(133,9)(134,8)(135,4)(136,3)(137,10)(138,11)(139,10)(140,12)(141,9)(142,10)(143,7)(144,7)(145,11)(146,12)(147,8)(148,7)(149,7)(150,12)(151,9)(152,8)(153,3)(154,12)(155,4)(156,10)(157,11)(158,8)(159,6)(160,6)(161,3)(162,5)(163,1)(164,9)(165,6)(166,2)(167,5)(168,2)(169,3)(170,3)(171,1)(172,3)(173,4)(174,3)(175,3)(176,2)(177,5)(178,6)(179,4)(180,6)(181,5)(182,2)(183,3)(184,1)(185,1)(186,3)(187,2)(188,2)(189,3)(190,2)(191,1)(192,2)(193,3)(194,2)(195,2)(196,4)(197,1)(198,1)(199,3)(200,4)(201,2)(202,1)(203,2)(204,5)(205,2)(206,6)(207,4)(208,5)(209,1)(210,4)(211,2)(212,3)(213,2)(214,2)(215,1)(216,3)(217,2)(218,1)(219,1)(220,5)(221,3)(222,2)(223,4)(224,3)(225,0)(226,2)(227,2)(228,1)(229,7)(230,5)(231,4)(232,1)(233,1)(234,7)(235,3)(236,1)(237,3)(238,2)(239,2)(240,2)(241,2)(242,0)(243,0)(244,0)(245,2)(246,2)(247,1)(248,2)(249,1)(250,3)(251,3)(252,1)(253,2)(254,4)(255,3)(256,3)(257,6)(258,2)(259,1)(260,0)(261,2)(262,1)(263,1)(264,1)(265,1)(266,0)(267,3)(268,2)(269,1)(270,1)(271,3)(272,2)(273,4)(274,3)(275,1)(276,0)(277,3)(278,2)(279,3)(280,2)(281,2)(282,4)(283,1)(284,2)(285,1)(286,0)(287,0)(288,4)(289,1)(290,1)(291,1)(292,1)(293,2)(294,2)(295,1)(296,2)(297,2)(298,1)(299,1)(300,0)(301,1)(302,1)(303,0)(304,1)(305,1)(306,3)(307,3)(308,2)(309,1)(310,1)(311,0)(312,3)(313,2)(314,2)(315,0)(316,2)(317,1)(318,3)(319,2)(320,2)(321,1)(322,0)(323,2)(324,1)(325,1)(326,1)(327,2)(328,5)(329,1)(330,3)(331,3)(332,1)(333,1)(334,3)(335,0)(336,1)(337,2)(338,1)(339,0)(340,1)(341,0)(342,1)(343,0)(344,1)(345,0)
}; % plot of the fifth histogram
\legend{30000 ppl,  27500 ppl, 25000 ppl, 22500 ppl, 20000 ppl} % legend of the plot

\end{axis}
\end{tikzpicture}

%% file: plots/num_ppl/evac_time.tex
\begin{tikzpicture}[scale=0.75]

% Set the axis labels and title
\begin{axis}[    xlabel={Time [s]},    ylabel={s},    title={Evacuation Time - Number of people},   legend style={at={(0,1)}, anchor=north west},]

\addplot[color=Red] coordinates{(0,10.62986111111111)(1,15.727734693877553)(2,9.368637931034483)(3,15.22831168831169)(4,12.367571428571427)(5,13.338355140186913)(6,14.773444444444443)(7,18.32413157894737)(8,18.902127659574468)(9,9.083776223776225)(10,15.11126712328767)(11,12.744193798449613)(12,12.92048125)(13,9.373718309859154)(14,12.826136363636362)(15,18.60796183206107)(16,12.43954411764706)(17,21.695402684563764)(18,22.56217567567567)(19,16.46952482269504)(20,19.57757861635219)(21,19.817468354430382)(22,21.314134615384614)(23,18.145424999999996)(24,26.492242038216556)(25,21.134309352517977)(26,26.491434210526315)(27,24.082681528662416)(28,21.601419354838704)(29,28.56121794871795)(30,24.63017730496453)(31,17.664461538461534)(32,24.422581196581195)(33,22.944792)(34,22.858400000000007)(35,16.93062121212121)(36,17.814236641221374)(37,15.025598214285713)(38,17.664540983606557)(39,25.233440860215048)(40,19.793484210526316)(41,26.587858823529416)(42,25.155422222222224)(43,17.696734375000002)(44,23.974026666666667)(45,22.692577464788734)(46,29.30796296296296)(47,26.558735294117646)(48,24.884522727272724)(49,29.339448717948724)(50,23.867025)(51,22.03658461538462)(52,25.43017721518987)(53,12.094113924050635)(54,26.361250000000002)(55,19.733014705882358)(56,24.01207142857142)(57,38.21717808219177)(58,29.540386666666667)(59,25.371315789473684)(60,19.5608115942029)(61,28.641567901234563)(62,31.80753424657535)(63,30.179454545454544)(64,27.12356000000001)(65,19.3669)(66,32.347238095238104)(67,21.053804347826087)(68,20.51407692307692)(69,14.538283333333334)(70,32.970274193548384)(71,26.060107142857138)(72,24.830468085106386)(73,34.090174999999995)(74,38.51774647887324)(75,39.8634693877551)(76,30.05728846153846)(77,27.907816326530615)(78,22.37160416666667)(79,34.12148837209302)(80,24.95844680851064)(81,28.91156818181818)(82,20.021871794871792)(83,25.08910256410257)(84,39.08485)(85,24.44025)(86,37.518205882352945)(87,37.697692307692314)(88,22.824117647058824)(89,30.431625000000004)(90,24.356050000000003)(91,17.761727272727274)(92,29.440733333333334)(93,27.936904761904763)(94,34.52564705882353)(95,23.24042857142857)(96,22.891117647058824)(97,21.236565217391302)(98,64.1192)(99,16.12783333333333)(100,15.9225)(101,22.234409090909093)(102,44.35045)(103,43.76866666666667)(104,44.862545454545455)(105,29.95113043478261)(106,39.0484)(107,31.04089473684211)(108,35.02321428571428)(109,32.565333333333335)(110,25.81984210526316)(111,29.257148148148147)(112,30.083384615384617)(113,30.6874375)(114,28.895470588235295)(115,39.4188)(116,22.43253846153846)(117,14.749769230769232)(118,27.396857142857147)(119,22.362636363636362)(120,40.72311764705882)(121,32.68183333333334)(122,14.61085)(123,46.684529411764714)(124,10.197444444444445)(125,32.87753333333333)(126,35.23492857142857)(127,38.56805555555556)(128,23.126117647058823)(129,9.1777)(130,26.222266666666666)(131,28.103928571428572)(132,30.272846153846153)(133,41.18808333333333)(134,18.3186875)(135,39.3868)(136,28.138571428571428)(137,43.788111111111114)(138,26.056473684210527)(139,22.579769230769234)(140,38.115692307692306)(141,49.836357142857146)(142,24.678812500000003)(143,36.735)(144,45.099363636363634)(145,30.391923076923074)(146,26.123842105263158)(147,30.407846153846155)(148,23.26435294117647)(149,53.38013333333333)(150,43.96644444444445)(151,33.13953333333333)(152,32.72188888888889)(153,16.099999999999998)(154,8.343363636363636)(155,49.525625)(156,30.484461538461538)(157,46.10938461538462)(158,19.3337)(159,19.3353)(160,12.892266666666668)(161,27.629428571428573)(162,21.493000000000002)(163,27.63828571428572)(164,21.49966666666667)(165,24.605416666666667)(166,55.42411111111111)(167,62.369125)(168,66.21066666666667)(169,30.592333333333332)(170,29.5529)(171,49.695499999999996)(172,48.43425)(173,18.3554)(174,60.181200000000004)(175,24.227375000000002)(176,49.3195)(177,9.1777)(178,15.296166666666666)(179,42.300428571428576)(180,18.3554)(181,42.320142857142855)(182,66.42816666666666)(183,32.93522222222222)(184,49.41883333333333)(185,21.577333333333332)(186,24.277124999999998)(187,38.849599999999995)(188,13.111)(189,49.47016666666667)(190,32.38833333333333)(191,48.5875)(192,64.79733333333333)(193,22.94425)(194,13.111)(195,18.3554)(196,64.86066666666666)(197,49.585499999999996)(198,32.443333333333335)(199,15.296166666666666)(200,42.52814285714286)(201,59.5424)(202,62.988125)(203,9.1777)(204,22.94425)(205,64.97666666666667)(206,74.53675)(207,32.49966666666666)(208,42.61185714285715)(209,44.620444444444445)(210,50.209625)(211,65.03266666666667)(212,13.111)(213,66.98016666666666)(214,97.588)(215,19.5214)(216,48.81125)(217,39.0564)(218,48.8245)(219,49.811)(220,22.94425)(221,24.422)(222,56.260999999999996)(223,65.15033333333334)(224,18.3554)(225,15.296166666666666)(226,42.759)(227,50.39675)(228,59.8966)(229,44.83277777777778)(230,59.9302)(231,59.9366)(232,48.94025)(233,48.945750000000004)(234,49.98583333333334)(235,48.96925)(236,18.3554)(237,27.987714285714283)(238,18.3554)(239,48.99025)(240,32.663833333333336)(241,67.42266666666667)(242,45.8885)(243,22.94425)(244,49.02475)(245,60.102999999999994)(246,98.0915)(247,42.95442857142857)(248,91.777)(249,30.592333333333332)(250,39.263)(251,32.727)(252,22.94425)(253,39.283)(254,22.94425)(255,30.592333333333332)(256,43.03185714285714)(257,49.13275)(258,22.94425)(259,30.592333333333332)(260,49.14625)(261,32.76416666666667)(262,67.72366666666667)(263,22.94425)(264,30.592333333333332)(265,100.541)(266,39.3464)(267,30.592333333333332)(268,32.7975)(269,15.296166666666666)(270,60.37140000000001)(271,30.592333333333332)(272,32.816833333333335)(273,65.64966666666668)(274,45.8885)(275,39.4064)(276,30.592333333333332)(277,65.69500000000001)(278,58.249857142857145)(279,45.8885)(280,75.62975)(281,65.72833333333334)(282,22.94425)(283,75.66425)(284,32.881)(285,45.8885)(286,65.78266666666667)(287,45.8885)(288,49.3465)(289,45.8885)(290,22.94425)(291,65.82133333333333)(292,22.94425)(293,32.9195)(294,75.82225)(295,45.8885)(296,98.801)(297,45.8885)(298,39.5298)(299,65.88833333333334)(300,101.19833333333334)(301,22.94425)(302,45.8885)(303,45.8885)(304,75.93825)(305,60.75940000000001)(306,22.94425)(307,102.487)(308,65.95533333333333)(309,32.98383333333334)(310,82.0592)(311,76.05725)(312,99.009)(313,91.777)(314,66.02166666666666)(315,30.592333333333332)(316,76.11525)(317,45.8885)(318,30.592333333333332)(319,30.592333333333332)(320,99.1405)(321,30.592333333333332)(322,30.592333333333332)(323,45.8885)(324,30.592333333333332)(325,22.94425)(326,91.777)(327,30.592333333333332)(328,91.777)(329,30.592333333333332)(330,30.592333333333332)(331,49.632999999999996)(332,30.592333333333332)(333,91.777)(334,30.592333333333332)(335,22.94425)(336,91.777)(337,30.592333333333332)(338,91.777)
}; % plot of the first histogram
\addplot[color=BurntOrange] coordinates{(0,8.97402857142857)(1,6.979263157894737)(2,9.308942307692307)(3,10.1985625)(4,12.159463157894736)(5,9.668938775510203)(6,9.077774193548386)(7,10.020714285714284)(8,9.962428571428573)(9,9.547881818181816)(10,7.648358778625953)(11,8.645412213740459)(12,14.9583488372093)(13,11.916141891891892)(14,9.334352)(15,12.130514492753623)(16,14.259536764705885)(17,12.41976377952756)(18,10.048308270676694)(19,16.2846884057971)(20,12.71110655737705)(21,14.377737931034485)(22,16.430523437499996)(23,15.462817518248178)(24,13.460457364341083)(25,11.766867132867134)(26,12.577614285714285)(27,21.59993103448276)(28,21.001089552238803)(29,15.905304054054051)(30,21.010143835616432)(31,16.5812)(32,16.284315384615383)(33,18.968053571428577)(34,14.80885)(35,16.03176811594203)(36,17.691905882352945)(37,15.744982456140352)(38,13.67699)(39,11.916048543689323)(40,14.122439024390246)(41,13.774772151898732)(42,17.520407407407404)(43,20.688783333333333)(44,23.685911764705878)(45,14.467563380281693)(46,20.25297959183673)(47,16.930080459770114)(48,22.34705263157895)(49,19.480017543859653)(50,17.274958904109592)(51,21.35824137931035)(52,14.020797101449274)(53,21.136533333333333)(54,14.174448275862067)(55,15.523036585365853)(56,23.368754098360657)(57,11.979666666666668)(58,19.81278947368421)(59,10.931565217391302)(60,25.54980882352942)(61,18.911781250000004)(62,20.27857746478874)(63,22.271440000000002)(64,20.62515254237288)(65,16.025543859649122)(66,25.867035714285702)(67,22.01403846153846)(68,16.554033333333333)(69,25.498148148148147)(70,22.969649999999998)(71,19.534352941176476)(72,23.019299999999998)(73,23.6485)(74,21.988523809523805)(75,24.162568627450973)(76,25.701755555555554)(77,26.219811320754715)(78,19.935064516129035)(79,10.448108108108109)(80,18.420809523809524)(81,20.06848148148148)(82,10.102695652173912)(83,2.76775)(84,17.233703703703704)(85,19.407149999999998)(86,12.944875000000001)(87,13.51504347826087)(88,11.10995238095238)(89,9.154470588235295)(90,19.47225)(91,7.422476190476191)(92,30.336666666666662)(93,24.654473684210526)(94,16.57048484848485)(95,16.452473684210524)(96,26.064333333333334)(97,19.55915)(98,14.239636363636363)(99,6.026846153846154)(100,13.068944444444444)(101,32.31605882352942)(102,8.722555555555555)(103,19.643549999999998)(104,12.09769230769231)(105,22.481571428571424)(106,15.746)(107,10.5028)(108,35.02711111111111)(109,23.6526)(110,8.763055555555557)(111,6.582833333333333)(112,8.783)(113,13.185583333333334)(114,27.709400000000006)(115,45.26628571428572)(116,29.728249999999996)(117,26.44033333333333)(118,26.461333333333336)(119,12.218615384615383)(120,14.44681818181818)(121,15.899599999999998)(122,14.46072727272727)(123,17.688888888888886)(124,39.8235)(125,21.730363636363634)(126,7.9719)(127,15.953)(128,21.763272727272728)(129,21.774545454545457)(130,24.581384615384614)(131,14.104529411764705)(132,30.757384615384616)(133,26.677866666666667)(134,13.347999999999999)(135,13.359499999999999)(136,10.0235)(137,12.339076923076922)(138,14.591636363636365)(139,57.33714285714286)(140,14.600727272727273)(141,22.959142857142858)(142,8.932)(143,14.621636363636362)(144,26.819)(145,10.061)(146,33.562916666666666)(147,15.114)(148,30.235125)(149,32.2628)(150,6.207999999999999)(151,13.456000000000001)(152,17.946)(153,8.0836)(154,26.96666666666667)(155,17.983333333333334)(156,13.491833333333332)(157,18.00488888888889)(158,28.947499999999998)(159,27.111666666666665)(160,29.591272727272727)(161,54.279333333333334)(162,16.2888)(163,8.147)(164,27.175)(165,9.06688888888889)(166,20.40575)(167,11.667428571428571)(168,11.67057142857143)(169,16.3532)(170,40.898)(171,40.908)(172,16.3736)(173,13.6565)(174,40.995)(175,20.5085)(176,35.177571428571426)(177,16.4404)(178,16.4508)(179,27.441666666666666)(180,13.729333333333335)(181,27.476)(182,27.506666666666664)(183,41.2795)(184,16.526400000000002)(185,55.18266666666667)(186,55.202)(187,27.608333333333334)(188,27.622)(189,27.669666666666668)(190,83.059)(191,33.2524)(192,55.46)(193,33.2932)(194,16.654400000000003)(195,83.345)(196,27.795333333333332)(197,55.678)(198,27.853333333333335)(199,27.863333333333333)(200,20.9205)(201,41.852999999999994)(202,83.753)(203,20.961)(204,42.039)(205,21.0245)(206,42.063)(207,28.05633333333333)(208,16.845599999999997)(209,16.860400000000002)(210,21.083)(211,42.2275)(212,42.25)(213,42.279)(214,28.209)(215,12.095428571428572)(216,28.240333333333336)(217,42.3965)(218,33.929199999999994)(219,42.444500000000005)(220,42.458)(221,33.9744)(222,28.32133333333333)(223,42.493)(224,56.68333333333334)(225,56.79333333333333)(226,63.96300000000001)(227,56.87733333333333)(228,17.069399999999998)(229,12.195857142857141)(230,42.706)(231,28.497666666666664)(232,42.7655)(233,42.8295)(234,34.2952)(235,21.4385)(236,24.509999999999998)(237,21.45025)(238,28.607666666666663)(239,85.854)(240,42.9615)(241,36.837857142857146)(242,57.31333333333333)(243,21.4965)(244,43.009)(245,86.039)(246,21.5215)(247,43.101)(248,17.2446)(249,64.734)(250,57.554833333333335)(251,43.259)(252,86.586)(253,28.874666666666666)(254,86.724)(255,86.968)(256,87.058)
}; % plot of the second histogram
\addplot[color=Cyan] coordinates{(0,12.569615384615386)(1,8.311625)(2,8.41766)(3,8.441513157894738)(4,5.780733333333333)(5,12.62866666666667)(6,8.65990291262136)(7,8.924593023255813)(8,8.201162393162393)(9,4.533152173913043)(10,7.311311926605503)(11,9.081366666666668)(12,8.821291262135922)(13,8.92335922330097)(14,7.742251798561149)(15,13.313499999999998)(16,10.581029197080293)(17,10.937186567164176)(18,10.972887931034483)(19,8.122070175438598)(20,10.412163636363633)(21,7.866625850340136)(22,13.723853448275861)(23,10.382165467625901)(24,14.1336)(25,14.344077519379844)(26,14.97266666666667)(27,17.808157024793395)(28,14.104246268656714)(29,17.99029197080292)(30,13.247529411764708)(31,12.70309482758621)(32,10.104681415929203)(33,11.177796296296297)(34,15.402433333333338)(35,12.215657894736847)(36,10.967376237623762)(37,12.55890476190476)(38,9.594837209302325)(39,12.298441558441557)(40,9.333957142857145)(41,12.874193181818178)(42,11.433117647058825)(43,13.338634920634922)(44,15.065249999999999)(45,14.1175)(46,17.660527272727276)(47,17.26426865671642)(48,17.579818181818183)(49,18.188359374999997)(50,13.538254237288136)(51,14.503882352941176)(52,19.406137254901964)(53,19.946785714285713)(54,21.049723076923076)(55,14.07433870967742)(56,9.698931034482758)(57,24.003148936170216)(58,19.038151515151515)(59,19.24329166666667)(60,11.35732)(61,20.36240677966102)(62,12.429882352941176)(63,11.45950819672131)(64,12.743380000000002)(65,16.940755102040818)(66,15.695490566037737)(67,16.027846153846156)(68,10.485408163265305)(69,7.145333333333332)(70,11.27455)(71,24.978193548387097)(72,16.506531914893614)(73,14.239699999999997)(74,10.453096774193549)(75,23.075644444444443)(76,14.596285714285715)(77,16.285979166666664)(78,16.781428571428574)(79,14.004595238095234)(80,14.0245)(81,10.236999999999998)(82,13.116239999999998)(83,13.677291666666669)(84,18.261777777777777)(85,17.94881818181818)(86,4.7055)(87,4.1204375)(88,19.410764705882357)(89,7.7724705882352945)(90,16.534375)(91,8.8272)(92,11.691529411764705)(93,17.295739130434782)(94,15.315923076923077)(95,27.67725)(96,28.003894736842096)(97,12.098363636363636)(98,30.73846153846154)(99,23.801642857142856)(100,24.25109090909091)(101,14.831111111111113)(102,4.771214285714286)(103,21.128315789473685)(104,17.621684210526315)(105,29.79888888888889)(106,14.38292857142857)(107,19.76088235294118)(108,26.9089)(109,7.482444444444444)(110,4.812357142857143)(111,14.450142857142856)(112,4.8225)(113,18.431818181818183)(114,25.38175)(115,10.699421052631578)(116,14.537214285714287)(117,19.969117647058823)(118,24.262142857142855)(119,20.3996)(120,17.013666666666666)(121,12.38290909090909)(122,6.1979999999999995)(123,34.1135)(124,31.039636363636365)(125,24.40342857142857)(126,11.39175)(127,13.694399999999998)(128,5.710083333333333)(129,42.8475)(130,28.24447058823529)(131,11.440833333333332)(132,29.445142857142855)(133,12.499272727272727)(134,11.473666666666666)(135,43.072625)(136,30.648888888888887)(137,19.725142857142856)(138,23.035066666666662)(139,23.056)(140,15.383111111111113)(141,6.927200000000001)(142,26.01025)(143,8.67425)(144,30.865333333333336)(145,10.693538461538461)(146,13.910333333333334)(147,23.195999999999998)(148,29.850285714285718)(149,8.708187500000001)(150,20.9155)(151,16.108615384615383)(152,19.958571428571428)(153,8.739125)(154,17.4895)(155,26.253375)(156,9.34)(157,17.52025)(158,17.54575)(159,17.553)(160,14.073400000000001)(161,23.513)(162,35.2975)(163,70.65)(164,23.573666666666668)(165,35.401)(166,17.7135)(167,17.7335)(168,23.672)(169,19.37618181818182)(170,42.670399999999994)(171,28.4552)(172,10.165571428571429)(173,17.8)(174,18.994933333333332)(175,14.266399999999999)(176,10.196)(177,17.89525)(178,23.888666666666666)(179,14.339400000000001)(180,14.35)(181,35.9145)(182,43.119600000000005)(183,72.095)(184,18.02775)(185,20.614571428571427)(186,20.628285714285713)(187,8.025555555555556)(188,28.9252)(189,36.2125)(190,18.11425)(191,12.082333333333333)(192,18.14725)(193,14.527600000000001)(194,18.195)(195,48.58733333333333)(196,36.474000000000004)(197,73.207)(198,24.413)(199,24.44333333333333)(200,48.925333333333334)(201,36.70416666666667)(202,24.480999999999998)(203,49.095333333333336)(204,18.42125)(205,14.742799999999999)(206,36.873)(207,73.787)(208,55.378499999999995)(209,36.944)(210,73.93766666666666)(211,37.1225)(212,49.536)(213,24.775333333333332)(214,18.589)(215,37.1985)(216,74.448)(217,37.2545)(218,49.897333333333336)(219,74.866)(220,56.203500000000005)(221,74.989)(222,75.017)(223,37.5285)(224,37.5365)(225,75.119)(226,37.6595)(227,56.50574999999999)(228,37.71)(229,18.87875)(230,25.192333333333334)(231,25.214333333333332)(232,50.446)(233,75.732)(234,75.732)(235,18.9675)(236,50.590666666666664)(237,25.305666666666667)(238,37.977)(239,50.64666666666667)(240,45.6006)(241,76.067)(242,76.195)(243,76.311)(244,76.458)(245,25.627)
}; % plot of the third histogram
\addplot[color=SeaGreen] coordinates{(0,2.6138214285714283)(1,7.619128205128206)(2,6.116048387096774)(3,5.489809523809524)(4,6.6645394736842105)(5,7.66720895522388)(6,9.335088888888889)(7,4.836293577981652)(8,6.830344444444444)(9,5.37887012987013)(10,9.03158064516129)(11,5.629380530973452)(12,10.06399099099099)(13,6.474206611570248)(14,9.007270491803279)(15,9.743300813008132)(16,7.00464347826087)(17,9.68195238095238)(18,5.62290579710145)(19,9.211909090909092)(20,9.545065420560748)(21,9.62103738317757)(22,10.352528455284553)(23,12.961845454545454)(24,13.018719298245614)(25,12.44058415841584)(26,11.950852941176464)(27,15.134465346534652)(28,14.37840495867769)(29,14.821034782608697)(30,11.45489075630252)(31,10.663432098765433)(32,5.391894736842106)(33,12.34028)(34,10.765468749999998)(35,16.062309523809525)(36,10.034028846153847)(37,10.47915)(38,10.526415384615383)(39,10.983272727272727)(40,10.475943661971831)(41,17.786681159420294)(42,20.315879310344826)(43,13.42776923076923)(44,9.87592957746479)(45,9.74504)(46,13.583071428571426)(47,16.632237288135595)(48,11.291253968253969)(49,11.430152777777778)(50,12.797196428571429)(51,13.564433962264152)(52,11.880196428571429)(53,14.585852459016392)(54,12.386755555555556)(55,13.9816875)(56,15.294295454545455)(57,12.732811320754713)(58,21.688288461538466)(59,12.686918367346939)(60,8.395148148148149)(61,17.234107142857145)(62,10.349454545454545)(63,13.2553488372093)(64,6.0784468085106385)(65,12.263833333333334)(66,15.765700000000004)(67,7.657066666666666)(68,16.662710526315788)(69,13.182400000000001)(70,14.054054054054054)(71,19.290939393939393)(72,16.346564102564102)(73,7.634552631578948)(74,21.811375)(75,13.443230769230771)(76,22.58296774193548)(77,16.12041379310345)(78,8.670518518518518)(79,16.747357142857144)(80,10.870481481481484)(81,12.595928571428573)(82,9.292684210526316)(83,9.42372)(84,12.819347826086958)(85,10.426058823529411)(86,8.454428571428572)(87,14.81825)(88,13.989823529411765)(89,14.012705882352941)(90,17.538411764705884)(91,17.054714285714287)(92,14.94075)(93,4.9895000000000005)(94,16.346636363636364)(95,9.999166666666667)(96,18.480615384615383)(97,12.887285714285714)(98,9.258923076923077)(99,12.0491)(100,15.074583333333335)(101,27.461909090909092)(102,10.996272727272727)(103,8.64942857142857)(104,20.197)(105,10.108)(106,20.23806666666667)(107,16.210666666666665)(108,14.045461538461536)(109,22.856749999999998)(110,30.5005)(111,4.361642857142857)(112,17.46542857142857)(113,15.302)(114,12.2548)(115,15.33125)(116,7.679125)(117,10.247333333333334)(118,21.98107142857143)(119,11.200181818181818)(120,11.557437499999999)(121,18.518099999999997)(122,10.295666666666667)(123,9.509384615384615)(124,13.251857142857144)(125,20.629133333333336)(126,6.5187368421052625)(127,22.148214285714285)(128,24.8364)(129,13.806222222222223)(130,34.974875)(131,19.142)(132,18.698)(133,27.731555555555556)(134,26.030666666666665)(135,4.465214285714286)(136,34.13945454545455)(137,31.3415)(138,16.731199999999998)(139,23.551875)(140,31.428)(141,22.87890909090909)(142,15.737)(143,16.815466666666666)(144,27.03857142857143)(145,9.021857142857142)(146,18.071142857142856)(147,15.8345)(148,18.10557142857143)(149,21.147333333333332)(150,14.113777777777777)(151,25.4216)(152,12.718700000000002)(153,12.7372)(154,23.905124999999998)(155,7.970875)(156,10.6365)(157,7.98375)(158,27.40757142857143)(159,12.804400000000001)(160,12.8248)(161,12.8466)(162,42.88333333333333)(163,19.3134)(164,25.8188)(165,12.9174)(166,32.333)(167,43.205333333333336)(168,25.9356)(169,25.951600000000003)(170,65.021)(171,16.2655)(172,26.037200000000002)(173,32.592)(174,43.53133333333333)(175,65.338)(176,65.429)(177,16.38)(178,26.2408)(179,16.408)(180,32.928)(181,22.007333333333335)(182,52.855999999999995)(183,44.15266666666667)(184,16.5705)(185,44.214666666666666)(186,66.454)(187,22.15833333333333)(188,44.336666666666666)(189,33.282)(190,22.20566666666667)(191,33.339)(192,22.236333333333334)(193,26.699599999999997)(194,33.419)(195,16.75675)(196,33.6205)(197,67.271)(198,50.498999999999995)(199,44.936)(200,67.625)(201,16.942)(202,22.605333333333334)(203,16.9725)(204,22.646)(205,22.66633333333333)(206,22.674000000000003)(207,45.401999999999994)(208,68.264)(209,22.772000000000002)(210,17.09175)(211,34.2095)(212,68.45)(213,17.1355)(214,34.3375)(215,68.756)(216,68.867)(217,34.453)(218,17.24225)(219,23.016666666666666)(220,23.043666666666667)(221,51.8925)(222,52.007999999999996)(223,69.599)(224,23.253666666666664)(225,34.9005)(226,27.932799999999997)(227,23.35166666666667)(228,35.0425)(229,35.058)(230,70.136)(231,70.164)(232,70.205)(233,23.436666666666667)(234,35.211)(235,35.2365)(236,17.6285)(237,23.521666666666665)(238,35.2825)(239,53.17725)(240,35.4615)
}; % plot of the fourth histogram
\addplot[color=LimeGreen] coordinates{(0,2.29075)(1,3.64674074074074)(2,9.476666666666668)(3,5.273745098039216)(4,7.770263157894737)(5,3.499753623188406)(6,7.774111111111112)(7,4.807049382716049)(8,5.604831325301205)(9,4.9266296296296295)(10,3.4105979381443303)(11,5.213021505376344)(12,6.080505376344084)(13,11.6319756097561)(14,6.548952830188678)(15,7.235298969072164)(16,6.756818181818181)(17,6.258537037037037)(18,10.13869230769231)(19,8.004860465116279)(20,5.8989111111111106)(21,10.80862626262626)(22,12.60250980392157)(23,7.267619565217391)(24,10.358166666666666)(25,8.898686046511628)(26,10.601415841584156)(27,9.315000000000001)(28,13.806609090909092)(29,9.15468181818182)(30,13.328764150943396)(31,11.883215517241378)(32,9.42565789473684)(33,10.56453086419753)(34,10.61232098765432)(35,11.371821428571431)(36,8.361440860215055)(37,4.178954545454546)(38,11.26476923076923)(39,12.246958333333332)(40,9.981785714285715)(41,8.56394366197183)(42,10.829192307692306)(43,8.184434782608696)(44,5.494046511627908)(45,6.510333333333333)(46,13.400031250000001)(47,16.2544)(48,15.002333333333334)(49,9.080283018867924)(50,6.5022884615384635)(51,14.359203703703704)(52,12.464692307692307)(53,13.927619047619048)(54,13.52440425531915)(55,13.295187500000003)(56,16.821536585365852)(57,13.102)(58,15.374758620689654)(59,18.785837837837843)(60,7.802254901960784)(61,18.12369696969697)(62,15.103837209302327)(63,10.871717391304347)(64,22.152488372093025)(65,14.3525)(66,16.78553846153846)(67,13.025870967741936)(68,13.044967741935482)(69,14.239968750000001)(70,16.91191111111111)(71,13.070457142857142)(72,15.089592592592592)(73,16.02894285714286)(74,13.714560975609757)(75,19.832387096774195)(76,11.59116129032258)(77,10.643551724137929)(78,17.774310344827587)(79,14.51778125)(80,9.526894736842106)(81,14.793999999999999)(82,8.652166666666666)(83,10.3948)(84,6.50925)(85,16.583772727272727)(86,8.696)(87,8.709083333333334)(88,6.54075)(89,19.047818181818183)(90,29.97614285714286)(91,7.004266666666666)(92,8.7635)(93,17.555500000000002)(94,7.534285714285714)(95,11.729999999999999)(96,6.60475)(97,16.273923076923076)(98,10.5922)(99,10.6102)(100,8.857083333333334)(101,7.603428571428571)(102,7.614714285714285)(103,6.301588235294118)(104,23.834666666666667)(105,10.7374)(106,13.451500000000001)(107,15.386571428571429)(108,12.436384615384615)(109,15.418142857142856)(110,27.0095)(111,6.763125)(112,4.922909090909091)(113,18.062)(114,9.039)(115,15.512428571428572)(116,6.043)(117,34.0285)(118,16.3442)(119,10.9068)(120,7.795857142857143)(121,27.3046)(122,9.946)(123,13.692)(124,31.329714285714285)(125,18.293333333333333)(126,24.422444444444444)(127,18.345999999999997)(128,7.869714285714286)(129,6.892125)(130,18.402333333333335)(131,6.144333333333333)(132,13.8425)(133,27.704500000000003)(134,36.976)(135,16.6533)(136,20.204)(137,22.2534)(138,23.208333333333332)(139,18.578444444444443)(140,22.316399999999998)(141,7.977428571428571)(142,23.949857142857145)(143,20.351636363636366)(144,4.664333333333333)(145,35.04375)(146,8.017857142857142)(147,9.368666666666668)(148,12.510666666666667)(149,14.0875)(150,18.807)(151,14.12)(152,28.2715)(153,16.9893)(154,10.307636363636364)(155,7.0945)(156,9.469)(157,57.029)(158,12.688888888888888)(159,9.527)(160,22.916800000000002)(161,19.130666666666666)(162,57.414)(163,14.3865)(164,19.197666666666667)(165,19.215)(166,23.0768)(167,9.621)(168,48.223000000000006)(169,11.5878)(170,38.833999999999996)(171,29.181)(172,29.243)(173,58.486)(174,29.2915)(175,19.548333333333336)(176,58.832)(177,19.642)(178,14.74625)(179,29.524)(180,19.760666666666665)(181,11.8794)(182,19.889)(183,29.919)(184,19.969666666666665)(185,20.061)(186,30.127)(187,20.094666666666665)(188,30.178)(189,30.188)(190,60.455)(191,34.564571428571426)(192,24.2012)(193,15.13975)(194,17.337714285714288)(195,20.25433333333333)(196,40.556000000000004)(197,30.482)(198,30.7115)(199,61.458)(200,41.028)(201,61.593)(202,30.8225)(203,20.598)(204,41.21866666666667)(205,30.929999999999996)(206,31.028)(207,62.124)(208,62.404)(209,20.808333333333334)(210,31.236)(211,20.849333333333334)(212,62.568)(213,41.806666666666665)(214,31.355)(215,62.883)(216,31.471)(217,31.506)(218,31.562)(219,31.6855)(220,63.513)(221,63.547)(222,31.794)(223,63.667)(224,31.8495)(225,31.8595)(226,64.045)(227,64.096)(228,21.377333333333336)(229,21.388666666666666)(230,64.351)(231,42.96866666666667)(232,32.2605)(233,32.351)(234,64.866)(235,32.541)(236,13.0594)(237,21.80033333333333)(238,65.442)(239,65.491)
}; % plot of the fifth histogram
\legend{30000 ppl,  27500 ppl, 25000 ppl, 22500 ppl, 20000 ppl} % legend of the plot

\end{axis}
\end{tikzpicture}

%% file: plots/num_ppl/avg_speed.tex
\begin{tikzpicture}[scale=0.75]

% Set the axis labels and title
\begin{axis}[    xlabel={Time [s]},    ylabel={Speed [m/s]},    title={Average Speed - Number of people},   legend style={at={(1,1)}, anchor=north east},]

\addplot[color=Red] coordinates{(0,4.018260258854219)(1,4.01683637463454)(2,4.0137139374238116)(3,4.0090585902782525)(4,4.003251470079557)(5,3.995790950543669)(6,3.9874532168953842)(7,3.9792316056542227)(8,3.970156971375808)(9,3.9617774430427413)(10,3.952640885676943)(11,3.941687637621894)(12,3.9304683727667795)(13,3.9185783122606153)(14,3.9066509115462527)(15,3.8923243712055506)(16,3.8787919224268306)(17,3.8653819809123706)(18,3.8502560729220505)(19,3.8352451478631235)(20,3.820024906600249)(21,3.803303919556741)(22,3.7838252406643393)(23,3.7623183655632033)(24,3.7391844572461315)(25,3.715972871842844)(26,3.687568223165555)(27,3.6597445094707695)(28,3.631654676258993)(29,3.5997956403269753)(30,3.5685067232837935)(31,3.5399323429916163)(32,3.5114294422432186)(33,3.48869277440706)(34,3.4682306940371457)(35,3.4459915611814345)(36,3.4270360013981125)(37,3.4054811866859622)(38,3.381824981301421)(39,3.3585853705672046)(40,3.335791886773647)(41,3.3162051282051284)(42,3.290909090909091)(43,3.2722330941509026)(44,3.2500556916908)(45,3.226704871060172)(46,3.2055294117647057)(47,3.1816420440784694)(48,3.156304401890077)(49,3.1252408477842004)(50,3.097454256165473)(51,3.0704032809295967)(52,3.0381690140845072)(53,3.004499274310595)(54,2.9660558726344246)(55,2.9292145296491774)(56,2.881555591415201)(57,2.840227767543125)(58,2.796880415944541)(59,2.755391804457225)(60,2.6957091999250515)(61,2.6416309012875536)(62,2.5701970443349755)(63,2.510497000856898)(64,2.445591204846309)(65,2.388862891904539)(66,2.332846359873387)(67,2.279028585884139)(68,2.2358934169278997)(69,2.1836900568951503)(70,2.1424148606811144)(71,2.0976608187134502)(72,2.0601571946795647)(73,2.0168961201501876)(74,1.9803479381443299)(75,1.9298657718120806)(76,1.884481558803062)(77,1.836231884057971)(78,1.7633472169632716)(79,1.7005136309758988)(80,1.6478988168094655)(81,1.6041315345699831)(82,1.5497588776852258)(83,1.509009009009009)(84,1.4680259499536608)(85,1.4485849056603775)(86,1.4184808901790034)(87,1.4089787863838183)(88,1.3973908680381335)(89,1.3944162436548224)(90,1.3858024691358024)(91,1.3733681462140992)(92,1.3678590496529632)(93,1.3580786026200873)(94,1.3541781959048147)(95,1.3408197641774284)(96,1.3335230506545248)(97,1.3254608294930876)(98,1.3230140186915889)(99,1.314540059347181)(100,1.3100961538461537)(101,1.3007290400972054)(102,1.2912801484230056)(103,1.2858040201005025)(104,1.2786989795918366)(105,1.2721970187945562)(106,1.2636783124588002)(107,1.257313829787234)(108,1.2510148849797023)(109,1.2472602739726026)(110,1.2427385892116183)(111,1.2352112676056337)(112,1.2287769784172662)(113,1.2219796215429404)(114,1.2179675994108983)(115,1.2137518684603887)(116,1.2099697885196374)(117,1.2068702290076336)(118,1.2009273570324575)(119,1.1954828660436136)(120,1.190779014308426)(121,1.1835084882780922)(122,1.1783960720130933)(123,1.1731890091590342)(124,1.1632825719120135)(125,1.1547008547008546)(126,1.1484375)(127,1.148375768217735)(128,1.1411974977658623)(129,1.1397459165154264)(130,1.1374885426214483)(131,1.1347583643122676)(132,1.130885122410546)(133,1.1296472831267874)(134,1.1256038647342994)(135,1.1218074656188606)(136,1.1157684630738522)(137,1.1124620060790273)(138,1.1094069529652353)(139,1.1042752867570387)(140,1.0976645435244161)(141,1.0958019375672767)(142,1.092896174863388)(143,1.0880713489409142)(144,1.0843644544431945)(145,1.082004555808656)(146,1.0776361529548089)(147,1.0725326991676576)(148,1.065217391304348)(149,1.0605686032138442)(150,1.0554156171284634)(151,1.0523627075351214)(152,1.0507152145643692)(153,1.0498687664041995)(154,1.0493333333333332)(155,1.0407055630936228)(156,1.0397805212620028)(157,1.0363128491620113)(158,1.0256045519203414)(159,1.0244604316546762)(160,1.0204678362573099)(161,1.017910447761194)(162,1.0165912518853695)(163,1.0137614678899083)(164,1.0138888888888888)(165,1.0078369905956113)(166,1.0047923322683705)(167,1.0032310177705976)(168,1)(169,1)(170,0.9983388704318937)(171,0.9983136593591906)(172,0.9982935153583617)(173,0.9982847341337907)(174,0.998272884283247)(175,0.9982456140350877)(176,0.9964476021314387)(177,0.996415770609319)(178,0.9963570127504554)(179,0.9963235294117647)(180,0.9962825278810409)(181,0.9962546816479401)(182,0.9962121212121212)(183,0.9961759082217974)(184,0.996116504854369)(185,0.996078431372549)(186,0.9960159362549801)(187,0.9959595959595959)(188,0.9959266802443992)(189,0.9958847736625515)(190,0.9958419958419958)(191,0.9957983193277311)(192,0.9957716701902748)(193,0.9957537154989384)(194,0.9957264957264957)(195,0.9956709956709957)(196,0.9956331877729258)(197,0.9956140350877193)(198,0.9955654101995566)(199,0.9955156950672646)(200,0.9954648526077098)(201,0.9954128440366973)(202,0.9953810623556582)(203,0.9953051643192489)(204,0.9952038369304557)(205,0.9951690821256038)(206,0.9951456310679612)(207,0.9951100244498777)(208,0.995049504950495)(209,0.9949748743718593)(210,0.9948849104859335)(211,0.9947916666666666)(212,0.9947643979057592)(213,0.9946808510638298)(214,0.9946091644204852)(215,0.9945945945945946)(216,0.9944598337950139)(217,0.994413407821229)(218,0.9943502824858758)(219,0.9943019943019943)(220,0.9942196531791907)(221,0.9941690962099126)(222,0.9940476190476191)(223,0.9939024390243902)(224,0.9938650306748467)(225,0.9938080495356038)(226,0.9937106918238994)(227,0.9935897435897436)(228,0.9934426229508196)(229,0.9933554817275747)(230,0.9931740614334471)(231,0.993103448275862)(232,0.993006993006993)(233,0.9929328621908127)(234,0.9928571428571429)(235,0.9927272727272727)(236,0.9926470588235294)(237,0.9925373134328358)(238,0.9923664122137404)(239,0.9922480620155039)(240,0.9921875)(241,0.9920318725099602)(242,0.991869918699187)(243,0.9918367346938776)(244,0.9917355371900827)(245,0.9916317991631799)(246,0.9914893617021276)(247,0.9914529914529915)(248,0.9912280701754386)(249,0.9912280701754386)(250,0.9911504424778761)(251,0.990990990990991)(252,0.9907834101382489)(253,0.9906542056074766)(254,0.9904761904761905)(255,0.9903381642512077)(256,0.9902439024390244)(257,0.9899497487437185)(258,0.9949238578680203)(259,0.9948453608247423)(260,0.9947916666666666)(261,0.9947368421052631)(262,0.9945945945945946)(263,0.9944444444444445)(264,0.9943502824858758)(265,0.9942857142857143)(266,0.9942196531791907)(267,0.9940828402366864)(268,0.9940119760479041)(269,0.9938271604938271)(270,0.9936305732484076)(271,0.9934640522875817)(272,0.9933774834437086)(273,0.9931506849315068)(274,0.9930555555555556)(275,0.993006993006993)(276,0.9928057553956835)(277,0.9927007299270073)(278,0.9925925925925926)(279,0.9922480620155039)(280,0.9921875)(281,0.992)(282,0.991869918699187)(283,0.9916666666666667)(284,0.9914529914529915)(285,0.9910714285714286)(286,0.9819819819819819)(287,0.9908256880733946)(288,0.9907407407407407)(289,0.9904761904761905)(290,0.9903846153846154)(291,0.9900990099009901)(292,0.98989898989899)(293,0.9895833333333334)(294,0.989010989010989)(295,1)(296,1)(297,1)(298,1)(299,1)(300,1)(301,1)(302,1)(303,1)(304,1)(305,1)(306,1)(307,1)(308,1)(309,1)(310,1)(311,1)(312,1)(313,1)(314,1)(315,1)(316,1)(317,1)(318,1)(319,1)(320,1)(321,1)(322,1)(323,1)(324,1)(325,1)(326,1)(327,1)(328,1)(329,1)(330,1)(331,1)(332,1)(333,1)(334,1)(335,1)(336,1)(337,1)(338,1)
}; % plot of the first histogram
\addplot[color=BurntOrange] coordinates{(0,4.019484073411902)(1,4.017131327953045)(2,4.014037680088659)(3,4.010251630941286)(4,4.004147031102733)(5,3.997592019263846)(6,3.9904144675566595)(7,3.9825606628526335)(8,3.974289961033222)(9,3.964688050469051)(10,3.955371693619155)(11,3.9451422712125095)(12,3.9346182422096936)(13,3.9239883746926)(14,3.9099136346931274)(15,3.8967125094197437)(16,3.8833664037167885)(17,3.868460388639761)(18,3.854197651161598)(19,3.8403246205733557)(20,3.825466579861111)(21,3.8096937921802665)(22,3.7896162268065)(23,3.7709683171848063)(24,3.748078460308676)(25,3.72289003370858)(26,3.6980149047022355)(27,3.669248166678089)(28,3.640202221589291)(29,3.605253394672405)(30,3.5740611691831203)(31,3.5457179652285897)(32,3.519606207241782)(33,3.496713371389033)(34,3.474324566112006)(35,3.452100528609849)(36,3.42765895674054)(37,3.40343862455018)(38,3.385322997416021)(39,3.3666702207058323)(40,3.3446564464541746)(41,3.319945541184479)(42,3.303330228225431)(43,3.281578947368421)(44,3.2590501902073874)(45,3.2377833753148613)(46,3.215137376879212)(47,3.1924003211131926)(48,3.160902047592695)(49,3.1319276250178087)(50,3.103726525821596)(51,3.0719043293975172)(52,3.040544771446462)(53,3.0093684380552417)(54,2.9736534934133734)(55,2.9406179872259623)(56,2.898600143575018)(57,2.861198738170347)(58,2.8095146379044684)(59,2.7608478304339132)(60,2.706616729088639)(61,2.6470076169749728)(62,2.579700272479564)(63,2.5206258890469417)(64,2.454905847373637)(65,2.3894491854150504)(66,2.335127860026918)(67,2.2827122443261416)(68,2.235961594413733)(69,2.1906060606060604)(70,2.1477878882962034)(71,2.100032583903552)(72,2.066711590296496)(73,2.023410202655486)(74,1.9776496034607065)(75,1.9400149031296572)(76,1.8836123005060335)(77,1.8216612377850163)(78,1.7598122866894197)(79,1.707855251544572)(80,1.6500457456541628)(81,1.5996231747527083)(82,1.5426205552849488)(83,1.4995)(84,1.4806122448979593)(85,1.4537859007832898)(86,1.4331379861481086)(87,1.4165307232191409)(88,1.4065265486725664)(89,1.40437464946719)(90,1.3959044368600682)(91,1.3943254198031267)(92,1.391764705882353)(93,1.3846153846153846)(94,1.3780193236714975)(95,1.3719362745098038)(96,1.3682559598494355)(97,1.3587786259541985)(98,1.3497091144149969)(99,1.3440577806959948)(100,1.3413493653974615)(101,1.3310719131614654)(102,1.3237113402061855)(103,1.3161200279134682)(104,1.3093883357041252)(105,1.3051337671728127)(106,1.2968405584129317)(107,1.285820895522388)(108,1.2792452830188679)(109,1.275229357798165)(110,1.2647058823529411)(111,1.2659323367427222)(112,1.2552166934189406)(113,1.2495921696574226)(114,1.2444260941370768)(115,1.2391666666666667)(116,1.233558178752108)(117,1.2246127366609294)(118,1.2223190932868353)(119,1.2178919397697077)(120,1.2116591928251121)(121,1.2076502732240437)(122,1.2040627885503232)(123,1.19644527595884)(124,1.1926345609065157)(125,1.191204588910134)(126,1.1847195357833655)(127,1.1820837390457644)(128,1.177690029615005)(129,1.1748251748251748)(130,1.1710526315789473)(131,1.1676891615541922)(132,1.1626442812172089)(133,1.1627659574468086)(134,1.1596091205211727)(135,1.1546961325966851)(136,1.1501128668171559)(137,1.1436781609195403)(138,1.1402085747392816)(139,1.134818288393904)(140,1.1288782816229117)(141,1.1248484848484848)(142,1.1214723926380368)(143,1.1161048689138577)(144,1.1125158027812896)(145,1.1112531969309463)(146,1.106631989596879)(147,1.1003963011889035)(148,1.0963855421686748)(149,1.0874316939890711)(150,1.0822873082287308)(151,1.0786838340486409)(152,1.072674418604651)(153,1.069219440353461)(154,1.061654135338346)(155,1.0519083969465648)(156,1.0434108527131782)(157,1.0362776025236593)(158,1.0305466237942122)(159,1.030844155844156)(160,1.0279605263157894)(161,1.0283333333333333)(162,1.026936026936027)(163,1.0256410256410255)(164,1.0175438596491229)(165,1.0142857142857142)(166,1.0108892921960073)(167,1.0073529411764706)(168,1.007421150278293)(169,1.0056285178236397)(170,1.0056603773584907)(171,1.0057803468208093)(172,1.003883495145631)(173,1.0039525691699605)(174,1.002016129032258)(175,1.0020408163265306)(176,1)(177,1)(178,1)(179,1)(180,1)(181,1)(182,1)(183,1)(184,1)(185,1)(186,1)(187,1)(188,1)(189,1)(190,1)(191,1)(192,1)(193,1)(194,1)(195,1)(196,1)(197,1)(198,1)(199,1)(200,1)(201,1)(202,1)(203,1)(204,1)(205,1)(206,1)(207,1)(208,0.997275204359673)(209,1)(210,1)(211,1)(212,1)(213,1)(214,1)(215,1)(216,1)(217,1)(218,1)(219,1)(220,1)(221,1)(222,1)(223,1)(224,1)(225,1)(226,1)(227,1)(228,1)(229,1)(230,1)(231,1)(232,1)(233,1)(234,1)(235,1)(236,1)(237,1)(238,1)(239,1)(240,1)(241,1)(242,1)(243,1)(244,1)(245,1)(246,1)(247,1)(248,1)(249,1)(250,1)(251,0.995850622406639)(252,1)(253,1)(254,1)(255,1)(256,1)(257,1)(258,1)(259,1)(260,1)(261,1)(262,1)(263,1)(264,1)(265,1)(266,1)(267,1)(268,1)(269,1)(270,1)(271,1)(272,1)(273,1)(274,1)(275,1)(276,1)(277,1)(278,1)(279,1)(280,1)(281,1)(282,1)(283,1)(284,1)(285,1)(286,1)(287,1)(288,1)(289,1)(290,1)(291,1)(292,1)(293,1)(294,1)(295,1)(296,1)(297,1)(298,1)(299,1)(300,1)(301,1)(302,1)(303,1)(304,1)(305,1)(306,1)(307,1)(308,1)(309,1)(310,1)(311,1)(312,1)(313,1)(314,1)(315,1)(316,1)(317,1)(318,1)(319,1)(320,1)(321,1)(322,1)(323,1)(324,1)(325,1)(326,1)(327,1)(328,1)(329,1)(330,1)(331,1)(332,1)(333,1)(334,1)(335,1)(336,1)(337,1)(338,1)(339,1)(340,1)(341,1)(342,1)(343,1)(344,1)(345,1)(346,1)(347,1)
}; % plot of the second histogram
\addplot[color=Cyan] coordinates{(0,4.019152780566151)(1,4.0174653113907715)(2,4.015038816404504)(3,4.011110200065596)(4,4.005439747529275)(5,3.9995374264087467)(6,3.99244138873468)(7,3.9843050302160776)(8,3.9765497101898144)(9,3.967484080747866)(10,3.9588140657320157)(11,3.94888178913738)(12,3.939763666058219)(13,3.9289080911233305)(14,3.9149236429615444)(15,3.9016122364613475)(16,3.8886590282198177)(17,3.875347923374993)(18,3.8608441521946832)(19,3.8454429941088137)(20,3.8285544318857823)(21,3.8115516397454723)(22,3.7913741857964967)(23,3.7746984023475707)(24,3.752027575020276)(25,3.7296652396659415)(26,3.702880058330295)(27,3.6718465736810186)(28,3.63899629132802)(29,3.6103672061118868)(30,3.5815184900503887)(31,3.554679191775966)(32,3.5305578531384985)(33,3.506208213944604)(34,3.482536855644603)(35,3.4622709122555544)(36,3.4396891302033428)(37,3.4205885594621406)(38,3.4013651877133104)(39,3.3800447111424874)(40,3.355151515151515)(41,3.3339977575682074)(42,3.3209338122113903)(43,3.3006613756613756)(44,3.275735294117647)(45,3.2585844428871757)(46,3.237300330982875)(47,3.2166715932604197)(48,3.19101805492338)(49,3.163616455498201)(50,3.135235332043843)(51,3.1057005152069137)(52,3.071969696969697)(53,3.0371621621621623)(54,3.0003669051550172)(55,2.9666857034075766)(56,2.9254437869822487)(57,2.887936313533374)(58,2.8405919661733616)(59,2.792739273927393)(60,2.742948864939234)(61,2.6862745098039214)(62,2.6360696517412934)(63,2.5733194913054764)(64,2.505697232772653)(65,2.445073612684032)(66,2.383933845245127)(67,2.3273791191869417)(68,2.282289002557545)(69,2.2462711302618494)(70,2.216649607642443)(71,2.1775733993632826)(72,2.133211678832117)(73,2.092838196286472)(74,2.0563657863618445)(75,2.014291547570437)(76,1.9513444302176697)(77,1.8819569120287254)(78,1.8129123468426014)(79,1.738423645320197)(80,1.6801848049281314)(81,1.6156716417910448)(82,1.5644983461962514)(83,1.5349492671927847)(84,1.5)(85,1.4755162241887905)(86,1.453558504221954)(87,1.4403669724770642)(88,1.427594779366066)(89,1.4218158890290038)(90,1.417146513115803)(91,1.4098254686489982)(92,1.4026229508196721)(93,1.3921438082556592)(94,1.3869653767820773)(95,1.3821305841924398)(96,1.3774373259052926)(97,1.3676366217175302)(98,1.3613505747126438)(99,1.3544395924308588)(100,1.3429416112342942)(101,1.3388182498130141)(102,1.3318216175359032)(103,1.3184615384615384)(104,1.3066666666666666)(105,1.2973621103117505)(106,1.2929782082324455)(107,1.2844262295081967)(108,1.2779632721202003)(109,1.2748735244519394)(110,1.272572402044293)(111,1.270293609671848)(112,1.2688266199649738)(113,1.264)(114,1.257918552036199)(115,1.2541133455210238)(116,1.249767008387698)(117,1.2407582938388626)(118,1.2353506243996157)(119,1.2294117647058824)(120,1.2235528942115768)(121,1.21840242669363)(122,1.2137718396711203)(123,1.2133194588969822)(124,1.2069693769799366)(125,1.20403825717322)(126,1.1984897518878102)(127,1.1960569550930997)(128,1.1906354515050168)(129,1.1837655016910935)(130,1.1837899543378996)(131,1.1840277777777777)(132,1.1789473684210525)(133,1.1744324970131421)(134,1.1670702179176755)(135,1.1627906976744187)(136,1.157960199004975)(137,1.149746192893401)(138,1.1437740693196405)(139,1.1391417425227568)(140,1.1339522546419099)(141,1.1326116373477673)(142,1.1275720164609053)(143,1.1199442119944212)(144,1.1136363636363635)(145,1.110632183908046)(146,1.1075581395348837)(147,1.107669616519174)(148,1.1069277108433735)(149,1.1003086419753085)(150,1.0923317683881064)(151,1.0858505564387917)(152,1.0800653594771241)(153,1.0787269681742044)(154,1.07679180887372)(155,1.062937062937063)(156,1.056637168141593)(157,1.0487364620938628)(158,1.0422794117647058)(159,1.0373831775700935)(160,1.0326923076923078)(161,1.0291828793774318)(162,1.023715415019763)(163,1.0141129032258065)(164,1.0101626016260163)(165,1.0102669404517455)(166,1.0082815734989647)(167,1.0084566596194504)(168,1.0063965884861408)(169,1.006423982869379)(170,1.002169197396963)(171,1.0021881838074398)(172,1.0021929824561404)(173,1.0022075055187638)(174,1.0022222222222221)(175,1)(176,1)(177,1)(178,1)(179,1)(180,1)(181,1)(182,1)(183,1)(184,1)(185,1)(186,1)(187,1)(188,1)(189,1)(190,1)(191,1)(192,1)(193,1)(194,1)(195,1)(196,1)(197,1)(198,1)(199,1)(200,1)(201,1)(202,1)(203,1)(204,1)(205,1)(206,1)(207,1)(208,1)(209,1)(210,1)(211,1)(212,1)(213,1)(214,1)(215,1)(216,1)(217,1)(218,1)(219,1)(220,1)(221,1)(222,1)(223,1)(224,1)(225,1)(226,1)(227,1)(228,1)(229,1)(230,1)(231,1)(232,1)(233,1)(234,1)(235,1)(236,1)(237,1)(238,1)(239,1)(240,1)(241,1)(242,1)(243,1)(244,1)(245,1)(246,1)(247,1)(248,1)(249,1)(250,1)(251,1)(252,1)(253,1)(254,1)(255,1)(256,1)(257,1)(258,1)(259,1)(260,1)(261,1)(262,1)(263,1)(264,1)(265,1)(266,1)(267,1)(268,1)(269,1)(270,1)(271,1)(272,1)(273,1)(274,1)(275,1)(276,1)(277,1)(278,1)(279,1)(280,1)(281,1)(282,1)(283,1)(284,1)(285,1)(286,1)(287,1)(288,1)(289,1)(290,1)(291,1)(292,1)(293,1)(294,1)(295,1)(296,1)(297,1)(298,1)(299,1)(300,1)(301,1)(302,1)(303,1)(304,1)(305,1)(306,1)(307,1)(308,1)(309,1)(310,1)(311,1)(312,1)(313,1)(314,1)(315,1)(316,1)(317,1)(318,1)(319,1)(320,1)(321,1)(322,1)(323,1)(324,1)(325,1)(326,1)(327,1)(328,1)(329,1)(330,1)(331,1)(332,1)(333,1)(334,1)(335,1)(336,1)(337,1)(338,1)(339,1)(340,1)(341,1)(342,1)(343,1)(344,1)(345,1)(346,1)(347,1)(348,1)(349,1)(350,1)(351,1)(352,1)(353,1)
}; % plot of the third histogram
\addplot[color=SeaGreen] coordinates{(0,4.018062617072518)(1,4.015831726241198)(2,4.012468940591823)(3,4.008486950173389)(4,4.003323025799603)(5,3.9972862957937583)(6,3.9896882721915987)(7,3.9797300566010354)(8,3.9723522744073727)(9,3.963832455480311)(10,3.954122340425532)(11,3.942901960784314)(12,3.931357839459865)(13,3.9192761173567314)(14,3.904643962848297)(15,3.8916902793037647)(16,3.8786206077872745)(17,3.8646767991210402)(18,3.8520007538161947)(19,3.836215271031669)(20,3.8208100186517453)(21,3.804655314474046)(22,3.7856737990647584)(23,3.763836017569546)(24,3.7431685716448415)(25,3.717594409986653)(26,3.695949243533431)(27,3.6678194854491775)(28,3.6411939775910365)(29,3.609302325581395)(30,3.5791820855868677)(31,3.5514589905362777)(32,3.52727087368743)(33,3.507583737097114)(34,3.48664915736485)(35,3.467189810076197)(36,3.447121034077556)(37,3.4276484213092773)(38,3.408832807570978)(39,3.39)(40,3.3689710610932475)(41,3.349016370890081)(42,3.329746566614753)(43,3.303392051244723)(44,3.2825956937799043)(45,3.264243917462273)(46,3.23858592263792)(47,3.220810942843185)(48,3.198019801980198)(49,3.1684886343918097)(50,3.138400860523485)(51,3.104286509556504)(52,3.0782775119617223)(53,3.043245387819877)(54,3.007591300779647)(55,2.9668014471164077)(56,2.9328785211267605)(57,2.892230576441103)(58,2.847975372957613)(59,2.7945745992601725)(60,2.7368150244404426)(61,2.68163484807744)(62,2.615536166619758)(63,2.5632855890090616)(64,2.511070670306339)(65,2.453683212140373)(66,2.389588859416446)(67,2.3343706777316737)(68,2.2883369330453562)(69,2.2494398805078415)(70,2.1991443018280825)(71,2.171887550200803)(72,2.1423251965246175)(73,2.093113902122131)(74,2.038914027149321)(75,1.9835603569751057)(76,1.9301416707376648)(77,1.8801422041645506)(78,1.8076718167288226)(79,1.7502768549280177)(80,1.7015437392795882)(81,1.6488413547237077)(82,1.6040515653775322)(83,1.5706214689265536)(84,1.528497409326425)(85,1.4973297730307076)(86,1.4735054347826086)(87,1.4482997918112421)(88,1.4377199155524278)(89,1.428368794326241)(90,1.4183526011560694)(91,1.4085231447465099)(92,1.4)(93,1.396525679758308)(94,1.388593155893536)(95,1.3878648233486943)(96,1.3766536964980545)(97,1.36600790513834)(98,1.3618895116092875)(99,1.3590576766856215)(100,1.3553173948887056)(101,1.347789824854045)(102,1.3423499577345732)(103,1.3290488431876606)(104,1.3185763888888888)(105,1.3089788732394365)(106,1.3033807829181494)(107,1.2974683544303798)(108,1.2943882244710212)(109,1.2887850467289719)(110,1.285037878787879)(111,1.2758620689655173)(112,1.2674418604651163)(113,1.2625368731563422)(114,1.2619047619047619)(115,1.2573158425832491)(116,1.2533197139938712)(117,1.2481977342945416)(118,1.2403746097814776)(119,1.2333685322069694)(120,1.2296137339055795)(121,1.2239130434782608)(122,1.2228381374722839)(123,1.2168539325842698)(124,1.2100456621004567)(125,1.2074159907300115)(126,1.1971664698937425)(127,1.1949458483754514)(128,1.1851851851851851)(129,1.1836477987421383)(130,1.1814720812182742)(131,1.1776061776061777)(132,1.1771653543307086)(133,1.1704697986577182)(134,1.1680327868852458)(135,1.163888888888889)(136,1.15625)(137,1.1499272197962154)(138,1.1471025260029717)(139,1.1437216338880485)(140,1.1381987577639752)(141,1.1340694006309149)(142,1.126602564102564)(143,1.1223491027732464)(144,1.1166666666666667)(145,1.1177474402730376)(146,1.1157167530224525)(147,1.1124780316344465)(148,1.1057347670250897)(149,1.1018181818181818)(150,1.0961182994454715)(151,1.0936329588014981)(152,1.09021113243762)(153,1.087890625)(154,1.0796812749003983)(155,1.0686868686868687)(156,1.060041407867495)(157,1.0529661016949152)(158,1.0475161987041037)(159,1.0397350993377483)(160,1.0335570469798658)(161,1.0296803652968036)(162,1.0255813953488373)(163,1.0238663484486874)(164,1.0216346153846154)(165,1.0148148148148148)(166,1.0100250626566416)(167,1.0076335877862594)(168,1.005128205128205)(169,1)(170,1)(171,1)(172,1)(173,1)(174,1)(175,1)(176,1)(177,1)(178,1)(179,1)(180,1)(181,1)(182,1)(183,1)(184,1)(185,1)(186,1)(187,1)(188,1)(189,1)(190,1)(191,1)(192,1)(193,1)(194,1)(195,1)(196,1)(197,1)(198,1)(199,1)(200,1)(201,1)(202,1)(203,1)(204,1)(205,1)(206,1)(207,1)(208,1)(209,1)(210,1)(211,1)(212,1)(213,1)(214,1)(215,1)(216,1)(217,1)(218,1)(219,1)(220,1)(221,1)(222,1)(223,1)(224,1)(225,1)(226,1)(227,1)(228,1)(229,1)(230,1)(231,1)(232,1)(233,1)(234,1)(235,1)(236,1)(237,1)(238,1)(239,1)(240,1)(241,1)(242,1)(243,1)(244,1)(245,1)(246,1)(247,1)(248,1)(249,1)(250,1)(251,1)(252,1)(253,1)(254,1)(255,1)(256,1)(257,1)(258,1)(259,1)(260,1)(261,1)(262,1)(263,1)(264,1)(265,1)(266,1)(267,1)(268,1)(269,1)(270,1)(271,1)(272,1)(273,1)(274,1)(275,1)(276,1)(277,1)(278,1)(279,1)(280,1)(281,1)(282,1)(283,1)(284,1)(285,1)(286,1)(287,1)(288,1)(289,1)(290,1)(291,1)(292,1)(293,1)(294,1)(295,1)(296,1)(297,1)(298,1)(299,1)(300,1)(301,1)(302,1)(303,1)(304,1)(305,1)(306,1)(307,1)(308,1)(309,1)(310,1)(311,1)(312,1)(313,1)(314,1)(315,1)(316,1)(317,1)(318,1)(319,1)(320,1)(321,1)(322,1)(323,1)(324,1)(325,1)(326,1)(327,1)(328,1)(329,1)(330,1)(331,1)(332,1)(333,1)(334,1)(335,1)(336,1)(337,1)(338,1)(339,1)(340,1)(341,1)(342,1)(343,1)(344,1)(345,1)(346,1)
}; % plot of the fourth histogram
\addplot[color=LimeGreen] coordinates{(0,4.019325368938862)(1,4.016863576693932)(2,4.013826758845059)(3,4.00934003900236)(4,4.003841561542854)(5,3.997156996946404)(6,3.9894629867351306)(7,3.981352235550709)(8,3.973234118169702)(9,3.9646127755794236)(10,3.9554558011049723)(11,3.9430137309210913)(12,3.930722891566265)(13,3.92052040942163)(14,3.9067426400759735)(15,3.8924808117601146)(16,3.8788365095285857)(17,3.864309210526316)(18,3.8491710758377424)(19,3.8359584815271828)(20,3.8203847300924547)(21,3.804297774366846)(22,3.786211258697027)(23,3.7646914589293012)(24,3.7414274828549656)(25,3.71802300061452)(26,3.6924895435533736)(27,3.664998111069135)(28,3.637183761521867)(29,3.6094372382800533)(30,3.576097404677988)(31,3.550721420643729)(32,3.5256958078300036)(33,3.5013760918990067)(34,3.479404466501241)(35,3.457818930041152)(36,3.437933796049119)(37,3.4202216066481994)(38,3.401090230956821)(39,3.379774948178857)(40,3.359413202933985)(41,3.3409162221524946)(42,3.316871761658031)(43,3.297791064607208)(44,3.2748208802456498)(45,3.2552893862563383)(46,3.2367193108399137)(47,3.2128167462357693)(48,3.1931281857655276)(49,3.1626365054602186)(50,3.13028594442207)(51,3.103548454036107)(52,3.0725477570294055)(53,3.0403190075321223)(54,3.000229726625316)(55,2.9648706384998813)(56,2.9222030130896517)(57,2.882142857142857)(58,2.8386840010612895)(59,2.801806734191076)(60,2.749218971882988)(61,2.695060632948832)(62,2.6365868631062)(63,2.5684887459807073)(64,2.516075016744809)(65,2.4623168178646195)(66,2.4101633393829403)(67,2.3547169811320754)(68,2.304262807978099)(69,2.271919191919192)(70,2.2363028021748224)(71,2.1986062717770034)(72,2.164934478084049)(73,2.1266355140186914)(74,2.0787937743190663)(75,2.0209077001529834)(76,1.9745762711864407)(77,1.9290038525041278)(78,1.8676975945017182)(79,1.799640071985603)(80,1.7261306532663316)(81,1.680921052631579)(82,1.6268758526603002)(83,1.5946890286512927)(84,1.5551575931232091)(85,1.5324107793153678)(86,1.5160089352196575)(87,1.5026375282592312)(88,1.4883720930232558)(89,1.483568075117371)(90,1.4758128469468677)(91,1.469224620303757)(92,1.463038180341186)(93,1.4586404586404587)(94,1.454771784232365)(95,1.449116904962153)(96,1.444160272804774)(97,1.4362532523850824)(98,1.4210992907801419)(99,1.4066726780883678)(100,1.3958143767060964)(101,1.3885291396854764)(102,1.383536014967259)(103,1.374762808349146)(104,1.370334928229665)(105,1.3570731707317074)(106,1.3476523476523476)(107,1.3444444444444446)(108,1.3357438016528926)(109,1.3256302521008403)(110,1.32237539766702)(111,1.3116178067318132)(112,1.3025302530253025)(113,1.294314381270903)(114,1.287015945330296)(115,1.2767441860465116)(116,1.2699530516431925)(117,1.2660332541567696)(118,1.259927797833935)(119,1.256723716381418)(120,1.2533992583436342)(121,1.2474874371859297)(122,1.2439949431099873)(123,1.2388250319284801)(124,1.2311688311688311)(125,1.2272126816380449)(126,1.221024258760108)(127,1.215846994535519)(128,1.209141274238227)(129,1.2)(130,1.1968616262482168)(131,1.1924746743849493)(132,1.1903367496339678)(133,1.1837037037037037)(134,1.1849624060150377)(135,1.180428134556575)(136,1.1738461538461538)(137,1.1679626749611198)(138,1.1629746835443038)(139,1.1594202898550725)(140,1.160392798690671)(141,1.154103852596315)(142,1.1457975986277873)(143,1.1398601398601398)(144,1.1352313167259787)(145,1.1283905967450272)(146,1.1291512915129152)(147,1.1245283018867924)(148,1.117533718689788)(149,1.1041257367387034)(150,1.1)(151,1.098360655737705)(152,1.0960334029227556)(153,1.0895522388059702)(154,1.084051724137931)(155,1.0798226164079823)(156,1.0741573033707865)(157,1.064516129032258)(158,1.0660377358490567)(159,1.0578313253012048)(160,1.0515970515970516)(161,1.045)(162,1.0378787878787878)(163,1.0332480818414322)(164,1.0283505154639174)(165,1.0238095238095237)(166,1.018918918918919)(167,1.016304347826087)(168,1.0110803324099722)(169,1.005586592178771)(170,1.0028169014084507)(171,1.0028409090909092)(172,1)(173,1)(174,1)(175,1)(176,1)(177,1)(178,1)(179,1)(180,1)(181,1)(182,1)(183,1)(184,1)(185,1)(186,1)(187,1)(188,1)(189,1)(190,1)(191,1)(192,1)(193,1)(194,1)(195,1)(196,1)(197,1)(198,1)(199,1)(200,1)(201,1)(202,1)(203,1)(204,1)(205,1)(206,1)(207,1)(208,1)(209,1)(210,1)(211,1)(212,1)(213,1)(214,1)(215,1)(216,1)(217,1)(218,1)(219,1)(220,1)(221,1)(222,1)(223,1)(224,1)(225,1)(226,1)(227,1)(228,1)(229,1)(230,1)(231,1)(232,1)(233,1)(234,1)(235,1)(236,1)(237,1)(238,1)(239,1)(240,1)(241,1)(242,1)(243,1)(244,1)(245,1)(246,1)(247,1)(248,1)(249,1)(250,1)(251,1)(252,1)(253,1)(254,1)(255,1)(256,1)(257,1)(258,1)(259,1)(260,1)(261,1)(262,1)(263,1)(264,1)(265,1)(266,1)(267,1)(268,1)(269,1)(270,1)(271,1)(272,1)(273,1)(274,1)(275,1)(276,1)(277,1)(278,1)(279,1)(280,1)(281,1)(282,1)(283,1)(284,1)(285,1)(286,1)(287,1)(288,1)(289,1)(290,1)(291,1)(292,1)(293,1)(294,1)(295,1)(296,1)(297,1)(298,1)(299,1)(300,1)(301,1)(302,1)(303,1)(304,1)(305,1)(306,1)(307,1)(308,1)(309,1)(310,1)(311,1)(312,1)(313,1)(314,1)(315,1)(316,1)(317,1)(318,1)(319,1)(320,1)(321,1)(322,1)(323,1)(324,1)(325,1)(326,1)(327,1)(328,1)(329,1)(330,1)(331,1)(332,1)(333,1)(334,1)(335,1)(336,1)(337,1)(338,1)(339,1)(340,1)(341,1)(342,1)(343,1)(344,1)
}; % plot of the fifth histogram
\legend{30000 ppl,  27500 ppl, 25000 ppl, 22500 ppl, 20000 ppl} % legend of the plot

\end{axis}
\end{tikzpicture}

%% file: plots/aware_fraction/avg_speed.tex
\begin{tikzpicture}[scale=0.75]

% Set the axis labels and title
\begin{axis}[    
xlabel={Time [s]},    ylabel={Speed [m/s]},    title={Average Speed - Mobile application},   legend style={at={(1,1)}, anchor=north east},]

\addplot[color=blue] coordinates{(0,4.018260258854219)(1,4.01683637463454)(2,4.0137139374238116)(3,4.0090585902782525)(4,4.003251470079557)(5,3.995790950543669)(6,3.9874532168953842)(7,3.9792316056542227)(8,3.970156971375808)(9,3.9617774430427413)(10,3.952640885676943)(11,3.941687637621894)(12,3.9304683727667795)(13,3.9185783122606153)(14,3.9066509115462527)(15,3.8923243712055506)(16,3.8787919224268306)(17,3.8653819809123706)(18,3.8502560729220505)(19,3.8352451478631235)(20,3.820024906600249)(21,3.803303919556741)(22,3.7838252406643393)(23,3.7623183655632033)(24,3.7391844572461315)(25,3.715972871842844)(26,3.687568223165555)(27,3.6597445094707695)(28,3.631654676258993)(29,3.5997956403269753)(30,3.5685067232837935)(31,3.5399323429916163)(32,3.5114294422432186)(33,3.48869277440706)(34,3.4682306940371457)(35,3.4459915611814345)(36,3.4270360013981125)(37,3.4054811866859622)(38,3.381824981301421)(39,3.3585853705672046)(40,3.335791886773647)(41,3.3162051282051284)(42,3.290909090909091)(43,3.2722330941509026)(44,3.2500556916908)(45,3.226704871060172)(46,3.2055294117647057)(47,3.1816420440784694)(48,3.156304401890077)(49,3.1252408477842004)(50,3.097454256165473)(51,3.0704032809295967)(52,3.0381690140845072)(53,3.004499274310595)(54,2.9660558726344246)(55,2.9292145296491774)(56,2.881555591415201)(57,2.840227767543125)(58,2.796880415944541)(59,2.755391804457225)(60,2.6957091999250515)(61,2.6416309012875536)(62,2.5701970443349755)(63,2.510497000856898)(64,2.445591204846309)(65,2.388862891904539)(66,2.332846359873387)(67,2.279028585884139)(68,2.2358934169278997)(69,2.1836900568951503)(70,2.1424148606811144)(71,2.0976608187134502)(72,2.0601571946795647)(73,2.0168961201501876)(74,1.9803479381443299)(75,1.9298657718120806)(76,1.884481558803062)(77,1.836231884057971)(78,1.7633472169632716)(79,1.7005136309758988)(80,1.6478988168094655)(81,1.6041315345699831)(82,1.5497588776852258)(83,1.509009009009009)(84,1.4680259499536608)(85,1.4485849056603775)(86,1.4184808901790034)(87,1.4089787863838183)(88,1.3973908680381335)(89,1.3944162436548224)(90,1.3858024691358024)(91,1.3733681462140992)(92,1.3678590496529632)(93,1.3580786026200873)(94,1.3541781959048147)(95,1.3408197641774284)(96,1.3335230506545248)(97,1.3254608294930876)(98,1.3230140186915889)(99,1.314540059347181)(100,1.3100961538461537)(101,1.3007290400972054)(102,1.2912801484230056)(103,1.2858040201005025)(104,1.2786989795918366)(105,1.2721970187945562)(106,1.2636783124588002)(107,1.257313829787234)(108,1.2510148849797023)(109,1.2472602739726026)(110,1.2427385892116183)(111,1.2352112676056337)(112,1.2287769784172662)(113,1.2219796215429404)(114,1.2179675994108983)(115,1.2137518684603887)(116,1.2099697885196374)(117,1.2068702290076336)(118,1.2009273570324575)(119,1.1954828660436136)(120,1.190779014308426)(121,1.1835084882780922)(122,1.1783960720130933)(123,1.1731890091590342)(124,1.1632825719120135)(125,1.1547008547008546)(126,1.1484375)(127,1.148375768217735)(128,1.1411974977658623)(129,1.1397459165154264)(130,1.1374885426214483)(131,1.1347583643122676)(132,1.130885122410546)(133,1.1296472831267874)(134,1.1256038647342994)(135,1.1218074656188606)(136,1.1157684630738522)(137,1.1124620060790273)(138,1.1094069529652353)(139,1.1042752867570387)(140,1.0976645435244161)(141,1.0958019375672767)(142,1.092896174863388)(143,1.0880713489409142)(144,1.0843644544431945)(145,1.082004555808656)(146,1.0776361529548089)(147,1.0725326991676576)(148,1.065217391304348)(149,1.0605686032138442)(150,1.0554156171284634)(151,1.0523627075351214)(152,1.0507152145643692)(153,1.0498687664041995)(154,1.0493333333333332)(155,1.0407055630936228)(156,1.0397805212620028)(157,1.0363128491620113)(158,1.0256045519203414)(159,1.0244604316546762)(160,1.0204678362573099)(161,1.017910447761194)(162,1.0165912518853695)(163,1.0137614678899083)(164,1.0138888888888888)(165,1.0078369905956113)(166,1.0047923322683705)(167,1.0032310177705976)(168,1)(169,1)(170,0.9983388704318937)(171,0.9983136593591906)(172,0.9982935153583617)(173,0.9982847341337907)(174,0.998272884283247)(175,0.9982456140350877)(176,0.9964476021314387)(177,0.996415770609319)(178,0.9963570127504554)(179,0.9963235294117647)(180,0.9962825278810409)(181,0.9962546816479401)(182,0.9962121212121212)(183,0.9961759082217974)(184,0.996116504854369)(185,0.996078431372549)(186,0.9960159362549801)(187,0.9959595959595959)(188,0.9959266802443992)(189,0.9958847736625515)(190,0.9958419958419958)(191,0.9957983193277311)(192,0.9957716701902748)(193,0.9957537154989384)(194,0.9957264957264957)(195,0.9956709956709957)(196,0.9956331877729258)(197,0.9956140350877193)(198,0.9955654101995566)(199,0.9955156950672646)(200,0.9954648526077098)(201,0.9954128440366973)(202,0.9953810623556582)(203,0.9953051643192489)(204,0.9952038369304557)(205,0.9951690821256038)(206,0.9951456310679612)(207,0.9951100244498777)(208,0.995049504950495)(209,0.9949748743718593)(210,0.9948849104859335)(211,0.9947916666666666)(212,0.9947643979057592)(213,0.9946808510638298)(214,0.9946091644204852)(215,0.9945945945945946)(216,0.9944598337950139)(217,0.994413407821229)(218,0.9943502824858758)(219,0.9943019943019943)(220,0.9942196531791907)(221,0.9941690962099126)(222,0.9940476190476191)(223,0.9939024390243902)(224,0.9938650306748467)(225,0.9938080495356038)(226,0.9937106918238994)(227,0.9935897435897436)(228,0.9934426229508196)(229,0.9933554817275747)(230,0.9931740614334471)(231,0.993103448275862)(232,0.993006993006993)(233,0.9929328621908127)(234,0.9928571428571429)(235,0.9927272727272727)(236,0.9926470588235294)(237,0.9925373134328358)(238,0.9923664122137404)(239,0.9922480620155039)(240,0.9921875)(241,0.9920318725099602)(242,0.991869918699187)(243,0.9918367346938776)(244,0.9917355371900827)(245,0.9916317991631799)(246,0.9914893617021276)(247,0.9914529914529915)(248,0.9912280701754386)(249,0.9912280701754386)(250,0.9911504424778761)(251,0.990990990990991)(252,0.9907834101382489)(253,0.9906542056074766)(254,0.9904761904761905)(255,0.9903381642512077)(256,0.9902439024390244)(257,0.9899497487437185)(258,0.9949238578680203)(259,0.9948453608247423)(260,0.9947916666666666)(261,0.9947368421052631)(262,0.9945945945945946)(263,0.9944444444444445)(264,0.9943502824858758)(265,0.9942857142857143)(266,0.9942196531791907)(267,0.9940828402366864)(268,0.9940119760479041)(269,0.9938271604938271)(270,0.9936305732484076)(271,0.9934640522875817)(272,0.9933774834437086)(273,0.9931506849315068)(274,0.9930555555555556)(275,0.993006993006993)(276,0.9928057553956835)(277,0.9927007299270073)(278,0.9925925925925926)(279,0.9922480620155039)(280,0.9921875)(281,0.992)(282,0.991869918699187)(283,0.9916666666666667)(284,0.9914529914529915)(285,0.9910714285714286)(286,0.9819819819819819)(287,0.9908256880733946)(288,0.9907407407407407)(289,0.9904761904761905)(290,0.9903846153846154)(291,0.9900990099009901)(292,0.98989898989899)(293,0.9895833333333334)(294,0.989010989010989)(295,1)(296,1)(297,1)(298,1)(299,1)(300,1)(301,1)(302,1)(303,1)(304,1)(305,1)(306,1)(307,1)(308,1)(309,1)(310,1)(311,1)(312,1)(313,1)(314,1)(315,1)(316,1)(317,1)(318,1)(319,1)(320,1)(321,1)(322,1)(323,1)(324,1)(325,1)(326,1)(327,1)(328,1)(329,1)(330,1)(331,1)(332,1)(333,1)(334,1)(335,1)(336,1)(337,1)(338,1)
}; % plot of the first histogram
\addplot[color=teal] coordinates{(0,4.017729903321848)(1,4.014674205708132)(2,4.011521908570267)(3,4.006234584817758)(4,4.0002431062026815)(5,3.9931646818406032)(6,3.9855540022224614)(7,3.9772228721848495)(8,3.967070797784763)(9,3.9559943366624575)(10,3.9450299729655605)(11,3.932961342378869)(12,3.9201037293158802)(13,3.9073448712536933)(14,3.8939663785239564)(15,3.8770907134209227)(16,3.8600074094655925)(17,3.845639507116248)(18,3.8292202354563813)(19,3.8102684256099426)(20,3.788863109048724)(21,3.765793780687398)(22,3.742210431248589)(23,3.716538664477927)(24,3.688913268726763)(25,3.6565817072395603)(26,3.627684175751569)(27,3.591556108347153)(28,3.549207038887682)(29,3.509792014574161)(30,3.4723546234509057)(31,3.434421463576709)(32,3.406182540367844)(33,3.3755501661726397)(34,3.35020979020979)(35,3.3223652375883095)(36,3.295452253705758)(37,3.2676100628930818)(38,3.2371403661726244)(39,3.2121791985592076)(40,3.1861411463783282)(41,3.1635099496523615)(42,3.1389265885256017)(43,3.1213272543059776)(44,3.100312256049961)(45,3.0717248762210625)(46,3.0529209621993125)(47,3.03125)(48,3.0073656845753898)(49,2.979476502082094)(50,2.945530081551008)(51,2.9091343093570976)(52,2.8762295081967215)(53,2.844380403458213)(54,2.8039593552908197)(55,2.7623456790123457)(56,2.7229932483120782)(57,2.6874878758486904)(58,2.6413765345139866)(59,2.606237006237006)(60,2.5660499139414803)(61,2.5092010771992816)(62,2.4519140989729227)(63,2.387740209194843)(64,2.340634441087613)(65,2.2712177121771218)(66,2.226063100137174)(67,2.1831345826235093)(68,2.1352785145888595)(69,2.094698130554704)(70,2.0596026490066226)(71,2.0273259596616784)(72,1.9952861952861953)(73,1.947829131652661)(74,1.9122870605291773)(75,1.870712401055409)(76,1.817149569303054)(77,1.7493837304847988)(78,1.7029324266893329)(79,1.638852097130243)(80,1.5705039297272307)(81,1.5398187887458274)(82,1.5090998524348254)(83,1.471955533097524)(84,1.4403956272774596)(85,1.408728046833422)(86,1.3929539295392954)(87,1.3847850055126791)(88,1.374015748031496)(89,1.3642611683848798)(90,1.3572669368847712)(91,1.3513831665685698)(92,1.3480597014925373)(93,1.3385731559854896)(94,1.3259668508287292)(95,1.314410480349345)(96,1.303684879288437)(97,1.2961290322580645)(98,1.2907662082514735)(99,1.2855245683930943)(100,1.28)(101,1.2778541953232463)(102,1.272156315422191)(103,1.2623762376237624)(104,1.2573687994248741)(105,1.2532751091703056)(106,1.2459618208516887)(107,1.2391629297458895)(108,1.2303625377643506)(109,1.2237547892720306)(110,1.220576773187841)(111,1.2167721518987342)(112,1.2115384615384615)(113,1.2078239608801955)(114,1.2034739454094292)(115,1.200503355704698)(116,1.19693094629156)(117,1.1934084995663488)(118,1.1894273127753303)(119,1.1829596412556054)(120,1.1790909090909092)(121,1.1735537190082646)(122,1.1721234798877456)(123,1.1723809523809523)(124,1.1695736434108528)(125,1.1641938674579624)(126,1.1566265060240963)(127,1.1557377049180328)(128,1.1522419186652764)(129,1.153028692879915)(130,1.1503267973856208)(131,1.143646408839779)(132,1.1428571428571428)(133,1.1370328425821064)(134,1.131791907514451)(135,1.125)(136,1.1174377224199288)(137,1.1123188405797102)(138,1.11015911872705)(139,1.1032338308457712)(140,1.0998719590268886)(141,1.0984455958549222)(142,1.0981432360742707)(143,1.0956873315363882)(144,1.093150684931507)(145,1.0846047156726768)(146,1.0829817158931083)(147,1.0786838340486409)(148,1.0725689404934688)(149,1.063893016344725)(150,1.0526315789473684)(151,1.0507936507936508)(152,1.0468497576736673)(153,1.0412541254125414)(154,1.04)(155,1.0320945945945945)(156,1.030612244897959)(157,1.027681660899654)(158,1.019538188277087)(159,1.0179856115107915)(160,1.0163339382940109)(161,1.014732965009208)(162,1.011214953271028)(163,1.0076045627376427)(164,1.001930501930502)(165,1.0019455252918288)(166,1.0019723865877712)(167,1.0019880715705765)(168,1.001996007984032)(169,1.0020120724346075)(170,1.0020325203252032)(171,1)(172,0.9979123173277662)(173,0.9978947368421053)(174,0.997872340425532)(175,0.9978632478632479)(176,0.9978401727861771)(177,0.9978260869565218)(178,0.9978021978021978)(179,0.9977678571428571)(180,0.9977272727272727)(181,0.9977064220183486)(182,0.9976905311778291)(183,0.9976635514018691)(184,0.9976415094339622)(185,0.9976019184652278)(186,0.9975845410628019)(187,0.9975429975429976)(188,0.9975247524752475)(189,0.9974489795918368)(190,0.9974160206718347)(191,0.9974093264248705)(192,0.9973890339425587)(193,0.9973753280839895)(194,0.9973544973544973)(195,0.9973045822102425)(196,0.9972602739726028)(197,0.9972222222222222)(198,0.9971910112359551)(199,0.9971751412429378)(200,0.997134670487106)(201,0.9971264367816092)(202,0.9971181556195965)(203,0.9971098265895953)(204,0.9970588235294118)(205,0.9970149253731343)(206,0.9969788519637462)(207,0.996969696969697)(208,0.9969512195121951)(209,0.9969325153374233)(210,0.9969230769230769)(211,0.9969040247678018)(212,0.9968944099378882)(213,0.9968553459119497)(214,0.9968454258675079)(215,0.9968354430379747)(216,0.9968253968253968)(217,0.9968253968253968)(218,0.9968152866242038)(219,0.9968051118210862)(220,0.9967845659163987)(221,0.9967637540453075)(222,0.9966996699669967)(223,0.9966442953020134)(224,1)(225,1)(226,1)(227,1)(228,1)(229,1)(230,1)(231,1)(232,1)(233,1)(234,1)(235,1)(236,1)(237,1)(238,1)(239,1)(240,1)(241,1)(242,1)(243,1)(244,1)(245,1)(246,1)(247,1)(248,1)(249,1)(250,1)(251,1)(252,1)(253,1)(254,1)(255,1)(256,1)(257,1)(258,1)(259,1)(260,1)(261,1)(262,1)(263,1)(264,1)(265,1)(266,1)(267,1)(268,1)(269,1)(270,1)(271,1)(272,1)(273,1)(274,1)(275,1)(276,1)(277,1)(278,1)(279,1)(280,1)(281,1)(282,1)(283,1)(284,1)(285,1)(286,1)(287,1)(288,1)(289,1)(290,1)(291,1)(292,1)(293,1)(294,1)(295,1)(296,1)(297,1)(298,1)(299,1)(300,1)(301,1)(302,1)(303,1)(304,1)(305,1)(306,1)(307,1)(308,1)(309,1)(310,1)(311,1)(312,1)(313,1)(314,1)(315,1)(316,1)(317,1)(318,1)(319,1)(320,1)(321,1)(322,1.0416666666666667)(323,1.0434782608695652)(324,1.0434782608695652)(325,1.05)(326,1.05)(327,1.0526315789473684)(328,1.0588235294117647)(329,1.0625)(330,1.0625)(331,1.0625)(332,1.0714285714285714)(333,1.0833333333333333)(334,1.0833333333333333)(335,1.0909090909090908)(336,1.1111111111111112)(337,1.1111111111111112)(338,1.1428571428571428)(339,1.1428571428571428)(340,1.2)(341,1.2)(342,1.2)(343,1.2)(344,1.2)(345,1.2)(346,1.25)(347,1.25)(348,1.25)(349,1.25)(350,1.3333333333333333)
}; % plot of the second histogram
\addplot[color=cyan] coordinates{(0,4.018613370827893)(1,4.015305778437058)(2,4.0113165228029635)(3,4.006529209621993)(4,3.9994074590449635)(5,3.9892770913222555)(6,3.9794105919454816)(7,3.96845694799659)(8,3.956911278994723)(9,3.9453140171670484)(10,3.9334024979051114)(11,3.920080453164765)(12,3.90635423263669)(13,3.890735102218356)(14,3.872369131138694)(15,3.8526443090376294)(16,3.833614335950672)(17,3.811919272654611)(18,3.7929055276901598)(19,3.7692679780715896)(20,3.74646397148903)(21,3.717325756697205)(22,3.691677706852945)(23,3.6606380306455666)(24,3.6287268199986893)(25,3.593700894992143)(26,3.5502984108722226)(27,3.5108580907857037)(28,3.459683196688689)(29,3.411468390563345)(30,3.353657452670874)(31,3.3098145213906234)(32,3.2687762476804374)(33,3.2307927450580802)(34,3.191539365452409)(35,3.161909009812667)(36,3.123874137326003)(37,3.0897560975609757)(38,3.05359643539147)(39,3.0219547678878453)(40,2.9974043715846994)(41,2.976425748164879)(42,2.9542607428987617)(43,2.9324040767386093)(44,2.911751116587094)(45,2.8865733375755647)(46,2.8620294599018004)(47,2.8389816219861745)(48,2.8093080006972286)(49,2.7837934743635713)(50,2.764781966001478)(51,2.7375047691720718)(52,2.7048536058164667)(53,2.6711273317112734)(54,2.6421404682274248)(55,2.6150194216659473)(56,2.574827662886369)(57,2.524924646417807)(58,2.4783965434469515)(59,2.442622950819672)(60,2.3931535269709543)(61,2.3509809191077666)(62,2.3081411503195333)(63,2.254408788667245)(64,2.19880059970015)(65,2.1501090003114296)(66,2.1006146878033)(67,2.048969942586964)(68,2.0101754385964914)(69,1.9737704918032788)(70,1.939268200678989)(71,1.9111369809856422)(72,1.8785800726099233)(73,1.8448060075093868)(74,1.7953063885267275)(75,1.7573033707865169)(76,1.7116279069767442)(77,1.669893514036786)(78,1.6201005025125628)(79,1.567539267015707)(80,1.5244831338411318)(81,1.4769144144144144)(82,1.4392794886693783)(83,1.409934171154997)(84,1.3854679802955665)(85,1.359771573604061)(86,1.3483365949119375)(87,1.3439363817097416)(88,1.3353658536585367)(89,1.326896551724138)(90,1.3247025892232331)(91,1.3222143364088006)(92,1.3152566883586407)(93,1.3105224429727742)(94,1.307576894223556)(95,1.3061068702290077)(96,1.2989930286599536)(97,1.2921259842519686)(98,1.2864038616251006)(99,1.283739837398374)(100,1.2716666666666667)(101,1.2619863013698631)(102,1.2602262837249782)(103,1.2513274336283187)(104,1.2506762849413886)(105,1.239889705882353)(106,1.2352389878163075)(107,1.2298195631528965)(108,1.227229146692234)(109,1.2253658536585366)(110,1.218344965104686)(111,1.2110091743119267)(112,1.2094455852156059)(113,1.207075962539022)(114,1.2027455121436115)(115,1.1997840172786176)(116,1.1978142076502731)(117,1.1955801104972377)(118,1.1845637583892616)(119,1.1829545454545454)(120,1.1802525832376578)(121,1.1777777777777778)(122,1.1744324970131421)(123,1.1743341404358354)(124,1.1695331695331694)(125,1.16625)(126,1.160668380462725)(127,1.154557463672391)(128,1.1524064171122994)(129,1.1452282157676348)(130,1.1404494382022472)(131,1.138373751783167)(132,1.1364296081277214)(133,1.1314623338257017)(134,1.1300448430493273)(135,1.1271056661562022)(136,1.1251956181533647)(137,1.1201923076923077)(138,1.1155115511551155)(139,1.116554054054054)(140,1.1143344709897611)(141,1.1120689655172413)(142,1.1081560283687943)(143,1.1023765996343693)(144,1.1003787878787878)(145,1.0891472868217054)(146,1.0836653386454183)(147,1.0843621399176955)(148,1.079002079002079)(149,1.0743099787685775)(150,1.0695652173913044)(151,1.0662251655629138)(152,1.054421768707483)(153,1.0468384074941453)(154,1.0404761904761906)(155,1.038647342995169)(156,1.0270935960591132)(157,1.0225)(158,1.0204081632653061)(159,1.0130548302872062)(160,1.0026737967914439)(161,1)(162,0.9945054945054945)(163,0.9944903581267218)(164,0.9943661971830986)(165,0.9914040114613181)(166,0.9853801169590644)(167,0.9852507374631269)(168,0.9851190476190477)(169,0.9819277108433735)(170,0.9787878787878788)(171,0.9785276073619632)(172,0.9720496894409938)(173,0.9716981132075472)(174,0.9716981132075472)(175,0.9716981132075472)(176,0.9714285714285714)(177,0.9711538461538461)(178,0.9707792207792207)(179,0.9704918032786886)(180,0.9700996677740864)(181,0.97)(182,0.9697986577181208)(183,0.9694915254237289)(184,0.9690721649484536)(185,0.9686411149825784)(186,0.968421052631579)(187,0.9679715302491103)(188,0.967391304347826)(189,0.9672727272727273)(190,0.9669117647058824)(191,0.9669117647058824)(192,0.9666666666666667)(193,0.9662921348314607)(194,0.9660377358490566)(195,0.9653846153846154)(196,0.9653846153846154)(197,0.96484375)(198,0.9647058823529412)(199,0.9644268774703557)(200,0.9637096774193549)(201,0.9635627530364372)(202,0.963265306122449)(203,0.963265306122449)(204,0.9628099173553719)(205,0.9626556016597511)(206,0.9625)(207,0.9621848739495799)(208,0.9615384615384616)(209,0.9613733905579399)(210,0.960352422907489)(211,0.9598214285714286)(212,0.9594594594594594)(213,0.9638009049773756)(214,0.963302752293578)(215,0.9629629629629629)(216,0.9629629629629629)(217,0.9622641509433962)(218,0.9620853080568721)(219,0.9611650485436893)(220,0.9609756097560975)(221,0.9609756097560975)(222,0.9605911330049262)(223,0.9597989949748744)(224,0.9593908629441624)(225,0.958974358974359)(226,0.9585492227979274)(227,0.9576719576719577)(228,0.9565217391304348)(229,0.9555555555555556)(230,0.9545454545454546)(231,0.9540229885057471)(232,0.9526627218934911)(233,0.9520958083832335)(234,0.9515151515151515)(235,0.9506172839506173)(236,0.9506172839506173)(237,0.95)(238,0.948051948051948)(239,0.9477124183006536)(240,0.9470198675496688)(241,0.9466666666666667)(242,0.9455782312925171)(243,0.9440559440559441)(244,0.9440559440559441)(245,0.9428571428571428)(246,0.95)(247,0.948905109489051)(248,0.948905109489051)(249,0.9481481481481482)(250,0.9473684210526315)(251,0.946969696969697)(252,0.946969696969697)(253,0.9465648854961832)(254,0.952755905511811)(255,0.9523809523809523)(256,0.9512195121951219)(257,0.9504132231404959)(258,0.95)(259,0.9495798319327731)(260,0.9495798319327731)(261,0.9491525423728814)(262,0.9482758620689655)(263,0.9478260869565217)(264,0.9473684210526315)(265,0.9473684210526315)(266,0.9469026548672567)(267,0.9469026548672567)(268,0.9469026548672567)(269,0.9459459459459459)(270,0.9459459459459459)(271,0.9454545454545454)(272,0.944954128440367)(273,0.944954128440367)(274,0.9423076923076923)(275,0.941747572815534)(276,0.9405940594059405)(277,0.9393939393939394)(278,0.9393939393939394)(279,0.9375)(280,0.9361702127659575)(281,0.9340659340659341)(282,0.9325842696629213)(283,0.9318181818181818)(284,0.9318181818181818)(285,0.9302325581395349)(286,0.9294117647058824)(287,0.927710843373494)(288,0.926829268292683)(289,0.9259259259259259)(290,0.9230769230769231)(291,0.9210526315789473)(292,0.9333333333333333)(293,0.9315068493150684)(294,0.9305555555555556)(295,0.9305555555555556)(296,0.9305555555555556)(297,0.9295774647887324)(298,0.9285714285714286)(299,0.9285714285714286)(300,0.9285714285714286)(301,0.9253731343283582)(302,0.9230769230769231)(303,0.921875)(304,0.9180327868852459)(305,0.9090909090909091)(306,0.9090909090909091)(307,0.9090909090909091)(308,0.9038461538461539)(309,0.9)(310,0.9)(311,0.8936170212765957)(312,0.8913043478260869)(313,0.8837209302325582)(314,0.8780487804878049)(315,0.875)(316,0.8717948717948718)(317,0.8717948717948718)(318,0.8648648648648649)(319,0.8648648648648649)(320,0.8648648648648649)(321,0.8611111111111112)(322,0.8571428571428571)(323,0.8529411764705882)(324,0.84375)(325,0.84375)(326,0.84375)(327,0.8387096774193549)(328,0.8333333333333334)(329,0.8214285714285714)(330,0.8148148148148148)(331,0.8)(332,0.7916666666666666)(333,0.7619047619047619)(334,0.7619047619047619)(335,0.7368421052631579)(336,0.7368421052631579)(337,0.7368421052631579)(338,0.7368421052631579)(339,0.7368421052631579)(340,0.7368421052631579)(341,0.7058823529411765)(342,0.6875)(343,0.6875)(344,0.6666666666666666)(345,0.6428571428571429)(346,0.6153846153846154)(347,0.6153846153846154)(348,0.6153846153846154)(349,0.6153846153846154)(350,0.5833333333333334)
}; % plot of the third histogram
\addplot[magenta] coordinates{(0,4.021356363392917)(1,4.01823206281805)(2,4.013415506145936)(3,4.007791491415569)(4,3.999579934889908)(5,3.990681899321671)(6,3.979163625748066)(7,3.9683915561367455)(8,3.955587942011199)(9,3.941675880587585)(10,3.928167581858857)(11,3.914682206942814)(12,3.898810296344365)(13,3.8819245889921374)(14,3.8611098255195078)(15,3.8371376288163344)(16,3.8143607705779337)(17,3.7913396059676665)(18,3.770281211205268)(19,3.743930603278319)(20,3.7125441696113075)(21,3.679443794991694)(22,3.643860327277413)(23,3.6055300162249866)(24,3.564524130111261)(25,3.5199194270367054)(26,3.4744313725490197)(27,3.4151934623082054)(28,3.3504523682810006)(29,3.2790411218495357)(30,3.211433756805808)(31,3.156998403406067)(32,3.102022471910112)(33,3.053290795226279)(34,3.011531308121513)(35,2.9651708844247326)(36,2.9142936134918442)(37,2.8652451407020596)(38,2.8233323180018277)(39,2.783171004311033)(40,2.7488002647691543)(41,2.727272727272727)(42,2.704225352112676)(43,2.684658576344865)(44,2.664490861618799)(45,2.6414732399769805)(46,2.619640387275242)(47,2.5967413441955194)(48,2.5839766325891924)(49,2.559209112400602)(50,2.535975204781935)(51,2.512189564821144)(52,2.4850623382733477)(53,2.461090909090909)(54,2.429967426710098)(55,2.4078573274747996)(56,2.3696695095948828)(57,2.3386388583973656)(58,2.30788876276958)(59,2.270698766881973)(60,2.2237122828405975)(61,2.1766750948166877)(62,2.1217105263157894)(63,2.073278800272665)(64,2.025859015232023)(65,1.9897247706422019)(66,1.9376434583014537)(67,1.891773162939297)(68,1.8616977225672877)(69,1.826311263972485)(70,1.8139946855624447)(71,1.7776236708275543)(72,1.7417345471969334)(73,1.7013888888888888)(74,1.6668377823408624)(75,1.6329787234042554)(76,1.5956858407079646)(77,1.549425287356322)(78,1.501194743130227)(79,1.4669987546699876)(80,1.433828276307295)(81,1.3952095808383234)(82,1.3620689655172413)(83,1.3390191897654584)(84,1.316096139839767)(85,1.296711509715994)(86,1.2820121951219512)(87,1.2696804364770071)(88,1.259904912836767)(89,1.2522231204527081)(90,1.2497945768282663)(91,1.2458053691275168)(92,1.241880341880342)(93,1.2387152777777777)(94,1.2343059239610963)(95,1.2243243243243243)(96,1.222323879231473)(97,1.2161410018552876)(98,1.2136832239925024)(99,1.2086124401913876)(100,1.2029268292682926)(101,1.2013820335636722)(102,1.1959798994974875)(103,1.195097037793667)(104,1.191866527632951)(105,1.1841269841269841)(106,1.185145317545748)(107,1.1836065573770491)(108,1.1720670391061452)(109,1.1683731513083049)(110,1.1672433679354095)(111,1.1672555948174324)(112,1.164071856287425)(113,1.1610169491525424)(114,1.1599015990159902)(115,1.1568381430363865)(116,1.1554140127388535)(117,1.151948051948052)(118,1.1532364597093792)(119,1.1507402422611037)(120,1.1493150684931508)(121,1.1472222222222221)(122,1.1454802259887005)(123,1.1455331412103746)(124,1.1413843888070692)(125,1.1400894187779433)(126,1.135258358662614)(127,1.130503144654088)(128,1.1254019292604502)(129,1.1233552631578947)(130,1.1142857142857143)(131,1.1065292096219932)(132,1.1003521126760563)(133,1.0934065934065933)(134,1.0900562851782365)(135,1.091254752851711)(136,1.09021113243762)(137,1.0901960784313725)(138,1.0873015873015872)(139,1.0835030549898168)(140,1.0798319327731092)(141,1.0765864332603938)(142,1.072234762979684)(143,1.066820276497696)(144,1.0665083135391924)(145,1.0635696821515892)(146,1.0643564356435644)(147,1.0583756345177664)(148,1.052219321148825)(149,1.048780487804878)(150,1.0413223140495869)(151,1.0337078651685394)(152,1.0288184438040346)(153,1.0238095238095237)(154,1.0152439024390243)(155,1.0125786163522013)(156,1.012779552715655)(157,1)(158,1)(159,0.9965397923875432)(160,0.9893992932862191)(161,0.9783393501805054)(162,0.9664179104477612)(163,0.9543726235741445)(164,0.9498069498069498)(165,0.9404761904761905)(166,0.9338842975206612)(167,0.9324894514767933)(168,0.927038626609442)(169,0.925764192139738)(170,0.9241071428571429)(171,0.9285714285714286)(172,0.9279279279279279)(173,0.926605504587156)(174,0.926605504587156)(175,0.9255813953488372)(176,0.9248826291079812)(177,0.9245283018867925)(178,0.9245283018867925)(179,0.9245283018867925)(180,0.9238095238095239)(181,0.9230769230769231)(182,0.9211822660098522)(183,0.92)(184,0.9187817258883249)(185,0.9187817258883249)(186,0.9179487179487179)(187,0.9179487179487179)(188,0.9175257731958762)(189,0.9162303664921466)(190,0.9153439153439153)(191,0.9148936170212766)(192,0.9180327868852459)(193,0.9166666666666666)(194,0.9162011173184358)(195,0.9152542372881356)(196,0.9147727272727273)(197,0.9132947976878613)(198,0.9122807017543859)(199,0.9107142857142857)(200,0.9096385542168675)(201,0.9085365853658537)(202,0.9074074074074074)(203,0.906832298136646)(204,0.906832298136646)(205,0.906832298136646)(206,0.90625)(207,0.9044585987261147)(208,0.9032258064516129)(209,0.9032258064516129)(210,0.9025974025974026)(211,0.9025974025974026)(212,0.9019607843137255)(213,0.9019607843137255)(214,0.9019607843137255)(215,0.9013157894736842)(216,0.9013157894736842)(217,0.8993288590604027)(218,0.8986486486486487)(219,0.8979591836734694)(220,0.8979591836734694)(221,0.896551724137931)(222,0.8958333333333334)(223,0.8958333333333334)(224,0.8943661971830986)(225,0.8905109489051095)(226,0.8880597014925373)(227,0.8880597014925373)(228,0.8872180451127819)(229,0.8854961832061069)(230,0.8837209302325582)(231,0.890625)(232,0.890625)(233,0.890625)(234,0.890625)(235,0.8976377952755905)(236,0.904)(237,0.912)(238,0.912)(239,0.912)(240,0.9112903225806451)(241,0.9112903225806451)(242,0.9090909090909091)(243,0.9090909090909091)(244,0.907563025210084)(245,0.9067796610169492)(246,0.9043478260869565)(247,0.9043478260869565)(248,0.9122807017543859)(249,0.911504424778761)(250,0.9107142857142857)(251,0.9074074074074074)(252,0.9065420560747663)(253,0.9065420560747663)(254,0.9056603773584906)(255,0.9047619047619048)(256,0.9038461538461539)(257,0.9038461538461539)(258,0.9019607843137255)(259,0.9019607843137255)(260,0.9019607843137255)(261,0.9019607843137255)(262,0.900990099009901)(263,0.900990099009901)(264,0.898989898989899)(265,0.898989898989899)(266,0.8979591836734694)(267,0.8969072164948454)(268,0.8969072164948454)(269,0.8958333333333334)(270,0.8936170212765957)(271,0.8888888888888888)(272,0.8876404494382022)(273,0.8850574712643678)(274,0.8953488372093024)(275,0.8953488372093024)(276,0.9058823529411765)(277,0.9047619047619048)(278,0.9036144578313253)(279,0.9024390243902439)(280,0.9024390243902439)(281,0.9)(282,0.8961038961038961)(283,0.8961038961038961)(284,0.8947368421052632)(285,0.8918918918918919)(286,0.8888888888888888)(287,0.8873239436619719)(288,0.8873239436619719)(289,0.8857142857142857)(290,0.8823529411764706)(291,0.8805970149253731)(292,0.8805970149253731)(293,0.8805970149253731)(294,0.8787878787878788)(295,0.8769230769230769)(296,0.8769230769230769)(297,0.8709677419354839)(298,0.8688524590163934)(299,0.8666666666666667)(300,0.864406779661017)(301,0.864406779661017)(302,0.864406779661017)(303,0.8596491228070176)(304,0.8545454545454545)(305,0.8545454545454545)(306,0.8518518518518519)(307,0.8490566037735849)(308,0.8461538461538461)(309,0.8461538461538461)(310,0.8461538461538461)(311,0.8461538461538461)(312,0.8431372549019608)(313,0.8297872340425532)(314,0.8222222222222222)(315,0.8222222222222222)(316,0.8181818181818182)(317,0.813953488372093)(318,0.8095238095238095)(319,0.8095238095238095)(320,0.8095238095238095)(321,0.8048780487804879)(322,0.8048780487804879)(323,0.8)(324,0.8)(325,0.8)(326,0.7948717948717948)(327,0.7714285714285715)(328,0.7714285714285715)(329,0.7575757575757576)(330,0.7575757575757576)(331,0.75)(332,0.7419354838709677)(333,0.7741935483870968)(334,0.7666666666666667)(335,0.7586206896551724)(336,0.7407407407407407)(337,0.7407407407407407)(338,0.7407407407407407)(339,0.7407407407407407)(340,0.7307692307692307)(341,0.7307692307692307)(342,0.7307692307692307)(343,0.72)(344,0.7083333333333334)(345,0.7083333333333334)(346,0.6956521739130435)
}; % plot of the fourth histogram
\addplot[color=red] coordinates{(0,4.019215961969803)(1,4.016321575504148)(2,4.011530720158292)(3,4.005190311418685)(4,3.997006515231555)(5,3.986225499010969)(6,3.9744119165253298)(7,3.9627164788625877)(8,3.947914230019493)(9,3.9327663347978556)(10,3.9188616535105942)(11,3.901035623737699)(12,3.882106576367849)(13,3.8635231152532987)(14,3.8364773820981712)(15,3.814926120711245)(16,3.787900989062745)(17,3.7616218867513216)(18,3.733192557927177)(19,3.694097138735273)(20,3.6621604627766597)(21,3.622941059734074)(22,3.583339132855453)(23,3.528814307794215)(24,3.480233009708738)(25,3.419847959015039)(26,3.3660714285714284)(27,3.28810861423221)(28,3.2133652981720107)(29,3.1142888094464305)(30,3.0302009524915787)(31,2.9344016024036055)(32,2.858288770053476)(33,2.7848911651728554)(34,2.722849826781142)(35,2.651293588301462)(36,2.5847644958340417)(37,2.52081090778615)(38,2.4498388014413046)(39,2.39042320683489)(40,2.336728330908335)(41,2.300709830070983)(42,2.2737169517884914)(43,2.249081726354454)(44,2.2328316086547506)(45,2.2190062711046794)(46,2.2069051167411824)(47,2.1843047034764824)(48,2.1599788806758182)(49,2.141573645521372)(50,2.1135531135531136)(51,2.0970562518216265)(52,2.077824519230769)(53,2.0550231839258113)(54,2.029765674477517)(55,1.9977146588312111)(56,1.968655207280081)(57,1.9618717504332757)(58,1.938059434300036)(59,1.9003703703703703)(60,1.864575363427697)(61,1.8489378442171518)(62,1.8234577922077921)(63,1.785503581963759)(64,1.7501096010521702)(65,1.7264840182648402)(66,1.6950757575757576)(67,1.6666666666666667)(68,1.6458015267175572)(69,1.6125265392781316)(70,1.5791227096057745)(71,1.5558126084441875)(72,1.5301568154402896)(73,1.4962358845671266)(74,1.4683626875407698)(75,1.4529972752043596)(76,1.4250177683013503)(77,1.3929903057419837)(78,1.3705426356589148)(79,1.311740890688259)(80,1.2940685045948204)(81,1.2670111972437554)(82,1.2406749555950267)(83,1.2247497725204732)(84,1.2015065913371)(85,1.1973051010587104)(86,1.1865745310957552)(87,1.1737373737373737)(88,1.1731958762886598)(89,1.166841552990556)(90,1.1584051724137931)(91,1.157488986784141)(92,1.155056179775281)(93,1.1557093425605536)(94,1.151764705882353)(95,1.1502403846153846)(96,1.147921760391198)(97,1.1455223880597014)(98,1.139240506329114)(99,1.1391752577319587)(100,1.1336828309305373)(101,1.132)(102,1.129251700680272)(103,1.1253443526170799)(104,1.1274509803921569)(105,1.1214689265536724)(106,1.114777618364419)(107,1.1075110456553756)(108,1.0970149253731343)(109,1.091324200913242)(110,1.088646967340591)(111,1.0850393700787402)(112,1.0861244019138756)(113,1.0834697217675942)(114,1.0836120401337792)(115,1.0797962648556876)(116,1.0738831615120275)(117,1.0660869565217392)(118,1.0676156583629892)(119,1.0672727272727274)(120,1.063551401869159)(121,1.0632183908045978)(122,1.0626223091976517)(123,1.0561122244488979)(124,1.049792531120332)(125,1.0467091295116773)(126,1.0453563714902807)(127,1.0443458980044347)(128,1.045766590389016)(129,1.044496487119438)(130,1.0381861575178997)(131,1.0338983050847457)(132,1.0297029702970297)(133,1.0227848101265822)(134,1.0130548302872062)(135,1.0131926121372032)(136,1.0161290322580645)(137,1.0137362637362637)(138,1.0113636363636365)(139,1.0087719298245614)(140,1)(141,0.9874213836477987)(142,0.987012987012987)(143,0.9764309764309764)(144,0.9680851063829787)(145,0.9672727272727273)(146,0.9571984435797666)(147,0.9512195121951219)(148,0.9356223175965666)(149,0.9282511210762332)(150,0.926605504587156)(151,0.9255813953488372)(152,0.9186602870813397)(153,0.914572864321608)(154,0.9119170984455959)(155,0.8983957219251337)(156,0.8956043956043956)(157,0.8932584269662921)(158,0.8901734104046243)(159,0.8888888888888888)(160,0.8875739644970414)(161,0.8862275449101796)(162,0.8848484848484849)(163,0.8703703703703703)(164,0.8598726114649682)(165,0.8496732026143791)(166,0.831081081081081)(167,0.8125)(168,0.8028169014084507)(169,0.8014184397163121)(170,0.8)(171,0.781021897810219)(172,0.781021897810219)(173,0.7777777777777778)(174,0.7761194029850746)(175,0.7744360902255639)(176,0.7727272727272727)(177,0.7692307692307693)(178,0.7674418604651163)(179,0.7674418604651163)(180,0.7637795275590551)(181,0.7619047619047619)(182,0.7619047619047619)(183,0.7698412698412699)(184,0.768)(185,0.768)(186,0.7661290322580645)(187,0.7642276422764228)(188,0.7583333333333333)(189,0.7583333333333333)(190,0.7583333333333333)(191,0.7563025210084033)(192,0.7542372881355932)(193,0.7542372881355932)(194,0.75)(195,0.7543859649122807)(196,0.75)(197,0.75)(198,0.7477477477477478)(199,0.7477477477477478)(200,0.7477477477477478)(201,0.7431192660550459)(202,0.7407407407407407)(203,0.7407407407407407)(204,0.7407407407407407)(205,0.7407407407407407)(206,0.7407407407407407)(207,0.75)(208,0.75)(209,0.7476635514018691)(210,0.7403846153846154)(211,0.7352941176470589)(212,0.7450980392156863)(213,0.7425742574257426)(214,0.7425742574257426)(215,0.7425742574257426)(216,0.7425742574257426)(217,0.7373737373737373)(218,0.7346938775510204)(219,0.7319587628865979)(220,0.7422680412371134)(221,0.75)(222,0.7446808510638298)(223,0.7419354838709677)(224,0.7391304347826086)(225,0.7391304347826086)(226,0.7391304347826086)(227,0.7362637362637363)(228,0.7303370786516854)(229,0.7303370786516854)(230,0.7241379310344828)(231,0.735632183908046)(232,0.735632183908046)(233,0.7471264367816092)(234,0.7380952380952381)(235,0.7349397590361446)(236,0.7283950617283951)(237,0.7283950617283951)(238,0.725)(239,0.725)(240,0.717948717948718)(241,0.7307692307692307)(242,0.7435897435897436)(243,0.7564102564102564)(244,0.7692307692307693)(245,0.7692307692307693)(246,0.7692307692307693)(247,0.7692307692307693)(248,0.7662337662337663)(249,0.7662337662337663)(250,0.7631578947368421)(251,0.76)(252,0.7567567567567568)(253,0.7534246575342466)(254,0.7534246575342466)(255,0.7534246575342466)(256,0.7638888888888888)(257,0.7638888888888888)(258,0.7638888888888888)(259,0.7638888888888888)(260,0.7638888888888888)(261,0.7638888888888888)(262,0.7571428571428571)(263,0.7571428571428571)(264,0.75)(265,0.75)(266,0.75)(267,0.75)(268,0.7647058823529411)(269,0.7647058823529411)(270,0.7575757575757576)(271,0.7575757575757576)(272,0.7538461538461538)(273,0.7538461538461538)(274,0.7538461538461538)(275,0.7538461538461538)(276,0.7538461538461538)(277,0.7538461538461538)(278,0.7538461538461538)(279,0.7538461538461538)(280,0.75)(281,0.746031746031746)(282,0.7419354838709677)(283,0.7419354838709677)(284,0.7377049180327869)(285,0.7377049180327869)(286,0.7377049180327869)(287,0.7377049180327869)(288,0.7377049180327869)(289,0.7333333333333333)(290,0.7333333333333333)(291,0.7192982456140351)(292,0.7090909090909091)(293,0.7272727272727273)(294,0.7272727272727273)(295,0.7272727272727273)(296,0.7272727272727273)(297,0.7222222222222222)(298,0.7169811320754716)(299,0.7169811320754716)(300,0.7169811320754716)(301,0.7169811320754716)(302,0.7169811320754716)(303,0.7169811320754716)(304,0.7169811320754716)(305,0.7115384615384616)(306,0.7115384615384616)(307,0.7058823529411765)(308,0.7058823529411765)(309,0.7058823529411765)(310,0.7058823529411765)(311,0.7)(312,0.7)(313,0.6938775510204082)(314,0.6938775510204082)(315,0.6938775510204082)(316,0.6938775510204082)(317,0.6938775510204082)(318,0.6938775510204082)(319,0.6875)(320,0.6875)(321,0.6739130434782609)(322,0.6739130434782609)(323,0.6590909090909091)(324,0.6511627906976745)(325,0.6511627906976745)(326,0.6511627906976745)(327,0.6511627906976745)(328,0.6511627906976745)(329,0.6511627906976745)(330,0.6511627906976745)(331,0.6511627906976745)(332,0.6511627906976745)(333,0.6511627906976745)(334,0.6341463414634146)(335,0.6341463414634146)(336,0.6341463414634146)(337,0.625)
}; % plot of the fifth histogram
\addplot[color=orange] coordinates{(0,4.019127055907279)(1,4.015734737979471)(2,4.010115162491884)(3,4.0032618502324935)(4,3.9936305732484074)(5,3.9831698577581527)(6,3.968988212546016)(7,3.955189582535427)(8,3.9381831142087758)(9,3.9226995871315866)(10,3.9057600473312766)(11,3.8863277826468012)(12,3.8640178815801476)(13,3.8407361613087314)(14,3.813603142246308)(15,3.7851847984965548)(16,3.754093871515417)(17,3.721947289676603)(18,3.685492667555448)(19,3.6478927203065132)(20,3.6058918266004563)(21,3.5581197188099125)(22,3.500151011778919)(23,3.44141406062545)(24,3.36712117314124)(25,3.2950774589788248)(26,3.2022549698348333)(27,3.1057599485144265)(28,2.995346672871103)(29,2.853095756954237)(30,2.7147669848399776)(31,2.5841100076394192)(32,2.4671941971645235)(33,2.3630427109071657)(34,2.2512810780034163)(35,2.1548807825555327)(36,2.0264433671220803)(37,1.9000715478177916)(38,1.7767857142857142)(39,1.6815974096060442)(40,1.5966913861950942)(41,1.539009196084248)(42,1.4930705266399753)(43,1.4636075949367089)(44,1.448075056615982)(45,1.4280026542800266)(46,1.4163543441226576)(47,1.4001399090591116)(48,1.3839798850574712)(49,1.369933677229182)(50,1.3575207860922147)(51,1.3398285268901013)(52,1.3263749498193496)(53,1.3157243087082129)(54,1.300639658848614)(55,1.2932461873638343)(56,1.2834751455441111)(57,1.277036355269213)(58,1.2705382436260624)(59,1.266183574879227)(60,1.2564741035856575)(61,1.245389344262295)(62,1.2372435560231456)(63,1.2300163132137032)(64,1.216704288939052)(65,1.2014092777451557)(66,1.1865942028985508)(67,1.174375)(68,1.1588845654993516)(69,1.144489247311828)(70,1.1275261324041812)(71,1.1064916119620716)(72,1.0903614457831325)(73,1.071651090342679)(74,1.049553208773355)(75,1.029109589041096)(76,1.014209591474245)(77,1)(78,0.9884726224783862)(79,0.9772277227722772)(80,0.9631147540983607)(81,0.9508547008547008)(82,0.9476553980370774)(83,0.9432703003337041)(84,0.9347079037800687)(85,0.9289871944121071)(86,0.9228028503562945)(87,0.9215922798552473)(88,0.9191542288557214)(89,0.9183417085427136)(90,0.9168797953964194)(91,0.9156939040207522)(92,0.9140211640211641)(93,0.9126344086021505)(94,0.9103448275862069)(95,0.909217877094972)(96,0.9075391180654339)(97,0.9080459770114943)(98,0.9062957540263543)(99,0.9046199701937406)(100,0.9021406727828746)(101,0.9023622047244094)(102,0.9014308426073132)(103,0.8998384491114702)(104,0.8978583196046128)(105,0.8950930626057529)(106,0.8929188255613126)(107,0.8902654867256637)(108,0.8898916967509025)(109,0.8880733944954129)(110,0.8880597014925373)(111,0.8888888888888888)(112,0.8867562380038387)(113,0.8854368932038835)(114,0.8822355289421158)(115,0.8850806451612904)(116,0.8829568788501027)(117,0.8828451882845189)(118,0.8829787234042553)(119,0.8796498905908097)(120,0.8764044943820225)(121,0.8735632183908046)(122,0.868421052631579)(123,0.8661800486618005)(124,0.8669950738916257)(125,0.8654822335025381)(126,0.8626943005181347)(127,0.8647214854111406)(128,0.8668478260869565)(129,0.8623595505617978)(130,0.8559077809798271)(131,0.8567251461988304)(132,0.8567164179104477)(133,0.8523076923076923)(134,0.8466453674121406)(135,0.8421052631578947)(136,0.8383838383838383)(137,0.8303886925795053)(138,0.8195488721804511)(139,0.8117647058823529)(140,0.8048780487804879)(141,0.7957446808510639)(142,0.788546255506608)(143,0.7788018433179723)(144,0.7692307692307693)(145,0.7635467980295566)(146,0.7512953367875648)(147,0.7419354838709677)(148,0.7251461988304093)(149,0.7080745341614907)(150,0.6928104575163399)(151,0.6824324324324325)(152,0.6666666666666666)(153,0.6594202898550725)(154,0.6209677419354839)(155,0.6115702479338843)(156,0.6050420168067226)(157,0.5803571428571429)(158,0.5523809523809524)(159,0.5392156862745098)(160,0.53)(161,0.5104166666666666)(162,0.4891304347826087)(163,0.4659090909090909)(164,0.45348837209302323)(165,0.4470588235294118)(166,0.4268292682926829)(167,0.4050632911392405)(168,0.4025974025974026)(169,0.4025974025974026)(170,0.39473684210526316)
}; % plot of the first histogram
\legend{50\%, 60\%, 70\%, 80\%, 90\%, 100\%} % legend of the plot

\end{axis}
\end{tikzpicture}

%% file: plots/aware_fraction/evac_speed.tex
\begin{tikzpicture}[scale=0.75]

% Set the axis labels and title
\begin{axis}[    xlabel={Time [s]},    ylabel={ppl/s},    title={Evacuation Speed - Mobile application},   legend style={at={(1,1)}, anchor=north east},]

\addplot[color=blue] coordinates{(0,36)(1,49)(2,58)(3,77)(4,70)(5,107)(6,108)(7,114)(8,141)(9,143)(10,146)(11,129)(12,160)(13,142)(14,176)(15,131)(16,136)(17,149)(18,148)(19,141)(20,159)(21,158)(22,156)(23,160)(24,157)(25,139)(26,152)(27,157)(28,155)(29,156)(30,141)(31,130)(32,117)(33,125)(34,130)(35,132)(36,131)(37,112)(38,122)(39,93)(40,95)(41,85)(42,90)(43,64)(44,75)(45,71)(46,81)(47,68)(48,88)(49,78)(50,80)(51,65)(52,79)(53,79)(54,80)(55,68)(56,84)(57,73)(58,75)(59,76)(60,69)(61,81)(62,73)(63,77)(64,75)(65,70)(66,63)(67,46)(68,52)(69,60)(70,62)(71,56)(72,47)(73,40)(74,71)(75,49)(76,52)(77,49)(78,48)(79,43)(80,47)(81,44)(82,39)(83,39)(84,20)(85,32)(86,34)(87,26)(88,17)(89,16)(90,20)(91,33)(92,30)(93,21)(94,17)(95,21)(96,17)(97,23)(98,20)(99,18)(100,12)(101,22)(102,20)(103,18)(104,22)(105,23)(106,10)(107,19)(108,14)(109,12)(110,19)(111,27)(112,13)(113,16)(114,17)(115,15)(116,13)(117,13)(118,7)(119,22)(120,17)(121,12)(122,20)(123,17)(124,9)(125,15)(126,14)(127,18)(128,17)(129,10)(130,15)(131,14)(132,13)(133,12)(134,16)(135,10)(136,14)(137,9)(138,19)(139,13)(140,13)(141,14)(142,16)(143,8)(144,11)(145,13)(146,19)(147,13)(148,17)(149,15)(150,9)(151,15)(152,9)(153,12)(154,11)(155,8)(156,13)(157,13)(158,10)(159,10)(160,15)(161,7)(162,9)(163,7)(164,9)(165,12)(166,9)(167,8)(168,6)(169,3)(170,10)(171,8)(172,4)(173,5)(174,10)(175,8)(176,6)(177,10)(178,6)(179,7)(180,5)(181,7)(182,6)(183,9)(184,6)(185,9)(186,8)(187,5)(188,7)(189,6)(190,6)(191,4)(192,3)(193,4)(194,7)(195,5)(196,3)(197,6)(198,6)(199,6)(200,7)(201,5)(202,8)(203,10)(204,4)(205,3)(206,4)(207,6)(208,7)(209,9)(210,8)(211,3)(212,7)(213,6)(214,2)(215,10)(216,4)(217,5)(218,4)(219,6)(220,4)(221,8)(222,9)(223,3)(224,5)(225,6)(226,7)(227,8)(228,5)(229,9)(230,5)(231,5)(232,4)(233,4)(234,6)(235,4)(236,5)(237,7)(238,5)(239,4)(240,6)(241,6)(242,2)(243,4)(244,4)(245,5)(246,2)(247,7)(248,1)(249,3)(250,5)(251,6)(252,4)(253,5)(254,4)(255,3)(256,7)(257,4)(258,4)(259,3)(260,4)(261,6)(262,6)(263,4)(264,3)(265,3)(266,5)(267,3)(268,6)(269,6)(270,5)(271,3)(272,6)(273,3)(274,2)(275,5)(276,3)(277,3)(278,7)(279,2)(280,4)(281,3)(282,4)(283,4)(284,6)(285,2)(286,3)(287,2)(288,4)(289,2)(290,4)(291,3)(292,4)(293,6)(294,4)(295,2)(296,2)(297,2)(298,5)(299,3)(300,3)(301,4)(302,2)(303,2)(304,4)(305,5)(306,4)(307,4)(308,3)(309,6)(310,5)(311,4)(312,2)(313,1)(314,3)(315,3)(316,4)(317,2)(318,3)(319,3)(320,2)(321,3)(322,3)(323,2)(324,3)(325,4)(326,1)(327,3)(328,1)(329,3)(330,3)(331,4)(332,3)(333,1)(334,3)(335,4)(336,1)(337,3)(338,1)
}; % plot of the first histogram
\addplot[color=teal] coordinates{(0,39)(1,40)(2,69)(3,98)(4,102)(5,124)(6,167)(7,160)(8,134)(9,154)(10,142)(11,148)(12,149)(13,156)(14,185)(15,175)(16,175)(17,163)(18,177)(19,144)(20,162)(21,155)(22,160)(23,186)(24,166)(25,163)(26,192)(27,172)(28,175)(29,180)(30,152)(31,135)(32,131)(33,122)(34,122)(35,117)(36,130)(37,100)(38,90)(39,101)(40,90)(41,85)(42,79)(43,82)(44,70)(45,69)(46,68)(47,70)(48,88)(49,73)(50,67)(51,60)(52,72)(53,60)(54,73)(55,58)(56,72)(57,66)(58,69)(59,68)(60,76)(61,63)(62,56)(63,62)(64,66)(65,82)(66,55)(67,52)(68,51)(69,38)(70,52)(71,57)(72,54)(73,56)(74,57)(75,44)(76,53)(77,41)(78,47)(79,49)(80,40)(81,37)(82,31)(83,40)(84,32)(85,29)(86,28)(87,24)(88,24)(89,13)(90,23)(91,20)(92,15)(93,23)(94,24)(95,25)(96,20)(97,22)(98,22)(99,29)(100,20)(101,19)(102,16)(103,20)(104,14)(105,11)(106,21)(107,14)(108,16)(109,20)(110,19)(111,16)(112,21)(113,16)(114,17)(115,18)(116,22)(117,17)(118,21)(119,15)(120,12)(121,21)(122,21)(123,16)(124,21)(125,13)(126,20)(127,17)(128,21)(129,22)(130,11)(131,11)(132,15)(133,18)(134,10)(135,15)(136,14)(137,14)(138,12)(139,23)(140,13)(141,19)(142,13)(143,14)(144,10)(145,11)(146,12)(147,11)(148,17)(149,26)(150,17)(151,13)(152,15)(153,8)(154,10)(155,6)(156,11)(157,15)(158,9)(159,6)(160,9)(161,10)(162,10)(163,7)(164,6)(165,9)(166,6)(167,4)(168,6)(169,7)(170,9)(171,8)(172,6)(173,7)(174,4)(175,7)(176,5)(177,8)(178,9)(179,10)(180,6)(181,5)(182,8)(183,6)(184,9)(185,6)(186,9)(187,6)(188,14)(189,7)(190,3)(191,5)(192,4)(193,5)(194,9)(195,8)(196,7)(197,6)(198,4)(199,7)(200,3)(201,3)(202,3)(203,8)(204,7)(205,6)(206,3)(207,4)(208,4)(209,3)(210,4)(211,3)(212,6)(213,3)(214,3)(215,3)(216,2)(217,3)(218,3)(219,4)(220,4)(221,8)(222,7)(223,6)(224,5)(225,6)(226,5)(227,4)(228,8)(229,4)(230,5)(231,5)(232,4)(233,9)(234,4)(235,6)(236,4)(237,5)(238,7)(239,6)(240,6)(241,6)(242,4)(243,8)(244,6)(245,3)(246,8)(247,6)(248,4)(249,6)(250,3)(251,5)(252,4)(253,3)(254,8)(255,8)(256,4)(257,4)(258,3)(259,5)(260,4)(261,3)(262,6)(263,5)(264,5)(265,4)(266,3)(267,5)(268,4)(269,4)(270,5)(271,5)(272,3)(273,5)(274,8)(275,3)(276,6)(277,7)(278,7)(279,5)(280,3)(281,3)(282,5)(283,5)(284,5)(285,6)(286,5)(287,4)(288,4)(289,4)(290,4)(291,4)(292,7)(293,3)(294,3)(295,5)(296,4)(297,4)(298,5)(299,5)(300,3)(301,2)(302,4)(303,4)(304,5)(305,6)(306,3)(307,4)(308,6)(309,3)(310,3)(311,5)(312,6)(313,3)(314,5)(315,6)(316,3)(317,2)(318,4)(319,7)(320,5)(321,2)(322,3)(323,2)(324,5)(325,2)(326,3)(327,4)(328,3)(329,2)(330,2)(331,4)(332,4)(333,2)(334,3)(335,4)(336,2)(337,4)(338,3)(339,4)(340,2)(341,2)(342,2)(343,2)(344,2)(345,3)(346,2)(347,2)(348,2)(349,3)(350,2)
}; % plot of the second histogram
\addplot[color=cyan] coordinates{(0,33)(1,46)(2,88)(3,83)(4,144)(5,127)(6,163)(7,168)(8,153)(9,179)(10,155)(11,164)(12,177)(13,164)(14,202)(15,168)(16,179)(17,176)(18,199)(19,174)(20,180)(21,169)(22,177)(23,170)(24,170)(25,184)(26,169)(27,157)(28,196)(29,180)(30,160)(31,142)(32,127)(33,124)(34,140)(35,126)(36,115)(37,115)(38,97)(39,94)(40,105)(41,94)(42,85)(43,71)(44,80)(45,76)(46,78)(47,78)(48,65)(49,79)(50,82)(51,59)(52,71)(53,64)(54,77)(55,55)(56,71)(57,67)(58,66)(59,69)(60,57)(61,60)(62,59)(63,55)(64,62)(65,56)(66,64)(67,59)(68,58)(69,54)(70,41)(71,58)(72,39)(73,48)(74,42)(75,42)(76,56)(77,48)(78,41)(79,47)(80,37)(81,36)(82,31)(83,36)(84,36)(85,36)(86,20)(87,27)(88,21)(89,24)(90,21)(91,20)(92,21)(93,27)(94,26)(95,19)(96,22)(97,23)(98,16)(99,26)(100,30)(101,22)(102,19)(103,25)(104,18)(105,23)(106,17)(107,12)(108,21)(109,22)(110,23)(111,10)(112,15)(113,17)(114,20)(115,14)(116,12)(117,12)(118,16)(119,13)(120,19)(121,19)(122,14)(123,13)(124,16)(125,22)(126,22)(127,11)(128,28)(129,13)(130,14)(131,14)(132,13)(133,12)(134,21)(135,18)(136,19)(137,18)(138,18)(139,9)(140,9)(141,19)(142,19)(143,21)(144,15)(145,17)(146,19)(147,8)(148,12)(149,15)(150,11)(151,15)(152,15)(153,12)(154,9)(155,11)(156,9)(157,11)(158,11)(159,10)(160,6)(161,10)(162,6)(163,12)(164,10)(165,10)(166,7)(167,7)(168,7)(169,6)(170,8)(171,7)(172,8)(173,4)(174,4)(175,7)(176,7)(177,8)(178,7)(179,8)(180,5)(181,6)(182,7)(183,8)(184,8)(185,6)(186,8)(187,9)(188,5)(189,7)(190,4)(191,7)(192,7)(193,6)(194,9)(195,4)(196,8)(197,5)(198,6)(199,9)(200,5)(201,6)(202,4)(203,8)(204,5)(205,5)(206,6)(207,8)(208,5)(209,10)(210,7)(211,6)(212,5)(213,7)(214,6)(215,4)(216,8)(217,5)(218,9)(219,5)(220,4)(221,6)(222,8)(223,6)(224,6)(225,6)(226,8)(227,9)(228,8)(229,8)(230,6)(231,9)(232,6)(233,6)(234,7)(235,4)(236,6)(237,10)(238,5)(239,6)(240,5)(241,7)(242,8)(243,4)(244,7)(245,4)(246,7)(247,4)(248,6)(249,6)(250,5)(251,4)(252,5)(253,8)(254,5)(255,7)(256,6)(257,5)(258,5)(259,4)(260,5)(261,6)(262,5)(263,5)(264,4)(265,5)(266,4)(267,4)(268,6)(269,4)(270,5)(271,5)(272,4)(273,9)(274,5)(275,6)(276,6)(277,4)(278,7)(279,6)(280,7)(281,6)(282,5)(283,4)(284,6)(285,5)(286,6)(287,5)(288,5)(289,7)(290,6)(291,5)(292,6)(293,5)(294,4)(295,4)(296,5)(297,5)(298,4)(299,4)(300,7)(301,6)(302,6)(303,7)(304,10)(305,4)(306,4)(307,7)(308,6)(309,4)(310,7)(311,5)(312,7)(313,6)(314,5)(315,5)(316,4)(317,6)(318,4)(319,4)(320,5)(321,5)(322,5)(323,6)(324,4)(325,4)(326,5)(327,5)(328,6)(329,5)(330,6)(331,6)(332,7)(333,4)(334,6)(335,4)(336,4)(337,4)(338,4)(339,4)(340,6)(341,5)(342,4)(343,5)(344,5)(345,5)(346,4)(347,4)(348,4)(349,5)(350,4)
}; % plot of the third histogram
\addplot[color=magenta] coordinates{(0,52)(1,59)(2,97)(3,92)(4,137)(5,139)(6,146)(7,156)(8,177)(9,191)(10,194)(11,196)(12,172)(13,189)(14,203)(15,176)(16,172)(17,177)(18,176)(19,181)(20,198)(21,209)(22,184)(23,180)(24,188)(25,162)(26,185)(27,204)(28,194)(29,178)(30,172)(31,131)(32,140)(33,140)(34,138)(35,150)(36,129)(37,124)(38,142)(39,94)(40,98)(41,71)(42,92)(43,82)(44,75)(45,80)(46,76)(47,65)(48,74)(49,71)(50,65)(51,75)(52,72)(53,68)(54,69)(55,58)(56,53)(57,63)(58,70)(59,64)(60,65)(61,59)(62,55)(63,64)(64,72)(65,62)(66,68)(67,52)(68,58)(69,53)(70,77)(71,53)(72,48)(73,46)(74,51)(75,53)(76,46)(77,52)(78,60)(79,48)(80,38)(81,52)(82,39)(83,34)(84,37)(85,28)(86,35)(87,29)(88,31)(89,29)(90,35)(91,30)(92,26)(93,28)(94,28)(95,27)(96,21)(97,22)(98,30)(99,26)(100,21)(101,25)(102,26)(103,29)(104,21)(105,25)(106,23)(107,25)(108,24)(109,22)(110,26)(111,23)(112,17)(113,22)(114,27)(115,22)(116,25)(117,24)(118,21)(119,20)(120,20)(121,22)(122,24)(123,23)(124,17)(125,21)(126,30)(127,21)(128,24)(129,21)(130,21)(131,21)(132,29)(133,21)(134,17)(135,15)(136,21)(137,16)(138,22)(139,23)(140,28)(141,21)(142,17)(143,22)(144,22)(145,16)(146,19)(147,21)(148,23)(149,15)(150,17)(151,18)(152,19)(153,17)(154,20)(155,15)(156,21)(157,13)(158,17)(159,16)(160,14)(161,16)(162,14)(163,14)(164,17)(165,19)(166,15)(167,13)(168,15)(169,15)(170,10)(171,12)(172,14)(173,10)(174,13)(175,12)(176,11)(177,10)(178,10)(179,12)(180,12)(181,15)(182,13)(183,13)(184,10)(185,12)(186,10)(187,11)(188,13)(189,12)(190,11)(191,15)(192,13)(193,11)(194,12)(195,11)(196,14)(197,12)(198,13)(199,12)(200,12)(201,12)(202,11)(203,10)(204,10)(205,11)(206,13)(207,12)(208,10)(209,11)(210,10)(211,11)(212,10)(213,10)(214,11)(215,10)(216,13)(217,11)(218,11)(219,10)(220,12)(221,11)(222,11)(223,12)(224,15)(225,13)(226,10)(227,11)(228,12)(229,12)(230,11)(231,10)(232,10)(233,10)(234,11)(235,12)(236,10)(237,10)(238,10)(239,11)(240,11)(241,13)(242,10)(243,12)(244,11)(245,13)(246,10)(247,11)(248,11)(249,11)(250,14)(251,11)(252,10)(253,11)(254,11)(255,11)(256,10)(257,12)(258,10)(259,10)(260,10)(261,11)(262,10)(263,12)(264,10)(265,11)(266,11)(267,10)(268,11)(269,12)(270,14)(271,11)(272,12)(273,11)(274,10)(275,11)(276,11)(277,11)(278,11)(279,10)(280,12)(281,13)(282,10)(283,11)(284,13)(285,12)(286,11)(287,10)(288,11)(289,12)(290,11)(291,10)(292,10)(293,12)(294,11)(295,10)(296,13)(297,11)(298,11)(299,11)(300,10)(301,10)(302,12)(303,12)(304,10)(305,11)(306,11)(307,11)(308,10)(309,10)(310,10)(311,11)(312,14)(313,12)(314,10)(315,11)(316,11)(317,11)(318,10)(319,10)(320,11)(321,10)(322,11)(323,10)(324,10)(325,11)(326,14)(327,10)(328,12)(329,10)(330,12)(331,11)(332,10)(333,11)(334,11)(335,12)(336,10)(337,10)(338,10)(339,11)(340,10)(341,10)(342,11)(343,11)(344,10)(345,11)(346,10)
}; % plot of the fourth histogram
\addplot[color=red] coordinates{(0,49)(1,69)(2,97)(3,131)(4,138)(5,152)(6,165)(7,197)(8,194)(9,168)(10,188)(11,187)(12,186)(13,203)(14,235)(15,218)(16,205)(17,215)(18,193)(19,196)(20,193)(21,236)(22,208)(23,193)(24,187)(25,210)(26,189)(27,204)(28,187)(29,213)(30,180)(31,155)(32,141)(33,136)(34,144)(35,130)(36,113)(37,115)(38,103)(39,102)(40,93)(41,82)(42,88)(43,74)(44,79)(45,86)(46,76)(47,80)(48,81)(49,78)(50,86)(51,73)(52,70)(53,49)(54,57)(55,67)(56,70)(57,69)(58,62)(59,59)(60,58)(61,62)(62,66)(63,69)(64,77)(65,58)(66,60)(67,61)(68,58)(69,65)(70,65)(71,66)(72,58)(73,55)(74,62)(75,58)(76,57)(77,51)(78,47)(79,38)(80,41)(81,40)(82,34)(83,42)(84,35)(85,37)(86,33)(87,34)(88,28)(89,37)(90,32)(91,31)(92,36)(93,30)(94,30)(95,27)(96,27)(97,25)(98,27)(99,26)(100,28)(101,28)(102,22)(103,26)(104,18)(105,25)(106,30)(107,18)(108,26)(109,28)(110,19)(111,23)(112,28)(113,26)(114,22)(115,20)(116,20)(117,28)(118,27)(119,28)(120,26)(121,24)(122,25)(123,28)(124,23)(125,21)(126,25)(127,29)(128,24)(129,22)(130,20)(131,23)(132,22)(133,25)(134,18)(135,21)(136,22)(137,25)(138,24)(139,26)(140,24)(141,24)(142,23)(143,29)(144,21)(145,32)(146,26)(147,25)(148,23)(149,19)(150,17)(151,19)(152,24)(153,20)(154,20)(155,19)(156,18)(157,19)(158,16)(159,16)(160,16)(161,16)(162,17)(163,18)(164,18)(165,18)(166,17)(167,16)(168,15)(169,15)(170,16)(171,14)(172,16)(173,15)(174,15)(175,15)(176,16)(177,15)(178,14)(179,16)(180,15)(181,14)(182,14)(183,15)(184,14)(185,15)(186,15)(187,17)(188,14)(189,14)(190,15)(191,15)(192,14)(193,16)(194,16)(195,16)(196,14)(197,15)(198,14)(199,14)(200,16)(201,15)(202,14)(203,14)(204,14)(205,14)(206,14)(207,14)(208,15)(209,17)(210,16)(211,14)(212,15)(213,14)(214,14)(215,14)(216,16)(217,15)(218,15)(219,14)(220,15)(221,16)(222,15)(223,15)(224,14)(225,14)(226,15)(227,16)(228,14)(229,16)(230,14)(231,14)(232,14)(233,17)(234,15)(235,16)(236,14)(237,15)(238,14)(239,16)(240,14)(241,14)(242,14)(243,14)(244,14)(245,14)(246,14)(247,15)(248,14)(249,15)(250,15)(251,15)(252,15)(253,14)(254,14)(255,15)(256,14)(257,14)(258,14)(259,14)(260,14)(261,16)(262,14)(263,16)(264,14)(265,14)(266,14)(267,14)(268,14)(269,16)(270,14)(271,15)(272,14)(273,14)(274,14)(275,14)(276,14)(277,14)(278,14)(279,15)(280,15)(281,15)(282,14)(283,15)(284,14)(285,14)(286,14)(287,14)(288,15)(289,14)(290,17)(291,16)(292,14)(293,14)(294,14)(295,14)(296,15)(297,15)(298,14)(299,14)(300,14)(301,14)(302,14)(303,14)(304,15)(305,14)(306,15)(307,14)(308,14)(309,14)(310,15)(311,14)(312,15)(313,14)(314,14)(315,14)(316,14)(317,14)(318,15)(319,14)(320,16)(321,14)(322,16)(323,15)(324,14)(325,14)(326,14)(327,14)(328,14)(329,14)(330,14)(331,14)(332,14)(333,16)(334,14)(335,14)(336,15)(337,14)
}; % plot of the fifth histogram
\addplot[color=orange] coordinates{(0,52)(1,62)(2,113)(3,125)(4,159)(5,171)(6,198)(7,212)(8,219)(9,209)(10,202)(11,225)(12,228)(13,216)(14,276)(15,256)(16,218)(17,204)(18,216)(19,205)(20,223)(21,208)(22,189)(23,212)(24,223)(25,200)(26,244)(27,190)(28,240)(29,197)(30,198)(31,157)(32,143)(33,118)(34,153)(35,112)(36,155)(37,115)(38,102)(39,104)(40,85)(41,90)(42,78)(43,71)(44,72)(45,72)(46,73)(47,73)(48,75)(49,66)(50,79)(51,77)(52,70)(53,77)(54,64)(55,70)(56,71)(57,64)(58,67)(59,71)(60,63)(61,62)(62,78)(63,81)(64,79)(65,60)(66,66)(67,71)(68,68)(69,61)(70,65)(71,57)(72,51)(73,63)(74,73)(75,56)(76,62)(77,53)(78,46)(79,50)(80,55)(81,40)(82,37)(83,43)(84,31)(85,35)(86,32)(87,45)(88,29)(89,35)(90,31)(91,35)(92,32)(93,39)(94,29)(95,34)(96,27)(97,33)(98,33)(99,37)(100,39)(101,26)(102,30)(103,32)(104,36)(105,32)(106,34)(107,32)(108,29)(109,29)(110,25)(111,30)(112,26)(113,34)(114,25)(115,29)(116,29)(117,28)(118,33)(119,32)(120,30)(121,37)(122,27)(123,26)(124,32)(125,28)(126,29)(127,29)(128,32)(129,29)(130,25)(131,27)(132,30)(133,32)(134,30)(135,27)(136,35)(137,38)(138,31)(139,29)(140,31)(141,28)(142,30)(143,30)(144,25)(145,31)(146,27)(147,35)(148,30)(149,28)(150,25)(151,27)(152,23)(153,34)(154,23)(155,22)(156,27)(157,27)(158,23)(159,22)(160,24)(161,24)(162,24)(163,22)(164,21)(165,23)(166,23)(167,22)(168,20)(169,21)(170,20)
}; % plot of the first histogram
\legend{50\%, 60\%, 70\%, 80\%, 90\%, 100\%} % legend of the plot

\end{axis}
\end{tikzpicture}

%% file: plots/aware_fraction/evac_time.tex
\begin{tikzpicture}[scale=0.75]

% Set the axis labels and title
\begin{axis}[    xlabel={Time[s]},    ylabel=s,    title={Evacuation Time - Mobile application},   legend style={at={(0,1)}, anchor=north west},]

\addplot[color=blue] coordinates{(0,14.621111111111112)(1,21.697020408163265)(2,12.988568965517242)(3,21.20374025974026)(4,17.298999999999996)(5,18.72670093457944)(6,20.815324074074073)(7,25.92685964912281)(8,26.854276595744675)(9,12.942587412587413)(10,21.614027397260273)(11,18.277720930232558)(12,18.63773125)(13,13.578985915492959)(14,18.643670454545454)(15,27.12715267175572)(16,18.186411764705884)(17,31.842456375838925)(18,33.29658783783784)(19,24.40582269503546)(20,29.145754716981145)(21,29.63033544303798)(22,31.991769230769233)(23,27.3524375)(24,40.14587898089173)(25,32.18358273381294)(26,40.58761842105263)(27,37.13652229299363)(28,33.53238064516129)(29,44.58303846153845)(30,38.65984397163121)(31,27.88193076923077)(32,38.6388547008547)(33,36.406560000000006)(34,36.33049230769232)(35,26.952878787878785)(36,28.439572519083963)(37,24.036339285714288)(38,28.29709836065574)(39,40.51297849462367)(40,31.82252631578947)(41,42.78911764705882)(42,40.51531111111112)(43,28.50540625)(44,38.68698666666668)(45,36.65795774647888)(46,47.42553086419755)(47,43.012779411764704)(48,40.33793181818181)(49,47.63915384615385)(50,38.7935)(51,35.84513846153846)(52,41.45332911392405)(53,19.701658227848103)(54,43.053887499999995)(55,32.23135294117646)(56,39.29688095238095)(57,62.63065753424656)(58,48.43674666666665)(59,41.60738157894736)(60,32.067)(61,47.03828395061729)(62,52.258109589041105)(63,49.60945454545455)(64,44.60352)(65,31.833400000000008)(66,53.25117460317459)(67,34.59643478260869)(68,33.757153846153855)(69,23.930983333333334)(70,54.46982258064516)(71,43.016035714285714)(72,40.97108510638297)(73,56.29729999999999)(74,63.725704225352096)(75,65.93969387755102)(76,49.72440384615384)(77,46.16885714285715)(78,36.99604166666666)(79,56.50720930232558)(80,41.32500000000001)(81,47.916522727272735)(82,33.10261538461539)(83,41.53492307692307)(84,64.6359)(85,40.44046875000001)(86,62.23455882352941)(87,62.47111538461537)(88,37.5364705882353)(89,50.1839375)(90,40.16595)(91,29.346545454545456)(92,48.77139999999999)(93,46.15304761904762)(94,57.049176470588236)(95,38.33942857142857)(96,37.66529411764706)(97,35.04130434782609)(98,106.46504999999999)(99,26.427444444444443)(100,25.862583333333333)(101,36.71877272727273)(102,73.57625)(103,72.5751111111111)(104,74.53690909090909)(105,49.64473913043478)(106,64.35050000000001)(107,51.38763157894737)(108,57.895071428571434)(109,53.698)(110,42.69636842105264)(111,48.58274074074074)(112,49.61884615384615)(113,50.7490625)(114,47.77876470588235)(115,65.29893333333332)(116,36.82130769230769)(117,23.98069230769231)(118,44.54314285714286)(119,36.9675)(120,67.52076470588236)(121,53.8715)(122,23.96585)(123,77.4184705882353)(124,16.076777777777778)(125,54.294333333333334)(126,58.18964285714286)(127,63.894499999999994)(128,38.0934705882353)(129,14.469100000000001)(130,43.197)(131,46.29828571428571)(132,49.87253846153846)(133,68.04675)(134,30.048937499999997)(135,64.8982)(136,46.36942857142857)(137,72.16055555555555)(138,43.051315789473676)(139,37.0353076923077)(140,62.95484615384616)(141,82.54107142857143)(142,40.6415625)(143,60.227125)(144,74.4719090909091)(145,50.06092307692307)(146,43.14678947368421)(147,50.087)(148,38.31117647058824)(149,88.4662)(150,72.39644444444444)(151,54.71766666666667)(152,53.65011111111111)(153,26.15341666666667)(154,13.153727272727274)(155,81.556625)(156,50.20738461538462)(157,76.27699999999999)(158,31.410800000000002)(159,31.414800000000003)(160,20.945933333333333)(161,44.88857142857143)(162,34.92011111111111)(163,44.906)(164,34.93333333333333)(165,40.34791666666666)(166,91.54477777777778)(167,103.01437500000002)(168,109.082)(169,48.230333333333334)(170,48.48630000000001)(171,81.891875)(172,78.7175)(173,28.938200000000002)(174,99.612)(175,39.380125)(176,80.92216666666666)(177,14.469100000000001)(178,24.115166666666667)(179,69.41385714285715)(180,28.938200000000002)(181,69.43414285714286)(182,109.471)(183,54.03344444444444)(184,81.06083333333333)(185,35.063222222222215)(186,39.4525)(187,63.133799999999994)(188,20.670142857142856)(189,81.14550000000001)(190,52.63733333333334)(191,78.97125)(192,105.30533333333334)(193,36.17275)(194,20.670142857142856)(195,28.938200000000002)(196,105.36966666666667)(197,81.26816666666667)(198,52.699999999999996)(199,24.115166666666667)(200,69.699)(201,97.5864)(202,103.922375)(203,14.469100000000001)(204,36.17275)(205,105.49166666666667)(206,122.09525)(207,52.76616666666666)(208,69.80442857142857)(209,73.41866666666667)(210,82.6135)(211,105.60633333333332)(212,20.670142857142856)(213,110.24016666666667)(214,158.495)(215,31.7041)(216,79.27799999999999)(217,63.4366)(218,79.30850000000001)(219,81.65116666666667)(220,36.17275)(221,39.675625)(222,92.85455555555556)(223,105.82600000000001)(224,28.938200000000002)(225,24.115166666666667)(226,70.08185714285715)(227,82.97125)(228,98.18140000000001)(229,73.79511111111111)(230,98.22139999999999)(231,98.2278)(232,79.49199999999999)(233,79.49600000000001)(234,81.89816666666665)(235,79.521)(236,28.938200000000002)(237,45.44942857142858)(238,28.938200000000002)(239,79.561)(240,53.0445)(241,110.91950000000001)(242,72.3455)(243,36.17275)(244,79.60425000000001)(245,98.44179999999999)(246,159.24)(247,70.33842857142858)(248,144.691)(249,48.230333333333334)(250,63.7264)(251,53.1105)(252,36.17275)(253,63.753)(254,36.17275)(255,48.230333333333334)(256,70.44328571428571)(257,79.74199999999999)(258,36.17275)(259,48.230333333333334)(260,79.79124999999999)(261,53.19416666666666)(262,111.40116666666667)(263,36.17275)(264,48.230333333333334)(265,164.713)(266,63.8952)(267,48.230333333333334)(268,53.2575)(269,24.115166666666667)(270,98.90899999999999)(271,48.230333333333334)(272,53.283166666666666)(273,106.58666666666666)(274,72.3455)(275,63.9722)(276,48.230333333333334)(277,106.64433333333334)(278,95.7867142857143)(279,72.3455)(280,123.83924999999999)(281,106.68433333333333)(282,36.17275)(283,123.89025)(284,53.36116666666667)(285,72.3455)(286,106.75266666666668)(287,72.3455)(288,80.07900000000001)(289,72.3455)(290,36.17275)(291,106.806)(292,36.17275)(293,53.415166666666664)(294,124.08824999999999)(295,72.3455)(296,160.29149999999998)(297,72.3455)(298,64.1288)(299,106.901)(300,165.59833333333333)(301,36.17275)(302,72.3455)(303,72.3455)(304,124.29625)(305,99.457)(306,36.17275)(307,168.487)(308,107.04699999999998)(309,53.5285)(310,134.8616)(311,124.45724999999999)(312,160.64100000000002)(313,144.691)(314,107.11766666666666)(315,48.230333333333334)(316,124.56475)(317,72.3455)(318,48.230333333333334)(319,48.230333333333334)(320,160.8075)(321,48.230333333333334)(322,48.230333333333334)(323,72.3455)(324,48.230333333333334)(325,36.17275)(326,144.691)(327,48.230333333333334)(328,144.691)(329,48.230333333333334)(330,48.230333333333334)(331,80.518)(332,48.230333333333334)(333,144.691)(334,48.230333333333334)(335,36.17275)(336,144.691)(337,48.230333333333334)(338,144.691)
}; % plot of the first histogram
\addplot[color=teal] coordinates{(0,7.2344871794871795)(1,9.479)(2,13.87455072463768)(3,17.73542857142857)(4,17.188970588235296)(5,17.423580645161294)(6,16.59222754491018)(7,20.582381250000005)(8,21.748649253731344)(9,17.10745454545455)(10,12.949401408450706)(11,15.98537837837838)(12,27.094557046979865)(13,18.693365384615387)(14,20.45281081081081)(15,20.01874857142857)(16,20.187857142857144)(17,19.92655828220859)(18,19.714050847457628)(19,19.073631944444443)(20,21.176080246913582)(21,22.336529032258067)(22,29.667612499999997)(23,22.683295698924734)(24,27.02490361445783)(25,28.467730061349698)(26,29.306348958333334)(27,35.82717441860467)(28,42.5878)(29,30.380261111111118)(30,24.19919736842106)(31,24.747918518518524)(32,24.70663358778626)(33,19.472459016393437)(34,35.00874590163934)(35,30.196923076923074)(36,31.184830769230775)(37,26.73534)(38,22.7722)(39,30.5810099009901)(40,25.81611111111111)(41,21.327929411764703)(42,27.94745569620253)(43,38.12337804878049)(44,37.3112)(45,32.259594202898555)(46,40.530779411764705)(47,26.32118571428572)(48,29.990920454545456)(49,36.267616438356164)(50,31.69567164179104)(51,31.017949999999992)(52,37.00252777777778)(53,46.72653333333333)(54,38.50823287671233)(55,23.132396551724135)(56,35.48379166666666)(57,28.58954545454545)(58,27.394391304347824)(59,37.77860294117646)(60,46.3342894736842)(61,34.45939682539683)(62,38.83869642857143)(63,48.33303225806453)(64,31.036106060606066)(65,38.3675731707317)(66,32.38976363636364)(67,36.95626923076923)(68,37.71798039215686)(69,36.20626315789473)(70,21.201384615384615)(71,33.888508771929835)(72,46.05579629629629)(73,39.529357142857144)(74,38.89140350877193)(75,44.14054545454546)(76,49.78256603773584)(77,27.128292682926833)(78,44.40855319148937)(79,48.32838775510205)(80,27.884149999999998)(81,37.71251351351351)(82,45.06509677419355)(83,27.961900000000004)(84,17.482062499999998)(85,28.974206896551724)(86,20.012714285714285)(87,35.07841666666666)(88,29.247041666666664)(89,43.19446153846154)(90,36.698565217391305)(91,42.23524999999999)(92,37.519466666666666)(93,36.77913043478261)(94,41.155833333333334)(95,45.22164)(96,21.130899999999997)(97,57.900090909090906)(98,57.946409090909086)(99,24.389655172413793)(100,28.289949999999997)(101,29.802315789473685)(102,62.1415625)(103,49.764799999999994)(104,40.551785714285714)(105,38.64081818181818)(106,47.515190476190476)(107,20.171)(108,35.5956875)(109,28.4894)(110,45.13052631578947)(111,62.611625000000004)(112,61.42447619047618)(113,26.69475)(114,42.0895294117647)(115,47.79072222222222)(116,45.690909090909095)(117,42.17535294117647)(118,41.066761904761904)(119,28.5772)(120,35.7475)(121,34.24209523809524)(122,27.358619047619044)(123,35.9308125)(124,68.91923809523809)(125,55.455384615384624)(126,21.523500000000002)(127,68.19458823529412)(128,48.308476190476185)(129,52.76909090909091)(130,52.47927272727273)(131,65.80545454545455)(132,67.79866666666666)(133,64.67366666666666)(134,43.235400000000006)(135,58.174)(136,30.910714285714285)(137,72.89628571428571)(138,36.1045)(139,44.43356521739131)(140,56.01153846153847)(141,38.342315789473695)(142,56.07399999999999)(143,20.476071428571426)(144,72.9934)(145,52.954)(146,23.920833333333334)(147,39.562909090909095)(148,25.60694117647059)(149,39.555461538461536)(150,25.634588235294117)(151,44.95246153846154)(152,48.88726666666666)(153,17.393)(154,13.9144)(155,48.012)(156,26.195909090909087)(157,68.94186666666666)(158,15.460444444444445)(159,97.8875)(160,48.67466666666667)(161,13.9144)(162,43.862)(163,41.27928571428572)(164,23.19066666666667)(165,32.132777777777775)(166,23.19066666666667)(167,34.786)(168,48.244)(169,62.849428571428575)(170,48.91466666666666)(171,55.05)(172,48.31666666666666)(173,41.428428571428576)(174,34.786)(175,41.45885714285714)(176,58.06079999999999)(177,36.311)(178,49.101111111111116)(179,59.362300000000005)(180,73.74900000000001)(181,88.52919999999999)(182,55.34725)(183,73.85233333333333)(184,49.25022222222222)(185,23.19066666666667)(186,32.37166666666667)(187,23.19066666666667)(188,20.819928571428573)(189,19.877714285714287)(190,46.38133333333334)(191,27.8288)(192,111.141)(193,58.387)(194,66.43744444444445)(195,36.520875000000004)(196,41.75428571428571)(197,23.19066666666667)(198,34.786)(199,19.877714285714287)(200,46.38133333333334)(201,46.38133333333334)(202,97.68966666666667)(203,55.89725)(204,19.877714285714287)(205,23.19066666666667)(206,97.83466666666668)(207,112.0445)(208,34.786)(209,97.91766666666668)(210,34.786)(211,46.38133333333334)(212,48.99816666666667)(213,46.38133333333334)(214,46.38133333333334)(215,98.07433333333334)(216,69.572)(217,46.38133333333334)(218,46.38133333333334)(219,73.6495)(220,34.786)(221,17.393)(222,86.66099999999999)(223,49.189166666666665)(224,59.049400000000006)(225,23.19066666666667)(226,27.8288)(227,73.89225)(228,56.51474999999999)(229,34.786)(230,90.51280000000001)(231,59.1846)(232,34.786)(233,102.6558888888889)(234,113.324)(235,23.19066666666667)(236,74.09700000000001)(237,59.2904)(238,64.83885714285715)(239,49.43683333333333)(240,49.447333333333326)(241,75.727)(242,74.21875)(243,17.393)(244,128.47)(245,46.38133333333334)(246,96.462875)(247,102.29966666666667)(248,113.9385)(249,23.19066666666667)(250,46.38133333333334)(251,122.99179999999998)(252,34.786)(253,46.38133333333334)(254,57.12125)(255,57.137)(256,34.786)(257,34.786)(258,46.38133333333334)(259,27.8288)(260,74.67099999999999)(261,46.38133333333334)(262,49.7945)(263,59.768600000000006)(264,59.7856)(265,74.74775)(266,46.38133333333334)(267,59.818000000000005)(268,34.786)(269,34.786)(270,59.86279999999999)(271,27.8288)(272,46.38133333333334)(273,59.9172)(274,37.46625)(275,46.38133333333334)(276,49.980333333333334)(277,65.83771428571428)(278,42.869285714285716)(279,92.2212)(280,100.12566666666667)(281,46.38133333333334)(282,60.1078)(283,27.8288)(284,60.1444)(285,50.131)(286,60.1722)(287,34.786)(288,34.786)(289,34.786)(290,75.29650000000001)(291,75.30924999999999)(292,66.21257142857142)(293,46.38133333333334)(294,46.38133333333334)(295,60.29200000000001)(296,75.39025000000001)(297,116.015)(298,92.824)(299,92.84079999999999)(300,46.38133333333334)(301,69.572)(302,75.45425)(303,75.47375)(304,60.388400000000004)(305,77.471)(306,46.38133333333334)(307,34.786)(308,77.55566666666668)(309,100.76966666666665)(310,46.38133333333334)(311,27.8288)(312,104.90516666666667)(313,100.88466666666666)(314,93.26280000000001)(315,23.19066666666667)(316,46.38133333333334)(317,69.572)(318,34.786)(319,43.30985714285714)(320,60.6656)(321,69.572)(322,46.38133333333334)(323,69.572)(324,93.6292)(325,69.572)(326,46.38133333333334)(327,34.786)(328,46.38133333333334)(329,69.572)(330,69.572)(331,34.786)(332,34.786)(333,69.572)(334,46.38133333333334)(335,76.05225)(336,69.572)(337,34.786)(338,46.38133333333334)(339,34.786)(340,69.572)(341,69.572)(342,69.572)(343,69.572)(344,69.572)(345,46.38133333333334)(346,69.572)(347,69.572)(348,69.572)(349,46.38133333333334)(350,69.572)
}; % plot of the second histogram
\addplot[color=cyan] coordinates{(0,8.304818181818183)(1,16.042260869565215)(2,15.855534090909094)(3,18.064939759036147)(4,21.025368055555553)(5,14.299417322834648)(6,18.883865030674855)(7,17.884095238095238)(8,16.60700653594771)(9,11.551145251396647)(10,13.433154838709678)(11,18.274335365853663)(12,25.025474576271183)(13,23.490018292682926)(14,16.156351485148516)(15,22.634333333333327)(16,22.5277094972067)(17,21.272142045454547)(18,21.06737185929648)(19,26.806178160919544)(20,26.219038888888885)(21,26.964402366863908)(22,27.303926553672316)(23,26.81178235294118)(24,27.686488235294114)(25,31.313059782608697)(26,29.70074556213018)(27,31.628343949044584)(28,38.688224489795914)(29,40.50733888888891)(30,27.432450000000006)(31,34.50904929577464)(32,36.94477952755906)(33,34.16421774193548)(34,30.41474285714286)(35,30.05533333333333)(36,28.852782608695648)(37,25.756886956521743)(38,38.32958762886597)(39,29.075765957446798)(40,28.478133333333336)(41,26.585861702127662)(42,39.82317647058824)(43,30.083732394366198)(44,44.158137499999995)(45,38.237210526315785)(46,30.843769230769233)(47,32.54048717948718)(48,37.17683076923077)(49,33.93001265822784)(50,29.651402439024395)(51,36.92791525423729)(52,36.19608450704225)(53,24.090265624999997)(54,36.88803896103896)(55,42.29665454545455)(56,34.699098591549294)(57,34.89061194029851)(58,43.3865606060606)(59,30.18447826086956)(60,27.4009649122807)(61,56.71843333333333)(62,28.784762711864413)(63,42.89645454545455)(64,59.38519354838709)(65,44.61901785714285)(66,34.950625)(67,37.99428813559322)(68,27.259706896551727)(69,39.14655555555555)(70,32.139219512195126)(71,43.4066724137931)(72,30.4428717948718)(73,33.10431249999999)(74,41.06623809523809)(75,41.10669047619048)(76,40.444017857142846)(77,38.84070833333334)(78,32.43165853658537)(79,39.75999999999999)(80,28.74048648648649)(81,29.55933333333333)(82,34.372064516129036)(83,33.36313888888889)(84,40.896555555555544)(85,22.136305555555552)(86,39.87405)(87,49.65281481481481)(88,31.573666666666668)(89,38.96745833333333)(90,31.634285714285717)(91,46.85509999999999)(92,51.141333333333336)(93,29.731703703703708)(94,30.895038461538466)(95,13.521315789473682)(96,24.12409090909091)(97,29.039347826086956)(98,50.367125)(99,31.008807692307695)(100,45.19816666666666)(101,36.68968181818182)(102,42.50847368421053)(103,37.83916)(104,44.95122222222222)(105,35.201913043478264)(106,47.655529411764704)(107,33.034333333333336)(108,45.219857142857144)(109,43.19681818181818)(110,59.402782608695645)(111,39.736599999999996)(112,35.7518)(113,39.726882352941175)(114,47.68275)(115,18.511071428571427)(116,10.009666666666666)(117,44.808833333333325)(118,16.2176875)(119,52.14246153846154)(120,28.354052631578952)(121,21.030578947368422)(122,28.556214285714287)(123,30.767692307692307)(124,25.019125000000003)(125,56.44340909090909)(126,56.49545454545454)(127,23.687363636363635)(128,49.46064285714287)(129,30.88230769230769)(130,48.80157142857143)(131,38.78171428571428)(132,20.08669230769231)(133,33.535833333333336)(134,39.37771428571428)(135,45.962722222222226)(136,58.49747368421051)(137,46.03261111111112)(138,38.17888888888889)(139,44.88388888888889)(140,44.90266666666666)(141,51.20357894736842)(142,43.748)(143,32.82647619047619)(144,45.99173333333333)(145,48.986117647058826)(146,21.33468421052632)(147,32.8615)(148,33.82150000000001)(149,17.541600000000003)(150,23.935000000000002)(151,36.65560000000001)(152,17.5684)(153,33.92250000000001)(154,13.346222222222222)(155,23.982000000000003)(156,45.28977777777778)(157,37.06436363636363)(158,24.00190909090909)(159,12.0116)(160,68.03966666666668)(161,26.425599999999996)(162,20.019333333333332)(163,10.009666666666666)(164,26.4563)(165,26.4686)(166,17.15942857142857)(167,17.15942857142857)(168,37.85814285714286)(169,20.019333333333332)(170,69.399125)(171,37.910000000000004)(172,33.181375)(173,30.029)(174,30.029)(175,37.97128571428571)(176,17.15942857142857)(177,33.252625)(178,38.01314285714285)(179,33.273625)(180,24.0232)(181,68.76800000000001)(182,17.15942857142857)(183,51.626374999999996)(184,33.3395)(185,44.47083333333333)(186,33.3635)(187,46.010444444444445)(188,24.0232)(189,38.18285714285714)(190,30.029)(191,38.225)(192,17.15942857142857)(193,20.019333333333332)(194,62.626555555555555)(195,30.029)(196,15.0145)(197,53.653800000000004)(198,20.019333333333332)(199,13.346222222222222)(200,24.0232)(201,44.781666666666666)(202,30.029)(203,15.0145)(204,24.0232)(205,24.0232)(206,20.019333333333332)(207,33.66325)(208,24.0232)(209,56.8292)(210,59.862)(211,44.940333333333335)(212,24.0232)(213,38.55457142857143)(214,69.96266666666666)(215,30.029)(216,33.77)(217,24.0232)(218,30.037555555555556)(219,54.083000000000006)(220,30.029)(221,45.101)(222,15.0145)(223,20.019333333333332)(224,45.1465)(225,20.019333333333332)(226,33.885625)(227,63.698555555555544)(228,33.9075)(229,33.930625)(230,45.25116666666667)(231,30.17533333333333)(232,20.019333333333332)(233,20.019333333333332)(234,38.847142857142856)(235,30.029)(236,20.019333333333332)(237,27.2217)(238,54.4636)(239,45.396166666666666)(240,54.484)(241,38.931285714285714)(242,53.126)(243,30.029)(244,38.963428571428565)(245,30.029)(246,38.985285714285716)(247,30.029)(248,71.02)(249,20.019333333333332)(250,54.6644)(251,30.029)(252,24.0232)(253,34.19925)(254,24.0232)(255,61.05757142857143)(256,20.019333333333332)(257,54.81420000000001)(258,24.0232)(259,30.029)(260,24.0232)(261,20.019333333333332)(262,24.0232)(263,54.925)(264,30.029)(265,24.0232)(266,30.029)(267,30.029)(268,20.019333333333332)(269,30.029)(270,55.0304)(271,24.0232)(272,30.029)(273,13.346222222222222)(274,24.0232)(275,20.019333333333332)(276,20.019333333333332)(277,30.029)(278,17.15942857142857)(279,20.019333333333332)(280,61.74228571428571)(281,46.038000000000004)(282,24.0232)(283,30.029)(284,72.16133333333333)(285,55.322799999999994)(286,46.1325)(287,24.0232)(288,55.3874)(289,39.58542857142857)(290,46.1935)(291,24.0232)(292,46.211999999999996)(293,55.472)(294,30.029)(295,30.029)(296,24.0232)(297,24.0232)(298,30.029)(299,30.029)(300,17.15942857142857)(301,46.33633333333333)(302,46.36683333333334)(303,17.15942857142857)(304,43.6616)(305,30.029)(306,30.029)(307,39.821285714285715)(308,72.97466666666666)(309,30.029)(310,17.15942857142857)(311,24.0232)(312,62.656)(313,20.019333333333332)(314,24.0232)(315,55.8996)(316,30.029)(317,20.019333333333332)(318,30.029)(319,30.029)(320,24.0232)(321,24.0232)(322,24.0232)(323,20.019333333333332)(324,30.029)(325,30.029)(326,24.0232)(327,24.0232)(328,20.019333333333332)(329,24.0232)(330,73.49833333333333)(331,20.019333333333332)(332,40.11442857142857)(333,30.029)(334,20.019333333333332)(335,30.029)(336,30.029)(337,30.029)(338,30.029)(339,30.029)(340,20.019333333333332)(341,24.0232)(342,30.029)(343,24.0232)(344,24.0232)(345,24.0232)(346,30.029)(347,30.029)(348,30.029)(349,24.0232)(350,30.029)
}; % plot of the third histogram
\addplot[color=magenta] coordinates{(0,16.09971153846154)(1,14.36415254237288)(2,18.99049484536083)(3,24.29785869565217)(4,29.36102189781022)(5,22.83799280575539)(6,15.84625342465754)(7,16.642096153846154)(8,22.211677966101693)(9,21.430827225130887)(10,18.54215463917526)(11,25.40992857142857)(12,29.24852325581395)(13,34.82560317460318)(14,26.62827586206897)(15,25.42711363636364)(16,23.81046511627907)(17,24.89794350282486)(18,26.065590909090908)(19,31.14390607734807)(20,32.408176767676764)(21,41.51763157894737)(22,36.33009782608695)(23,36.66896666666667)(24,46.0049574468085)(25,47.38537037037037)(26,55.14731351351352)(27,55.12931862745095)(28,53.5579175257732)(29,51.80767977528091)(30,41.80491860465117)(31,50.30583969465649)(32,44.97268571428572)(33,43.971450000000004)(34,45.96115217391304)(35,42.41916000000001)(36,51.99513953488371)(37,42.25753225806451)(38,33.474781690140844)(39,36.59858510638298)(40,38.54299999999999)(41,48.59312676056337)(42,53.758108695652176)(43,48.23758536585366)(44,43.92942666666666)(45,45.366724999999995)(46,54.37130263157895)(47,45.66670769230769)(48,42.387662162162165)(49,51.2599295774648)(50,43.17127692307692)(51,53.03477333333333)(52,43.65806944444445)(53,43.7830294117647)(54,50.44626086956521)(55,57.15663793103447)(56,78.40167924528302)(57,55.346444444444444)(58,49.83632857142857)(59,54.53887499999999)(60,66.66118461538461)(61,42.11242372881355)(62,57.44505454545453)(63,49.385140625000005)(64,39.23193055555556)(65,59.179548387096766)(66,36.61123529411765)(67,47.88773076923077)(68,48.76741379310345)(69,50.19837735849057)(70,54.32029870129871)(71,53.41603773584906)(72,34.33016666666667)(73,57.8956304347826)(74,72.15378431372551)(75,63.064603773584906)(76,46.90421739130435)(77,54.539269230769236)(78,55.75760000000001)(79,48.510749999999994)(80,56.83742105263158)(81,48.07567307692307)(82,59.769128205128204)(83,33.58817647058824)(84,44.65335135135134)(85,65.09267857142858)(86,56.951114285714276)(87,45.26734482758621)(88,42.35103225806452)(89,45.275999999999996)(90,52.119657142857136)(91,43.777866666666675)(92,50.52469230769231)(93,34.73885714285714)(94,59.11003571428572)(95,61.31318518518518)(96,62.582904761904764)(97,51.98527272727272)(98,66.59866666666667)(99,57.14346153846154)(100,78.89576190476191)(101,66.28132000000001)(102,63.74026923076922)(103,51.260344827586216)(104,46.3462380952381)(105,73.1724)(106,64.65382608695653)(107,52.63688)(108,69.10962500000001)(109,90.9775909090909)(110,63.812615384615384)(111,64.69191304347827)(112,67.35976470588236)(113,67.65027272727274)(114,48.773259259259255)(115,44.26263636363637)(116,59.554120000000005)(117,83.50829166666665)(118,54.55357142857144)(119,57.2863)(120,74.4813)(121,52.08945454545455)(122,47.75291666666667)(123,49.836173913043474)(124,77.55176470588235)(125,62.79261904761905)(126,61.191066666666664)(127,79.2277619047619)(128,62.14908333333333)(129,71.03342857142856)(130,46.40804761904762)(131,79.26171428571428)(132,75.24668965517242)(133,71.06419047619048)(134,67.49211764705882)(135,53.47586666666667)(136,79.32180952380952)(137,82.52687500000002)(138,44.31568181818182)(139,64.936)(140,65.6942142857143)(141,79.3707619047619)(142,67.53176470588235)(143,75.77490909090909)(144,83.65699999999998)(145,93.41312500000001)(146,69.55715789473685)(147,87.68333333333332)(148,57.471304347826084)(149,65.02726666666666)(150,57.378882352941176)(151,63.82533333333335)(152,51.34426315789474)(153,77.78776470588234)(154,48.779399999999995)(155,65.04213333333334)(156,79.51014285714284)(157,61.70292307692308)(158,47.184588235294115)(159,60.984500000000004)(160,57.29557142857143)(161,71.849625)(162,57.29557142857143)(163,69.71057142857143)(164,57.411058823529416)(165,42.21778947368421)(166,65.0706)(167,75.08376923076924)(168,76.67479999999999)(169,65.07533333333333)(170,80.2138)(171,66.84483333333334)(172,57.29557142857143)(173,80.2138)(174,75.11492307692308)(175,66.84483333333334)(176,72.92163636363637)(177,80.2138)(178,80.2138)(179,66.84483333333334)(180,81.3925)(181,53.47586666666667)(182,88.56953846153846)(183,61.70292307692308)(184,80.2138)(185,66.84483333333334)(186,80.2138)(187,72.92163636363637)(188,88.60138461538463)(189,95.996)(190,88.82772727272727)(191,53.47586666666667)(192,61.70292307692308)(193,72.92163636363637)(194,66.84483333333334)(195,72.92163636363637)(196,57.29557142857143)(197,66.84483333333334)(198,88.66476923076924)(199,66.84483333333334)(200,66.84483333333334)(201,66.84483333333334)(202,72.92163636363637)(203,80.2138)(204,80.2138)(205,72.92163636363637)(206,75.20646153846154)(207,81.47708333333334)(208,80.2138)(209,72.92163636363637)(210,80.2138)(211,72.92163636363637)(212,80.2138)(213,80.2138)(214,72.92163636363637)(215,80.2138)(216,75.2346153846154)(217,72.92163636363637)(218,72.92163636363637)(219,80.2138)(220,66.84483333333334)(221,88.93481818181819)(222,88.93927272727272)(223,96.21316666666667)(224,88.73526666666666)(225,88.838)(226,80.2138)(227,72.92163636363637)(228,81.55658333333334)(229,66.84483333333334)(230,72.92163636363637)(231,80.2138)(232,80.2138)(233,80.2138)(234,72.92163636363637)(235,81.58525)(236,80.2138)(237,80.2138)(238,80.2138)(239,72.92163636363637)(240,72.92163636363637)(241,61.70292307692308)(242,80.2138)(243,81.61225)(244,72.92163636363637)(245,75.3406923076923)(246,80.21379999999999)(247,72.92163636363637)(248,89.0480909090909)(249,72.92163636363637)(250,69.96942857142857)(251,72.92163636363637)(252,80.2138)(253,72.92163636363637)(254,72.92163636363637)(255,89.06672727272728)(256,80.2138)(257,66.84483333333334)(258,80.2138)(259,80.2138)(260,80.2138)(261,72.92163636363637)(262,80.2138)(263,66.84483333333334)(264,80.21379999999999)(265,72.92163636363637)(266,89.10372727272727)(267,80.2138)(268,89.10836363636363)(269,66.84483333333334)(270,70.01957142857142)(271,72.92163636363637)(272,81.69608333333333)(273,89.125)(274,80.21379999999999)(275,89.13618181818183)(276,89.13872727272728)(277,72.92163636363637)(278,72.92163636363637)(279,80.2138)(280,96.60333333333334)(281,102.91515384615384)(282,80.2138)(283,89.16172727272728)(284,89.19384615384615)(285,81.73575000000001)(286,72.92163636363637)(287,80.2138)(288,89.17627272727273)(289,66.84483333333334)(290,72.92163636363637)(291,80.2138)(292,80.2138)(293,96.67716666666666)(294,89.19381818181817)(295,80.2138)(296,75.47646153846155)(297,72.92163636363637)(298,72.92163636363637)(299,72.92163636363637)(300,80.2138)(301,80.2138)(302,66.84483333333334)(303,81.78258333333333)(304,80.2138)(305,89.22218181818182)(306,89.22363636363637)(307,72.92163636363637)(308,80.2138)(309,80.2138)(310,80.2138)(311,72.92163636363637)(312,95.75000000000001)(313,96.762)(314,80.2138)(315,72.92163636363637)(316,89.2459090909091)(317,72.92163636363637)(318,80.2138)(319,80.2138)(320,72.92163636363637)(321,80.2138)(322,89.25663636363637)(323,80.2138)(324,80.2138)(325,72.92163636363637)(326,82.98242857142857)(327,80.2138)(328,66.84483333333334)(329,80.2138)(330,66.84483333333334)(331,72.92163636363637)(332,80.2138)(333,72.92163636363637)(334,72.92163636363637)(335,96.87366666666668)(336,80.2138)(337,80.2138)(338,80.2138)(339,72.92163636363637)(340,80.2138)(341,80.2138)(342,72.92163636363637)(343,72.92163636363637)(344,80.2138)(345,72.92163636363637)(346,80.2138)
}; % plot of the fourth histogram
\addplot[color=red] coordinates{(0,11.379632653061224)(1,18.472347826086956)(2,20.64680412371134)(3,27.577274809160304)(4,21.127710144927534)(5,23.2415197368421)(6,23.43253333333333)(7,22.100015228426393)(8,21.1150206185567)(9,18.17713690476191)(10,29.51657978723404)(11,28.344860962566848)(12,38.071306451612905)(13,33.65610837438423)(14,27.281029787234043)(15,42.99953211009173)(16,34.164829268292685)(17,36.609525581395346)(18,34.18444559585492)(19,40.618994897959176)(20,47.54727461139896)(21,46.85754661016947)(22,47.76181730769231)(23,48.19938860103627)(24,55.40957219251337)(25,60.93206666666664)(26,63.434343915343916)(27,59.821519607843136)(28,68.30449197860965)(29,54.45697652582161)(30,61.73506111111112)(31,50.537741935483865)(32,42.0496879432624)(33,43.67139705882354)(34,54.69199999999999)(35,57.963446153846164)(36,52.75934513274336)(37,36.625017391304354)(38,47.77211650485436)(39,37.91924509803922)(40,45.4135376344086)(41,42.92017073170732)(42,44.044261363636366)(43,47.63041891891892)(44,71.5329240506329)(45,53.389511627906984)(46,48.78359210526315)(47,50.80759999999999)(48,56.776790123456806)(49,45.31635897435896)(50,36.98741860465117)(51,48.46180821917809)(52,35.31021428571429)(53,54.087428571428575)(54,59.012)(55,58.20700000000001)(56,53.19414285714284)(57,59.148811594202904)(58,57.21551612903227)(59,57.125983050847466)(60,33.49427586206896)(61,45.74835483870968)(62,48.4055)(63,48.90471014492755)(64,46.16279220779221)(65,55.130172413793105)(66,56.28578333333334)(67,49.50249180327869)(68,39.7175)(69,49.22903076923078)(70,51.99687692307692)(71,64.79880303030302)(72,64.47381034482757)(73,64.74074545454543)(74,63.22914516129031)(75,52.13287931034482)(76,53.054228070175434)(77,45.22560784313726)(78,52.8995744680851)(79,60.71281578947369)(80,47.5130243902439)(81,44.2141)(82,41.448470588235296)(83,50.67323809523809)(84,55.678514285714286)(85,81.83570270270273)(86,69.97133333333333)(87,52.047029411764704)(88,69.644)(89,57.580216216216215)(90,55.3295625)(91,45.493870967741934)(92,39.17888888888889)(93,59.03933333333334)(94,59.044399999999996)(95,65.60962962962964)(96,58.93125925925926)(97,63.64904000000001)(98,65.6282962962963)(99,47.32265384615384)(100,63.29685714285715)(101,50.39778571428572)(102,72.35886363636365)(103,75.1334230769231)(104,68.36638888888889)(105,63.689080000000004)(106,53.07733333333333)(107,68.3715)(108,61.2493076923077)(109,63.33899999999999)(110,64.77668421052631)(111,77.11652173913043)(112,56.885071428571415)(113,89.10950000000001)(114,72.40763636363637)(115,61.544399999999996)(116,61.545500000000004)(117,69.83603571428571)(118,52.29959259259259)(119,63.37428571428571)(120,75.22130769230769)(121,73.9455)(122,78.24276)(123,76.34071428571428)(124,85.06147826086955)(125,84.532)(126,63.7594)(127,73.73558620689654)(128,51.302416666666666)(129,64.21609090909091)(130,61.565549999999995)(131,53.53660869565218)(132,72.47595454545454)(133,92.83116)(134,88.58683333333333)(135,58.63938095238096)(136,80.74363636363636)(137,85.58832000000001)(138,66.45354166666667)(139,61.344730769230765)(140,58.88666666666668)(141,74.03895833333334)(142,85.16713043478262)(143,67.55168965517242)(144,84.63276190476192)(145,49.85890625)(146,82.35915384615386)(147,56.54836)(148,93.11965217391307)(149,55.253894736842106)(150,72.46282352941178)(151,74.41852631578948)(152,58.916666666666664)(153,61.59715)(154,61.597950000000004)(155,64.8415789473684)(156,58.32355555555556)(157,64.84357894736843)(158,65.614)(159,65.614)(160,65.614)(161,65.614)(162,61.75435294117648)(163,58.32355555555556)(164,58.323555555555544)(165,68.4558888888889)(166,61.75435294117648)(167,65.614)(168,69.98826666666666)(169,69.98826666666666)(170,65.61399999999999)(171,74.98742857142857)(172,65.614)(173,69.98826666666668)(174,69.98826666666666)(175,69.98826666666668)(176,77.038)(177,69.98826666666668)(178,74.98742857142858)(179,88.4725)(180,69.98826666666668)(181,74.98742857142858)(182,74.98742857142858)(183,69.98826666666666)(184,74.98742857142857)(185,69.98826666666668)(186,69.98826666666666)(187,61.75435294117648)(188,74.98742857142857)(189,74.98742857142858)(190,69.98826666666668)(191,82.20253333333334)(192,74.98742857142858)(193,77.0668125)(194,77.06875000000001)(195,77.07075)(196,74.98742857142857)(197,69.98826666666666)(198,74.98742857142858)(199,74.98742857142857)(200,65.61399999999999)(201,69.98826666666666)(202,74.98742857142858)(203,74.98742857142858)(204,74.98742857142858)(205,74.98742857142858)(206,74.98742857142858)(207,74.98742857142857)(208,69.98826666666666)(209,83.35611764705881)(210,88.56775)(211,74.98742857142858)(212,82.23446666666665)(213,74.98742857142858)(214,74.98742857142857)(215,74.98742857142857)(216,65.61399999999999)(217,82.2426)(218,69.98826666666668)(219,74.98742857142857)(220,69.98826666666666)(221,88.602625)(222,69.98826666666668)(223,69.98826666666666)(224,74.98742857142858)(225,74.98742857142858)(226,69.98826666666668)(227,65.614)(228,74.98742857142858)(229,77.1221875)(230,74.98742857142858)(231,74.98742857142858)(232,74.98742857142858)(233,83.42670588235295)(234,69.98826666666668)(235,77.12943750000001)(236,74.98742857142858)(237,69.98826666666668)(238,74.98742857142858)(239,77.13568749999999)(240,74.98742857142858)(241,74.98742857142858)(242,74.98742857142857)(243,74.98742857142858)(244,74.98742857142857)(245,74.98742857142858)(246,74.98742857142858)(247,82.29140000000001)(248,74.98742857142857)(249,82.29533333333333)(250,69.98826666666668)(251,69.98826666666668)(252,82.29913333333333)(253,74.98742857142858)(254,74.98742857142858)(255,69.98826666666668)(256,74.98742857142858)(257,74.98742857142857)(258,74.98742857142858)(259,74.98742857142858)(260,74.98742857142858)(261,77.16793750000001)(262,74.98742857142858)(263,77.17125)(264,74.98742857142858)(265,74.98742857142858)(266,74.98742857142858)(267,74.98742857142858)(268,74.98742857142858)(269,77.18243749999999)(270,74.98742857142857)(271,69.98826666666666)(272,74.98742857142857)(273,74.98742857142858)(274,74.98742857142857)(275,74.98742857142858)(276,74.98742857142857)(277,74.98742857142858)(278,74.98742857142857)(279,69.98826666666668)(280,69.98826666666668)(281,69.98826666666668)(282,74.98742857142858)(283,69.98826666666668)(284,74.98742857142858)(285,74.98742857142857)(286,74.98742857142857)(287,74.98742857142858)(288,69.98826666666668)(289,74.98742857142858)(290,72.67535294117647)(291,65.61399999999999)(292,74.98742857142858)(293,74.98742857142858)(294,74.98742857142858)(295,74.98742857142858)(296,69.98826666666668)(297,69.98826666666668)(298,74.98742857142858)(299,74.98742857142858)(300,74.98742857142858)(301,74.98742857142858)(302,74.98742857142858)(303,74.98742857142857)(304,69.98826666666666)(305,74.98742857142858)(306,69.98826666666668)(307,74.98742857142858)(308,74.98742857142858)(309,74.98742857142858)(310,82.39246666666666)(311,74.98742857142858)(312,82.3956)(313,74.98742857142858)(314,74.98742857142858)(315,74.98742857142858)(316,74.98742857142857)(317,74.98742857142858)(318,82.4044)(319,74.98742857142858)(320,77.257375)(321,74.98742857142858)(322,65.614)(323,69.98826666666666)(324,74.98742857142858)(325,74.98742857142857)(326,74.98742857142857)(327,74.98742857142858)(328,74.98742857142857)(329,74.98742857142858)(330,74.98742857142858)(331,74.98742857142857)(332,74.98742857142857)(333,65.61399999999999)(334,74.98742857142858)(335,74.98742857142858)(336,69.98826666666668)(337,74.98742857142858)
}; % plot of the fifth histogram
\addplot[color=orange] coordinates{(0,14.332038461538463)(1,13.951193548387096)(2,24.206256637168142)(3,21.274736)(4,17.616559748427676)(5,17.87612865497076)(6,17.90971717171717)(7,21.793377358490563)(8,18.092652968036536)(9,18.573636363636364)(10,14.699985148514854)(11,16.52824444444444)(12,31.92035526315789)(13,27.806615740740735)(14,17.96785507246377)(15,28.931937500000004)(16,32.034908256880726)(17,24.528044117647056)(18,20.343958333333344)(19,29.89253658536586)(20,36.51619730941704)(21,35.86129807692308)(22,43.41105820105818)(23,36.42224528301886)(24,41.093964125560525)(25,42.07581999999999)(26,40.50117213114754)(27,50.2669315789474)(28,46.64321666666663)(29,49.36393908629444)(30,50.160070707070716)(31,47.97475796178346)(32,43.721755244755265)(33,47.011737288135585)(34,44.103300653594765)(35,47.20477678571429)(36,40.89118064516129)(37,41.016600000000004)(38,30.252882352941178)(39,42.617961538461536)(40,48.7025411764706)(41,37.72672222222222)(42,53.16534615384615)(43,50.00625352112676)(44,63.973680555555546)(45,61.95295833333333)(46,42.58327397260275)(47,44.682863013698636)(48,49.57034666666668)(49,58.67775757575759)(50,52.90605063291139)(51,42.48831168831169)(52,68.4474857142857)(53,50.41772727272727)(54,65.4345)(55,57.6763)(56,67.60733802816901)(57,65.53364062499999)(58,46.70807462686567)(59,57.01877464788732)(60,47.30182539682539)(61,75.27893548387097)(62,50.04226923076923)(63,65.31935802469137)(64,59.263708860759486)(65,57.52038333333334)(66,64.01200000000001)(67,50.82649295774649)(68,64.4484411764706)(69,56.670213114754105)(70,65.1023076923077)(71,66.11975438596491)(72,64.81464705882352)(73,62.33950793650792)(74,51.702246575342464)(75,59.12512499999999)(76,55.93454838709678)(77,62.5377358490566)(78,72.08867391304348)(79,53.878199999999985)(80,54.68807272727272)(81,43.96935)(82,68.68564864864864)(83,66.3910930232558)(84,56.749322580645156)(85,63.701228571428565)(86,69.67990625000002)(87,63.49633333333333)(88,60.67155172413793)(89,68.21631428571429)(90,66.89632258064516)(91,63.75197142857144)(92,69.7355625)(93,65.29494871794871)(94,71.55010344827586)(95,74.94126470588235)(96,65.1888888888889)(97,58.126090909090905)(98,82.0557878787879)(99,47.581243243243236)(100,69.47384615384615)(101,73.81288461538462)(102,58.70159999999999)(103,55.034749999999995)(104,62.143027777777775)(105,74.91956249999998)(106,56.489088235294105)(107,79.92271875)(108,82.71765517241379)(109,60.75465517241379)(110,76.85332)(111,64.05016666666667)(112,67.77161538461539)(113,65.91091176470589)(114,76.87755999999999)(115,66.27789655172414)(116,71.7924827586207)(117,80.08303571428573)(118,67.9600303030303)(119,60.086656250000004)(120,74.78463333333333)(121,73.64697297297297)(122,77.17729629629629)(123,73.98592307692307)(124,65.13925)(125,80.19767857142857)(126,60.81275862068965)(127,77.48162068965517)(128,75.29296875)(129,71.96006896551725)(130,83.48448)(131,83.30388888888886)(132,74.98176666666667)(133,65.247)(134,58.814)(135,83.34314814814815)(136,64.29705714285714)(137,93.34292105263155)(138,93.52954838709678)(139,77.62979310344828)(140,67.39551612903226)(141,63.03135714285714)(142,69.66)(143,75.08563333333332)(144,83.61084000000001)(145,77.92838709677419)(146,77.44222222222223)(147,64.40071428571427)(148,64.28223333333334)(149,63.055821428571434)(150,70.62772000000001)(151,83.53514814814814)(152,76.77504347826087)(153,56.74261764705883)(154,76.78326086956523)(155,72.84204545454546)(156,59.352777777777774)(157,59.35277777777778)(158,69.675)(159,87.74268181818181)(160,73.603)(161,66.77187500000001)(162,66.771875)(163,72.84204545454546)(164,76.31071428571428)(165,69.675)(166,69.675)(167,72.84204545454544)(168,80.12625)(169,76.31071428571428)(170,80.12625)
}; % plot of the first histogram
\legend{50\%, 60\%, 70\%, 80\%, 90\%, 100\%} % legend of the plot

\end{axis}
\end{tikzpicture}

%% file: plots/aware_fraction/injury_levels.tex
\begin{tikzpicture}[scale=0.75]
\begin{axis}[
ybar, % style of the histogram: clustered columns
width=\columnwidth, % width of the plot
height=8cm, % height of the plot
title={Injury Levels  - Mobile application},
ylabel={Amount}, % label for the y-axis
symbolic x coords={HEALTHY,MINOR,MODERATE,SERIOUS}, % labels for the x-ticks
xtick=data, % position the x-ticks at the data points
ymin=0, % minimum value of the y-axis
legend style={at={(1,1)}, anchor=north east}, % position of the legend
legend cell align=left, % alignment of the legend cells
ymajorgrids=true, % display major grids
bar width=0.15cm % width of the bars
]
\addplot[color=blue, fill] coordinates{(HEALTHY,16459)(MINOR,10648)(MODERATE,2090)(SERIOUS,659)
}; % plot of the first histogram
\addplot[color=teal, fill] coordinates{(HEALTHY,16679)(MINOR,10612)(MODERATE,1914)(SERIOUS,616)
}; % plot of the second histogram
\addplot[color=cyan, fill] coordinates{(HEALTHY,16787)(MINOR,10648)(MODERATE,1698)(SERIOUS,641)
}; % plot of the third histogram
\addplot[color=magenta, fill] coordinates{(HEALTHY,16404)(MINOR,11031)(MODERATE,1661)(SERIOUS,606)
}; % plot of the fourth histogram
\addplot[color=red, fill] coordinates{(HEALTHY,15966)(MINOR,11413)(MODERATE,1572)(SERIOUS,637)
}; % plot of the fifth histogram
\addplot[color=orange, fill] coordinates{(HEALTHY,15051)(MINOR,12203)(MODERATE,1511)(SERIOUS,663)
}; % plot of the first histogram
\legend{50\%, 60\%, 70\%, 80\%, 90\%, 100\%} % legend of the plot
\end{axis}
\end{tikzpicture}

\begin{tikzpicture}[scale=0.75]
\begin{axis}[
ybar, % style of the histogram: clustered columns
width=\columnwidth, % width of the plot
height=8cm, % height of the plot
ylabel={Amount}, % label for the y-axis
symbolic x coords={SEVERE,CRITICAL,FATAL}, % labels for the x-ticks
xtick=data, % position the x-ticks at the data points
ymin=0, % minimum value of the y-axis
legend style={at={(1,1)}, anchor=north east}, % position of the legend
legend cell align=left, % alignment of the legend cells
ymajorgrids=true, % display major grids
bar width=0.15cm % width of the bars
]
\addplot[color=blue, fill] coordinates{(SEVERE,128)(CRITICAL,12)(FATAL,4)
}; % plot of the first histogram
\addplot[color=teal, fill] coordinates{(SEVERE,154)(CRITICAL,22)(FATAL,3)
}; % plot of the second histogram
\addplot[color=cyan, fill] coordinates{(SEVERE,168)(CRITICAL,46)(FATAL,12)
}; % plot of the third histogram
\addplot[color=magenta, fill] coordinates{(SEVERE,212)(CRITICAL,63)(FATAL,23)
}; % plot of the fourth histogram
\addplot[color=red, fill] coordinates{(SEVERE,274)(CRITICAL,98)(FATAL,40)
}; % plot of the fifth histogram
\addplot[color=orange, fill] coordinates{(SEVERE,347)(CRITICAL,149)(FATAL,76)
}; % plot of the first histogram
\legend{50\%, 60\%, 70\%, 80\%, 90\%, 100\%} % legend of the plot
\end{axis}
\end{tikzpicture}

%% file: plots/accessible_exits/evac_time.tex
\begin{tikzpicture}[scale=0.75]

% Set the axis labels and title
\begin{axis}[    xlabel={Time [s]},    ylabel={[s]},    title={Evacuation Time - Accessible exits},   legend style={at={(0,1)}, anchor=north west},]

\addplot[color=teal] coordinates{(0,15.556666666666665)(1,23.16391836734694)(2,13.852810344827585)(3,22.63127272727273)(4,18.463985714285716)(5,20.001037383177568)(6,22.226064814814816)(7,27.689096491228067)(8,28.66907092198581)(9,13.831083916083914)(10,23.103787671232876)(11,19.561604651162792)(12,19.887487500000002)(13,14.484359154929578)(14,19.874511363636366)(15,28.932358778625957)(16,19.36854411764706)(17,33.87943624161075)(18,35.3582027027027)(19,25.89837588652482)(20,30.872006289308178)(21,31.34072784810127)(22,33.80363461538463)(23,28.884793750000007)(24,42.31208917197453)(25,33.87415827338129)(26,42.5736447368421)(27,38.802745222929936)(28,34.837258064516135)(29,46.07985897435897)(30,39.701049645390064)(31,28.435230769230774)(32,39.27123931623933)(33,36.85504000000001)(34,36.646307692307694)(35,27.116174242424236)(36,28.511809160305344)(37,24.03791071428571)(38,28.242122950819677)(39,40.32548387096776)(40,31.61782105263158)(41,42.4488705882353)(42,40.1374)(43,28.22809375)(44,38.200559999999996)(45,36.13688732394367)(46,46.648506172839504)(47,42.25192647058824)(48,39.564715909090914)(49,46.63125641025641)(50,37.9084625)(51,35.00495384615384)(52,40.38381012658227)(53,19.219506329113923)(54,41.853412500000005)(55,31.332779411764704)(56,38.106750000000005)(57,60.62353424657533)(58,46.820600000000006)(59,40.19535526315789)(60,30.993434782608702)(61,45.34391358024693)(62,50.33997260273975)(63,47.74953246753246)(64,42.907266666666665)(65,30.63455714285714)(66,51.12879365079366)(67,33.29182608695652)(68,32.42013461538462)(69,22.97895)(70,52.04737096774194)(71,41.142339285714286)(72,39.209361702127666)(73,53.807624999999994)(74,60.74547887323943)(75,62.86424489795917)(76,47.39007692307692)(77,44.002142857142864)(78,35.27662499999999)(79,53.77230232558139)(80,39.34293617021277)(81,45.556204545454555)(82,31.56874358974359)(83,39.51779487179487)(84,61.59369999999999)(85,38.50996875)(86,59.05370588235296)(87,59.358153846153854)(88,36.06482352941176)(89,48.02624999999999)(90,38.4283)(91,27.999424242424244)(92,46.3338)(93,44.01933333333333)(94,54.38976470588235)(95,36.63895238095238)(96,36.12658823529412)(97,33.467565217391304)(98,100.70870000000001)(99,25.494333333333334)(100,25.281583333333334)(101,35.018772727272726)(102,69.67710000000001)(103,68.79149999999998)(104,70.46359090909091)(105,47.08973913043478)(106,61.5523)(107,48.82084210526316)(108,55.131857142857136)(109,51.328583333333334)(110,40.64305263157895)(111,45.95688888888889)(112,47.41253846153847)(113,48.2928125)(114,45.46129411764706)(115,61.955999999999996)(116,35.42046153846154)(117,23.39676923076923)(118,43.456714285714284)(119,35.166045454545454)(120,63.931411764705885)(121,51.45033333333333)(122,23.049500000000002)(123,73.19864705882352)(124,16.418444444444443)(125,51.64973333333333)(126,55.34928571428572)(127,60.479888888888894)(128,36.37047058823529)(129,14.776599999999998)(130,41.233666666666664)(131,44.18557142857143)(132,47.59323076923077)(133,64.65283333333333)(134,28.865875000000003)(135,61.901199999999996)(136,44.221785714285716)(137,68.80011111111111)(138,40.86731578947368)(139,35.553384615384616)(140,59.74969230769231)(141,77.96171428571428)(142,38.734750000000005)(143,57.814499999999995)(144,70.67472727272727)(145,47.70007692307692)(146,40.932315789473684)(147,47.72338461538462)(148,36.49823529411765)(149,83.409)(150,68.97066666666666)(151,51.903533333333336)(152,51.47266666666667)(153,25.461166666666667)(154,13.433272727272726)(155,77.65775)(156,47.797923076923084)(157,72.10084615384615)(158,30.571299999999997)(159,30.5751)(160,20.385466666666666)(161,43.687571428571424)(162,33.98166666666666)(163,43.695142857142855)(164,33.98866666666667)(165,38.675916666666666)(166,86.726)(167,97.58225000000002)(168,103.75466666666667)(169,49.25533333333333)(170,46.4368)(171,77.849375)(172,76.53399999999999)(173,29.553199999999997)(174,94.0038)(175,38.281499999999994)(176,77.467)(177,14.776599999999998)(178,24.627666666666666)(179,66.43542857142857)(180,29.553199999999997)(181,66.45285714285714)(182,103.99416666666666)(183,51.69955555555555)(184,77.56166666666667)(185,34.06666666666667)(186,38.329625)(187,61.333600000000004)(188,21.10942857142857)(189,77.618)(190,51.128166666666665)(191,76.70425)(192,102.27933333333333)(193,36.9415)(194,21.10942857142857)(195,29.553199999999997)(196,102.31766666666665)(197,77.69533333333334)(198,51.173)(199,24.627666666666666)(200,66.63342857142858)(201,93.3006)(202,98.20175)(203,14.776599999999998)(204,36.9415)(205,102.45566666666666)(206,116.7575)(207,51.238166666666665)(208,66.73371428571429)(209,69.65211111111111)(210,78.370625)(211,102.517)(212,21.10942857142857)(213,104.55383333333334)(214,153.82549999999998)(215,30.7682)(216,76.9245)(217,61.5458)(218,76.93799999999999)(219,77.967)(220,36.9415)(221,38.4815)(222,87.58511111111112)(223,102.64366666666666)(224,29.553199999999997)(225,24.627666666666666)(226,66.90542857142857)(227,78.59)(228,93.69919999999999)(229,69.89377777777776)(230,93.7336)(231,93.7408)(232,77.071)(233,77.07900000000001)(234,78.14933333333333)(235,77.09625)(236,29.553199999999997)(237,44.06171428571429)(238,29.553199999999997)(239,77.12125)(240,51.422)(241,105.02166666666666)(242,73.883)(243,36.9415)(244,77.16749999999999)(245,93.9336)(246,154.3775)(247,67.11514285714286)(248,147.766)(249,49.25533333333333)(250,61.77719999999999)(251,51.48616666666667)(252,36.9415)(253,61.794200000000004)(254,36.9415)(255,49.25533333333333)(256,67.19657142857143)(257,77.2755)(258,36.9415)(259,49.25533333333333)(260,77.3005)(261,51.53366666666667)(262,105.38816666666666)(263,36.9415)(264,49.25533333333333)(265,157.00266666666667)(266,61.8806)(267,49.25533333333333)(268,51.57866666666666)(269,24.627666666666666)(270,94.2544)(271,49.25533333333333)(272,51.59649999999999)(273,103.20366666666666)(274,73.883)(275,61.937599999999996)(276,49.25533333333333)(277,103.25766666666668)(278,90.55428571428571)(279,73.883)(280,117.98400000000001)(281,103.291)(282,36.9415)(283,118.01849999999999)(284,51.656)(285,73.883)(286,103.335)(287,73.883)(288,77.513)(289,73.883)(290,36.9415)(291,103.37866666666666)(292,36.9415)(293,51.7035)(294,118.18799999999999)(295,73.883)(296,155.153)(297,73.883)(298,62.071799999999996)(299,103.45833333333333)(300,157.682)(301,36.9415)(302,73.883)(303,73.883)(304,118.3275)(305,94.67080000000001)(306,36.9415)(307,159.07225)(308,103.546)(309,51.776666666666664)(310,127.2986)(311,118.412)(312,155.361)(313,147.766)(314,103.59199999999998)(315,49.25533333333333)(316,118.462)(317,73.883)(318,49.25533333333333)(319,49.25533333333333)(320,155.45999999999998)(321,49.25533333333333)(322,49.25533333333333)(323,73.883)(324,49.25533333333333)(325,36.9415)(326,147.766)(327,49.25533333333333)(328,147.766)(329,49.25533333333333)(330,49.25533333333333)(331,77.82875)(332,49.25533333333333)(333,147.766)(334,49.25533333333333)(335,36.9415)(336,147.766)(337,49.25533333333333)(338,147.766)
}; % plot of the first histogram
\addplot[color=blue] coordinates{(0,9.507027777777779)(1,14.240306122448981)(2,8.612172413793104)(3,14.164558441558443)(4,11.683800000000002)(5,12.71729906542056)(6,14.239629629629633)(7,17.81368421052632)(8,18.57599290780142)(9,8.989993006993007)(10,15.126527397260274)(11,12.850961240310083)(12,13.158325000000001)(13,9.606570422535214)(14,13.229829545454548)(15,19.335099236641227)(16,12.96830882352941)(17,22.807114093959722)(18,23.854871621621623)(19,17.52663120567376)(20,20.959018867924524)(21,21.281417721518988)(22,23.026685897435897)(23,19.718243750000006)(24,28.903171974522316)(25,23.1748345323741)(26,29.223618421052624)(27,26.67037579617834)(28,24.123193548387093)(29,32.012519230769236)(30,27.729574468085104)(31,20.00826153846154)(32,27.7178547008547)(33,26.131776000000002)(34,26.108576923076914)(35,19.366053030303025)(36,20.420282442748096)(37,17.284401785714287)(38,20.386245901639345)(39,29.20772043010753)(40,22.943263157894737)(41,30.85708235294118)(42,29.246677777777776)(43,20.622890625)(44,28.039266666666663)(45,26.57819718309859)(46,34.38925925925926)(47,31.212176470588226)(48,29.32945454545455)(49,34.712153846153846)(50,28.304150000000003)(51,26.15869230769231)(52,30.266835443037973)(53,14.377683544303798)(54,31.483187499999996)(55,23.573808823529415)(56,28.785047619047617)(57,45.940150684931496)(58,35.577999999999996)(59,30.56517105263158)(60,23.550695652173918)(61,34.593641975308635)(62,38.46008219178083)(63,36.553675324675325)(64,32.92026666666666)(65,23.5336)(66,39.39447619047619)(67,25.571891304347822)(68,24.950384615384614)(69,17.67435)(70,40.29593548387096)(71,31.829392857142857)(72,30.31014893617021)(73,41.70312500000001)(74,47.30171830985915)(75,49.07422448979591)(76,37.030153846153844)(77,34.37648979591837)(78,27.540895833333334)(79,42.10853488372092)(80,30.781127659574473)(81,35.69218181818182)(82,24.627384615384614)(83,30.931461538461534)(84,48.101200000000006)(85,30.11065625)(86,46.45041176470589)(87,46.64530769230768)(88,27.89194117647059)(89,37.431125)(90,29.97285)(91,21.935272727272725)(92,36.583666666666666)(93,34.538476190476196)(94,42.69629411764706)(95,28.66647619047619)(96,28.103411764705886)(97,26.208565217391307)(98,80.10735)(99,19.653555555555556)(100,19.076666666666668)(101,27.489181818181816)(102,55.3366)(103,54.608)(104,56.157272727272726)(105,37.362086956521736)(106,48.2384)(107,38.694210526315786)(108,43.52142857142858)(109,40.29458333333333)(110,32.125157894736844)(111,36.70474074074074)(112,37.260769230769235)(113,38.228375)(114,36.00164705882353)(115,49.300666666666665)(116,27.56846153846154)(117,17.786615384615384)(118,33.04557142857143)(119,27.918818181818178)(120,51.20341176470589)(121,40.64733333333333)(122,18.0045)(123,58.9214705882353)(124,11.542444444444445)(125,41.22486666666667)(126,44.204571428571434)(127,48.7675)(128,28.88964705882353)(129,10.388200000000001)(130,32.79566666666666)(131,35.15371428571429)(132,37.87807692307692)(133,51.862750000000005)(134,22.7100625)(135,49.334199999999996)(136,35.25378571428571)(137,54.90022222222222)(138,32.88747368421053)(139,28.073384615384615)(140,48.184461538461534)(141,63.48599999999999)(142,31.0609375)(143,45.762)(144,57.146727272727276)(145,38.28053846153846)(146,33.14031578947368)(147,38.34123076923077)(148,29.333941176470585)(149,68.37480000000001)(150,55.473444444444446)(151,42.0904)(152,40.86377777777778)(153,19.6655)(154,9.443818181818182)(155,62.58875)(156,38.533692307692306)(157,58.94784615384616)(158,23.6463)(159,23.6559)(160,15.786666666666667)(161,33.845)(162,26.33866666666667)(163,33.90828571428572)(164,26.381222222222224)(165,30.927000000000003)(166,71.00844444444445)(167,79.92875)(168,84.311)(169,34.62733333333333)(170,37.244099999999996)(171,63.37125)(172,59.5765)(173,20.776400000000002)(174,77.7086)(175,29.823875)(176,62.24366666666666)(177,10.388200000000001)(178,17.313666666666666)(179,53.47128571428571)(180,20.776400000000002)(181,53.62685714285714)(182,85.25033333333333)(183,41.75366666666667)(184,62.69016666666667)(185,26.67577777777778)(186,30.019125)(187,48.0456)(188,14.840285714285715)(189,62.846666666666664)(190,40.092)(191,60.155750000000005)(192,80.23366666666666)(193,25.9705)(194,14.840285714285715)(195,20.776400000000002)(196,80.39133333333334)(197,63.129)(198,40.25116666666667)(199,17.313666666666666)(200,54.28685714285714)(201,76.0298)(202,82.17912500000001)(203,10.388200000000001)(204,25.9705)(205,80.824)(206,95.297)(207,40.44083333333334)(208,54.507000000000005)(209,57.84188888888889)(210,65.09812500000001)(211,80.98933333333333)(212,14.840285714285715)(213,86.92866666666664)(214,121.5905)(215,24.3244)(216,60.854)(217,48.701)(218,60.9075)(219,63.96516666666667)(220,25.9705)(221,30.5025)(222,73.86066666666666)(223,81.38766666666668)(224,20.776400000000002)(225,17.313666666666666)(226,54.99942857142857)(227,65.717)(228,77.0408)(229,58.48188888888889)(230,77.12960000000001)(231,77.1436)(232,61.22475)(233,61.250750000000004)(234,64.37433333333333)(235,61.27825)(236,20.776400000000002)(237,35.026428571428575)(238,20.776400000000002)(239,61.326750000000004)(240,40.893)(241,88.08216666666668)(242,51.941)(243,25.9705)(244,61.4055)(245,77.5208)(246,122.9165)(247,55.41857142857143)(248,103.882)(249,34.62733333333333)(250,49.2356)(251,41.04)(252,25.9705)(253,49.294)(254,25.9705)(255,34.62733333333333)(256,55.62714285714286)(257,61.67375)(258,25.9705)(259,34.62733333333333)(260,61.73175)(261,41.1545)(262,88.88266666666668)(263,25.9705)(264,34.62733333333333)(265,130.16133333333332)(266,49.4426)(267,34.62733333333333)(268,41.2175)(269,17.313666666666666)(270,78.1904)(271,34.62733333333333)(272,41.25633333333334)(273,82.52966666666667)(274,51.941)(275,49.5504)(276,34.62733333333333)(277,82.638)(278,76.60642857142857)(279,51.941)(280,98.132)(281,82.76)(282,25.9705)(283,98.233)(284,41.43516666666667)(285,51.941)(286,82.90400000000001)(287,51.941)(288,62.2005)(289,51.941)(290,25.9705)(291,83.01033333333334)(292,25.9705)(293,41.5235)(294,98.619)(295,51.941)(296,124.67450000000001)(297,51.941)(298,49.8872)(299,83.165)(300,131.73466666666667)(301,25.9705)(302,51.941)(303,51.941)(304,98.886)(305,79.12960000000001)(306,25.9705)(307,135.42925)(308,83.28833333333334)(309,41.64933333333334)(310,108.41300000000001)(311,99.024)(312,125.0485)(313,103.882)(314,83.40933333333334)(315,34.62733333333333)(316,99.193)(317,51.941)(318,34.62733333333333)(319,34.62733333333333)(320,125.269)(321,34.62733333333333)(322,34.62733333333333)(323,51.941)(324,34.62733333333333)(325,25.9705)(326,103.882)(327,34.62733333333333)(328,103.882)(329,34.62733333333333)(330,34.62733333333333)(331,62.74675)(332,34.62733333333333)(333,103.882)(334,34.62733333333333)(335,25.9705)(336,103.882)(337,34.62733333333333)(338,103.882)
}; % plot of the second histogram
\addplot[color=cyan] coordinates{(0,9.170416666666668)(1,13.77189795918367)(2,8.343517241379311)(3,13.75775324675325)(4,11.312614285714284)(5,12.370626168224298)(6,13.820953703703706)(7,17.412964912280696)(8,18.109510638297866)(9,8.798923076923076)(10,14.776719178082194)(11,12.572023255813953)(12,12.864518749999998)(13,9.403485915492958)(14,12.9770625)(15,18.90576335877862)(16,12.733323529411766)(17,22.309932885906044)(18,23.382141891891894)(19,17.155092198581556)(20,20.470415094339625)(21,20.85700632911392)(22,22.55228205128206)(23,19.274106250000003)(24,28.32208917197452)(25,22.693597122302158)(26,28.57655921052632)(27,26.19112738853503)(28,23.584729032258068)(29,31.356038461538457)(30,27.205517730496457)(31,19.583553846153848)(32,27.165632478632485)(33,25.602736000000004)(34,25.545492307692307)(35,18.963007575757576)(36,20.038244274809163)(37,16.945401785714285)(38,19.955057377049183)(39,28.57288172043011)(40,22.459126315789476)(41,30.236247058823526)(42,28.690799999999996)(43,20.20171875)(44,27.428853333333326)(45,25.99053521126761)(46,33.64477777777777)(47,30.542249999999996)(48,28.695522727272728)(49,33.931730769230775)(50,27.626725)(51,25.521446153846156)(52,29.514367088607592)(53,14.031898734177215)(54,30.7231625)(55,23.009441176470585)(56,28.10314285714286)(57,44.85787671232877)(58,34.703653333333335)(59,29.80951315789473)(60,22.975942028985507)(61,33.73517283950616)(62,37.50120547945204)(63,35.67105194805195)(64,32.12928)(65,22.950557142857143)(66,38.479174603174606)(67,24.96641304347826)(68,24.36113461538461)(69,17.249133333333337)(70,39.33577419354838)(71,31.062232142857138)(72,29.606553191489358)(73,40.737049999999996)(74,46.21487323943662)(75,47.841122448979604)(76,36.09986538461539)(77,33.513612244897956)(78,26.865604166666667)(79,41.07344186046513)(80,30.02453191489361)(81,34.81429545454545)(82,24.02102564102564)(83,30.168230769230775)(84,46.922399999999996)(85,29.352906250000004)(86,45.27917647058824)(87,45.45696153846154)(88,27.187470588235296)(89,36.455625000000005)(90,29.196349999999995)(91,21.369000000000003)(92,35.64363333333334)(93,33.664857142857144)(94,41.62335294117647)(95,27.961285714285715)(96,27.41394117647059)(97,25.560347826086957)(98,78.09555)(99,19.18411111111111)(100,18.63383333333333)(101,26.813863636363635)(102,53.942049999999995)(103,53.208)(104,54.71236363636364)(105,36.411434782608694)(106,47.0109)(107,37.703736842105265)(108,42.436928571428574)(109,39.32)(110,31.34352631578947)(111,35.800481481481484)(112,36.38215384615384)(113,37.31875)(114,35.14464705882352)(115,48.14073333333334)(116,26.92453846153846)(117,17.381307692307693)(118,32.29985714285714)(119,27.251954545454545)(120,49.93535294117647)(121,39.633583333333334)(122,17.56595)(123,57.43282352941177)(124,11.301222222222222)(125,40.19766666666667)(126,43.09492857142857)(127,47.506499999999996)(128,28.178941176470587)(129,10.1711)(130,31.9706)(131,34.297928571428564)(132,36.95230769230769)(133,50.580833333333324)(134,22.1574375)(135,48.127700000000004)(136,34.393857142857144)(137,53.529222222222224)(138,32.0508947368421)(139,27.345153846153842)(140,46.88823076923077)(141,61.72285714285715)(142,30.2103125)(143,44.546125)(144,55.58463636363637)(145,37.27192307692308)(146,32.25268421052632)(147,37.33030769230769)(148,28.602294117647062)(149,66.64026666666666)(150,54.085777777777786)(151,41.0672)(152,39.88922222222223)(153,19.205916666666667)(154,9.246454545454545)(155,61.079375)(156,37.605384615384615)(157,57.488615384615386)(158,23.0988)(159,23.105900000000002)(160,15.409333333333333)(161,33.029999999999994)(162,25.704666666666665)(163,33.07085714285714)(164,25.736444444444444)(165,30.158083333333334)(166,69.23477777777778)(167,77.987875)(168,82.28433333333332)(169,33.903666666666666)(170,36.3335)(171,61.81599999999999)(172,58.192750000000004)(173,20.3422)(174,75.7741)(175,29.12975)(176,60.75049999999999)(177,10.1711)(178,16.951833333333333)(179,52.15500000000001)(180,20.3422)(181,52.20157142857142)(182,82.91516666666666)(183,40.655)(184,61.033500000000004)(185,26.00511111111111)(186,29.272624999999998)(187,46.858399999999996)(188,14.530142857142858)(189,61.20149999999999)(190,39.119)(191,58.698499999999996)(192,78.28500000000001)(193,25.42775)(194,14.530142857142858)(195,20.3422)(196,78.41933333333333)(197,61.49516666666667)(198,39.254)(199,16.951833333333333)(200,52.80328571428571)(201,73.93879999999999)(202,79.763125)(203,10.1711)(204,25.42775)(205,78.70166666666667)(206,92.65675)(207,39.386833333333335)(208,53.04557142857142)(209,56.31877777777778)(210,63.525999999999996)(211,79.09366666666666)(212,14.530142857142858)(213,84.86333333333333)(214,118.80199999999999)(215,23.766699999999997)(216,59.456500000000005)(217,47.5794)(218,59.489999999999995)(219,62.39266666666666)(220,25.42775)(221,29.778)(222,72.00666666666667)(223,79.455)(224,20.3422)(225,16.951833333333333)(226,53.718714285714285)(227,64.285125)(228,75.37939999999999)(229,57.21111111111111)(230,75.46099999999998)(231,75.49719999999999)(232,59.938)(233,59.953500000000005)(234,63.00899999999999)(235,59.98675)(236,20.3422)(237,34.30414285714286)(238,20.3422)(239,60.065)(240,40.05350000000001)(241,86.3125)(242,50.8555)(243,25.42775)(244,60.15325)(245,75.9526)(246,120.42949999999999)(247,54.31842857142857)(248,101.711)(249,33.903666666666666)(250,48.2692)(251,40.236)(252,25.42775)(253,48.3134)(254,25.42775)(255,33.903666666666666)(256,54.56699999999999)(257,60.47225)(258,25.42775)(259,33.903666666666666)(260,60.53675)(261,40.35783333333333)(262,87.21233333333333)(263,25.42775)(264,33.903666666666666)(265,127.70633333333335)(266,48.4952)(267,33.903666666666666)(268,40.434666666666665)(269,16.951833333333333)(270,76.74940000000001)(271,33.903666666666666)(272,40.47016666666667)(273,80.96066666666667)(274,50.8555)(275,48.6226)(276,33.903666666666666)(277,81.099)(278,75.25828571428572)(279,50.8555)(280,96.36724999999998)(281,81.22033333333333)(282,25.42775)(283,96.51325)(284,40.6575)(285,50.8555)(286,81.35199999999999)(287,50.8555)(288,61.06725)(289,50.8555)(290,25.42775)(291,81.48333333333333)(292,25.42775)(293,40.765833333333326)(294,96.89525)(295,50.8555)(296,122.36850000000001)(297,50.8555)(298,48.9754)(299,81.63933333333334)(300,129.40633333333335)(301,25.42775)(302,50.8555)(303,50.8555)(304,97.18425)(305,77.779)(306,25.42775)(307,133.22975)(308,81.85566666666666)(309,40.940666666666665)(310,106.7404)(311,97.46125)(312,122.91550000000001)(313,101.711)(314,82.038)(315,33.903666666666666)(316,97.68525)(317,50.8555)(318,33.903666666666666)(319,33.903666666666666)(320,123.24549999999999)(321,33.903666666666666)(322,33.903666666666666)(323,50.8555)(324,33.903666666666666)(325,25.42775)(326,101.711)(327,33.903666666666666)(328,101.711)(329,33.903666666666666)(330,33.903666666666666)(331,61.748000000000005)(332,33.903666666666666)(333,101.711)(334,33.903666666666666)(335,25.42775)(336,101.711)(337,33.903666666666666)(338,101.711)
}; % plot of the third histogram
\addplot[color=violet] coordinates{(0,10.715595744680853)(1,13.529145454545452)(2,15.504265625)(3,22.607908045977013)(4,16.227169642857145)(5,17.315492424242425)(6,20.480923566878978)(7,26.78297916666666)(8,24.578243421052633)(9,21.77669230769231)(10,20.210007194244604)(11,22.16704827586207)(12,26.04784126984127)(13,18.188766129032256)(14,23.006200000000003)(15,17.341177777777776)(16,24.14757462686567)(17,22.13182608695652)(18,21.865711267605633)(19,27.5218085106383)(20,25.859020979020983)(21,27.495625)(22,31.33818493150684)(23,24.969086956521735)(24,30.052656250000005)(25,27.906029197080294)(26,34.57661635220127)(27,32.99297333333334)(28,41.3465487804878)(29,40.6315125)(30,33.77786805555556)(31,35.92196825396827)(32,33.24586507936508)(33,31.288936363636363)(34,35.09563779527559)(35,31.93029411764706)(36,33.339999999999996)(37,32.85487619047619)(38,33.18524418604652)(39,22.55117582417582)(40,28.06737037037037)(41,24.583723684210526)(42,24.704513157894734)(43,25.22793670886076)(44,29.40030379746835)(45,24.696315068493153)(46,37.412594594594594)(47,26.063038461538458)(48,37.72341558441558)(49,18.689962962962966)(50,27.706931372549015)(51,33.33734722222223)(52,31.979)(53,41.339357142857146)(54,28.438364864864862)(55,26.611591549295778)(56,32.10208219178082)(57,38.613677777777774)(58,31.917133333333332)(59,25.769240506329115)(60,29.07209302325581)(61,23.612810344827583)(62,30.044324999999994)(63,35.9984328358209)(64,35.54166153846153)(65,26.243584905660377)(66,24.582461538461537)(67,11.890979591836732)(68,26.769041666666666)(69,21.278045454545452)(70,38.34869387755102)(71,37.79177358490566)(72,25.493823529411767)(73,38.85326229508197)(74,24.650867924528303)(75,25.337255319148934)(76,21.05705882352941)(77,16.771350877192983)(78,30.708205128205126)(79,24.611641025641024)(80,30.686297872340422)(81,21.891787878787877)(82,17.73626470588235)(83,16.792333333333332)(84,7.118970588235294)(85,31.444518518518517)(86,28.010846153846156)(87,34.02232)(88,33.17790909090909)(89,19.22342105263158)(90,19.50664)(91,12.842421052631579)(92,42.7403)(93,36.6854)(94,16.88641379310345)(95,11.67352380952381)(96,16.371199999999998)(97,24.579200000000004)(98,17.57357142857143)(99,43.466058823529416)(100,38.5346875)(101,44.87427272727273)(102,17.64242857142857)(103,41.234)(104,28.131363636363634)(105,33.788000000000004)(106,27.563222222222223)(107,27.585111111111114)(108,12.420399999999999)(109,26.647)(110,49.767)(111,41.497)(112,46.72925)(113,46.760125)(114,19.714052631578948)(115,26.780357142857145)(116,50.06280000000001)(117,12.5298)(118,8.955)(119,50.183066666666654)(120,48.34315384615385)(121,31.444350000000004)(122,44.95128571428571)(123,41.995)(124,22.934363636363635)(125,56.13644444444445)(126,19.963157894736838)(127,29.76870588235294)(128,31.6455)(129,25.348266666666667)(130,16.908666666666665)(131,38.0707)(132,21.170416666666668)(133,7.94625)(134,45.44842857142857)(135,47.7565)(136,51.0057)(137,42.559916666666666)(138,49.14392307692307)(139,51.1378)(140,30.135882352941174)(141,40.048)(142,27.4785)(143,42.772777777777776)(144,38.5344)(145,25.721299999999996)(146,38.6051)(147,42.919444444444444)(148,36.82285714285714)(149,21.49266666666667)(150,16.12775)(151,11.736)(152,23.50772727272727)(153,36.96114285714285)(154,35.302181818181815)(155,21.596833333333333)(156,9.262857142857143)(157,25.973000000000003)(158,43.381)(159,21.703)(160,32.57475)(161,32.64825)(162,32.68475)(163,65.49033333333334)(164,18.732857142857142)(165,52.4836)(166,52.5264)(167,18.783142857142856)(168,18.79342857142857)(169,52.6948)(170,33.04025)(171,44.139)(172,53.041599999999995)(173,53.0708)(174,66.467)(175,44.43866666666667)(176,66.6955)(177,33.3645)(178,19.075142857142858)(179,44.57533333333333)(180,44.604000000000006)(181,16.748125)(182,44.68566666666666)(183,22.362333333333336)(184,26.8522)(185,26.879199999999997)(186,26.922800000000002)(187,22.4495)(188,26.975599999999996)(189,54.0692)(190,33.8135)(191,33.83375)(192,45.13233333333333)(193,38.73185714285715)(194,27.1402)(195,67.885)(196,19.424857142857142)(197,45.36600000000001)(198,19.462714285714288)(199,68.2055)(200,34.15275)(201,54.672799999999995)(202,39.11)(203,34.2485)(204,60.97155555555556)(205,68.64)(206,137.551)(207,45.91433333333333)(208,68.99)(209,34.51275)(210,46.038000000000004)(211,27.6344)(212,138.284)(213,27.6692)(214,23.079166666666666)(215,46.178666666666665)(216,19.80557142857143)(217,69.354)(218,34.69375)(219,46.28066666666667)(220,55.5612)(221,69.552)(222,46.546)(223,46.69383333333334)(224,60.06371428571429)(225,60.14742857142857)(226,56.1476)(227,35.1035)(228,46.828)(229,46.89066666666667)(230,140.713)(231,140.935)(232,112.8296)(233,47.029333333333334)(234,70.5745)(235,47.13933333333333)(236,70.746)(237,56.6256)(238,35.4365)(239,71.0755)(240,35.56575)(241,28.475)(242,95.06033333333335)(243,47.559666666666665)(244,35.7025)(245,95.22066666666667)(246,47.63433333333333)(247,47.65133333333333)(248,71.543)(249,47.72566666666666)(250,71.6395)(251,71.69)
}; % plot of the fifth histogram
\addplot[color=purple] coordinates{(0,10.182808510638298)(1,12.81029090909091)(2,14.657328125000003)(3,21.404517241379306)(4,15.415580357142858)(5,16.506409090909088)(6,19.550828025477703)(7,25.614479166666655)(8,23.514296052631575)(9,20.78842603550296)(10,19.208906474820143)(11,21.103855172413795)(12,24.889198412698413)(13,17.39931451612903)(14,22.057779999999998)(15,16.641133333333332)(16,23.20182835820896)(17,21.28224637681159)(18,21.117345070422534)(19,26.611886524822697)(20,25.010517482517482)(21,26.58616447368421)(22,30.356568493150686)(23,24.175920289855068)(24,29.141475000000007)(25,27.089277372262778)(26,33.603459119496854)(27,32.083313333333344)(28,40.22281707317076)(29,39.59560625000001)(30,32.910347222222214)(31,35.05869047619047)(32,32.439626984126974)(33,30.523309090909084)(34,34.29311023622048)(35,31.21843697478992)(36,32.556566037735855)(37,32.09997142857143)(38,32.4558488372093)(39,22.063571428571432)(40,27.427456790123454)(41,24.010986842105265)(42,24.13065789473684)(43,24.64786075949368)(44,28.743240506329112)(45,24.159616438356164)(46,36.61486486486487)(47,25.54532051282051)(48,37.01848051948051)(49,18.333160493827158)(50,27.16026470588235)(51,32.70312499999999)(52,31.39776388888889)(53,40.56003571428571)(54,27.883824324324323)(55,26.100042253521128)(56,31.47194520547945)(57,37.84681111111111)(58,31.285066666666662)(59,25.256113924050634)(60,28.460302325581402)(61,23.129862068965515)(62,29.454312500000004)(63,35.324462686567166)(64,34.84775384615385)(65,25.70518867924528)(66,24.09013461538461)(67,11.65210204081633)(68,26.23872916666667)(69,20.862590909090912)(70,37.58379591836734)(71,37.0236037735849)(72,24.973098039215685)(73,38.079967213114756)(74,24.147188679245282)(75,24.820297872340426)(76,20.638764705882355)(77,16.45261403508772)(78,30.125487179487184)(79,24.154615384615386)(80,30.10923404255319)(81,21.484333333333336)(82,17.40558823529412)(83,16.46147222222222)(84,6.987235294117647)(85,30.836)(86,27.48853846153846)(87,33.38963999999999)(88,32.556318181818185)(89,18.872789473684207)(90,19.1478)(91,12.61578947368421)(92,41.99235)(93,36.0457)(94,16.60437931034483)(95,11.478047619047619)(96,16.084600000000002)(97,24.1779)(98,17.284142857142857)(99,42.73952941176471)(100,37.91025)(101,44.13818181818181)(102,17.351642857142856)(103,40.52086666666666)(104,27.65909090909091)(105,33.222)(106,27.10238888888889)(107,27.125777777777778)(108,12.2152)(109,26.203857142857142)(110,48.95786666666666)(111,40.836866666666666)(112,45.998125)(113,46.0385)(114,19.412842105263156)(115,26.37035714285714)(116,49.263949999999994)(117,12.340399999999999)(118,8.820357142857143)(119,49.43026666666667)(120,47.608384615384615)(121,30.9646)(122,44.270714285714284)(123,41.366)(124,22.578727272727274)(125,55.285000000000004)(126,19.655157894736842)(127,29.314)(128,31.190833333333334)(129,24.9762)(130,16.662866666666666)(131,37.547200000000004)(132,20.8835)(133,7.8364375)(134,44.81185714285714)(135,47.143625)(136,50.3131)(137,41.962833333333336)(138,48.451153846153844)(139,50.4655)(140,29.707529411764707)(141,39.479375)(142,27.097071428571432)(143,42.17933333333333)(144,37.9778)(145,25.339)(146,38.0466)(147,42.302666666666674)(148,36.28542857142857)(149,21.19183333333333)(150,15.908)(151,11.577636363636364)(152,23.211363636363636)(153,36.50114285714285)(154,34.86445454545455)(155,21.31866666666667)(156,9.154285714285715)(157,25.6652)(158,42.86933333333334)(159,21.449166666666667)(160,32.195125000000004)(161,32.21275)(162,32.260625000000005)(163,64.59183333333334)(164,18.482714285714284)(165,51.779999999999994)(166,51.8366)(167,18.538)(168,18.550857142857144)(169,52.0432)(170,32.6055)(171,43.57566666666667)(172,52.39880000000001)(173,52.428399999999996)(174,65.60900000000001)(175,43.855333333333334)(176,65.856)(177,32.94475)(178,18.836714285714287)(179,44.02766666666667)(180,44.05733333333333)(181,16.544375)(182,44.167)(183,22.099500000000003)(184,26.532799999999998)(185,26.587799999999998)(186,26.6184)(187,22.192333333333334)(188,26.6796)(189,53.424)(190,33.40525)(191,33.4275)(192,44.617333333333335)(193,38.28771428571429)(194,26.8185)(195,67.109)(196,19.204857142857144)(197,44.855)(198,19.25114285714286)(199,67.46875)(200,33.834)(201,54.1712)(202,38.73142857142857)(203,33.93325)(204,60.388444444444445)(205,67.979)(206,136.168)(207,45.48566666666667)(208,68.2965)(209,34.1655)(210,45.57533333333333)(211,27.371199999999998)(212,136.949)(213,27.4086)(214,22.8685)(215,45.760333333333335)(216,19.646285714285714)(217,68.7975)(218,34.4165)(219,45.91166666666667)(220,55.1188)(221,69.021)(222,46.162)(223,46.28466666666667)(224,59.537142857142854)(225,59.62071428571429)(226,55.67719999999999)(227,34.82025)(228,46.455333333333336)(229,46.544999999999995)(230,139.69)(231,140.026)(232,112.1152)(233,46.737)(234,70.1555)(235,46.845333333333336)(236,70.3)(237,56.281600000000005)(238,35.2035)(239,70.5855)(240,35.33575)(241,28.3008)(242,94.44566666666667)(243,47.28933333333333)(244,35.4995)(245,94.68566666666668)(246,47.39033333333333)(247,47.410666666666664)(248,71.1765)(249,47.51933333333333)(250,71.334)(251,71.3595)
}; % plot of the first histogram
\addplot[color=magenta] coordinates{(0,11.13240425531915)(1,14.025236363636367)(2,16.090500000000002)(3,23.364862068965518)(4,16.802749999999996)(5,17.963886363636362)(6,21.190528662420377)(7,27.831951388888886)(8,25.609835526315784)(9,22.745514792899407)(10,21.15509352517986)(11,23.213062068965517)(12,27.306849206349206)(13,19.06620161290323)(14,24.082039999999996)(15,18.153000000000002)(16,25.234052238805972)(17,23.214644927536234)(18,22.915338028169018)(19,28.906588652482274)(20,27.110419580419585)(21,28.871467105263157)(22,32.98003424657534)(23,26.24909420289855)(24,31.600781249999994)(25,29.290182481751827)(26,36.27516981132076)(27,34.57371333333332)(28,43.30539024390244)(29,42.5739)(30,35.30202777777778)(31,37.6275)(32,34.785603174603175)(33,32.75018181818182)(34,36.79065354330709)(35,33.415478991596636)(36,34.86877358490567)(37,34.42609523809523)(38,34.78891860465116)(39,23.63426373626374)(40,29.45217283950617)(41,25.815289473684206)(42,25.937644736842106)(43,26.455582278481014)(44,30.895227848101264)(45,25.961493150684934)(46,39.33714864864864)(47,27.43107692307692)(48,39.81625974025974)(49,19.728938271604935)(50,29.241490196078438)(51,35.19463888888889)(52,33.873708333333326)(53,43.747089285714274)(54,30.07343243243243)(55,28.150732394366198)(56,34.00613698630137)(57,40.94976666666666)(58,33.83103333333333)(59,27.28856962025317)(60,30.753232558139537)(61,25.01977586206896)(62,31.856787500000003)(63,38.183835820895524)(64,37.62892307692308)(65,27.75722641509434)(66,26.018250000000002)(67,12.63369387755102)(68,28.4506875)(69,22.61959090909091)(70,40.71765306122449)(71,40.073094339622635)(72,27.015176470588234)(73,41.184672131147536)(74,26.13567924528302)(75,26.894851063829787)(76,22.366264705882354)(77,17.812649122807017)(78,32.59110256410257)(79,26.11569230769231)(80,32.541191489361694)(81,23.20748484848485)(82,18.813088235294117)(83,17.802722222222222)(84,7.548470588235295)(85,33.37837037037038)(86,29.751730769230768)(87,36.16168)(88,35.250727272727275)(89,20.44057894736842)(90,20.731440000000003)(91,13.648210526315792)(92,45.432849999999995)(93,38.971799999999995)(94,17.933379310344826)(95,12.408904761904761)(96,17.39466666666667)(97,26.134750000000004)(98,18.68257142857143)(99,46.214705882352945)(100,40.9915625)(101,47.73545454545454)(102,18.766428571428573)(103,43.852666666666664)(104,29.91868181818182)(105,35.936090909090915)(106,29.3085)(107,29.330777777777776)(108,13.2066)(109,28.334142857142858)(110,52.91753333333334)(111,44.126533333333334)(112,49.71575)(113,49.760125)(114,20.992421052631578)(115,28.514714285714287)(116,53.263200000000005)(117,13.3389)(118,9.540142857142857)(119,53.46006666666667)(120,51.48323076923077)(121,33.488150000000005)(122,47.876714285714286)(123,44.73577777777778)(124,24.429454545454544)(125,59.800055555555545)(126,21.269)(127,31.71788235294118)(128,33.72325)(129,27.003000000000004)(130,18.028266666666664)(131,40.625099999999996)(132,22.647499999999997)(133,8.50225)(134,48.65085714285714)(135,51.112875)(136,54.5467)(137,45.4905)(138,52.558461538461536)(139,54.69)(140,32.19094117647059)(141,42.7950625)(142,29.363714285714284)(143,45.702555555555556)(144,41.148900000000005)(145,27.472699999999996)(146,41.28465)(147,45.93555555555556)(148,39.47485714285715)(149,23.038333333333338)(150,17.287875)(151,12.58990909090909)(152,25.20836363636364)(153,39.63342857142857)(154,37.86563636363636)(155,23.152666666666665)(156,9.927571428571428)(157,27.8358)(158,46.44)(159,23.232)(160,34.87575)(161,34.9135)(162,34.98775)(163,70.126)(164,20.067857142857143)(165,56.221199999999996)(166,56.2504)(167,20.10685714285714)(168,20.13)(169,56.413599999999995)(170,35.31675)(171,47.12799999999999)(172,56.6188)(173,56.6348)(174,70.8395)(175,47.34466666666666)(176,71.0625)(177,35.574)(178,20.337714285714288)(179,47.537)(180,47.57666666666666)(181,17.856375)(182,47.62799999999999)(183,23.820833333333336)(184,28.615)(185,28.6336)(186,28.657999999999998)(187,23.895333333333337)(188,28.6946)(189,57.426)(190,35.9085)(191,35.923)(192,47.907000000000004)(193,41.09342857142857)(194,28.786)(195,72.0015)(196,20.607714285714284)(197,48.13466666666667)(198,20.649571428571427)(199,72.40525)(200,36.24325)(201,58.0052)(202,41.48514285714286)(203,36.3115)(204,64.59555555555556)(205,72.7035)(206,145.532)(207,48.559666666666665)(208,72.8815)(209,36.4485)(210,48.621)(211,29.179199999999998)(212,145.928)(213,29.188)(214,24.340333333333334)(215,48.693333333333335)(216,20.87742857142857)(217,73.082)(218,36.56175)(219,48.769666666666666)(220,58.552800000000005)(221,73.2375)(222,48.959666666666664)(223,49.014)(224,63.03128571428572)(225,63.05742857142856)(226,58.8744)(227,36.8035)(228,49.079)(229,49.13333333333333)(230,147.43)(231,147.563)(232,118.11679999999998)(233,49.22233333333333)(234,73.854)(235,49.263666666666666)(236,73.9095)(237,59.136)(238,36.9775)(239,74.0085)(240,37.01475)(241,29.6238)(242,98.77533333333334)(243,49.40166666666667)(244,37.06175)(245,98.842)(246,49.43233333333333)(247,49.43933333333334)(248,74.1855)(249,49.478)(250,74.243)(251,74.253)
}; % plot of the second histogram
\legend{false \#1,  false \#2, false \#3, true \#1, true \#2, true \#3} % legend of the plot

\end{axis}
\end{tikzpicture}

%% file: plots/accessible_exits/avg_speed.tex
\begin{tikzpicture}[scale=0.75]

% Set the axis labels and title
\begin{axis}[    xlabel={Time [s]},    ylabel={Speed [m/s]},    title={Average Speed - Accessible exits},   legend style={at={(1,1)}, anchor=north east},]

\addplot[color=teal] coordinates{(0,4.018260258854219)(1,4.01683637463454)(2,4.0137139374238116)(3,4.0090585902782525)(4,4.003251470079557)(5,3.995790950543669)(6,3.9874532168953842)(7,3.9792316056542227)(8,3.970156971375808)(9,3.9617774430427413)(10,3.952640885676943)(11,3.941687637621894)(12,3.9304683727667795)(13,3.9185783122606153)(14,3.9066509115462527)(15,3.8923243712055506)(16,3.8787919224268306)(17,3.8653819809123706)(18,3.8502560729220505)(19,3.8352451478631235)(20,3.820024906600249)(21,3.803303919556741)(22,3.7838252406643393)(23,3.7623183655632033)(24,3.7391844572461315)(25,3.715972871842844)(26,3.687568223165555)(27,3.6597445094707695)(28,3.631654676258993)(29,3.5997956403269753)(30,3.5685067232837935)(31,3.5399323429916163)(32,3.5114294422432186)(33,3.48869277440706)(34,3.4682306940371457)(35,3.4459915611814345)(36,3.4270360013981125)(37,3.4054811866859622)(38,3.381824981301421)(39,3.3585853705672046)(40,3.335791886773647)(41,3.3162051282051284)(42,3.290909090909091)(43,3.2722330941509026)(44,3.2500556916908)(45,3.226704871060172)(46,3.2055294117647057)(47,3.1816420440784694)(48,3.156304401890077)(49,3.1252408477842004)(50,3.097454256165473)(51,3.0704032809295967)(52,3.0381690140845072)(53,3.004499274310595)(54,2.9660558726344246)(55,2.9292145296491774)(56,2.881555591415201)(57,2.840227767543125)(58,2.796880415944541)(59,2.755391804457225)(60,2.6957091999250515)(61,2.6416309012875536)(62,2.5701970443349755)(63,2.510497000856898)(64,2.445591204846309)(65,2.388862891904539)(66,2.332846359873387)(67,2.279028585884139)(68,2.2358934169278997)(69,2.1836900568951503)(70,2.1424148606811144)(71,2.0976608187134502)(72,2.0601571946795647)(73,2.0168961201501876)(74,1.9803479381443299)(75,1.9298657718120806)(76,1.884481558803062)(77,1.836231884057971)(78,1.7633472169632716)(79,1.7005136309758988)(80,1.6478988168094655)(81,1.6041315345699831)(82,1.5497588776852258)(83,1.509009009009009)(84,1.4680259499536608)(85,1.4485849056603775)(86,1.4184808901790034)(87,1.4089787863838183)(88,1.3973908680381335)(89,1.3944162436548224)(90,1.3858024691358024)(91,1.3733681462140992)(92,1.3678590496529632)(93,1.3580786026200873)(94,1.3541781959048147)(95,1.3408197641774284)(96,1.3335230506545248)(97,1.3254608294930876)(98,1.3230140186915889)(99,1.314540059347181)(100,1.3100961538461537)(101,1.3007290400972054)(102,1.2912801484230056)(103,1.2858040201005025)(104,1.2786989795918366)(105,1.2721970187945562)(106,1.2636783124588002)(107,1.257313829787234)(108,1.2510148849797023)(109,1.2472602739726026)(110,1.2427385892116183)(111,1.2352112676056337)(112,1.2287769784172662)(113,1.2219796215429404)(114,1.2179675994108983)(115,1.2137518684603887)(116,1.2099697885196374)(117,1.2068702290076336)(118,1.2009273570324575)(119,1.1954828660436136)(120,1.190779014308426)(121,1.1835084882780922)(122,1.1783960720130933)(123,1.1731890091590342)(124,1.1632825719120135)(125,1.1547008547008546)(126,1.1484375)(127,1.148375768217735)(128,1.1411974977658623)(129,1.1397459165154264)(130,1.1374885426214483)(131,1.1347583643122676)(132,1.130885122410546)(133,1.1296472831267874)(134,1.1256038647342994)(135,1.1218074656188606)(136,1.1157684630738522)(137,1.1124620060790273)(138,1.1094069529652353)(139,1.1042752867570387)(140,1.0976645435244161)(141,1.0958019375672767)(142,1.092896174863388)(143,1.0880713489409142)(144,1.0843644544431945)(145,1.082004555808656)(146,1.0776361529548089)(147,1.0725326991676576)(148,1.065217391304348)(149,1.0605686032138442)(150,1.0554156171284634)(151,1.0523627075351214)(152,1.0507152145643692)(153,1.0498687664041995)(154,1.0493333333333332)(155,1.0407055630936228)(156,1.0397805212620028)(157,1.0363128491620113)(158,1.0256045519203414)(159,1.0244604316546762)(160,1.0204678362573099)(161,1.017910447761194)(162,1.0165912518853695)(163,1.0137614678899083)(164,1.0138888888888888)(165,1.0078369905956113)(166,1.0047923322683705)(167,1.0032310177705976)(168,1)(169,1)(170,0.9983388704318937)(171,0.9983136593591906)(172,0.9982935153583617)(173,0.9982847341337907)(174,0.998272884283247)(175,0.9982456140350877)(176,0.9964476021314387)(177,0.996415770609319)(178,0.9963570127504554)(179,0.9963235294117647)(180,0.9962825278810409)(181,0.9962546816479401)(182,0.9962121212121212)(183,0.9961759082217974)(184,0.996116504854369)(185,0.996078431372549)(186,0.9960159362549801)(187,0.9959595959595959)(188,0.9959266802443992)(189,0.9958847736625515)(190,0.9958419958419958)(191,0.9957983193277311)(192,0.9957716701902748)(193,0.9957537154989384)(194,0.9957264957264957)(195,0.9956709956709957)(196,0.9956331877729258)(197,0.9956140350877193)(198,0.9955654101995566)(199,0.9955156950672646)(200,0.9954648526077098)(201,0.9954128440366973)(202,0.9953810623556582)(203,0.9953051643192489)(204,0.9952038369304557)(205,0.9951690821256038)(206,0.9951456310679612)(207,0.9951100244498777)(208,0.995049504950495)(209,0.9949748743718593)(210,0.9948849104859335)(211,0.9947916666666666)(212,0.9947643979057592)(213,0.9946808510638298)(214,0.9946091644204852)(215,0.9945945945945946)(216,0.9944598337950139)(217,0.994413407821229)(218,0.9943502824858758)(219,0.9943019943019943)(220,0.9942196531791907)(221,0.9941690962099126)(222,0.9940476190476191)(223,0.9939024390243902)(224,0.9938650306748467)(225,0.9938080495356038)(226,0.9937106918238994)(227,0.9935897435897436)(228,0.9934426229508196)(229,0.9933554817275747)(230,0.9931740614334471)(231,0.993103448275862)(232,0.993006993006993)(233,0.9929328621908127)(234,0.9928571428571429)(235,0.9927272727272727)(236,0.9926470588235294)(237,0.9925373134328358)(238,0.9923664122137404)(239,0.9922480620155039)(240,0.9921875)(241,0.9920318725099602)(242,0.991869918699187)(243,0.9918367346938776)(244,0.9917355371900827)(245,0.9916317991631799)(246,0.9914893617021276)(247,0.9914529914529915)(248,0.9912280701754386)(249,0.9912280701754386)(250,0.9911504424778761)(251,0.990990990990991)(252,0.9907834101382489)(253,0.9906542056074766)(254,0.9904761904761905)(255,0.9903381642512077)(256,0.9902439024390244)(257,0.9899497487437185)(258,0.9949238578680203)(259,0.9948453608247423)(260,0.9947916666666666)(261,0.9947368421052631)(262,0.9945945945945946)(263,0.9944444444444445)(264,0.9943502824858758)(265,0.9942857142857143)(266,0.9942196531791907)(267,0.9940828402366864)(268,0.9940119760479041)(269,0.9938271604938271)(270,0.9936305732484076)(271,0.9934640522875817)(272,0.9933774834437086)(273,0.9931506849315068)(274,0.9930555555555556)(275,0.993006993006993)(276,0.9928057553956835)(277,0.9927007299270073)(278,0.9925925925925926)(279,0.9922480620155039)(280,0.9921875)(281,0.992)(282,0.991869918699187)(283,0.9916666666666667)(284,0.9914529914529915)(285,0.9910714285714286)(286,0.9819819819819819)(287,0.9908256880733946)(288,0.9907407407407407)(289,0.9904761904761905)(290,0.9903846153846154)(291,0.9900990099009901)(292,0.98989898989899)(293,0.9895833333333334)(294,0.989010989010989)(295,1)(296,1)(297,1)(298,1)(299,1)(300,1)(301,1)(302,1)(303,1)(304,1)(305,1)(306,1)(307,1)(308,1)(309,1)(310,1)(311,1)(312,1)(313,1)(314,1)(315,1)(316,1)(317,1)(318,1)(319,1)(320,1)(321,1)(322,1)(323,1)(324,1)(325,1)(326,1)(327,1)(328,1)(329,1)(330,1)(331,1)(332,1)(333,1)(334,1)(335,1)(336,1)(337,1)(338,1)
}; % plot of the first histogram
\addplot[color=blue] coordinates{(0,4.018260258854219)(1,4.01683637463454)(2,4.0137139374238116)(3,4.0090585902782525)(4,4.003251470079557)(5,3.995790950543669)(6,3.9874532168953842)(7,3.9792316056542227)(8,3.970156971375808)(9,3.9617774430427413)(10,3.952640885676943)(11,3.941687637621894)(12,3.9304683727667795)(13,3.9185783122606153)(14,3.9066509115462527)(15,3.8923243712055506)(16,3.8787919224268306)(17,3.8653819809123706)(18,3.8502560729220505)(19,3.8352451478631235)(20,3.820024906600249)(21,3.803303919556741)(22,3.7838252406643393)(23,3.7623183655632033)(24,3.7391844572461315)(25,3.715972871842844)(26,3.687568223165555)(27,3.6597445094707695)(28,3.631654676258993)(29,3.5997956403269753)(30,3.5685067232837935)(31,3.5399323429916163)(32,3.5114294422432186)(33,3.48869277440706)(34,3.4682306940371457)(35,3.4459915611814345)(36,3.4270360013981125)(37,3.4054811866859622)(38,3.381824981301421)(39,3.3585853705672046)(40,3.335791886773647)(41,3.3162051282051284)(42,3.290909090909091)(43,3.2722330941509026)(44,3.2500556916908)(45,3.226704871060172)(46,3.2055294117647057)(47,3.1816420440784694)(48,3.156304401890077)(49,3.1252408477842004)(50,3.097454256165473)(51,3.0704032809295967)(52,3.0381690140845072)(53,3.004499274310595)(54,2.9660558726344246)(55,2.9292145296491774)(56,2.881555591415201)(57,2.840227767543125)(58,2.796880415944541)(59,2.755391804457225)(60,2.6957091999250515)(61,2.6416309012875536)(62,2.5701970443349755)(63,2.510497000856898)(64,2.445591204846309)(65,2.388862891904539)(66,2.332846359873387)(67,2.279028585884139)(68,2.2358934169278997)(69,2.1836900568951503)(70,2.1424148606811144)(71,2.0976608187134502)(72,2.0601571946795647)(73,2.0168961201501876)(74,1.9803479381443299)(75,1.9298657718120806)(76,1.884481558803062)(77,1.836231884057971)(78,1.7633472169632716)(79,1.7005136309758988)(80,1.6478988168094655)(81,1.6041315345699831)(82,1.5497588776852258)(83,1.509009009009009)(84,1.4680259499536608)(85,1.4485849056603775)(86,1.4184808901790034)(87,1.4089787863838183)(88,1.3973908680381335)(89,1.3944162436548224)(90,1.3858024691358024)(91,1.3733681462140992)(92,1.3678590496529632)(93,1.3580786026200873)(94,1.3541781959048147)(95,1.3408197641774284)(96,1.3335230506545248)(97,1.3254608294930876)(98,1.3230140186915889)(99,1.314540059347181)(100,1.3100961538461537)(101,1.3007290400972054)(102,1.2912801484230056)(103,1.2858040201005025)(104,1.2786989795918366)(105,1.2721970187945562)(106,1.2636783124588002)(107,1.257313829787234)(108,1.2510148849797023)(109,1.2472602739726026)(110,1.2427385892116183)(111,1.2352112676056337)(112,1.2287769784172662)(113,1.2219796215429404)(114,1.2179675994108983)(115,1.2137518684603887)(116,1.2099697885196374)(117,1.2068702290076336)(118,1.2009273570324575)(119,1.1954828660436136)(120,1.190779014308426)(121,1.1835084882780922)(122,1.1783960720130933)(123,1.1731890091590342)(124,1.1632825719120135)(125,1.1547008547008546)(126,1.1484375)(127,1.148375768217735)(128,1.1411974977658623)(129,1.1397459165154264)(130,1.1374885426214483)(131,1.1347583643122676)(132,1.130885122410546)(133,1.1296472831267874)(134,1.1256038647342994)(135,1.1218074656188606)(136,1.1157684630738522)(137,1.1124620060790273)(138,1.1094069529652353)(139,1.1042752867570387)(140,1.0976645435244161)(141,1.0958019375672767)(142,1.092896174863388)(143,1.0880713489409142)(144,1.0843644544431945)(145,1.082004555808656)(146,1.0776361529548089)(147,1.0725326991676576)(148,1.065217391304348)(149,1.0605686032138442)(150,1.0554156171284634)(151,1.0523627075351214)(152,1.0507152145643692)(153,1.0498687664041995)(154,1.0493333333333332)(155,1.0407055630936228)(156,1.0397805212620028)(157,1.0363128491620113)(158,1.0256045519203414)(159,1.0244604316546762)(160,1.0204678362573099)(161,1.017910447761194)(162,1.0165912518853695)(163,1.0137614678899083)(164,1.0138888888888888)(165,1.0078369905956113)(166,1.0047923322683705)(167,1.0032310177705976)(168,1)(169,1)(170,0.9983388704318937)(171,0.9983136593591906)(172,0.9982935153583617)(173,0.9982847341337907)(174,0.998272884283247)(175,0.9982456140350877)(176,0.9964476021314387)(177,0.996415770609319)(178,0.9963570127504554)(179,0.9963235294117647)(180,0.9962825278810409)(181,0.9962546816479401)(182,0.9962121212121212)(183,0.9961759082217974)(184,0.996116504854369)(185,0.996078431372549)(186,0.9960159362549801)(187,0.9959595959595959)(188,0.9959266802443992)(189,0.9958847736625515)(190,0.9958419958419958)(191,0.9957983193277311)(192,0.9957716701902748)(193,0.9957537154989384)(194,0.9957264957264957)(195,0.9956709956709957)(196,0.9956331877729258)(197,0.9956140350877193)(198,0.9955654101995566)(199,0.9955156950672646)(200,0.9954648526077098)(201,0.9954128440366973)(202,0.9953810623556582)(203,0.9953051643192489)(204,0.9952038369304557)(205,0.9951690821256038)(206,0.9951456310679612)(207,0.9951100244498777)(208,0.995049504950495)(209,0.9949748743718593)(210,0.9948849104859335)(211,0.9947916666666666)(212,0.9947643979057592)(213,0.9946808510638298)(214,0.9946091644204852)(215,0.9945945945945946)(216,0.9944598337950139)(217,0.994413407821229)(218,0.9943502824858758)(219,0.9943019943019943)(220,0.9942196531791907)(221,0.9941690962099126)(222,0.9940476190476191)(223,0.9939024390243902)(224,0.9938650306748467)(225,0.9938080495356038)(226,0.9937106918238994)(227,0.9935897435897436)(228,0.9934426229508196)(229,0.9933554817275747)(230,0.9931740614334471)(231,0.993103448275862)(232,0.993006993006993)(233,0.9929328621908127)(234,0.9928571428571429)(235,0.9927272727272727)(236,0.9926470588235294)(237,0.9925373134328358)(238,0.9923664122137404)(239,0.9922480620155039)(240,0.9921875)(241,0.9920318725099602)(242,0.991869918699187)(243,0.9918367346938776)(244,0.9917355371900827)(245,0.9916317991631799)(246,0.9914893617021276)(247,0.9914529914529915)(248,0.9912280701754386)(249,0.9912280701754386)(250,0.9911504424778761)(251,0.990990990990991)(252,0.9907834101382489)(253,0.9906542056074766)(254,0.9904761904761905)(255,0.9903381642512077)(256,0.9902439024390244)(257,0.9899497487437185)(258,0.9949238578680203)(259,0.9948453608247423)(260,0.9947916666666666)(261,0.9947368421052631)(262,0.9945945945945946)(263,0.9944444444444445)(264,0.9943502824858758)(265,0.9942857142857143)(266,0.9942196531791907)(267,0.9940828402366864)(268,0.9940119760479041)(269,0.9938271604938271)(270,0.9936305732484076)(271,0.9934640522875817)(272,0.9933774834437086)(273,0.9931506849315068)(274,0.9930555555555556)(275,0.993006993006993)(276,0.9928057553956835)(277,0.9927007299270073)(278,0.9925925925925926)(279,0.9922480620155039)(280,0.9921875)(281,0.992)(282,0.991869918699187)(283,0.9916666666666667)(284,0.9914529914529915)(285,0.9910714285714286)(286,0.9819819819819819)(287,0.9908256880733946)(288,0.9907407407407407)(289,0.9904761904761905)(290,0.9903846153846154)(291,0.9900990099009901)(292,0.98989898989899)(293,0.9895833333333334)(294,0.989010989010989)(295,1)(296,1)(297,1)(298,1)(299,1)(300,1)(301,1)(302,1)(303,1)(304,1)(305,1)(306,1)(307,1)(308,1)(309,1)(310,1)(311,1)(312,1)(313,1)(314,1)(315,1)(316,1)(317,1)(318,1)(319,1)(320,1)(321,1)(322,1)(323,1)(324,1)(325,1)(326,1)(327,1)(328,1)(329,1)(330,1)(331,1)(332,1)(333,1)(334,1)(335,1)(336,1)(337,1)(338,1)
}; % plot of the second histogram
\addplot[color=cyan] coordinates{(0,4.018260258854219)(1,4.01683637463454)(2,4.0137139374238116)(3,4.0090585902782525)(4,4.003251470079557)(5,3.995790950543669)(6,3.9874532168953842)(7,3.9792316056542227)(8,3.970156971375808)(9,3.9617774430427413)(10,3.952640885676943)(11,3.941687637621894)(12,3.9304683727667795)(13,3.9185783122606153)(14,3.9066509115462527)(15,3.8923243712055506)(16,3.8787919224268306)(17,3.8653819809123706)(18,3.8502560729220505)(19,3.8352451478631235)(20,3.820024906600249)(21,3.803303919556741)(22,3.7838252406643393)(23,3.7623183655632033)(24,3.7391844572461315)(25,3.715972871842844)(26,3.687568223165555)(27,3.6597445094707695)(28,3.631654676258993)(29,3.5997956403269753)(30,3.5685067232837935)(31,3.5399323429916163)(32,3.5114294422432186)(33,3.48869277440706)(34,3.4682306940371457)(35,3.4459915611814345)(36,3.4270360013981125)(37,3.4054811866859622)(38,3.381824981301421)(39,3.3585853705672046)(40,3.335791886773647)(41,3.3162051282051284)(42,3.290909090909091)(43,3.2722330941509026)(44,3.2500556916908)(45,3.226704871060172)(46,3.2055294117647057)(47,3.1816420440784694)(48,3.156304401890077)(49,3.1252408477842004)(50,3.097454256165473)(51,3.0704032809295967)(52,3.0381690140845072)(53,3.004499274310595)(54,2.9660558726344246)(55,2.9292145296491774)(56,2.881555591415201)(57,2.840227767543125)(58,2.796880415944541)(59,2.755391804457225)(60,2.6957091999250515)(61,2.6416309012875536)(62,2.5701970443349755)(63,2.510497000856898)(64,2.445591204846309)(65,2.388862891904539)(66,2.332846359873387)(67,2.279028585884139)(68,2.2358934169278997)(69,2.1836900568951503)(70,2.1424148606811144)(71,2.0976608187134502)(72,2.0601571946795647)(73,2.0168961201501876)(74,1.9803479381443299)(75,1.9298657718120806)(76,1.884481558803062)(77,1.836231884057971)(78,1.7633472169632716)(79,1.7005136309758988)(80,1.6478988168094655)(81,1.6041315345699831)(82,1.5497588776852258)(83,1.509009009009009)(84,1.4680259499536608)(85,1.4485849056603775)(86,1.4184808901790034)(87,1.4089787863838183)(88,1.3973908680381335)(89,1.3944162436548224)(90,1.3858024691358024)(91,1.3733681462140992)(92,1.3678590496529632)(93,1.3580786026200873)(94,1.3541781959048147)(95,1.3408197641774284)(96,1.3335230506545248)(97,1.3254608294930876)(98,1.3230140186915889)(99,1.314540059347181)(100,1.3100961538461537)(101,1.3007290400972054)(102,1.2912801484230056)(103,1.2858040201005025)(104,1.2786989795918366)(105,1.2721970187945562)(106,1.2636783124588002)(107,1.257313829787234)(108,1.2510148849797023)(109,1.2472602739726026)(110,1.2427385892116183)(111,1.2352112676056337)(112,1.2287769784172662)(113,1.2219796215429404)(114,1.2179675994108983)(115,1.2137518684603887)(116,1.2099697885196374)(117,1.2068702290076336)(118,1.2009273570324575)(119,1.1954828660436136)(120,1.190779014308426)(121,1.1835084882780922)(122,1.1783960720130933)(123,1.1731890091590342)(124,1.1632825719120135)(125,1.1547008547008546)(126,1.1484375)(127,1.148375768217735)(128,1.1411974977658623)(129,1.1397459165154264)(130,1.1374885426214483)(131,1.1347583643122676)(132,1.130885122410546)(133,1.1296472831267874)(134,1.1256038647342994)(135,1.1218074656188606)(136,1.1157684630738522)(137,1.1124620060790273)(138,1.1094069529652353)(139,1.1042752867570387)(140,1.0976645435244161)(141,1.0958019375672767)(142,1.092896174863388)(143,1.0880713489409142)(144,1.0843644544431945)(145,1.082004555808656)(146,1.0776361529548089)(147,1.0725326991676576)(148,1.065217391304348)(149,1.0605686032138442)(150,1.0554156171284634)(151,1.0523627075351214)(152,1.0507152145643692)(153,1.0498687664041995)(154,1.0493333333333332)(155,1.0407055630936228)(156,1.0397805212620028)(157,1.0363128491620113)(158,1.0256045519203414)(159,1.0244604316546762)(160,1.0204678362573099)(161,1.017910447761194)(162,1.0165912518853695)(163,1.0137614678899083)(164,1.0138888888888888)(165,1.0078369905956113)(166,1.0047923322683705)(167,1.0032310177705976)(168,1)(169,1)(170,0.9983388704318937)(171,0.9983136593591906)(172,0.9982935153583617)(173,0.9982847341337907)(174,0.998272884283247)(175,0.9982456140350877)(176,0.9964476021314387)(177,0.996415770609319)(178,0.9963570127504554)(179,0.9963235294117647)(180,0.9962825278810409)(181,0.9962546816479401)(182,0.9962121212121212)(183,0.9961759082217974)(184,0.996116504854369)(185,0.996078431372549)(186,0.9960159362549801)(187,0.9959595959595959)(188,0.9959266802443992)(189,0.9958847736625515)(190,0.9958419958419958)(191,0.9957983193277311)(192,0.9957716701902748)(193,0.9957537154989384)(194,0.9957264957264957)(195,0.9956709956709957)(196,0.9956331877729258)(197,0.9956140350877193)(198,0.9955654101995566)(199,0.9955156950672646)(200,0.9954648526077098)(201,0.9954128440366973)(202,0.9953810623556582)(203,0.9953051643192489)(204,0.9952038369304557)(205,0.9951690821256038)(206,0.9951456310679612)(207,0.9951100244498777)(208,0.995049504950495)(209,0.9949748743718593)(210,0.9948849104859335)(211,0.9947916666666666)(212,0.9947643979057592)(213,0.9946808510638298)(214,0.9946091644204852)(215,0.9945945945945946)(216,0.9944598337950139)(217,0.994413407821229)(218,0.9943502824858758)(219,0.9943019943019943)(220,0.9942196531791907)(221,0.9941690962099126)(222,0.9940476190476191)(223,0.9939024390243902)(224,0.9938650306748467)(225,0.9938080495356038)(226,0.9937106918238994)(227,0.9935897435897436)(228,0.9934426229508196)(229,0.9933554817275747)(230,0.9931740614334471)(231,0.993103448275862)(232,0.993006993006993)(233,0.9929328621908127)(234,0.9928571428571429)(235,0.9927272727272727)(236,0.9926470588235294)(237,0.9925373134328358)(238,0.9923664122137404)(239,0.9922480620155039)(240,0.9921875)(241,0.9920318725099602)(242,0.991869918699187)(243,0.9918367346938776)(244,0.9917355371900827)(245,0.9916317991631799)(246,0.9914893617021276)(247,0.9914529914529915)(248,0.9912280701754386)(249,0.9912280701754386)(250,0.9911504424778761)(251,0.990990990990991)(252,0.9907834101382489)(253,0.9906542056074766)(254,0.9904761904761905)(255,0.9903381642512077)(256,0.9902439024390244)(257,0.9899497487437185)(258,0.9949238578680203)(259,0.9948453608247423)(260,0.9947916666666666)(261,0.9947368421052631)(262,0.9945945945945946)(263,0.9944444444444445)(264,0.9943502824858758)(265,0.9942857142857143)(266,0.9942196531791907)(267,0.9940828402366864)(268,0.9940119760479041)(269,0.9938271604938271)(270,0.9936305732484076)(271,0.9934640522875817)(272,0.9933774834437086)(273,0.9931506849315068)(274,0.9930555555555556)(275,0.993006993006993)(276,0.9928057553956835)(277,0.9927007299270073)(278,0.9925925925925926)(279,0.9922480620155039)(280,0.9921875)(281,0.992)(282,0.991869918699187)(283,0.9916666666666667)(284,0.9914529914529915)(285,0.9910714285714286)(286,0.9819819819819819)(287,0.9908256880733946)(288,0.9907407407407407)(289,0.9904761904761905)(290,0.9903846153846154)(291,0.9900990099009901)(292,0.98989898989899)(293,0.9895833333333334)(294,0.989010989010989)(295,1)(296,1)(297,1)(298,1)(299,1)(300,1)(301,1)(302,1)(303,1)(304,1)(305,1)(306,1)(307,1)(308,1)(309,1)(310,1)(311,1)(312,1)(313,1)(314,1)(315,1)(316,1)(317,1)(318,1)(319,1)(320,1)(321,1)(322,1)(323,1)(324,1)(325,1)(326,1)(327,1)(328,1)(329,1)(330,1)(331,1)(332,1)(333,1)(334,1)(335,1)(336,1)(337,1)(338,1)
}; % plot of the third histogram
\addplot[color=violet] coordinates{(0,4.0180047521836615)(1,4.0160051216389245)(2,4.012272232798477)(3,4.007080254339233)(4,3.999302576978066)(5,3.9898828541001063)(6,3.979365021902038)(7,3.9686609686609686)(8,3.959213759213759)(9,3.9499826060067256)(10,3.939175257731959)(11,3.9287510137875103)(12,3.91823690545364)(13,3.908609215378746)(14,3.897523892267593)(15,3.887556088675641)(16,3.8757229382030145)(17,3.863825412402449)(18,3.849822132487261)(19,3.837132262051916)(20,3.8210504885993486)(21,3.8046109510086454)(22,3.787081081081081)(23,3.7700951215441956)(24,3.7453237410071942)(25,3.7255985267034992)(26,3.7012987012987013)(27,3.678480529396793)(28,3.6454376411209988)(29,3.6182332252948073)(30,3.587706685837527)(31,3.56)(32,3.534195647122609)(33,3.513546008336005)(34,3.4914057958980322)(35,3.4717710750193347)(36,3.447455209911757)(37,3.425163488993276)(38,3.4044397865853657)(39,3.3822431742290315)(40,3.3586912479740683)(41,3.341138515974607)(42,3.320188094474725)(43,3.298157086441422)(44,3.2778403967538323)(45,3.2580719824094433)(46,3.237936772046589)(47,3.218070818070818)(48,3.190613990941117)(49,3.1672054829949565)(50,3.143142742902839)(51,3.11095776980403)(52,3.0780759349129316)(53,3.039710658399764)(54,3.002599388379205)(55,2.9584126984126984)(56,2.9164744645799012)(57,2.8737399624124382)(58,2.8340425531914892)(59,2.7874102705687465)(60,2.7292828302249568)(61,2.6695582005245106)(62,2.6048914189331644)(63,2.537086092715232)(64,2.4751609935602574)(65,2.421796407185629)(66,2.362615423009733)(67,2.3078512396694215)(68,2.263157894736842)(69,2.217511776115267)(70,2.173390557939914)(71,2.1337466784765278)(72,2.093137254901961)(73,2.0510462904248574)(74,2.0098781692459666)(75,1.9626712328767124)(76,1.9151278409090908)(77,1.8634020618556701)(78,1.8047564250095895)(79,1.7429028388644543)(80,1.6820448877805487)(81,1.6181818181818182)(82,1.5582959641255605)(83,1.5212373037857803)(84,1.4840855106888362)(85,1.4425989252564728)(86,1.4197098549274638)(87,1.4120040691759919)(88,1.4041343669250645)(89,1.3990561090718405)(90,1.3931806073521578)(91,1.390481341265549)(92,1.3824657534246576)(93,1.3731260410882842)(94,1.3667975322490185)(95,1.3602058319039452)(96,1.3478766724840023)(97,1.3343248066627007)(98,1.3272837265577737)(99,1.3211009174311927)(100,1.3162128712871286)(101,1.309284818067754)(102,1.3035487959442331)(103,1.2980769230769231)(104,1.293774319066148)(105,1.2833553500660502)(106,1.273881095524382)(107,1.2642663043478262)(108,1.2565157750342935)(109,1.25)(110,1.2443977591036415)(111,1.2352941176470589)(112,1.2327586206896552)(113,1.2265112891478513)(114,1.222466960352423)(115,1.217326362957431)(116,1.2149886449659348)(117,1.2093918398768284)(118,1.2079131109387122)(119,1.2042586750788644)(120,1.2)(121,1.1954582319545823)(122,1.187396351575456)(123,1.1825063078216989)(124,1.1789652247667515)(125,1.1733905579399142)(126,1.169844020797227)(127,1.1682819383259913)(128,1.1626235399820306)(129,1.1601097895699908)(130,1.155844155844156)(131,1.1497175141242937)(132,1.1483253588516746)(133,1.1453488372093024)(134,1.1404715127701375)(135,1.1341341341341342)(136,1.1262729124236253)(137,1.12217659137577)(138,1.116910229645094)(139,1.1164021164021165)(140,1.10752688172043)(141,1.10239651416122)(142,1.0968819599109132)(143,1.0920454545454545)(144,1.0870069605568446)(145,1.0833333333333333)(146,1.0807600950118765)(147,1.0734939759036144)(148,1.066831683168317)(149,1.0627352572145545)(150,1.0596446700507614)(151,1.054263565891473)(152,1.049738219895288)(153,1.044)(154,1.0420624151967435)(155,1.0413223140495869)(156,1.0362622036262203)(157,1.0311614730878187)(158,1.0286123032904149)(159,1.0234260614934114)(160,1.0192878338278932)(161,1.014992503748126)(162,1.0136157337367624)(163,1.0122137404580154)(164,1.0107858243451464)(165,1.0093603744149766)(166,1.0062992125984251)(167,1.0047770700636942)(168,1.0048387096774194)(169,1.0048543689320388)(170,1.0032733224222585)(171,1)(172,1)(173,1)(174,1)(175,1)(176,1)(177,1)(178,1)(179,1)(180,1)(181,1)(182,1)(183,1)(184,1)(185,1)(186,1)(187,1)(188,1)(189,1)(190,1)(191,1)(192,1)(193,1)(194,1)(195,1)(196,1)(197,1)(198,1)(199,1)(200,1)(201,1)(202,1)(203,1)(204,1)(205,1)(206,1)(207,1)(208,1)(209,1)(210,1)(211,1)(212,1)(213,1)(214,1)(215,1)(216,1)(217,1)(218,1)(219,1)(220,1)(221,1)(222,1)(223,1)(224,1)(225,1)(226,1)(227,1)(228,1)(229,1)(230,1)(231,1)(232,1)(233,1)(234,1)(235,1)(236,1)(237,1)(238,1)(239,1)(240,1)(241,1)(242,1)(243,1)(244,1)(245,1)(246,1)(247,1)(248,1)(249,1)(250,1)(251,1)(252,1)(253,1)(254,1)(255,1)(256,1)(257,1)(258,1)(259,1)(260,1)(261,1)(262,1)(263,1)(264,1)(265,1)(266,1)(267,1)(268,1)(269,1)(270,1)(271,1)(272,1)(273,1)(274,1)(275,1)(276,1)(277,1)(278,1)(279,1)(280,1)(281,1)(282,1)(283,1)(284,1)(285,1)(286,1)(287,1)(288,1)(289,1)(290,1)(291,1)(292,1)(293,1)(294,1)(295,1)(296,1)(297,1)(298,1)(299,1)(300,1)(301,1)(302,1)(303,1)(304,1)(305,1)(306,1)(307,1)(308,1)(309,1)(310,1)(311,1)(312,1)(313,1)(314,1)(315,1)(316,1)(317,1)(318,1)(319,1)(320,1)(321,1)(322,1)(323,1)(324,1)(325,1)(326,1)(327,1)(328,1)(329,1)(330,1)(331,1)(332,1)(333,1)(334,1)(335,1)(336,1)(337,1)(338,1)(339,1)(340,1)(341,1)(342,1)(343,1)
}; % plot of the fifth histogram
\addplot[color=purple] coordinates{(0,4.0180047521836615)(1,4.0160051216389245)(2,4.012272232798477)(3,4.007080254339233)(4,3.999302576978066)(5,3.9898828541001063)(6,3.979365021902038)(7,3.9686609686609686)(8,3.959213759213759)(9,3.9499826060067256)(10,3.939175257731959)(11,3.9287510137875103)(12,3.91823690545364)(13,3.908609215378746)(14,3.897523892267593)(15,3.887556088675641)(16,3.8757229382030145)(17,3.863825412402449)(18,3.849822132487261)(19,3.837132262051916)(20,3.8210504885993486)(21,3.8046109510086454)(22,3.787081081081081)(23,3.7700951215441956)(24,3.7453237410071942)(25,3.7255985267034992)(26,3.7012987012987013)(27,3.678480529396793)(28,3.6454376411209988)(29,3.6182332252948073)(30,3.587706685837527)(31,3.56)(32,3.534195647122609)(33,3.513546008336005)(34,3.4914057958980322)(35,3.4717710750193347)(36,3.447455209911757)(37,3.425163488993276)(38,3.4044397865853657)(39,3.3822431742290315)(40,3.3586912479740683)(41,3.341138515974607)(42,3.320188094474725)(43,3.298157086441422)(44,3.2778403967538323)(45,3.2580719824094433)(46,3.237936772046589)(47,3.218070818070818)(48,3.190613990941117)(49,3.1672054829949565)(50,3.143142742902839)(51,3.11095776980403)(52,3.0780759349129316)(53,3.039710658399764)(54,3.002599388379205)(55,2.9584126984126984)(56,2.9164744645799012)(57,2.8737399624124382)(58,2.8340425531914892)(59,2.7874102705687465)(60,2.7292828302249568)(61,2.6695582005245106)(62,2.6048914189331644)(63,2.537086092715232)(64,2.4751609935602574)(65,2.421796407185629)(66,2.362615423009733)(67,2.3078512396694215)(68,2.263157894736842)(69,2.217511776115267)(70,2.173390557939914)(71,2.1337466784765278)(72,2.093137254901961)(73,2.0510462904248574)(74,2.0098781692459666)(75,1.9626712328767124)(76,1.9151278409090908)(77,1.8634020618556701)(78,1.8047564250095895)(79,1.7429028388644543)(80,1.6820448877805487)(81,1.6181818181818182)(82,1.5582959641255605)(83,1.5212373037857803)(84,1.4840855106888362)(85,1.4425989252564728)(86,1.4197098549274638)(87,1.4120040691759919)(88,1.4041343669250645)(89,1.3990561090718405)(90,1.3931806073521578)(91,1.390481341265549)(92,1.3824657534246576)(93,1.3731260410882842)(94,1.3667975322490185)(95,1.3602058319039452)(96,1.3478766724840023)(97,1.3343248066627007)(98,1.3272837265577737)(99,1.3211009174311927)(100,1.3162128712871286)(101,1.309284818067754)(102,1.3035487959442331)(103,1.2980769230769231)(104,1.293774319066148)(105,1.2833553500660502)(106,1.273881095524382)(107,1.2642663043478262)(108,1.2565157750342935)(109,1.25)(110,1.2443977591036415)(111,1.2352941176470589)(112,1.2327586206896552)(113,1.2265112891478513)(114,1.222466960352423)(115,1.217326362957431)(116,1.2149886449659348)(117,1.2093918398768284)(118,1.2079131109387122)(119,1.2042586750788644)(120,1.2)(121,1.1954582319545823)(122,1.187396351575456)(123,1.1825063078216989)(124,1.1789652247667515)(125,1.1733905579399142)(126,1.169844020797227)(127,1.1682819383259913)(128,1.1626235399820306)(129,1.1601097895699908)(130,1.155844155844156)(131,1.1497175141242937)(132,1.1483253588516746)(133,1.1453488372093024)(134,1.1404715127701375)(135,1.1341341341341342)(136,1.1262729124236253)(137,1.12217659137577)(138,1.116910229645094)(139,1.1164021164021165)(140,1.10752688172043)(141,1.10239651416122)(142,1.0968819599109132)(143,1.0920454545454545)(144,1.0870069605568446)(145,1.0833333333333333)(146,1.0807600950118765)(147,1.0734939759036144)(148,1.066831683168317)(149,1.0627352572145545)(150,1.0596446700507614)(151,1.054263565891473)(152,1.049738219895288)(153,1.044)(154,1.0420624151967435)(155,1.0413223140495869)(156,1.0362622036262203)(157,1.0311614730878187)(158,1.0286123032904149)(159,1.0234260614934114)(160,1.0192878338278932)(161,1.014992503748126)(162,1.0136157337367624)(163,1.0122137404580154)(164,1.0107858243451464)(165,1.0093603744149766)(166,1.0062992125984251)(167,1.0047770700636942)(168,1.0048387096774194)(169,1.0048543689320388)(170,1.0032733224222585)(171,1)(172,1)(173,1)(174,1)(175,1)(176,1)(177,1)(178,1)(179,1)(180,1)(181,1)(182,1)(183,1)(184,1)(185,1)(186,1)(187,1)(188,1)(189,1)(190,1)(191,1)(192,1)(193,1)(194,1)(195,1)(196,1)(197,1)(198,1)(199,1)(200,1)(201,1)(202,1)(203,1)(204,1)(205,1)(206,1)(207,1)(208,1)(209,1)(210,1)(211,1)(212,1)(213,1)(214,1)(215,1)(216,1)(217,1)(218,1)(219,1)(220,1)(221,1)(222,1)(223,1)(224,1)(225,1)(226,1)(227,1)(228,1)(229,1)(230,1)(231,1)(232,1)(233,1)(234,1)(235,1)(236,1)(237,1)(238,1)(239,1)(240,1)(241,1)(242,1)(243,1)(244,1)(245,1)(246,1)(247,1)(248,1)(249,1)(250,1)(251,1)(252,1)(253,1)(254,1)(255,1)(256,1)(257,1)(258,1)(259,1)(260,1)(261,1)(262,1)(263,1)(264,1)(265,1)(266,1)(267,1)(268,1)(269,1)(270,1)(271,1)(272,1)(273,1)(274,1)(275,1)(276,1)(277,1)(278,1)(279,1)(280,1)(281,1)(282,1)(283,1)(284,1)(285,1)(286,1)(287,1)(288,1)(289,1)(290,1)(291,1)(292,1)(293,1)(294,1)(295,1)(296,1)(297,1)(298,1)(299,1)(300,1)(301,1)(302,1)(303,1)(304,1)(305,1)(306,1)(307,1)(308,1)(309,1)(310,1)(311,1)(312,1)(313,1)(314,1)(315,1)(316,1)(317,1)(318,1)(319,1)(320,1)(321,1)(322,1)(323,1)(324,1)(325,1)(326,1)(327,1)(328,1)(329,1)(330,1)(331,1)(332,1)(333,1)(334,1)(335,1)(336,1)(337,1)(338,1)(339,1)(340,1)(341,1)(342,1)(343,1)
}; % plot of the first histogram
\addplot[color=magenta] coordinates{(0,4.0180047521836615)(1,4.0160051216389245)(2,4.012272232798477)(3,4.007080254339233)(4,3.999302576978066)(5,3.9898828541001063)(6,3.979365021902038)(7,3.9686609686609686)(8,3.959213759213759)(9,3.9499826060067256)(10,3.939175257731959)(11,3.9287510137875103)(12,3.91823690545364)(13,3.908609215378746)(14,3.897523892267593)(15,3.887556088675641)(16,3.8757229382030145)(17,3.863825412402449)(18,3.849822132487261)(19,3.837132262051916)(20,3.8210504885993486)(21,3.8046109510086454)(22,3.787081081081081)(23,3.7700951215441956)(24,3.7453237410071942)(25,3.7255985267034992)(26,3.7012987012987013)(27,3.678480529396793)(28,3.6454376411209988)(29,3.6182332252948073)(30,3.587706685837527)(31,3.56)(32,3.534195647122609)(33,3.513546008336005)(34,3.4914057958980322)(35,3.4717710750193347)(36,3.447455209911757)(37,3.425163488993276)(38,3.4044397865853657)(39,3.3822431742290315)(40,3.3586912479740683)(41,3.341138515974607)(42,3.320188094474725)(43,3.298157086441422)(44,3.2778403967538323)(45,3.2580719824094433)(46,3.237936772046589)(47,3.218070818070818)(48,3.190613990941117)(49,3.1672054829949565)(50,3.143142742902839)(51,3.11095776980403)(52,3.0780759349129316)(53,3.039710658399764)(54,3.002599388379205)(55,2.9584126984126984)(56,2.9164744645799012)(57,2.8737399624124382)(58,2.8340425531914892)(59,2.7874102705687465)(60,2.7292828302249568)(61,2.6695582005245106)(62,2.6048914189331644)(63,2.537086092715232)(64,2.4751609935602574)(65,2.421796407185629)(66,2.362615423009733)(67,2.3078512396694215)(68,2.263157894736842)(69,2.217511776115267)(70,2.173390557939914)(71,2.1337466784765278)(72,2.093137254901961)(73,2.0510462904248574)(74,2.0098781692459666)(75,1.9626712328767124)(76,1.9151278409090908)(77,1.8634020618556701)(78,1.8047564250095895)(79,1.7429028388644543)(80,1.6820448877805487)(81,1.6181818181818182)(82,1.5582959641255605)(83,1.5212373037857803)(84,1.4840855106888362)(85,1.4425989252564728)(86,1.4197098549274638)(87,1.4120040691759919)(88,1.4041343669250645)(89,1.3990561090718405)(90,1.3931806073521578)(91,1.390481341265549)(92,1.3824657534246576)(93,1.3731260410882842)(94,1.3667975322490185)(95,1.3602058319039452)(96,1.3478766724840023)(97,1.3343248066627007)(98,1.3272837265577737)(99,1.3211009174311927)(100,1.3162128712871286)(101,1.309284818067754)(102,1.3035487959442331)(103,1.2980769230769231)(104,1.293774319066148)(105,1.2833553500660502)(106,1.273881095524382)(107,1.2642663043478262)(108,1.2565157750342935)(109,1.25)(110,1.2443977591036415)(111,1.2352941176470589)(112,1.2327586206896552)(113,1.2265112891478513)(114,1.222466960352423)(115,1.217326362957431)(116,1.2149886449659348)(117,1.2093918398768284)(118,1.2079131109387122)(119,1.2042586750788644)(120,1.2)(121,1.1954582319545823)(122,1.187396351575456)(123,1.1825063078216989)(124,1.1789652247667515)(125,1.1733905579399142)(126,1.169844020797227)(127,1.1682819383259913)(128,1.1626235399820306)(129,1.1601097895699908)(130,1.155844155844156)(131,1.1497175141242937)(132,1.1483253588516746)(133,1.1453488372093024)(134,1.1404715127701375)(135,1.1341341341341342)(136,1.1262729124236253)(137,1.12217659137577)(138,1.116910229645094)(139,1.1164021164021165)(140,1.10752688172043)(141,1.10239651416122)(142,1.0968819599109132)(143,1.0920454545454545)(144,1.0870069605568446)(145,1.0833333333333333)(146,1.0807600950118765)(147,1.0734939759036144)(148,1.066831683168317)(149,1.0627352572145545)(150,1.0596446700507614)(151,1.054263565891473)(152,1.049738219895288)(153,1.044)(154,1.0420624151967435)(155,1.0413223140495869)(156,1.0362622036262203)(157,1.0311614730878187)(158,1.0286123032904149)(159,1.0234260614934114)(160,1.0192878338278932)(161,1.014992503748126)(162,1.0136157337367624)(163,1.0122137404580154)(164,1.0107858243451464)(165,1.0093603744149766)(166,1.0062992125984251)(167,1.0047770700636942)(168,1.0048387096774194)(169,1.0048543689320388)(170,1.0032733224222585)(171,1)(172,1)(173,1)(174,1)(175,1)(176,1)(177,1)(178,1)(179,1)(180,1)(181,1)(182,1)(183,1)(184,1)(185,1)(186,1)(187,1)(188,1)(189,1)(190,1)(191,1)(192,1)(193,1)(194,1)(195,1)(196,1)(197,1)(198,1)(199,1)(200,1)(201,1)(202,1)(203,1)(204,1)(205,1)(206,1)(207,1)(208,1)(209,1)(210,1)(211,1)(212,1)(213,1)(214,1)(215,1)(216,1)(217,1)(218,1)(219,1)(220,1)(221,1)(222,1)(223,1)(224,1)(225,1)(226,1)(227,1)(228,1)(229,1)(230,1)(231,1)(232,1)(233,1)(234,1)(235,1)(236,1)(237,1)(238,1)(239,1)(240,1)(241,1)(242,1)(243,1)(244,1)(245,1)(246,1)(247,1)(248,1)(249,1)(250,1)(251,1)(252,1)(253,1)(254,1)(255,1)(256,1)(257,1)(258,1)(259,1)(260,1)(261,1)(262,1)(263,1)(264,1)(265,1)(266,1)(267,1)(268,1)(269,1)(270,1)(271,1)(272,1)(273,1)(274,1)(275,1)(276,1)(277,1)(278,1)(279,1)(280,1)(281,1)(282,1)(283,1)(284,1)(285,1)(286,1)(287,1)(288,1)(289,1)(290,1)(291,1)(292,1)(293,1)(294,1)(295,1)(296,1)(297,1)(298,1)(299,1)(300,1)(301,1)(302,1)(303,1)(304,1)(305,1)(306,1)(307,1)(308,1)(309,1)(310,1)(311,1)(312,1)(313,1)(314,1)(315,1)(316,1)(317,1)(318,1)(319,1)(320,1)(321,1)(322,1)(323,1)(324,1)(325,1)(326,1)(327,1)(328,1)(329,1)(330,1)(331,1)(332,1)(333,1)(334,1)(335,1)(336,1)(337,1)(338,1)(339,1)(340,1)(341,1)(342,1)(343,1)
}; % plot of the second histogram
\legend{false \#1,  false \#2, false \#3, true \#1, true \#2, true \#3} % legend of the plot

\end{axis}
\end{tikzpicture}

%% file: plots/accessible_exits/evac_speed.tex
\begin{tikzpicture}[scale=0.75]

% Set the axis labels and title
\begin{axis}[    xlabel={Time [s]},    ylabel={Speed [ppl/s]},    title={Evacuation Speed - Accessible exits},   legend style={at={(1,1)}, anchor=north east},]

\addplot[color=teal] coordinates{(0,36)(1,49)(2,58)(3,77)(4,70)(5,107)(6,108)(7,114)(8,141)(9,143)(10,146)(11,129)(12,160)(13,142)(14,176)(15,131)(16,136)(17,149)(18,148)(19,141)(20,159)(21,158)(22,156)(23,160)(24,157)(25,139)(26,152)(27,157)(28,155)(29,156)(30,141)(31,130)(32,117)(33,125)(34,130)(35,132)(36,131)(37,112)(38,122)(39,93)(40,95)(41,85)(42,90)(43,64)(44,75)(45,71)(46,81)(47,68)(48,88)(49,78)(50,80)(51,65)(52,79)(53,79)(54,80)(55,68)(56,84)(57,73)(58,75)(59,76)(60,69)(61,81)(62,73)(63,77)(64,75)(65,70)(66,63)(67,46)(68,52)(69,60)(70,62)(71,56)(72,47)(73,40)(74,71)(75,49)(76,52)(77,49)(78,48)(79,43)(80,47)(81,44)(82,39)(83,39)(84,20)(85,32)(86,34)(87,26)(88,17)(89,16)(90,20)(91,33)(92,30)(93,21)(94,17)(95,21)(96,17)(97,23)(98,20)(99,18)(100,12)(101,22)(102,20)(103,18)(104,22)(105,23)(106,10)(107,19)(108,14)(109,12)(110,19)(111,27)(112,13)(113,16)(114,17)(115,15)(116,13)(117,13)(118,7)(119,22)(120,17)(121,12)(122,20)(123,17)(124,9)(125,15)(126,14)(127,18)(128,17)(129,10)(130,15)(131,14)(132,13)(133,12)(134,16)(135,10)(136,14)(137,9)(138,19)(139,13)(140,13)(141,14)(142,16)(143,8)(144,11)(145,13)(146,19)(147,13)(148,17)(149,15)(150,9)(151,15)(152,9)(153,12)(154,11)(155,8)(156,13)(157,13)(158,10)(159,10)(160,15)(161,7)(162,9)(163,7)(164,9)(165,12)(166,9)(167,8)(168,6)(169,3)(170,10)(171,8)(172,4)(173,5)(174,10)(175,8)(176,6)(177,10)(178,6)(179,7)(180,5)(181,7)(182,6)(183,9)(184,6)(185,9)(186,8)(187,5)(188,7)(189,6)(190,6)(191,4)(192,3)(193,4)(194,7)(195,5)(196,3)(197,6)(198,6)(199,6)(200,7)(201,5)(202,8)(203,10)(204,4)(205,3)(206,4)(207,6)(208,7)(209,9)(210,8)(211,3)(212,7)(213,6)(214,2)(215,10)(216,4)(217,5)(218,4)(219,6)(220,4)(221,8)(222,9)(223,3)(224,5)(225,6)(226,7)(227,8)(228,5)(229,9)(230,5)(231,5)(232,4)(233,4)(234,6)(235,4)(236,5)(237,7)(238,5)(239,4)(240,6)(241,6)(242,2)(243,4)(244,4)(245,5)(246,2)(247,7)(248,1)(249,3)(250,5)(251,6)(252,4)(253,5)(254,4)(255,3)(256,7)(257,4)(258,4)(259,3)(260,4)(261,6)(262,6)(263,4)(264,3)(265,3)(266,5)(267,3)(268,6)(269,6)(270,5)(271,3)(272,6)(273,3)(274,2)(275,5)(276,3)(277,3)(278,7)(279,2)(280,4)(281,3)(282,4)(283,4)(284,6)(285,2)(286,3)(287,2)(288,4)(289,2)(290,4)(291,3)(292,4)(293,6)(294,4)(295,2)(296,2)(297,2)(298,5)(299,3)(300,3)(301,4)(302,2)(303,2)(304,4)(305,5)(306,4)(307,4)(308,3)(309,6)(310,5)(311,4)(312,2)(313,1)(314,3)(315,3)(316,4)(317,2)(318,3)(319,3)(320,2)(321,3)(322,3)(323,2)(324,3)(325,4)(326,1)(327,3)(328,1)(329,3)(330,3)(331,4)(332,3)(333,1)(334,3)(335,4)(336,1)(337,3)(338,1)
}; % plot of the first histogram
\addplot[color=blue] coordinates{(0,36)(1,49)(2,58)(3,77)(4,70)(5,107)(6,108)(7,114)(8,141)(9,143)(10,146)(11,129)(12,160)(13,142)(14,176)(15,131)(16,136)(17,149)(18,148)(19,141)(20,159)(21,158)(22,156)(23,160)(24,157)(25,139)(26,152)(27,157)(28,155)(29,156)(30,141)(31,130)(32,117)(33,125)(34,130)(35,132)(36,131)(37,112)(38,122)(39,93)(40,95)(41,85)(42,90)(43,64)(44,75)(45,71)(46,81)(47,68)(48,88)(49,78)(50,80)(51,65)(52,79)(53,79)(54,80)(55,68)(56,84)(57,73)(58,75)(59,76)(60,69)(61,81)(62,73)(63,77)(64,75)(65,70)(66,63)(67,46)(68,52)(69,60)(70,62)(71,56)(72,47)(73,40)(74,71)(75,49)(76,52)(77,49)(78,48)(79,43)(80,47)(81,44)(82,39)(83,39)(84,20)(85,32)(86,34)(87,26)(88,17)(89,16)(90,20)(91,33)(92,30)(93,21)(94,17)(95,21)(96,17)(97,23)(98,20)(99,18)(100,12)(101,22)(102,20)(103,18)(104,22)(105,23)(106,10)(107,19)(108,14)(109,12)(110,19)(111,27)(112,13)(113,16)(114,17)(115,15)(116,13)(117,13)(118,7)(119,22)(120,17)(121,12)(122,20)(123,17)(124,9)(125,15)(126,14)(127,18)(128,17)(129,10)(130,15)(131,14)(132,13)(133,12)(134,16)(135,10)(136,14)(137,9)(138,19)(139,13)(140,13)(141,14)(142,16)(143,8)(144,11)(145,13)(146,19)(147,13)(148,17)(149,15)(150,9)(151,15)(152,9)(153,12)(154,11)(155,8)(156,13)(157,13)(158,10)(159,10)(160,15)(161,7)(162,9)(163,7)(164,9)(165,12)(166,9)(167,8)(168,6)(169,3)(170,10)(171,8)(172,4)(173,5)(174,10)(175,8)(176,6)(177,10)(178,6)(179,7)(180,5)(181,7)(182,6)(183,9)(184,6)(185,9)(186,8)(187,5)(188,7)(189,6)(190,6)(191,4)(192,3)(193,4)(194,7)(195,5)(196,3)(197,6)(198,6)(199,6)(200,7)(201,5)(202,8)(203,10)(204,4)(205,3)(206,4)(207,6)(208,7)(209,9)(210,8)(211,3)(212,7)(213,6)(214,2)(215,10)(216,4)(217,5)(218,4)(219,6)(220,4)(221,8)(222,9)(223,3)(224,5)(225,6)(226,7)(227,8)(228,5)(229,9)(230,5)(231,5)(232,4)(233,4)(234,6)(235,4)(236,5)(237,7)(238,5)(239,4)(240,6)(241,6)(242,2)(243,4)(244,4)(245,5)(246,2)(247,7)(248,1)(249,3)(250,5)(251,6)(252,4)(253,5)(254,4)(255,3)(256,7)(257,4)(258,4)(259,3)(260,4)(261,6)(262,6)(263,4)(264,3)(265,3)(266,5)(267,3)(268,6)(269,6)(270,5)(271,3)(272,6)(273,3)(274,2)(275,5)(276,3)(277,3)(278,7)(279,2)(280,4)(281,3)(282,4)(283,4)(284,6)(285,2)(286,3)(287,2)(288,4)(289,2)(290,4)(291,3)(292,4)(293,6)(294,4)(295,2)(296,2)(297,2)(298,5)(299,3)(300,3)(301,4)(302,2)(303,2)(304,4)(305,5)(306,4)(307,4)(308,3)(309,6)(310,5)(311,4)(312,2)(313,1)(314,3)(315,3)(316,4)(317,2)(318,3)(319,3)(320,2)(321,3)(322,3)(323,2)(324,3)(325,4)(326,1)(327,3)(328,1)(329,3)(330,3)(331,4)(332,3)(333,1)(334,3)(335,4)(336,1)(337,3)(338,1)
}; % plot of the second histogram
\addplot[color=cyan] coordinates{(0,36)(1,49)(2,58)(3,77)(4,70)(5,107)(6,108)(7,114)(8,141)(9,143)(10,146)(11,129)(12,160)(13,142)(14,176)(15,131)(16,136)(17,149)(18,148)(19,141)(20,159)(21,158)(22,156)(23,160)(24,157)(25,139)(26,152)(27,157)(28,155)(29,156)(30,141)(31,130)(32,117)(33,125)(34,130)(35,132)(36,131)(37,112)(38,122)(39,93)(40,95)(41,85)(42,90)(43,64)(44,75)(45,71)(46,81)(47,68)(48,88)(49,78)(50,80)(51,65)(52,79)(53,79)(54,80)(55,68)(56,84)(57,73)(58,75)(59,76)(60,69)(61,81)(62,73)(63,77)(64,75)(65,70)(66,63)(67,46)(68,52)(69,60)(70,62)(71,56)(72,47)(73,40)(74,71)(75,49)(76,52)(77,49)(78,48)(79,43)(80,47)(81,44)(82,39)(83,39)(84,20)(85,32)(86,34)(87,26)(88,17)(89,16)(90,20)(91,33)(92,30)(93,21)(94,17)(95,21)(96,17)(97,23)(98,20)(99,18)(100,12)(101,22)(102,20)(103,18)(104,22)(105,23)(106,10)(107,19)(108,14)(109,12)(110,19)(111,27)(112,13)(113,16)(114,17)(115,15)(116,13)(117,13)(118,7)(119,22)(120,17)(121,12)(122,20)(123,17)(124,9)(125,15)(126,14)(127,18)(128,17)(129,10)(130,15)(131,14)(132,13)(133,12)(134,16)(135,10)(136,14)(137,9)(138,19)(139,13)(140,13)(141,14)(142,16)(143,8)(144,11)(145,13)(146,19)(147,13)(148,17)(149,15)(150,9)(151,15)(152,9)(153,12)(154,11)(155,8)(156,13)(157,13)(158,10)(159,10)(160,15)(161,7)(162,9)(163,7)(164,9)(165,12)(166,9)(167,8)(168,6)(169,3)(170,10)(171,8)(172,4)(173,5)(174,10)(175,8)(176,6)(177,10)(178,6)(179,7)(180,5)(181,7)(182,6)(183,9)(184,6)(185,9)(186,8)(187,5)(188,7)(189,6)(190,6)(191,4)(192,3)(193,4)(194,7)(195,5)(196,3)(197,6)(198,6)(199,6)(200,7)(201,5)(202,8)(203,10)(204,4)(205,3)(206,4)(207,6)(208,7)(209,9)(210,8)(211,3)(212,7)(213,6)(214,2)(215,10)(216,4)(217,5)(218,4)(219,6)(220,4)(221,8)(222,9)(223,3)(224,5)(225,6)(226,7)(227,8)(228,5)(229,9)(230,5)(231,5)(232,4)(233,4)(234,6)(235,4)(236,5)(237,7)(238,5)(239,4)(240,6)(241,6)(242,2)(243,4)(244,4)(245,5)(246,2)(247,7)(248,1)(249,3)(250,5)(251,6)(252,4)(253,5)(254,4)(255,3)(256,7)(257,4)(258,4)(259,3)(260,4)(261,6)(262,6)(263,4)(264,3)(265,3)(266,5)(267,3)(268,6)(269,6)(270,5)(271,3)(272,6)(273,3)(274,2)(275,5)(276,3)(277,3)(278,7)(279,2)(280,4)(281,3)(282,4)(283,4)(284,6)(285,2)(286,3)(287,2)(288,4)(289,2)(290,4)(291,3)(292,4)(293,6)(294,4)(295,2)(296,2)(297,2)(298,5)(299,3)(300,3)(301,4)(302,2)(303,2)(304,4)(305,5)(306,4)(307,4)(308,3)(309,6)(310,5)(311,4)(312,2)(313,1)(314,3)(315,3)(316,4)(317,2)(318,3)(319,3)(320,2)(321,3)(322,3)(323,2)(324,3)(325,4)(326,1)(327,3)(328,1)(329,3)(330,3)(331,4)(332,3)(333,1)(334,3)(335,4)(336,1)(337,3)(338,1)
}; % plot of the third histogram
\addplot[color=violet] coordinates{(0,47)(1,55)(2,64)(3,87)(4,112)(5,132)(6,157)(7,144)(8,152)(9,169)(10,139)(11,145)(12,126)(13,124)(14,100)(15,135)(16,134)(17,138)(18,142)(19,141)(20,143)(21,152)(22,146)(23,138)(24,160)(25,137)(26,159)(27,150)(28,164)(29,160)(30,144)(31,126)(32,126)(33,110)(34,127)(35,119)(36,106)(37,105)(38,86)(39,91)(40,81)(41,76)(42,76)(43,79)(44,79)(45,73)(46,74)(47,78)(48,77)(49,81)(50,102)(51,72)(52,72)(53,56)(54,74)(55,71)(56,73)(57,90)(58,60)(59,79)(60,86)(61,58)(62,80)(63,67)(64,65)(65,53)(66,52)(67,49)(68,48)(69,44)(70,49)(71,53)(72,51)(73,61)(74,53)(75,47)(76,34)(77,57)(78,39)(79,39)(80,47)(81,33)(82,34)(83,36)(84,34)(85,27)(86,26)(87,25)(88,22)(89,19)(90,25)(91,19)(92,20)(93,10)(94,29)(95,21)(96,30)(97,20)(98,14)(99,17)(100,16)(101,11)(102,14)(103,15)(104,22)(105,11)(106,18)(107,9)(108,10)(109,14)(110,15)(111,15)(112,16)(113,8)(114,19)(115,14)(116,20)(117,10)(118,14)(119,15)(120,13)(121,20)(122,14)(123,9)(124,11)(125,9)(126,18)(127,19)(128,17)(129,12)(130,15)(131,15)(132,10)(133,12)(134,16)(135,14)(136,8)(137,10)(138,12)(139,13)(140,10)(141,17)(142,16)(143,14)(144,9)(145,10)(146,10)(147,20)(148,9)(149,7)(150,12)(151,8)(152,11)(153,14)(154,11)(155,7)(156,11)(157,6)(158,14)(159,8)(160,5)(161,6)(162,6)(163,6)(164,8)(165,4)(166,6)(167,8)(168,2)(169,6)(170,7)(171,7)(172,10)(173,5)(174,10)(175,7)(176,7)(177,2)(178,5)(179,7)(180,5)(181,5)(182,8)(183,3)(184,3)(185,4)(186,5)(187,5)(188,6)(189,4)(190,4)(191,8)(192,3)(193,3)(194,2)(195,4)(196,7)(197,5)(198,3)(199,3)(200,2)(201,8)(202,3)(203,6)(204,5)(205,6)(206,5)(207,3)(208,5)(209,6)(210,4)(211,5)(212,1)(213,5)(214,4)(215,4)(216,6)(217,3)(218,7)(219,10)(220,2)(221,2)(222,2)(223,7)(224,6)(225,3)(226,7)(227,3)(228,4)(229,4)(230,1)(231,4)(232,5)(233,6)(234,7)(235,4)(236,3)(237,9)(238,2)(239,7)(240,1)(241,3)(242,2)(243,3)(244,4)(245,2)(246,4)(247,3)(248,5)(249,1)(250,5)(251,1)(252,6)(253,3)(254,7)(255,6)(256,4)(257,3)(258,5)(259,4)(260,2)(261,3)(262,5)(263,5)(264,3)(265,2)(266,3)(267,3)(268,2)(269,1)(270,2)(271,6)(272,7)(273,3)(274,7)(275,5)(276,4)(277,3)(278,2)(279,7)(280,3)(281,1)(282,5)(283,2)(284,1)(285,2)(286,5)(287,3)(288,2)(289,2)(290,2)(291,3)(292,2)(293,5)(294,4)(295,4)(296,3)(297,4)(298,2)(299,1)(300,2)(301,1)(302,4)(303,2)(304,5)(305,2)(306,3)(307,2)(308,3)(309,1)(310,4)(311,3)(312,3)(313,3)(314,3)(315,2)(316,2)(317,3)(318,0)(319,2)(320,2)(321,1)(322,2)(323,1)(324,2)(325,2)(326,0)(327,1)(328,1)(329,3)(330,0)(331,0)(332,0)(333,2)(334,0)(335,1)(336,1)(337,1)(338,1)(339,1)(340,1)(341,0)(342,0)(343,1)(344,0)
}; % plot of the fifth histogram
\addplot[color=purple] coordinates{(0,47)(1,55)(2,64)(3,87)(4,112)(5,132)(6,157)(7,144)(8,152)(9,169)(10,139)(11,145)(12,126)(13,124)(14,100)(15,135)(16,134)(17,138)(18,142)(19,141)(20,143)(21,152)(22,146)(23,138)(24,160)(25,137)(26,159)(27,150)(28,164)(29,160)(30,144)(31,126)(32,126)(33,110)(34,127)(35,119)(36,106)(37,105)(38,86)(39,91)(40,81)(41,76)(42,76)(43,79)(44,79)(45,73)(46,74)(47,78)(48,77)(49,81)(50,102)(51,72)(52,72)(53,56)(54,74)(55,71)(56,73)(57,90)(58,60)(59,79)(60,86)(61,58)(62,80)(63,67)(64,65)(65,53)(66,52)(67,49)(68,48)(69,44)(70,49)(71,53)(72,51)(73,61)(74,53)(75,47)(76,34)(77,57)(78,39)(79,39)(80,47)(81,33)(82,34)(83,36)(84,34)(85,27)(86,26)(87,25)(88,22)(89,19)(90,25)(91,19)(92,20)(93,10)(94,29)(95,21)(96,30)(97,20)(98,14)(99,17)(100,16)(101,11)(102,14)(103,15)(104,22)(105,11)(106,18)(107,9)(108,10)(109,14)(110,15)(111,15)(112,16)(113,8)(114,19)(115,14)(116,20)(117,10)(118,14)(119,15)(120,13)(121,20)(122,14)(123,9)(124,11)(125,9)(126,18)(127,19)(128,17)(129,12)(130,15)(131,15)(132,10)(133,12)(134,16)(135,14)(136,8)(137,10)(138,12)(139,13)(140,10)(141,17)(142,16)(143,14)(144,9)(145,10)(146,10)(147,20)(148,9)(149,7)(150,12)(151,8)(152,11)(153,14)(154,11)(155,7)(156,11)(157,6)(158,14)(159,8)(160,5)(161,6)(162,6)(163,6)(164,8)(165,4)(166,6)(167,8)(168,2)(169,6)(170,7)(171,7)(172,10)(173,5)(174,10)(175,7)(176,7)(177,2)(178,5)(179,7)(180,5)(181,5)(182,8)(183,3)(184,3)(185,4)(186,5)(187,5)(188,6)(189,4)(190,4)(191,8)(192,3)(193,3)(194,2)(195,4)(196,7)(197,5)(198,3)(199,3)(200,2)(201,8)(202,3)(203,6)(204,5)(205,6)(206,5)(207,3)(208,5)(209,6)(210,4)(211,5)(212,1)(213,5)(214,4)(215,4)(216,6)(217,3)(218,7)(219,10)(220,2)(221,2)(222,2)(223,7)(224,6)(225,3)(226,7)(227,3)(228,4)(229,4)(230,1)(231,4)(232,5)(233,6)(234,7)(235,4)(236,3)(237,9)(238,2)(239,7)(240,1)(241,3)(242,2)(243,3)(244,4)(245,2)(246,4)(247,3)(248,5)(249,1)(250,5)(251,1)(252,6)(253,3)(254,7)(255,6)(256,4)(257,3)(258,5)(259,4)(260,2)(261,3)(262,5)(263,5)(264,3)(265,2)(266,3)(267,3)(268,2)(269,1)(270,2)(271,6)(272,7)(273,3)(274,7)(275,5)(276,4)(277,3)(278,2)(279,7)(280,3)(281,1)(282,5)(283,2)(284,1)(285,2)(286,5)(287,3)(288,2)(289,2)(290,2)(291,3)(292,2)(293,5)(294,4)(295,4)(296,3)(297,4)(298,2)(299,1)(300,2)(301,1)(302,4)(303,2)(304,5)(305,2)(306,3)(307,2)(308,3)(309,1)(310,4)(311,3)(312,3)(313,3)(314,3)(315,2)(316,2)(317,3)(318,0)(319,2)(320,2)(321,1)(322,2)(323,1)(324,2)(325,2)(326,0)(327,1)(328,1)(329,3)(330,0)(331,0)(332,0)(333,2)(334,0)(335,1)(336,1)(337,1)(338,1)(339,1)(340,1)(341,0)(342,0)(343,1)(344,0)
}; % plot of the first histogram
\addplot[color=magenta] coordinates{(0,47)(1,55)(2,64)(3,87)(4,112)(5,132)(6,157)(7,144)(8,152)(9,169)(10,139)(11,145)(12,126)(13,124)(14,100)(15,135)(16,134)(17,138)(18,142)(19,141)(20,143)(21,152)(22,146)(23,138)(24,160)(25,137)(26,159)(27,150)(28,164)(29,160)(30,144)(31,126)(32,126)(33,110)(34,127)(35,119)(36,106)(37,105)(38,86)(39,91)(40,81)(41,76)(42,76)(43,79)(44,79)(45,73)(46,74)(47,78)(48,77)(49,81)(50,102)(51,72)(52,72)(53,56)(54,74)(55,71)(56,73)(57,90)(58,60)(59,79)(60,86)(61,58)(62,80)(63,67)(64,65)(65,53)(66,52)(67,49)(68,48)(69,44)(70,49)(71,53)(72,51)(73,61)(74,53)(75,47)(76,34)(77,57)(78,39)(79,39)(80,47)(81,33)(82,34)(83,36)(84,34)(85,27)(86,26)(87,25)(88,22)(89,19)(90,25)(91,19)(92,20)(93,10)(94,29)(95,21)(96,30)(97,20)(98,14)(99,17)(100,16)(101,11)(102,14)(103,15)(104,22)(105,11)(106,18)(107,9)(108,10)(109,14)(110,15)(111,15)(112,16)(113,8)(114,19)(115,14)(116,20)(117,10)(118,14)(119,15)(120,13)(121,20)(122,14)(123,9)(124,11)(125,9)(126,18)(127,19)(128,17)(129,12)(130,15)(131,15)(132,10)(133,12)(134,16)(135,14)(136,8)(137,10)(138,12)(139,13)(140,10)(141,17)(142,16)(143,14)(144,9)(145,10)(146,10)(147,20)(148,9)(149,7)(150,12)(151,8)(152,11)(153,14)(154,11)(155,7)(156,11)(157,6)(158,14)(159,8)(160,5)(161,6)(162,6)(163,6)(164,8)(165,4)(166,6)(167,8)(168,2)(169,6)(170,7)(171,7)(172,10)(173,5)(174,10)(175,7)(176,7)(177,2)(178,5)(179,7)(180,5)(181,5)(182,8)(183,3)(184,3)(185,4)(186,5)(187,5)(188,6)(189,4)(190,4)(191,8)(192,3)(193,3)(194,2)(195,4)(196,7)(197,5)(198,3)(199,3)(200,2)(201,8)(202,3)(203,6)(204,5)(205,6)(206,5)(207,3)(208,5)(209,6)(210,4)(211,5)(212,1)(213,5)(214,4)(215,4)(216,6)(217,3)(218,7)(219,10)(220,2)(221,2)(222,2)(223,7)(224,6)(225,3)(226,7)(227,3)(228,4)(229,4)(230,1)(231,4)(232,5)(233,6)(234,7)(235,4)(236,3)(237,9)(238,2)(239,7)(240,1)(241,3)(242,2)(243,3)(244,4)(245,2)(246,4)(247,3)(248,5)(249,1)(250,5)(251,1)(252,6)(253,3)(254,7)(255,6)(256,4)(257,3)(258,5)(259,4)(260,2)(261,3)(262,5)(263,5)(264,3)(265,2)(266,3)(267,3)(268,2)(269,1)(270,2)(271,6)(272,7)(273,3)(274,7)(275,5)(276,4)(277,3)(278,2)(279,7)(280,3)(281,1)(282,5)(283,2)(284,1)(285,2)(286,5)(287,3)(288,2)(289,2)(290,2)(291,3)(292,2)(293,5)(294,4)(295,4)(296,3)(297,4)(298,2)(299,1)(300,2)(301,1)(302,4)(303,2)(304,5)(305,2)(306,3)(307,2)(308,3)(309,1)(310,4)(311,3)(312,3)(313,3)(314,3)(315,2)(316,2)(317,3)(318,0)(319,2)(320,2)(321,1)(322,2)(323,1)(324,2)(325,2)(326,0)(327,1)(328,1)(329,3)(330,0)(331,0)(332,0)(333,2)(334,0)(335,1)(336,1)(337,1)(338,1)(339,1)(340,1)(341,0)(342,0)(343,1)(344,0)
}; % plot of the second histogram
\legend{false \#1,  false \#2, false \#3, true \#1, true \#2, true \#3} % legend of the plot

\end{axis}
\end{tikzpicture}

%% file: plots/accessible_exits/injury_levels.tex
\begin{tikzpicture}[scale=0.75]
\begin{axis}[
title={Injury Levels - Accessible exits},
ybar, % style of the histogram: clustered columns
width=\columnwidth, % width of the plot
height=8cm, % height of the plot
ylabel={Amount}, % label for the y-axis
symbolic x coords={HEALTHY,MINOR,MODERATE,SERIOUS}, % labels for the x-ticks
xtick=data, % position the x-ticks at the data points
ymin=0, % minimum value of the y-axis
legend style={at={(1,1)}, anchor=north east}, % position of the legend
legend cell align=left, % alignment of the legend cells
ymajorgrids=true, % display major grids
bar width=0.15cm % width of the bars
]
\addplot[color=blue, fill] coordinates{(HEALTHY,16538)(MINOR,10550)(MODERATE,2106)(SERIOUS,686)
}; % plot of the first histogram
\addplot[color=teal, fill] coordinates{(HEALTHY,16348)(MINOR,10830)(MODERATE,2035)(SERIOUS,617)
}; % plot of the second histogram
\addplot[color=cyan, fill] coordinates{(HEALTHY,16338)(MINOR,10800)(MODERATE,2068)(SERIOUS,627)
}; % plot of the third histogram
\addplot[color=violet, fill] coordinates{(HEALTHY,17470)(MINOR,10152)(MODERATE,1803)(SERIOUS,486)
}; % plot of the fifth histogram
\addplot[color=purple, fill] coordinates{(HEALTHY,17416)(MINOR,10160)(MODERATE,1826)(SERIOUS,513)
}; % plot of the first histogram
\addplot[color=magenta, fill] coordinates{(HEALTHY,17370)(MINOR,10134)(MODERATE,1916)(SERIOUS,482)
}; % plot of the second histogram
\legend{false \#1,  false \#2, false \#3, true \#1, true \#2, true \#3} % legend of the plot
\end{axis}
\end{tikzpicture}

\begin{tikzpicture}[scale=0.75]
\begin{axis}[
ybar, % style of the histogram: clustered columns
width=\columnwidth, % width of the plot
height=8cm, % height of the plot
ylabel={Amount}, % label for the y-axis
symbolic x coords={SEVERE,CRITICAL,FATAL}, % labels for the x-ticks
xtick=data, % position the x-ticks at the data points
ymin=0, % minimum value of the y-axis
legend style={at={(1,1)}, anchor=north east}, % position of the legend
legend cell align=left, % alignment of the legend cells
ymajorgrids=true, % display major grids
bar width=0.15cm % width of the bars
]
\addplot[color=blue, fill] coordinates{(SEVERE,113)(CRITICAL,6)(FATAL,1)
}; % plot of the first histogram
\addplot[color=teal, fill] coordinates{(SEVERE,146)(CRITICAL,24)(FATAL,0)
}; % plot of the second histogram
\addplot[color=cyan, fill] coordinates{(SEVERE,148)(CRITICAL,16)(FATAL,3)
}; % plot of the third histogram
\addplot[color=violet, fill] coordinates{(SEVERE,80)(CRITICAL,7)(FATAL,2)
}; % plot of the fifth histogram
\addplot[color=purple, fill] coordinates{(SEVERE,73)(CRITICAL,12)(FATAL,0)
}; % plot of the first histogram
\addplot[color=magenta, fill] coordinates{(SEVERE,91)(CRITICAL,7)(FATAL,0)
}; % plot of the second histogram
\legend{false \#1,  false \#2, false \#3, true \#1, true \#2, true \#3} % legend of the plot
\end{axis}
\end{tikzpicture}

%% file: plots/glass_bottles/injury_levels.tex
\begin{tikzpicture}[scale=0.75]
\begin{axis}[
title={Injury Levels - Glass bottles},
ybar, % style of the histogram: clustered columns
width=\columnwidth, % width of the plot
height=8cm, % height of the plot
ylabel={Amount}, % label for the y-axis
symbolic x coords={HEALTHY,MINOR,MODERATE,SERIOUS}, % labels for the x-ticks
xtick=data, % position the x-ticks at the data points
ymin=0, % minimum value of the y-axis
legend style={at={(1,1)}, anchor=north east}, % position of the legend
legend cell align=left, % alignment of the legend cells
ymajorgrids=true, % display major grids
bar width=0.15cm % width of the bars
]
\addplot[color=PineGreen, fill] coordinates{(HEALTHY,16515)(MINOR,10639)(MODERATE,2047)(SERIOUS,660)
}; % plot of the first histogram
\addplot[color=Green, fill] coordinates{(HEALTHY,16461)(MINOR,10690)(MODERATE,2074)(SERIOUS,634)
}; % plot of the second histogram
\addplot[color=LimeGreen, fill] coordinates{(HEALTHY,16438)(MINOR,10650)(MODERATE,2106)(SERIOUS,660)
}; % plot of the third histogram
\addplot[color=RedOrange, fill] coordinates{(HEALTHY,23840)(MINOR,5101)(MODERATE,824)(SERIOUS,211)
}; % plot of the fourth histogram
\addplot[color=Orange, fill] coordinates{(HEALTHY,23781)(MINOR,5170)(MODERATE,793)(SERIOUS,192)
}; % plot of the fifth histogram
\addplot[color=Melon, fill] coordinates{(HEALTHY,23902)(MINOR,5062)(MODERATE,791)(SERIOUS,210)
}; % plot of the first histogram
\legend{glass \#1,  glass \#2, glass \#3, no glass \#1,  no glass \#2, no glass \#3,} % legend of the plot
\end{axis}
\end{tikzpicture}

\begin{tikzpicture}[scale=0.75]
\begin{axis}[
ybar, % style of the histogram: clustered columns
width=\columnwidth, % width of the plot
height=8cm, % height of the plot
ylabel={Amount}, % label for the y-axis
symbolic x coords={SEVERE,CRITICAL,FATAL}, % labels for the x-ticks
xtick=data, % position the x-ticks at the data points
ymin=0, % minimum value of the y-axis
legend style={at={(1,1)}, anchor=north east}, % position of the legend
legend cell align=left, % alignment of the legend cells
ymajorgrids=true, % display major grids
bar width=0.15cm % width of the bars
]
\addplot[color=PineGreen, fill] coordinates{(SEVERE,124)(CRITICAL,13)(FATAL,2)
}; % plot of the first histogram
\addplot[color=Green, fill] coordinates{(SEVERE,131)(CRITICAL,9)(FATAL,1)
}; % plot of the second histogram
\addplot[color=LimeGreen, fill] coordinates{(SEVERE,129)(CRITICAL,15)(FATAL,2)
}; % plot of the third histogram
\addplot[color=RedOrange, fill] coordinates{(SEVERE,23)(CRITICAL,1)(FATAL,0)
}; % plot of the fourth histogram
\addplot[color=Orange, fill] coordinates{(SEVERE,58)(CRITICAL,6)(FATAL,0)
}; % plot of the fifth histogram
\addplot[color=Melon, fill] coordinates{(SEVERE,33)(CRITICAL,2)(FATAL,0)
}; % plot of the first histogram
\legend{glass \#1,  glass \#2, glass \#3, no glass \#1,  no glass \#2, no glass \#3,} % legend of the plot
\end{axis}
\end{tikzpicture}

%% file: plots/glass_bottles/avg_speed.tex
\begin{tikzpicture}[scale=0.75]

% Set the axis labels and title
\begin{axis}[    xlabel={Time [s]},    ylabel={Speed [m/s]},    title={Average Speed - Glass bottles},   legend style={at={(1,1)}, anchor=north east},]

\addplot[color=PineGreen] coordinates{(0,4.018462155924947)(1,4.016671708513999)(2,4.013646214276039)(3,4.008721825084653)(4,4.002215989751047)(5,3.9962810932182578)(6,3.988265087744329)(7,3.978970689717793)(8,3.970913257197768)(9,3.9626059122575787)(10,3.953357134630888)(11,3.9423695566888974)(12,3.931678761878032)(13,3.920307742944129)(14,3.905992651716711)(15,3.8921109666233202)(16,3.878932146829811)(17,3.8636322049405307)(18,3.846473420829006)(19,3.8320962499393585)(20,3.8168102375143156)(21,3.7979056516605922)(22,3.780514880795052)(23,3.7618138164356654)(24,3.739145164022359)(25,3.7136512388966807)(26,3.686848484848485)(27,3.6589795148925472)(28,3.632392254026211)(29,3.602261953135582)(30,3.570190274841438)(31,3.5404632751795924)(32,3.511787594415198)(33,3.487473326483838)(34,3.465756783262504)(35,3.4429587000677047)(36,3.4171207593601687)(37,3.3935290375352176)(38,3.3707009960533734)(39,3.3465865011636926)(40,3.3241917725953356)(41,3.301453158816861)(42,3.28106164745691)(43,3.2638542457434117)(44,3.2402178746109382)(45,3.221522610188895)(46,3.202728127939793)(47,3.1794128307357736)(48,3.1578489816194732)(49,3.131258803944167)(50,3.1082827749934054)(51,3.076158038147139)(52,3.0440639910188043)(53,3.0072202166064983)(54,2.9626966292134833)(55,2.9305233904585455)(56,2.8913078277573234)(57,2.8450190618266205)(58,2.798203178991016)(59,2.7443447037701976)(60,2.687007134810364)(61,2.623923257635082)(62,2.557029177718833)(63,2.5010611205432935)(64,2.4345422379478108)(65,2.383467278989667)(66,2.333810888252149)(67,2.282527881040892)(68,2.2359866905554133)(69,2.1947159841479524)(70,2.1524356869184453)(71,2.1211608903916597)(72,2.073542338118957)(73,2.024173806609547)(74,1.974920634920635)(75,1.919631093544137)(76,1.8744027303754267)(77,1.8203486303806475)(78,1.7602968460111317)(79,1.7029930928626247)(80,1.6448784376245515)(81,1.5865782932891466)(82,1.5382969619169875)(83,1.494931687968268)(84,1.4596188747731398)(85,1.4390356977283263)(86,1.4175355450236966)(87,1.4070327552986512)(88,1.3981481481481481)(89,1.3920314805705853)(90,1.3880597014925373)(91,1.3839959738298944)(92,1.3753830439223698)(93,1.365334717176959)(94,1.3593096234309623)(95,1.3526287838555497)(96,1.3448275862068966)(97,1.3371584699453551)(98,1.3298109010011123)(99,1.3291995490417137)(100,1.3245714285714285)(101,1.3138051044083527)(102,1.309551886792453)(103,1.3004189108318371)(104,1.2947941888619854)(105,1.2893772893772895)(106,1.2868649318463445)(107,1.2768361581920904)(108,1.267515923566879)(109,1.2621359223300972)(110,1.2540929927963327)(111,1.2451698867421719)(112,1.2387053270397843)(113,1.2344497607655502)(114,1.227681660899654)(115,1.2192362093352191)(116,1.2124910265613784)(117,1.205710102489019)(118,1.1986656782802076)(119,1.191904047976012)(120,1.1854961832061068)(121,1.1817478731631863)(122,1.175686274509804)(123,1.1727707006369428)(124,1.1670702179176755)(125,1.165302782324059)(126,1.1572379367720467)(127,1.1544991511035654)(128,1.1512027491408934)(129,1.1514360313315928)(130,1.147887323943662)(131,1.1410601976639712)(132,1.1339449541284403)(133,1.130232558139535)(134,1.1271186440677967)(135,1.1248808388941849)(136,1.1194605009633911)(137,1.115686274509804)(138,1.107)(139,1.104251012145749)(140,1.1007194244604317)(141,1.098635886673662)(142,1.0970149253731343)(143,1.093649085037675)(144,1.087051142546246)(145,1.0833333333333333)(146,1.079455164585698)(147,1.0774566473988438)(148,1.0762016412661195)(149,1.0681818181818181)(150,1.0665859564164648)(151,1.0626535626535627)(152,1.0590452261306533)(153,1.0574712643678161)(154,1.0568475452196382)(155,1.0484927916120577)(156,1.0482573726541555)(157,1.0462585034013605)(158,1.042995839112344)(159,1.0355113636363635)(160,1.0343839541547277)(161,1.0305232558139534)(162,1.0251107828655834)(163,1.0164917541229386)(164,1.013719512195122)(165,1.0109375)(166,1.0063492063492063)(167,1.0032258064516129)(168,1.0032679738562091)(169,1.00164744645799)(170,1.001669449081803)(171,1.0016863406408094)(172,1)(173,1)(174,1)(175,1)(176,1)(177,1)(178,1)(179,1)(180,1)(181,1)(182,1)(183,1)(184,1)(185,1)(186,1)(187,1)(188,1)(189,1)(190,1)(191,1)(192,1)(193,1)(194,1)(195,1)(196,1)(197,1)(198,1)(199,1)(200,1)(201,1)(202,1)(203,1)(204,1)(205,1)(206,1)(207,1)(208,1)(209,1)(210,1)(211,1)(212,1)(213,1)(214,1)(215,1)(216,1)(217,1)(218,1)(219,1)(220,1)(221,1)(222,1)(223,1)(224,1)(225,1)(226,1)(227,1)(228,1)(229,1)(230,1)(231,1)(232,1)(233,1)(234,1)(235,1)(236,1)(237,1)(238,1)(239,1)(240,1)(241,1)(242,1)(243,1)(244,1)(245,1)(246,1)(247,1)(248,1)(249,1)(250,1)(251,1)(252,1)(253,1)(254,1)(255,1)(256,1)(257,1)(258,1)(259,1)(260,1)(261,1)(262,1)(263,1)(264,1)(265,1)(266,1)(267,1)(268,1)(269,1)(270,1)(271,1)(272,1)(273,1)(274,1)(275,1)(276,1)(277,1)(278,1)(279,1)(280,1)(281,1)(282,1)(283,1)(284,1)(285,1)(286,1)(287,1.0081967213114753)(288,1.0084033613445378)(289,1.008695652173913)(290,1.008849557522124)(291,1.009090909090909)(292,1.0093457943925233)(293,1.0096153846153846)(294,1.01)(295,1.01)(296,1.010204081632653)(297,1.0104166666666667)(298,1.0108695652173914)(299,1.0112359550561798)(300,1.0113636363636365)(301,1.011764705882353)(302,1.0119047619047619)(303,1.0125)(304,1.0129870129870129)(305,1.0135135135135136)(306,1.0140845070422535)(307,1.0149253731343284)(308,1.0161290322580645)(309,1.0166666666666666)(310,1.0166666666666666)(311,1.018181818181818)(312,1.0192307692307692)(313,1.02)(314,1.0425531914893618)(315,1.0444444444444445)(316,1.0465116279069768)(317,1.0526315789473684)(318,1.0571428571428572)(319,1.0588235294117647)(320,1.064516129032258)(321,1.0769230769230769)(322,1.0833333333333333)(323,1.105263157894737)(324,1.1176470588235294)(325,1.125)(326,1.1333333333333333)(327,1.1666666666666667)(328,1.1666666666666667)(329,1.1666666666666667)(330,1.1818181818181819)(331,1.2)(332,1.25)(333,1.25)(334,1.25)(335,1.2857142857142858)(336,1.2857142857142858)(337,1.5)(338,1.5)(339,1.6666666666666667)(340,1.6666666666666667)(341,1.6666666666666667)(342,2)
}; % plot of the first histogram
\addplot[color=Green] coordinates{(0,4.019208245490748)(1,4.017361461592813)(2,4.01435390500694)(3,4.009068510026692)(4,4.00377633037694)(5,3.9974014116655545)(6,3.9899123119697726)(7,3.9812880765883376)(8,3.973362397014815)(9,3.964796686746988)(10,3.955594002306805)(11,3.9462813162648236)(12,3.9359704725988927)(13,3.924498090897894)(14,3.9110961441320087)(15,3.896245999481014)(16,3.8815075198006586)(17,3.8665324978273796)(18,3.850873475537976)(19,3.8361116495444856)(20,3.81800938404712)(21,3.8018693508627774)(22,3.78300439176676)(23,3.764532664412695)(24,3.7426155580608795)(25,3.7179756282432512)(26,3.694484585741811)(27,3.6672288855572215)(28,3.6344106957424716)(29,3.602565833896016)(30,3.5717194411289754)(31,3.5427322404371586)(32,3.5185269207501513)(33,3.4954488386691778)(34,3.4731513728031103)(35,3.451079197110943)(36,3.426059709287144)(37,3.4046394078887987)(38,3.3835820895522386)(39,3.3598689282960676)(40,3.3394276629570747)(41,3.3196897009288557)(42,3.2979147018757202)(43,3.2777897226402923)(44,3.2561892130857646)(45,3.234859675036928)(46,3.209212376933896)(47,3.1892543067100347)(48,3.1603328365623446)(49,3.1340153452685424)(50,3.107722772277228)(51,3.07120318287831)(52,3.0352441613588113)(53,2.997214076246334)(54,2.9632041187159297)(55,2.9215440138082536)(56,2.880441845354126)(57,2.836247478143914)(58,2.788397693517386)(59,2.73622977640429)(60,2.6783957623912222)(61,2.6247548058061985)(62,2.569560776302349)(63,2.508633553613302)(64,2.453578550853091)(65,2.3960739030023093)(66,2.339966434907696)(67,2.277501244400199)(68,2.227530991735537)(69,2.1854038100348805)(70,2.1354137642797437)(71,2.0948947216613787)(72,2.052788547569341)(73,2.0018587360594795)(74,1.9536530415191502)(75,1.9092122830440588)(76,1.8630705394190872)(77,1.813978494623656)(78,1.757201646090535)(79,1.7020534676481984)(80,1.637389202256245)(81,1.5898502495840265)(82,1.5411106328024107)(83,1.4988923349579086)(84,1.463013698630137)(85,1.4274988268418582)(86,1.3975961538461539)(87,1.3798717316230884)(88,1.3703148425787106)(89,1.360163015792155)(90,1.3534571723426212)(91,1.3482794577685089)(92,1.3436334576451785)(93,1.3405172413793103)(94,1.3340623291416074)(95,1.3266888150609082)(96,1.3228523301516002)(97,1.3185437997724687)(98,1.3140877598152425)(99,1.30697129466901)(100,1.300774270399047)(101,1.296318648159324)(102,1.292326431181486)(103,1.287313432835821)(104,1.2815047021943573)(105,1.2755102040816326)(106,1.2701482914248872)(107,1.264089121887287)(108,1.2583001328021248)(109,1.2491582491582491)(110,1.2454112848402448)(111,1.2386519944979368)(112,1.231145251396648)(113,1.2288732394366197)(114,1.2235714285714285)(115,1.2173913043478262)(116,1.214760147601476)(117,1.2119402985074628)(118,1.2066616199848599)(119,1.2040030792917629)(120,1.2018706157443493)(121,1.197172034564022)(122,1.187849720223821)(123,1.183739837398374)(124,1.1817434210526316)(125,1.1769616026711185)(126,1.1711864406779662)(127,1.1673856773080242)(128,1.1633187772925764)(129,1.1604609929078014)(130,1.1572832886505808)(131,1.1546860782529573)(132,1.1488372093023256)(133,1.1446124763705103)(134,1.1435406698564594)(135,1.1358748778103618)(136,1.1332675222112536)(137,1.126873126873127)(138,1.1233569261880687)(139,1.120162932790224)(140,1.1141975308641976)(141,1.1107628004179728)(142,1.1058201058201058)(143,1.105037513397642)(144,1.1055495103373232)(145,1.1040974529346623)(146,1.1037204058624577)(147,1.0942528735632184)(148,1.0851808634772462)(149,1.0802395209580837)(150,1.079268292682927)(151,1.070631970260223)(152,1.0617906683480454)(153,1.0561941251596425)(154,1.0506493506493506)(155,1.0501981505944518)(156,1.0456989247311828)(157,1.0397260273972602)(158,1.0361613351877608)(159,1.0308555399719495)(160,1.0297872340425531)(161,1.0261248185776488)(162,1.017725258493353)(163,1.0134730538922156)(164,1.0105421686746987)(165,1.0091603053435114)(166,1.0061728395061729)(167,1.0062402496099845)(168,1.0047169811320755)(169,1.0047543581616483)(170,1.0048076923076923)(171,1.0032310177705976)(172,1.0016313213703099)(173,1)(174,1)(175,1)(176,1)(177,1)(178,1)(179,1)(180,1)(181,1)(182,1)(183,1)(184,1)(185,1)(186,1)(187,1)(188,1)(189,1)(190,1)(191,1)(192,1)(193,1)(194,1)(195,1)(196,1)(197,1)(198,1)(199,1.0021413276231264)(200,1.002145922746781)(201,1.0021505376344086)(202,1.002169197396963)(203,1.0021929824561404)(204,1.0022222222222221)(205,1.0022371364653244)(206,1.0022421524663676)(207,1.002267573696145)(208,1.0022988505747126)(209,1.0023201856148491)(210,1.0023364485981308)(211,1.0023584905660377)(212,1.0023809523809524)(213,1.0023923444976077)(214,1.0024038461538463)(215,1.002433090024331)(216,1.0024630541871922)(217,1.0025188916876575)(218,1.0025445292620865)(219,1.0025839793281655)(220,1.0026246719160106)(221,1.002680965147453)(222,1.002710027100271)(223,1.002754820936639)(224,1.0027624309392265)(225,1.0027855153203342)(226,1.0027932960893855)(227,1.0028328611898016)(228,1.0028818443804035)(229,1.002906976744186)(230,1.0029411764705882)(231,1.0029585798816567)(232,1.0029940119760479)(233,1.0030211480362539)(234,1.0030674846625767)(235,1.0062305295950156)(236,1.0063291139240507)(237,1.0064102564102564)(238,1.0064935064935066)(239,1.0066225165562914)(240,1.006779661016949)(241,1.0069686411149825)(242,1.0070671378091873)(243,1.007168458781362)(244,1.0072992700729928)(245,1.0074074074074073)(246,1.0074349442379182)(247,1.0075471698113208)(248,1.0076335877862594)(249,1.0076923076923077)(250,1.0077519379844961)(251,1.0078125)(252,1.008)(253,1.008130081300813)(254,1.0081632653061225)(255,1.0082644628099173)(256,1.0084033613445378)(257,1.0085106382978724)(258,1.0086206896551724)(259,1.0089686098654709)(260,1.009090909090909)(261,1.0092592592592593)(262,1.0093457943925233)(263,1.0095238095238095)(264,1.0097560975609756)(265,1.00990099009901)(266,1.0100502512562815)(267,1.0103092783505154)(268,1.0105820105820107)(269,1.0108108108108107)(270,1.010989010989011)(271,1.0113636363636365)(272,1.0115606936416186)(273,1.011764705882353)(274,1.0119047619047619)(275,1.0121951219512195)(276,1.0124223602484472)(277,1.0126582278481013)(278,1.0128205128205128)(279,1.0135135135135136)(280,1.0138888888888888)(281,1.0144927536231885)(282,1.0150375939849625)(283,1.0151515151515151)(284,1.015267175572519)(285,1.016)(286,1.0161290322580645)(287,1.0163934426229508)(288,1.0172413793103448)(289,1.018018018018018)(290,1.0185185185185186)(291,1.0194174757281553)(292,1.02020202020202)(293,1.0208333333333333)(294,1.0210526315789474)(295,1.021505376344086)(296,1.0222222222222221)(297,1.0227272727272727)(298,1.0232558139534884)(299,1.0235294117647058)(300,1.0246913580246915)(301,1.0256410256410255)(302,1.0266666666666666)(303,1.027027027027027)(304,1.0277777777777777)(305,1.0289855072463767)(306,1.0298507462686568)(307,1.0298507462686568)(308,1.0307692307692307)(309,1.0307692307692307)(310,1.0327868852459017)(311,1.0350877192982457)(312,1.0363636363636364)(313,1.037037037037037)(314,1.04)(315,1.0425531914893618)(316,1.0444444444444445)(317,1.05)(318,1.0588235294117647)(319,1.064516129032258)(320,1.0714285714285714)(321,1.0769230769230769)(322,1.0869565217391304)(323,1.0869565217391304)(324,1.0952380952380953)(325,1.1111111111111112)(326,1.1176470588235294)(327,1.1176470588235294)(328,1.125)(329,1.1333333333333333)(330,1.1538461538461537)(331,1.1818181818181819)(332,1.2222222222222223)(333,1.2222222222222223)(334,1.25)(335,1.25)(336,1.25)(337,1.25)(338,1.25)(339,1.4)(340,1.4)(341,1.5)(342,1.6666666666666667)(343,1.6666666666666667)(344,2)(345,2)(346,3)
}; % plot of the second histogram
\addplot[color=LimeGreen] coordinates{(0,4.019136835061894)(1,4.017214134418182)(2,4.014381429344883)(3,4.010995390131467)(4,4.005559008355776)(5,4.000104858441104)(6,3.992185279909065)(7,3.9843196762771877)(8,3.97596260030921)(9,3.9665678586169375)(10,3.957813814732655)(11,3.947884134915678)(12,3.9363632729743774)(13,3.925173134450682)(14,3.9122777520911267)(15,3.8979450872042825)(16,3.884296788482835)(17,3.8712245269286756)(18,3.856078706957133)(19,3.8404609228098936)(20,3.8240003978516013)(21,3.8057081258662286)(22,3.787110406761754)(23,3.7674215552523873)(24,3.7467253839205057)(25,3.721245399848122)(26,3.694713870029098)(27,3.667400465438078)(28,3.6360958366064415)(29,3.6022137195955177)(30,3.570231501207215)(31,3.541285756414037)(32,3.5152418060967223)(33,3.4935445544554455)(34,3.474214145383104)(35,3.4505895326151497)(36,3.4328358208955225)(37,3.4112514768699445)(38,3.3894004886299567)(39,3.370183930300097)(40,3.3469631993617233)(41,3.327338129496403)(42,3.307895292379143)(43,3.289185378023647)(44,3.2668301382077574)(45,3.241498161764706)(46,3.214582098399526)(47,3.194227959084267)(48,3.1643268989721736)(49,3.1387739265390584)(50,3.106394339874516)(51,3.0768912773873502)(52,3.043360818879727)(53,3.0060134936931653)(54,2.9695446931627836)(55,2.935001573811772)(56,2.897113973585521)(57,2.859045692126117)(58,2.810296684118674)(59,2.754322111010009)(60,2.6956356736242886)(61,2.643937905285911)(62,2.583639330885353)(63,2.5250957039557633)(64,2.4577324162414023)(65,2.3913948646773076)(66,2.332531280076997)(67,2.274342928660826)(68,2.220427305888484)(69,2.1726831896551726)(70,2.1233455364686002)(71,2.0839140534262484)(72,2.0494011976047903)(73,2.007203257124961)(74,1.9603638726445745)(75,1.913484285231497)(76,1.858093903293623)(77,1.8016017473607573)(78,1.7481203007518797)(79,1.6856134849078792)(80,1.6203252032520326)(81,1.5790139064475348)(82,1.5445501730103806)(83,1.512638580931264)(84,1.4785974499089254)(85,1.4502347417840376)(86,1.4251554280248684)(87,1.4092909535452323)(88,1.397308075772682)(89,1.387113140537798)(90,1.3806085611139762)(91,1.3760504201680672)(92,1.3659711075441412)(93,1.3569482288828338)(94,1.3464305478693968)(95,1.340628507295174)(96,1.3327654741624078)(97,1.3256484149855907)(98,1.3214909726266744)(99,1.3171884229178972)(100,1.3083832335329342)(101,1.3026076409945422)(102,1.2917690417690417)(103,1.2849127182044888)(104,1.2808132147395173)(105,1.2768834513844172)(106,1.2682291666666667)(107,1.2635046113306982)(108,1.2580213903743316)(109,1.2539035980991176)(110,1.2450103234686856)(111,1.2424877707896576)(112,1.2386685552407932)(113,1.226848528356066)(114,1.2213294375456538)(115,1.2163958641063515)(116,1.2102486812358704)(117,1.2055003819709702)(118,1.2009273570324575)(119,1.195447409733124)(120,1.1883012820512822)(121,1.1849757673667205)(122,1.1763740771123872)(123,1.1727574750830565)(124,1.169463087248322)(125,1.166098807495741)(126,1.1612068965517242)(127,1.1577092511013216)(128,1.1544642857142857)(129,1.1459655485040798)(130,1.1425942962281508)(131,1.1380597014925373)(132,1.133649289099526)(133,1.127536231884058)(134,1.125)(135,1.1222879684418146)(136,1.1181181181181181)(137,1.110435663627153)(138,1.1017653167185877)(139,1.1009463722397477)(140,1.0972222222222223)(141,1.0944625407166124)(142,1.0898138006571743)(143,1.0845383759733036)(144,1.0810810810810811)(145,1.0785876993166288)(146,1.0775462962962963)(147,1.070921985815603)(148,1.0661853188929)(149,1.0625)(150,1.0586034912718205)(151,1.053231939163498)(152,1.0501285347043703)(153,1.0430809399477807)(154,1.0411686586985391)(155,1.0377358490566038)(156,1.0355677154582763)(157,1.0332871012482663)(158,1.0295774647887324)(159,1.026950354609929)(160,1.027220630372493)(161,1.0231884057971015)(162,1.0221238938053097)(163,1.0194610778443114)(164,1.0136363636363637)(165,1.0122137404580154)(166,1.0108359133126934)(167,1.0062893081761006)(168,1.0048)(169,1.0048309178743962)(170,1.0016313213703099)(171,1.001639344262295)(172,1.001669449081803)(173,1.0016920473773265)(174,1.0017035775127767)(175,1.0017123287671232)(176,1.0017391304347827)(177,1.0017574692442883)(178,1.0017699115044247)(179,1.00177304964539)(180,1.0017857142857143)(181,1.0018050541516246)(182,1)(183,1)(184,1)(185,1)(186,1)(187,1)(188,1)(189,1)(190,1)(191,1)(192,1)(193,1)(194,1)(195,1)(196,1)(197,1)(198,1)(199,1)(200,1)(201,1)(202,1)(203,1)(204,1)(205,1)(206,1)(207,1)(208,1)(209,1)(210,1)(211,1)(212,1)(213,1)(214,1)(215,1)(216,1)(217,1)(218,1)(219,1)(220,1)(221,1)(222,1)(223,1)(224,1)(225,1)(226,1)(227,1)(228,1)(229,1.002932551319648)(230,1.0029761904761905)(231,1.0030120481927711)(232,1.003058103975535)(233,1.0030959752321982)(234,1.003125)(235,1.0031446540880504)(236,1.0031948881789137)(237,1.0032258064516129)(238,1.0032362459546926)(239,1.0032894736842106)(240,1.0033444816053512)(241,1.0033783783783783)(242,1.0034129692832765)(243,1.0034129692832765)(244,1.0034843205574913)(245,1.0035211267605635)(246,1.00355871886121)(247,1.0036101083032491)(248,1.0036764705882353)(249,1.003717472118959)(250,1.0038022813688212)(251,1.0038610038610039)(252,1.003921568627451)(253,1.0039370078740157)(254,1.00398406374502)(255,1.0040650406504066)(256,1.0040983606557377)(257,1.0041322314049588)(258,1.004149377593361)(259,1.0042372881355932)(260,1.0043290043290043)(261,1.0043859649122806)(262,1.0046728971962617)(263,1.0046948356807512)(264,1.0048076923076923)(265,1.0049261083743843)(266,1.0050251256281406)(267,1.005050505050505)(268,1.0051546391752577)(269,1.0052083333333333)(270,1.0054347826086956)(271,1.005586592178771)(272,1.0056179775280898)(273,1.0057471264367817)(274,1.0058823529411764)(275,1.0059171597633136)(276,1.006060606060606)(277,1.0060975609756098)(278,1.0062111801242235)(279,1.00625)(280,1.0063694267515924)(281,1.0064516129032257)(282,1.0064516129032257)(283,1.0066666666666666)(284,1.0068493150684932)(285,1.0070422535211268)(286,1.0071942446043165)(287,1.0074074074074073)(288,1.0078125)(289,1.008)(290,1.008130081300813)(291,1.0082644628099173)(292,1.0084745762711864)(293,1.008695652173913)(294,1.0091743119266054)(295,1.0095238095238095)(296,1.0098039215686274)(297,1.0101010101010102)(298,1.0104166666666667)(299,1.0105263157894737)(300,1.0108695652173914)(301,1.0108695652173914)(302,1.011111111111111)(303,1.0113636363636365)(304,1.0114942528735633)(305,1.011764705882353)(306,1.0119047619047619)(307,1.0128205128205128)(308,1.0129870129870129)(309,1.0136986301369864)(310,1.0142857142857142)(311,1.0147058823529411)(312,1.0151515151515151)(313,1.015625)(314,1.032258064516129)(315,1.0350877192982457)(316,1.0392156862745099)(317,1.0408163265306123)(318,1.0408163265306123)(319,1.0425531914893618)(320,1.0454545454545454)(321,1.05)(322,1.054054054054054)(323,1.0555555555555556)(324,1.0555555555555556)(325,1.0625)(326,1.0666666666666667)(327,1.0666666666666667)(328,1.0714285714285714)(329,1.0740740740740742)(330,1.0869565217391304)(331,1.0909090909090908)(332,1.1)(333,1.105263157894737)(334,1.105263157894737)(335,1.1111111111111112)(336,1.1111111111111112)(337,1.1333333333333333)(338,1.1538461538461537)(339,1.1538461538461537)(340,1.1666666666666667)(341,1.2)(342,1.2222222222222223)(343,1.2857142857142858)(344,1.2857142857142858)(345,1.5)(346,1.5)(347,1.5)(348,1.5)(349,1.5)(350,1.6666666666666667)(351,1.6666666666666667)(352,1.6666666666666667)(353,1.6666666666666667)(354,1.6666666666666667)(355,1.6666666666666667)(356,1.6666666666666667)(357,2)
}; % plot of the third histogram
\addplot[color=RedOrange] coordinates{(0,4.019220466113046)(1,4.016968554306107)(2,4.013728813559322)(3,4.009113646486449)(4,4.00249757180519)(5,3.995638870326733)(6,3.9876120095676697)(7,3.9786630802224563)(8,3.970222222222222)(9,3.960589457774419)(10,3.9511112826053405)(11,3.940116691634471)(12,3.9293956265633825)(13,3.9168978510206616)(14,3.9036426233688952)(15,3.8913328096360305)(16,3.876407518729532)(17,3.8635609621233065)(18,3.8485136290245987)(19,3.8345467871976546)(20,3.8190049801297854)(21,3.802228556620886)(22,3.7829440754231842)(23,3.7646440621715804)(24,3.7436981994855674)(25,3.722327453533799)(26,3.6932300313595277)(27,3.665986611412177)(28,3.6310342525205317)(29,3.6007089241034196)(30,3.5730182816677507)(31,3.545932152506755)(32,3.5231690250349867)(33,3.501047711154094)(34,3.4801905717151453)(35,3.4584385063974237)(36,3.440552496163221)(37,3.420468327269335)(38,3.3968807517194612)(39,3.3729491796718687)(40,3.3509278350515466)(41,3.3345713073005094)(42,3.3157205240174674)(43,3.3016139878950908)(44,3.2823610631687954)(45,3.2593646277856805)(46,3.2381649585163492)(47,3.2156147850138295)(48,3.190791180285344)(49,3.1585873506109845)(50,3.1289786880708554)(51,3.0991711917690767)(52,3.0680674755844923)(53,3.0357634112792296)(54,2.9976262066782717)(55,2.9618546530746466)(56,2.910958904109589)(57,2.8707264957264957)(58,2.816235337925898)(59,2.765549312148808)(60,2.7100243506493507)(61,2.6503823279524212)(62,2.5934309809143365)(63,2.5351326198231736)(64,2.482364388226709)(65,2.428680010134279)(66,2.384168865435356)(67,2.344203891477117)(68,2.3017953833000857)(69,2.2638600652238363)(70,2.241579914268218)(71,2.1941008563273074)(72,2.1549110085695453)(73,2.1139154160982265)(74,2.061785714285714)(75,2.025536639526277)(76,1.9615384615384615)(77,1.8980839788014676)(78,1.845859872611465)(79,1.7898711683696136)(80,1.7351450944265316)(81,1.6870265151515151)(82,1.6341821743388834)(83,1.5953464845725847)(84,1.5750386797318205)(85,1.5625)(86,1.5538954108858059)(87,1.5479825517993457)(88,1.543598233995585)(89,1.5382452193475815)(90,1.5357548240635641)(91,1.5293440736478712)(92,1.5210280373831775)(93,1.5109662122110255)(94,1.5045099218280216)(95,1.501219512195122)(96,1.4966153846153847)(97,1.4831038798498122)(98,1.4735837046467217)(99,1.464935064935065)(100,1.4567656765676569)(101,1.4491298527443106)(102,1.4398368456832087)(103,1.4261954261954262)(104,1.415666901905434)(105,1.4113018597997138)(106,1.4072202166064982)(107,1.403225806451613)(108,1.3958177744585512)(109,1.391238670694864)(110,1.3865546218487395)(111,1.3843774168600154)(112,1.3796875)(113,1.3769716088328077)(114,1.3745019920318724)(115,1.3713592233009708)(116,1.3690671031096564)(117,1.3670256835128418)(118,1.3667790893760539)(119,1.3651202749140894)(120,1.3626852659110724)(121,1.3613817537643933)(122,1.3591359135913592)(123,1.3556370302474794)(124,1.354089219330855)(125,1.3503787878787878)(126,1.3474494706448508)(127,1.342156862745098)(128,1.3396414342629481)(129,1.3302658486707566)(130,1.3274428274428274)(131,1.3210526315789475)(132,1.3186695278969958)(133,1.3107960741548528)(134,1.3008948545861299)(135,1.2942528735632184)(136,1.2900584795321637)(137,1.2855436081242533)(138,1.2794117647058822)(139,1.2752179327521793)(140,1.27020202020202)(141,1.2667525773195876)(142,1.2641261498028908)(143,1.2663989290495314)(144,1.2579310344827586)(145,1.2514285714285713)(146,1.2419354838709677)(147,1.231578947368421)(148,1.2269938650306749)(149,1.2192429022082019)(150,1.203583061889251)(151,1.1871838111298483)(152,1.175043327556326)(153,1.1553571428571427)(154,1.1397058823529411)(155,1.1197718631178708)(156,1.1045364891518739)(157,1.0936863543788187)(158,1.0873180873180872)(159,1.0584415584415585)(160,1.0503282275711159)(161,1.0401785714285714)(162,1.036281179138322)(163,1.0275229357798166)(164,1.020979020979021)(165,1.0141509433962264)(166,1.01187648456057)(167,1.0095693779904307)(168,1.0072115384615385)(169,1.002433090024331)(170,1)(171,1)(172,1)(173,1)(174,1)(175,1)(176,1)(177,1)(178,1)(179,1)(180,1)(181,1)(182,1)(183,1)(184,1)(185,1)(186,1)(187,1)(188,1)(189,1)(190,1)(191,1)(192,1)(193,1)(194,1)(195,1)(196,1)(197,1)(198,1)(199,1)(200,1)(201,1)(202,1)(203,1)(204,1)(205,1)(206,1)(207,1)(208,1)(209,1)(210,1)(211,1)(212,1)(213,1)(214,1)(215,1)(216,1)(217,1)(218,1)(219,1)(220,1)(221,1)(222,1)(223,1)(224,1)(225,1)(226,1)(227,1)(228,1)(229,1)(230,1)(231,1)(232,1)(233,1)(234,1)(235,1)(236,1)(237,1)(238,1)(239,1)(240,1)(241,1)(242,1)(243,1)(244,1)(245,1)(246,1)(247,1)(248,1)(249,1)(250,1)(251,1)(252,1)(253,1)(254,1)(255,1)(256,1)(257,1)(258,1)(259,1)(260,1)(261,1)(262,1)(263,1)(264,1)(265,1)(266,1)(267,1)(268,1)(269,1)(270,1)(271,1)(272,1)(273,1)(274,1)(275,1)(276,1)(277,1)(278,1)(279,1)(280,1)(281,1)(282,1)(283,1)(284,1)(285,1)(286,1)(287,1)(288,1)(289,1)(290,1)(291,1)(292,1)(293,1)(294,1)(295,1)(296,1)(297,1)(298,1)(299,1)(300,1)(301,1)(302,1)(303,1)(304,1)(305,1)(306,1)(307,1)(308,1)(309,1)(310,1)(311,1)(312,1)(313,1)(314,1)(315,1)(316,1)(317,1)(318,1)(319,1)(320,1)(321,1)(322,1)(323,1)(324,1)(325,1)(326,1)(327,1)(328,1)(329,1)(330,1)(331,1)(332,1)(333,1)(334,1)(335,1)(336,1)(337,1)(338,1)(339,1)(340,1)(341,1)(342,1)(343,1)(344,1)
}; % plot of the fourth histogram
\addplot[color=Orange] coordinates{(0,4.018109392783022)(1,4.0158847681227705)(2,4.012343167175314)(3,4.007705479452055)(4,4.002494629616797)(5,3.9961722151987638)(6,3.987574978577549)(7,3.980984867728708)(8,3.9722170772365253)(9,3.9642479613409844)(10,3.9539684377049813)(11,3.943539802073887)(12,3.9323714331774946)(13,3.920049504950495)(14,3.907528156490812)(15,3.8946704920883324)(16,3.882106389619655)(17,3.86731510368875)(18,3.850989302281549)(19,3.8332276783537327)(20,3.816976393771974)(21,3.79757399823752)(22,3.777908531692966)(23,3.7571152252003315)(24,3.7341728677733257)(25,3.70519233059426)(26,3.6796302003081665)(27,3.652678857876163)(28,3.623003408407405)(29,3.588337624382437)(30,3.5554830287206265)(31,3.5300113250283127)(32,3.5107879924953096)(33,3.4872085492227978)(34,3.465303385853168)(35,3.442158486618429)(36,3.420981295744104)(37,3.3997003184116874)(38,3.378945323880742)(39,3.3591182364729457)(40,3.3379801734820322)(41,3.3208907741251323)(42,3.2996181123840698)(43,3.2848158131177)(44,3.2651716564559012)(45,3.24420952607198)(46,3.2196135974565907)(47,3.198309574870695)(48,3.169789227166276)(49,3.1400940228341168)(50,3.1215393133997784)(51,3.0925766695328174)(52,3.060633080695497)(53,3.037344398340249)(54,2.9996815286624203)(55,2.961977186311787)(56,2.9155509783728117)(57,2.8756014970593475)(58,2.83462389380531)(59,2.794066985645933)(60,2.7342279190867216)(61,2.6704378797402053)(62,2.605993829881005)(63,2.5450554528650646)(64,2.483527131782946)(65,2.4268754736044458)(66,2.374703557312253)(67,2.331055221432477)(68,2.29738933030647)(69,2.265505020673361)(70,2.234862385321101)(71,2.1959287531806617)(72,2.148603723404255)(73,2.091790786283339)(74,2.032561505065123)(75,1.9853107344632768)(76,1.9327301337529503)(77,1.8619264158743283)(78,1.7983592400690847)(79,1.7415578568212517)(80,1.6903648269410665)(81,1.6533269045323047)(82,1.6086309523809523)(83,1.5685071574642127)(84,1.5467362924281984)(85,1.5357142857142858)(86,1.5267712276906436)(87,1.52)(88,1.5166481687014428)(89,1.514623172103487)(90,1.5113636363636365)(91,1.5066206102475532)(92,1.500292568753657)(93,1.4940828402366864)(94,1.4900781719783525)(95,1.4810281517747859)(96,1.4718045112781954)(97,1.4673430564362715)(98,1.4617363344051446)(99,1.4526938239159002)(100,1.4415497661990648)(101,1.4364864864864866)(102,1.430622009569378)(103,1.4249134948096887)(104,1.416022487702038)(105,1.4133522727272727)(106,1.4044700793078586)(107,1.3953147877013177)(108,1.3890532544378698)(109,1.3778280542986425)(110,1.3733741392501913)(111,1.370255615801704)(112,1.3667711598746082)(113,1.3623304070231446)(114,1.3576051779935274)(115,1.3554463554463554)(116,1.3522064945878434)(117,1.3460884353741496)(118,1.3445017182130585)(119,1.3382737576285963)(120,1.332146037399822)(121,1.3297101449275361)(122,1.3275229357798166)(123,1.323666978484565)(124,1.321969696969697)(125,1.3169556840077072)(126,1.3151219512195123)(127,1.30990099009901)(128,1.3063872255489022)(129,1.3007063572149344)(130,1.295267489711934)(131,1.2897489539748954)(132,1.2864749733759318)(133,1.2782608695652173)(134,1.2749445676274944)(135,1.268878718535469)(136,1.266355140186916)(137,1.264218009478673)(138,1.256969696969697)(139,1.2518518518518518)(140,1.25)(141,1.2432432432432432)(142,1.2311756935270806)(143,1.2235772357723578)(144,1.2083916083916084)(145,1.2022630834512023)(146,1.190406976744186)(147,1.1773472429210134)(148,1.1671732522796352)(149,1.1567398119122256)(150,1.1430868167202572)(151,1.1335504885993486)(152,1.1202003338898163)(153,1.1160409556313993)(154,1.1120840630472855)(155,1.105357142857143)(156,1.0965391621129326)(157,1.0890538033395176)(158,1.0721062618595825)(159,1.0635838150289016)(160,1.0570866141732282)(161,1.0346938775510204)(162,1.022680412371134)(163,1.0210084033613445)(164,1.0129310344827587)(165,1.0130718954248366)(166,1.013129102844639)(167,1.008849557522124)(168,1.0089086859688197)(169,1.0044742729306488)(170,1.002267573696145)(171,1.0022831050228311)(172,1)(173,1)(174,1)(175,1)(176,1)(177,1)(178,1)(179,1)(180,1)(181,1)(182,1)(183,1)(184,1)(185,1)(186,1)(187,1)(188,1)(189,1)(190,1)(191,1)(192,1)(193,1)(194,1)(195,1)(196,1)(197,1)(198,1)(199,1)(200,1)(201,1)(202,1)(203,1)(204,1)(205,1)(206,1)(207,1)(208,1)(209,1)(210,1)(211,1)(212,1)(213,1)(214,1)(215,1)(216,1)(217,1)(218,1)(219,1)(220,1)(221,1)(222,1)(223,1)(224,1)(225,1)(226,1)(227,1)(228,1)(229,1)(230,1)(231,1)(232,1)(233,1)(234,1)(235,1)(236,1)(237,1)(238,1)(239,1)(240,1)(241,1)(242,1)(243,1)(244,1)(245,1)(246,1)(247,1)(248,1)(249,1)(250,1)(251,1)(252,1)(253,1)(254,1)(255,1)(256,1)(257,1)(258,1)(259,1)(260,1)(261,1)(262,1)(263,1)(264,1)(265,1)(266,1)(267,1)(268,1)(269,1)(270,1)(271,1)(272,1)(273,1)(274,1)(275,1)(276,1)(277,1)(278,1)(279,1)(280,1)(281,1)(282,1)(283,1)(284,1)(285,1)(286,1)(287,1)(288,1)(289,1)(290,1)(291,1)(292,1)(293,1)(294,1)(295,1)(296,1)(297,1)(298,1)(299,1)(300,1)(301,1)(302,1)(303,1)(304,1)(305,1)(306,1)(307,1)(308,1)(309,1)(310,1)(311,1)(312,1)(313,1)(314,1)(315,1)(316,1)(317,1)(318,1)(319,1)(320,1)(321,1)(322,1)(323,1)(324,1)(325,1)(326,1)(327,1)(328,1)(329,1)(330,1)(331,1)(332,1)(333,1)(334,1)(335,1)(336,1)(337,1)(338,1)(339,1)(340,1)(341,1)(342,1)(343,1)(344,1)(345,1)
}; % plot of the fifth histogram
\addplot[color=Melon] coordinates{(0,4.01959210966232)(1,4.017304526057592)(2,4.0138047707663675)(3,4.00830513688096)(4,4.003041299464317)(5,3.9961430575035064)(6,3.9886327192388555)(7,3.9806833448313426)(8,3.9721996002073)(9,3.96269445703066)(10,3.9516508978567293)(11,3.9412530063478295)(12,3.9287818262518663)(13,3.9154620386861616)(14,3.9008633521881513)(15,3.8876212955677945)(16,3.8753027720462905)(17,3.8638728989177986)(18,3.8465626037861176)(19,3.8313859692006846)(20,3.8128466270041343)(21,3.7935789364137786)(22,3.77327174088199)(23,3.7525801797802685)(24,3.728555855183889)(25,3.702261904761905)(26,3.6737356250772844)(27,3.641833258340297)(28,3.611747371241042)(29,3.578374607603767)(30,3.552063721940623)(31,3.5223362511319047)(32,3.4971428571428573)(33,3.4763527216678836)(34,3.4568067226890755)(35,3.437412831241283)(36,3.4152496382054993)(37,3.396709665358011)(38,3.382222650231125)(39,3.362175034729113)(40,3.3440596588006946)(41,3.3242561244874356)(42,3.3049269085002706)(43,3.2836468885672936)(44,3.2635552505147563)(45,3.2474141984015046)(46,3.2262826718296225)(47,3.1998501124156884)(48,3.1733969810347054)(49,3.1504589596913664)(50,3.127796842827728)(51,3.0977272727272727)(52,3.061149492870792)(53,3.028740875912409)(54,2.994654928470366)(55,2.9507203667321544)(56,2.907838802924673)(57,2.860333215171925)(58,2.805986696230599)(59,2.758687258687259)(60,2.7053499597747384)(61,2.6449168596085033)(62,2.5926738319806977)(63,2.5432801822323463)(64,2.494078635717669)(65,2.439544103072349)(66,2.3876288659793814)(67,2.3354821380607036)(68,2.2968619827825605)(69,2.2623092427296285)(70,2.218236173393124)(71,2.172531921519776)(72,2.11907098462545)(73,2.0791978246091096)(74,2.0307311903920877)(75,1.9692706405035172)(76,1.9147376543209877)(77,1.8639455782312926)(78,1.8129945855893377)(79,1.760312635692575)(80,1.7006772009029345)(81,1.650749063670412)(82,1.614498319731157)(83,1.5716410510659395)(84,1.545915778792491)(85,1.532676348547718)(86,1.528391167192429)(87,1.5234291799787008)(88,1.5206073752711498)(89,1.5158991228070176)(90,1.5066666666666666)(91,1.492342597844583)(92,1.4864553314121038)(93,1.4792761237594863)(94,1.4757969303423848)(95,1.471505698860228)(96,1.4654126213592233)(97,1.4594594594594594)(98,1.449874686716792)(99,1.4426751592356688)(100,1.4372574385510997)(101,1.4310907903331156)(102,1.4250663129973475)(103,1.417004048582996)(104,1.4145677331518038)(105,1.4058171745152355)(106,1.4001401541695866)(107,1.3902787705503932)(108,1.3830557566980448)(109,1.380952380952381)(110,1.3748137108792846)(111,1.3672230652503794)(112,1.362095531587057)(113,1.3604378420641126)(114,1.3579952267303104)(115,1.3562753036437247)(116,1.3530864197530865)(117,1.3542713567839195)(118,1.352542372881356)(119,1.3482758620689654)(120,1.3455497382198953)(121,1.3442188879082082)(122,1.3405211141060198)(123,1.3357664233576643)(124,1.3333333333333333)(125,1.3298774740810555)(126,1.3244274809160306)(127,1.3223938223938223)(128,1.3136094674556213)(129,1.312)(130,1.30668016194332)(131,1.3030927835051547)(132,1.2952380952380953)(133,1.2887931034482758)(134,1.287912087912088)(135,1.2871508379888268)(136,1.2840909090909092)(137,1.2790697674418605)(138,1.2777777777777777)(139,1.2750301568154403)(140,1.2730627306273063)(141,1.2673392181588903)(142,1.2571059431524547)(143,1.2490066225165564)(144,1.2411924119241193)(145,1.24)(146,1.2260371959942775)(147,1.2194403534609721)(148,1.209895052473763)(149,1.2076923076923076)(150,1.2006269592476488)(151,1.1861958266452648)(152,1.1647254575707155)(153,1.148972602739726)(154,1.1401050788091067)(155,1.1292639138240574)(156,1.1078066914498141)(157,1.0981132075471698)(158,1.0917782026768643)(159,1.0729783037475344)(160,1.0602409638554218)(161,1.0551020408163265)(162,1.0434782608695652)(163,1.0378151260504203)(164,1.0319829424307037)(165,1.025974025974026)(166,1.019736842105263)(167,1.0155902004454342)(168,1.0068181818181818)(169,1.0045977011494254)(170,1.0046189376443417)(171,1.0023364485981308)(172,1.0023474178403755)(173,1.0023809523809524)(174,1.0024038461538463)(175,1.002415458937198)(176,1.0024449877750612)(177,1.0024630541871922)(178,1.0024875621890548)(179,1.0025188916876575)(180,1)(181,1)(182,1)(183,1)(184,1)(185,1)(186,1)(187,1)(188,1)(189,1)(190,1)(191,1)(192,1)(193,1)(194,1)(195,1)(196,1)(197,1)(198,1)(199,1)(200,1)(201,1)(202,1)(203,1)(204,1)(205,1)(206,1)(207,1)(208,1)(209,1)(210,1)(211,1)(212,1)(213,1)(214,1)(215,1)(216,1)(217,1)(218,1)(219,1)(220,1)(221,1)(222,1)(223,1)(224,1)(225,1)(226,1)(227,1)(228,1)(229,1)(230,1)(231,1)(232,1)(233,1)(234,1)(235,1)(236,1)(237,1)(238,1)(239,1)(240,1)(241,1)(242,1)(243,1)(244,1)(245,1)(246,1)(247,1)(248,1)(249,1)(250,1)(251,1)(252,1)(253,1)(254,1)(255,1)(256,1)(257,1)(258,1)(259,1)(260,1)(261,1)(262,1)(263,1)(264,1)(265,1)(266,1)(267,1)(268,1)(269,1)(270,1)(271,1)(272,1)(273,1)(274,1)(275,1)(276,1)(277,1)(278,1)(279,1)(280,1)(281,1)(282,1)(283,1)(284,1)(285,1)(286,1)(287,1)(288,1)(289,1)(290,1)(291,1)(292,1)(293,1)(294,1)(295,1)(296,1)(297,1)(298,1)(299,1)(300,1)(301,1)(302,1)(303,1)(304,1)(305,1)(306,1)(307,1)(308,1)(309,1)(310,1)(311,1)(312,1)(313,1)(314,1)(315,1)(316,1)(317,1)(318,1)(319,1)(320,1)(321,1)(322,1)(323,1)(324,1)(325,1)(326,1)(327,1)(328,1)(329,1)(330,1)(331,1)(332,1)(333,1)(334,1)(335,1)(336,1)(337,1)(338,1)(339,1)(340,1)(341,1)
}; % plot of the first histogram
\legend{glass \#1,  glass \#2, glass \#3, no glass \#1,  no glass \#2, no glass \#3,} % legend of the plot

\end{axis}
\end{tikzpicture}

%% file: plots/glass_bottles/evac_speed.tex
\begin{tikzpicture}[scale=0.75]

% Set the axis labels and title
\begin{axis}[    xlabel={Time [s]},    ylabel={ppl/s},    title={Evacuation Speed - Glass bottles},   legend style={at={(1,1)}, anchor=north east},]

\addplot[color=PineGreen] coordinates{(0,28)(1,30)(2,65)(3,82)(4,97)(5,119)(6,138)(7,121)(8,129)(9,133)(10,136)(11,141)(12,161)(13,139)(14,142)(15,143)(16,185)(17,154)(18,149)(19,136)(20,156)(21,132)(22,161)(23,150)(24,145)(25,168)(26,144)(27,150)(28,153)(29,153)(30,153)(31,156)(32,134)(33,135)(34,105)(35,138)(36,90)(37,106)(38,102)(39,106)(40,105)(41,88)(42,79)(43,81)(44,101)(45,74)(46,79)(47,83)(48,90)(49,75)(50,71)(51,70)(52,64)(53,84)(54,74)(55,86)(56,68)(57,88)(58,73)(59,96)(60,71)(61,60)(62,71)(63,71)(64,61)(65,57)(66,59)(67,53)(68,40)(69,49)(70,48)(71,54)(72,59)(73,45)(74,45)(75,50)(76,62)(77,60)(78,38)(79,43)(80,44)(81,39)(82,34)(83,36)(84,26)(85,27)(86,28)(87,15)(88,15)(89,19)(90,19)(91,25)(92,20)(93,13)(94,21)(95,24)(96,19)(97,24)(98,21)(99,20)(100,20)(101,24)(102,19)(103,16)(104,11)(105,21)(106,12)(107,20)(108,19)(109,12)(110,18)(111,11)(112,17)(113,17)(114,24)(115,19)(116,21)(117,13)(118,11)(119,20)(120,17)(121,15)(122,16)(123,11)(124,13)(125,14)(126,22)(127,12)(128,15)(129,11)(130,22)(131,22)(132,12)(133,13)(134,11)(135,9)(136,17)(137,15)(138,11)(139,14)(140,18)(141,13)(142,7)(143,10)(144,17)(145,18)(146,15)(147,11)(148,15)(149,9)(150,10)(151,16)(152,12)(153,9)(154,7)(155,16)(156,11)(157,12)(158,15)(159,6)(160,9)(161,9)(162,7)(163,11)(164,15)(165,7)(166,8)(167,9)(168,4)(169,8)(170,6)(171,8)(172,6)(173,10)(174,6)(175,6)(176,2)(177,8)(178,3)(179,8)(180,4)(181,2)(182,7)(183,9)(184,2)(185,7)(186,2)(187,5)(188,5)(189,3)(190,3)(191,1)(192,2)(193,6)(194,4)(195,6)(196,4)(197,3)(198,6)(199,9)(200,8)(201,5)(202,11)(203,4)(204,4)(205,9)(206,1)(207,4)(208,7)(209,5)(210,7)(211,5)(212,5)(213,2)(214,7)(215,5)(216,2)(217,3)(218,4)(219,1)(220,3)(221,9)(222,6)(223,6)(224,5)(225,6)(226,3)(227,2)(228,3)(229,5)(230,2)(231,2)(232,3)(233,2)(234,3)(235,7)(236,7)(237,8)(238,4)(239,1)(240,4)(241,1)(242,7)(243,4)(244,7)(245,3)(246,3)(247,5)(248,2)(249,5)(250,3)(251,1)(252,1)(253,1)(254,1)(255,2)(256,3)(257,3)(258,3)(259,2)(260,4)(261,1)(262,3)(263,3)(264,4)(265,6)(266,3)(267,3)(268,2)(269,5)(270,3)(271,4)(272,2)(273,4)(274,1)(275,4)(276,4)(277,2)(278,8)(279,1)(280,4)(281,2)(282,2)(283,2)(284,1)(285,3)(286,6)(287,3)(288,4)(289,2)(290,3)(291,3)(292,3)(293,4)(294,0)(295,2)(296,2)(297,4)(298,3)(299,1)(300,3)(301,1)(302,5)(303,3)(304,3)(305,3)(306,4)(307,5)(308,2)(309,0)(310,5)(311,3)(312,2)(313,3)(314,2)(315,2)(316,5)(317,3)(318,1)(319,3)(320,5)(321,2)(322,5)(323,2)(324,1)(325,1)(326,3)(327,0)(328,0)(329,1)(330,1)(331,2)(332,0)(333,0)(334,1)(335,0)(336,3)(337,0)(338,1)(339,0)(340,0)(341,1)(342,0)
}; % plot of the first histogram
\addplot[color=Green] coordinates{(0,38)(1,38)(2,75)(3,82)(4,98)(5,105)(6,122)(7,147)(8,115)(9,140)(10,130)(11,122)(12,123)(13,124)(14,161)(15,166)(16,148)(17,163)(18,137)(19,127)(20,154)(21,147)(22,127)(23,177)(24,153)(25,134)(26,150)(27,151)(28,170)(29,151)(30,129)(31,148)(32,142)(33,103)(34,128)(35,137)(36,121)(37,115)(38,108)(39,121)(40,84)(41,78)(42,84)(43,85)(44,71)(45,94)(46,77)(47,82)(48,71)(49,95)(50,91)(51,76)(52,73)(53,77)(54,78)(55,77)(56,55)(57,75)(58,63)(59,62)(60,60)(61,92)(62,65)(63,65)(64,72)(65,57)(66,56)(67,58)(68,63)(69,58)(70,47)(71,43)(72,54)(73,55)(74,47)(75,49)(76,40)(77,62)(78,42)(79,45)(80,39)(81,42)(82,32)(83,37)(84,34)(85,30)(86,35)(87,15)(88,25)(89,22)(90,15)(91,28)(92,17)(93,16)(94,17)(95,20)(96,17)(97,21)(98,20)(99,22)(100,15)(101,9)(102,30)(103,8)(104,21)(105,12)(106,19)(107,19)(108,19)(109,11)(110,13)(111,20)(112,10)(113,15)(114,18)(115,21)(116,12)(117,17)(118,21)(119,15)(120,8)(121,16)(122,15)(123,12)(124,17)(125,14)(126,18)(127,10)(128,15)(129,8)(130,17)(131,21)(132,14)(133,11)(134,18)(135,9)(136,9)(137,13)(138,3)(139,8)(140,14)(141,10)(142,12)(143,15)(144,15)(145,16)(146,12)(147,10)(148,20)(149,14)(150,9)(151,9)(152,8)(153,11)(154,13)(155,11)(156,12)(157,10)(158,4)(159,7)(160,17)(161,11)(162,9)(163,3)(164,9)(165,7)(166,7)(167,4)(168,8)(169,7)(170,5)(171,5)(172,3)(173,9)(174,5)(175,3)(176,6)(177,6)(178,9)(179,7)(180,9)(181,7)(182,7)(183,6)(184,10)(185,5)(186,3)(187,2)(188,6)(189,5)(190,3)(191,6)(192,6)(193,3)(194,5)(195,3)(196,7)(197,5)(198,9)(199,1)(200,2)(201,4)(202,5)(203,6)(204,3)(205,1)(206,6)(207,6)(208,5)(209,3)(210,4)(211,4)(212,2)(213,2)(214,5)(215,5)(216,9)(217,4)(218,6)(219,6)(220,8)(221,4)(222,6)(223,1)(224,3)(225,1)(226,5)(227,7)(228,3)(229,4)(230,2)(231,4)(232,3)(233,6)(234,5)(235,5)(236,5)(237,4)(238,6)(239,7)(240,8)(241,4)(242,4)(243,5)(244,4)(245,1)(246,4)(247,3)(248,2)(249,2)(250,3)(251,6)(252,4)(253,1)(254,3)(255,4)(256,3)(257,3)(258,9)(259,3)(260,4)(261,2)(262,4)(263,5)(264,3)(265,3)(266,5)(267,5)(268,4)(269,3)(270,6)(271,3)(272,3)(273,2)(274,4)(275,3)(276,3)(277,2)(278,8)(279,4)(280,6)(281,5)(282,1)(283,2)(284,6)(285,1)(286,2)(287,6)(288,6)(289,3)(290,5)(291,4)(292,3)(293,1)(294,2)(295,3)(296,2)(297,2)(298,2)(299,4)(300,3)(301,3)(302,1)(303,2)(304,3)(305,2)(306,0)(307,2)(308,0)(309,5)(310,4)(311,2)(312,1)(313,4)(314,3)(315,2)(316,5)(317,6)(318,4)(319,3)(320,2)(321,3)(322,0)(323,2)(324,3)(325,1)(326,0)(327,1)(328,1)(329,2)(330,2)(331,2)(332,0)(333,1)(334,0)(335,0)(336,0)(337,0)(338,3)(339,0)(340,1)(341,1)(342,0)(343,1)(344,0)(345,1)(346,0)
}; % plot of the second histogram
\addplot[color=LimeGreen] coordinates{(0,37)(1,58)(2,61)(3,66)(4,69)(5,104)(6,107)(7,121)(8,129)(9,122)(10,116)(11,144)(12,143)(13,142)(14,143)(15,145)(16,156)(17,140)(18,155)(19,166)(20,169)(21,133)(22,153)(23,177)(24,157)(25,150)(26,134)(27,178)(28,161)(29,167)(30,139)(31,115)(32,141)(33,134)(34,128)(35,127)(36,108)(37,120)(38,112)(39,92)(40,93)(41,83)(42,85)(43,73)(44,83)(45,91)(46,82)(47,72)(48,78)(49,93)(50,77)(51,69)(52,67)(53,87)(54,71)(55,75)(56,71)(57,70)(58,68)(59,70)(60,60)(61,52)(62,76)(63,75)(64,59)(65,50)(66,58)(67,58)(68,47)(69,74)(70,49)(71,44)(72,67)(73,51)(74,56)(75,37)(76,46)(77,38)(78,52)(79,31)(80,44)(81,31)(82,35)(83,29)(84,40)(85,22)(86,35)(87,28)(88,28)(89,26)(90,29)(91,29)(92,26)(93,20)(94,22)(95,14)(96,17)(97,14)(98,20)(99,16)(100,15)(101,16)(102,16)(103,26)(104,17)(105,14)(106,14)(107,18)(108,21)(109,16)(110,18)(111,15)(112,14)(113,20)(114,13)(115,23)(116,16)(117,11)(118,16)(119,21)(120,10)(121,15)(122,15)(123,12)(124,14)(125,11)(126,22)(127,12)(128,14)(129,11)(130,12)(131,16)(132,15)(133,10)(134,9)(135,14)(136,10)(137,19)(138,11)(139,13)(140,15)(141,7)(142,14)(143,10)(144,8)(145,13)(146,17)(147,14)(148,13)(149,12)(150,12)(151,11)(152,11)(153,13)(154,10)(155,11)(156,11)(157,8)(158,4)(159,7)(160,7)(161,12)(162,9)(163,7)(164,4)(165,8)(166,9)(167,11)(168,4)(169,7)(170,3)(171,13)(172,8)(173,4)(174,3)(175,9)(176,7)(177,4)(178,1)(179,4)(180,7)(181,3)(182,8)(183,6)(184,4)(185,5)(186,5)(187,4)(188,5)(189,8)(190,3)(191,5)(192,9)(193,3)(194,3)(195,7)(196,5)(197,2)(198,2)(199,4)(200,2)(201,2)(202,3)(203,5)(204,5)(205,3)(206,9)(207,5)(208,3)(209,8)(210,6)(211,5)(212,3)(213,0)(214,7)(215,4)(216,3)(217,3)(218,3)(219,4)(220,6)(221,1)(222,8)(223,7)(224,4)(225,3)(226,5)(227,5)(228,3)(229,5)(230,4)(231,5)(232,4)(233,3)(234,2)(235,6)(236,3)(237,1)(238,5)(239,5)(240,3)(241,3)(242,1)(243,6)(244,3)(245,3)(246,4)(247,5)(248,3)(249,6)(250,4)(251,4)(252,1)(253,3)(254,5)(255,2)(256,2)(257,1)(258,5)(259,5)(260,3)(261,14)(262,1)(263,5)(264,5)(265,4)(266,1)(267,4)(268,2)(269,8)(270,5)(271,1)(272,4)(273,4)(274,1)(275,4)(276,1)(277,4)(278,1)(279,3)(280,2)(281,0)(282,5)(283,4)(284,4)(285,3)(286,4)(287,7)(288,3)(289,2)(290,3)(291,3)(292,3)(293,6)(294,4)(295,3)(296,3)(297,3)(298,1)(299,3)(300,0)(301,2)(302,2)(303,1)(304,2)(305,1)(306,6)(307,1)(308,5)(309,3)(310,2)(311,2)(312,2)(313,2)(314,5)(315,6)(316,2)(317,0)(318,2)(319,4)(320,4)(321,3)(322,1)(323,0)(324,4)(325,2)(326,0)(327,2)(328,1)(329,4)(330,1)(331,2)(332,1)(333,0)(334,1)(335,1)(336,3)(337,2)(338,0)(339,1)(340,2)(341,1)(342,2)(343,1)(344,3)(345,0)(346,0)(347,0)(348,0)(349,1)(350,0)(351,0)(352,0)(353,0)(354,0)(355,0)(356,1)(357,0)
}; % plot of the third histogram
\addplot[color=RedOrange] coordinates{(0,32)(1,38)(2,85)(3,62)(4,76)(5,114)(6,133)(7,138)(8,112)(9,127)(10,126)(11,151)(12,145)(13,141)(14,159)(15,156)(16,162)(17,151)(18,167)(19,148)(20,128)(21,152)(22,159)(23,147)(24,189)(25,149)(26,144)(27,168)(28,165)(29,139)(30,116)(31,118)(32,123)(33,125)(34,122)(35,123)(36,103)(37,115)(38,105)(39,85)(40,89)(41,83)(42,85)(43,71)(44,81)(45,67)(46,73)(47,84)(48,86)(49,69)(50,72)(51,71)(52,72)(53,63)(54,81)(55,67)(56,81)(57,85)(58,63)(59,76)(60,71)(61,58)(62,62)(63,82)(64,46)(65,61)(66,55)(67,60)(68,54)(69,46)(70,41)(71,53)(72,39)(73,56)(74,45)(75,55)(76,48)(77,48)(78,46)(79,30)(80,15)(81,28)(82,33)(83,20)(84,28)(85,19)(86,27)(87,19)(88,21)(89,11)(90,17)(91,17)(92,17)(93,16)(94,17)(95,9)(96,13)(97,20)(98,16)(99,20)(100,15)(101,14)(102,16)(103,17)(104,13)(105,9)(106,17)(107,17)(108,10)(109,8)(110,11)(111,10)(112,8)(113,8)(114,12)(115,9)(116,10)(117,17)(118,19)(119,11)(120,14)(121,16)(122,14)(123,13)(124,15)(125,12)(126,11)(127,14)(128,19)(129,12)(130,7)(131,16)(132,13)(133,15)(134,16)(135,11)(136,12)(137,16)(138,12)(139,10)(140,14)(141,11)(142,13)(143,17)(144,18)(145,9)(146,13)(147,10)(148,12)(149,14)(150,14)(151,11)(152,11)(153,11)(154,14)(155,15)(156,12)(157,9)(158,11)(159,3)(160,7)(161,7)(162,2)(163,6)(164,2)(165,3)(166,3)(167,1)(168,4)(169,2)(170,4)(171,1)(172,1)(173,5)(174,4)(175,2)(176,3)(177,4)(178,1)(179,6)(180,3)(181,2)(182,4)(183,3)(184,2)(185,3)(186,1)(187,1)(188,3)(189,4)(190,3)(191,4)(192,7)(193,3)(194,1)(195,2)(196,1)(197,1)(198,3)(199,3)(200,2)(201,3)(202,0)(203,1)(204,5)(205,3)(206,1)(207,1)(208,0)(209,2)(210,4)(211,2)(212,6)(213,2)(214,1)(215,2)(216,3)(217,5)(218,4)(219,1)(220,2)(221,1)(222,5)(223,1)(224,8)(225,4)(226,2)(227,2)(228,1)(229,2)(230,3)(231,2)(232,4)(233,6)(234,4)(235,7)(236,3)(237,1)(238,0)(239,3)(240,5)(241,3)(242,5)(243,2)(244,3)(245,4)(246,4)(247,2)(248,2)(249,4)(250,4)(251,3)(252,0)(253,5)(254,6)(255,3)(256,3)(257,2)(258,5)(259,1)(260,4)(261,8)(262,3)(263,2)(264,1)(265,1)(266,4)(267,3)(268,2)(269,1)(270,4)(271,3)(272,3)(273,0)(274,2)(275,1)(276,2)(277,0)(278,5)(279,1)(280,3)(281,3)(282,3)(283,1)(284,1)(285,5)(286,3)(287,1)(288,0)(289,0)(290,4)(291,0)(292,1)(293,3)(294,3)(295,4)(296,5)(297,2)(298,4)(299,0)(300,2)(301,1)(302,1)(303,3)(304,2)(305,1)(306,1)(307,2)(308,2)(309,2)(310,3)(311,2)(312,4)(313,1)(314,2)(315,0)(316,3)(317,2)(318,3)(319,1)(320,0)(321,1)(322,2)(323,3)(324,2)(325,0)(326,1)(327,1)(328,1)(329,1)(330,0)(331,0)(332,0)(333,0)(334,1)(335,0)(336,0)(337,0)(338,0)(339,0)(340,1)(341,0)(342,0)(343,0)(344,1)(345,0)
}; % plot of the fourth histogram
\addplot[color=Orange] coordinates{(0,36)(1,48)(2,82)(3,77)(4,77)(5,118)(6,99)(7,150)(8,101)(9,129)(10,138)(11,143)(12,142)(13,134)(14,186)(15,123)(16,152)(17,196)(18,130)(19,152)(20,162)(21,144)(22,159)(23,173)(24,152)(25,153)(26,163)(27,177)(28,153)(29,162)(30,156)(31,124)(32,106)(33,126)(34,115)(35,123)(36,116)(37,118)(38,84)(39,96)(40,83)(41,72)(42,95)(43,80)(44,69)(45,72)(46,72)(47,69)(48,65)(49,70)(50,72)(51,72)(52,77)(53,64)(54,81)(55,73)(56,63)(57,66)(58,61)(59,71)(60,64)(61,78)(62,76)(63,66)(64,55)(65,60)(66,60)(67,49)(68,61)(69,47)(70,48)(71,56)(72,50)(73,37)(74,47)(75,51)(76,53)(77,45)(78,36)(79,33)(80,34)(81,30)(82,27)(83,25)(84,25)(85,17)(86,17)(87,13)(88,16)(89,11)(90,13)(91,21)(92,9)(93,19)(94,18)(95,20)(96,11)(97,17)(98,22)(99,13)(100,12)(101,14)(102,12)(103,13)(104,10)(105,14)(106,10)(107,8)(108,14)(109,13)(110,10)(111,9)(112,16)(113,13)(114,11)(115,17)(116,16)(117,11)(118,10)(119,17)(120,15)(121,10)(122,14)(123,11)(124,10)(125,10)(126,9)(127,5)(128,7)(129,11)(130,10)(131,13)(132,11)(133,17)(134,20)(135,16)(136,10)(137,15)(138,9)(139,16)(140,14)(141,14)(142,13)(143,18)(144,4)(145,15)(146,11)(147,9)(148,15)(149,10)(150,6)(151,10)(152,12)(153,14)(154,8)(155,8)(156,5)(157,6)(158,5)(159,10)(160,12)(161,2)(162,8)(163,10)(164,5)(165,2)(166,5)(167,3)(168,2)(169,6)(170,3)(171,3)(172,4)(173,5)(174,4)(175,2)(176,2)(177,3)(178,3)(179,4)(180,2)(181,2)(182,1)(183,5)(184,3)(185,5)(186,6)(187,2)(188,3)(189,0)(190,2)(191,2)(192,2)(193,6)(194,6)(195,4)(196,3)(197,5)(198,0)(199,4)(200,6)(201,2)(202,2)(203,2)(204,2)(205,6)(206,5)(207,2)(208,2)(209,1)(210,2)(211,4)(212,5)(213,3)(214,1)(215,4)(216,2)(217,2)(218,1)(219,1)(220,3)(221,2)(222,5)(223,4)(224,1)(225,2)(226,3)(227,8)(228,4)(229,2)(230,5)(231,3)(232,1)(233,3)(234,2)(235,4)(236,6)(237,3)(238,3)(239,4)(240,5)(241,2)(242,4)(243,1)(244,0)(245,4)(246,4)(247,3)(248,3)(249,4)(250,3)(251,1)(252,5)(253,3)(254,4)(255,2)(256,2)(257,1)(258,2)(259,4)(260,3)(261,3)(262,1)(263,1)(264,4)(265,2)(266,2)(267,2)(268,2)(269,2)(270,5)(271,4)(272,5)(273,3)(274,3)(275,1)(276,3)(277,2)(278,3)(279,5)(280,3)(281,2)(282,1)(283,3)(284,6)(285,1)(286,1)(287,1)(288,1)(289,6)(290,2)(291,1)(292,2)(293,1)(294,3)(295,3)(296,4)(297,1)(298,1)(299,4)(300,3)(301,1)(302,2)(303,2)(304,0)(305,2)(306,6)(307,1)(308,2)(309,3)(310,1)(311,0)(312,5)(313,2)(314,3)(315,4)(316,2)(317,1)(318,0)(319,1)(320,7)(321,1)(322,0)(323,2)(324,1)(325,0)(326,1)(327,0)(328,1)(329,0)(330,1)(331,0)(332,0)(333,0)(334,0)(335,2)(336,0)(337,1)(338,2)(339,0)(340,0)(341,0)(342,1)(343,0)(344,0)(345,1)(346,0)
}; % plot of the fifth histogram
\addplot[color=Melon] coordinates{(0,31)(1,43)(2,66)(3,63)(4,77)(5,102)(6,132)(7,112)(8,122)(9,148)(10,136)(11,135)(12,158)(13,156)(14,162)(15,156)(16,147)(17,142)(18,170)(19,124)(20,142)(21,129)(22,154)(23,160)(24,155)(25,149)(26,138)(27,164)(28,139)(29,158)(30,154)(31,134)(32,114)(33,124)(34,130)(35,118)(36,103)(37,109)(38,100)(39,80)(40,86)(41,86)(42,67)(43,85)(44,77)(45,75)(46,84)(47,91)(48,64)(49,76)(50,78)(51,70)(52,72)(53,73)(54,72)(55,65)(56,67)(57,54)(58,74)(59,63)(60,80)(61,64)(62,49)(63,58)(64,63)(65,55)(66,65)(67,43)(68,55)(69,46)(70,54)(71,58)(72,48)(73,45)(74,53)(75,40)(76,42)(77,46)(78,48)(79,38)(80,41)(81,20)(82,32)(83,22)(84,33)(85,21)(86,14)(87,27)(88,15)(89,15)(90,22)(91,17)(92,12)(93,12)(94,17)(95,12)(96,10)(97,22)(98,15)(99,17)(100,9)(101,16)(102,17)(103,9)(104,18)(105,13)(106,18)(107,11)(108,11)(109,15)(110,17)(111,13)(112,15)(113,17)(114,19)(115,16)(116,16)(117,10)(118,16)(119,11)(120,10)(121,14)(122,14)(123,12)(124,15)(125,10)(126,9)(127,16)(128,10)(129,6)(130,15)(131,15)(132,12)(133,14)(134,12)(135,11)(136,16)(137,12)(138,15)(139,11)(140,14)(141,14)(142,13)(143,10)(144,10)(145,17)(146,14)(147,7)(148,14)(149,8)(150,9)(151,13)(152,13)(153,8)(154,7)(155,12)(156,4)(157,6)(158,9)(159,7)(160,5)(161,4)(162,6)(163,6)(164,5)(165,6)(166,5)(167,6)(168,4)(169,2)(170,4)(171,2)(172,6)(173,4)(174,2)(175,5)(176,3)(177,4)(178,5)(179,8)(180,0)(181,1)(182,4)(183,3)(184,2)(185,4)(186,1)(187,2)(188,1)(189,1)(190,1)(191,3)(192,5)(193,2)(194,4)(195,5)(196,4)(197,3)(198,6)(199,5)(200,1)(201,3)(202,4)(203,3)(204,3)(205,5)(206,0)(207,3)(208,2)(209,1)(210,0)(211,2)(212,3)(213,4)(214,2)(215,4)(216,3)(217,3)(218,1)(219,2)(220,0)(221,4)(222,1)(223,4)(224,3)(225,3)(226,3)(227,6)(228,4)(229,2)(230,3)(231,4)(232,4)(233,0)(234,2)(235,6)(236,2)(237,3)(238,3)(239,5)(240,3)(241,4)(242,9)(243,2)(244,3)(245,3)(246,1)(247,1)(248,1)(249,3)(250,5)(251,5)(252,2)(253,3)(254,3)(255,3)(256,6)(257,3)(258,1)(259,1)(260,4)(261,3)(262,4)(263,2)(264,1)(265,5)(266,2)(267,0)(268,3)(269,0)(270,5)(271,1)(272,4)(273,5)(274,0)(275,7)(276,4)(277,1)(278,1)(279,3)(280,5)(281,0)(282,1)(283,3)(284,1)(285,1)(286,1)(287,2)(288,5)(289,1)(290,4)(291,3)(292,2)(293,0)(294,4)(295,3)(296,4)(297,3)(298,1)(299,0)(300,3)(301,3)(302,3)(303,1)(304,7)(305,2)(306,2)(307,2)(308,1)(309,2)(310,1)(311,1)(312,1)(313,0)(314,1)(315,4)(316,1)(317,2)(318,1)(319,1)(320,1)(321,1)(322,0)(323,1)(324,3)(325,1)(326,2)(327,0)(328,3)(329,1)(330,1)(331,1)(332,1)(333,1)(334,0)(335,0)(336,0)(337,1)(338,0)(339,0)(340,1)(341,1)(342,0)
}; % plot of the first histogram
\legend{glass \#1,  glass \#2, glass \#3, no glass \#1,  no glass \#2, no glass \#3,} % legend of the plot

\end{axis}
\end{tikzpicture}

%% file: plots/glass_bottles/evac_time.tex
\begin{tikzpicture}[scale=0.75]

% Set the axis labels and title
\begin{axis}[    xlabel={Time [s]},    ylabel={s},    title={Evacuation Time - Glass bottles},   legend style={at={(0,1)}, anchor=north west},]

\addplot[color=PineGreen] coordinates{(0,27.283142857142856)(1,30.052566666666667)(2,23.221999999999994)(3,14.446097560975609)(4,11.321659793814433)(5,27.214739495798323)(6,23.913623188405797)(7,20.334628099173557)(8,20.451341085271316)(9,25.202992481203008)(10,21.14211029411765)(11,31.537638297872338)(12,25.481913043478265)(13,31.07048920863309)(14,30.83947183098592)(15,34.16776923076923)(16,34.797897297297304)(17,36.480753246753245)(18,42.37516107382551)(19,34.40525735294118)(20,33.49012820512821)(21,42.495636363636365)(22,48.43037267080745)(23,47.14800000000001)(24,50.61913103448275)(25,45.15317857142858)(26,46.13713194444444)(27,52.8968)(28,37.58199346405229)(29,39.190039215686284)(30,48.977457516339875)(31,46.22048076923076)(32,40.75864925373133)(33,36.50191111111111)(34,33.10545714285714)(35,51.83977536231885)(36,44.94265555555555)(37,46.933698113207534)(38,47.09440196078431)(39,36.7563679245283)(40,33.665152380952385)(41,55.11428409090908)(42,26.05582278481013)(43,41.671469135802475)(44,39.08615841584159)(45,58.556027027027035)(46,47.808506329113925)(47,61.55724096385543)(48,54.771277777777776)(49,55.738640000000004)(50,50.99449295774649)(51,60.0419)(52,56.90103125)(53,41.18778571428571)(54,52.09866216216217)(55,53.90111627906976)(56,56.95107352941176)(57,55.119909090909076)(58,58.63927397260273)(59,50.847864583333326)(60,55.11542253521127)(61,52.2856)(62,44.274563380281684)(63,66.52908450704224)(64,54.96795081967213)(65,58.910280701754395)(66,53.72803389830509)(67,56.29964150943397)(68,44.82835)(69,36.68659183673469)(70,58.38102083333332)(71,63.12125925925926)(72,78.30883050847459)(73,71.57657777777777)(74,62.747199999999985)(75,56.55237999999999)(76,68.52929032258065)(77,54.053)(78,37.39992105263158)(79,52.01441860465117)(80,37.01427272727273)(81,20.908692307692306)(82,36.007264705882356)(83,39.73727777777778)(84,47.23453846153846)(85,45.53170370370371)(86,58.61925)(87,27.386333333333333)(88,41.130266666666664)(89,21.670052631578947)(90,65.13621052631579)(91,41.30088)(92,31.004150000000003)(93,31.834)(94,19.726)(95,43.203291666666665)(96,43.68568421052631)(97,60.56812499999999)(98,89.12085714285715)(99,20.820100000000004)(100,31.261450000000004)(101,86.93029166666668)(102,65.96868421052632)(103,26.151625)(104,38.09463636363636)(105,59.91680952380952)(106,69.94850000000001)(107,31.499149999999997)(108,55.32184210526316)(109,35.06841666666667)(110,11.696944444444444)(111,37.21764705882352)(112,49.663117647058826)(113,35.20441666666667)(114,33.38521052631579)(115,60.462)(116,32.606307692307695)(117,96.48136363636364)(118,95.5616)(119,62.49288235294119)(120,42.5162)(121,39.8835)(122,96.79763636363637)(123,65.589)(124,106.66185714285716)(125,58.218545454545456)(126,71.20066666666666)(127,56.98720000000001)(128,38.896)(129,48.69654545454546)(130,68.2554090909091)(131,71.55308333333333)(132,39.10736363636364)(133,47.827333333333335)(134,25.340058823529407)(135,129.31406666666666)(136,19.604454545454544)(137,61.657357142857144)(138,36.001888888888885)(139,16.640307692307694)(140,92.783)(141,108.33040000000001)(142,102.0939411764706)(143,84.49038888888889)(144,57.962599999999995)(145,39.54809090909091)(146,29.02266666666667)(147,48.403)(148,43.587599999999995)(149,54.50725)(150,54.528083333333335)(151,48.49288888888889)(152,62.373714285714286)(153,81.892)(154,59.577636363636366)(155,36.42216666666666)(156,58.29593333333333)(157,36.45633333333333)(158,24.313111111111112)(159,24.32488888888889)(160,31.286857142857144)(161,99.585)(162,73.04566666666666)(163,31.322285714285716)(164,27.417)(165,24.38222222222222)(166,109.76724999999999)(167,27.450625)(168,73.22716666666666)(169,54.938625)(170,36.642)(171,43.9841)(172,73.34233333333333)(173,73.37266666666666)(174,110.127625)(175,73.44433333333333)(176,55.107)(177,55.12825)(178,110.2565)(179,63.04657142857143)(180,49.05522222222222)(181,110.4175)(182,63.10828571428572)(183,88.4288)(184,73.762)(185,110.7915)(186,55.4485)(187,147.93483333333333)(188,98.7882222222222)(189,27.795375)(190,88.9742)(191,80.91436363636363)(192,55.64825)(193,111.4915)(194,31.864714285714285)(195,44.6286)(196,31.887571428571427)(197,89.31219999999999)(198,44.6704)(199,111.7265)(200,31.93642857142857)(201,89.457)(202,111.8605)(203,149.18866666666665)(204,24.901444444444444)(205,37.364333333333335)(206,37.3805)(207,44.8686)(208,112.21133333333334)(209,112.276)(210,89.8922)(211,225.043)(212,32.16685714285715)(213,32.17457142857143)(214,28.16025)(215,56.33575)(216,64.45242857142857)(217,169.238)(218,64.488)(219,75.256)(220,75.27433333333333)(221,150.72566666666668)(222,226.323)(223,113.1915)(224,113.3065)(225,75.601)(226,75.621)(227,56.7295)(228,113.48650000000002)(229,151.357)(230,151.396)(231,113.574)(232,136.3218)(233,75.75433333333334)(234,113.6565)(235,113.689)(236,56.85575)(237,227.485)(238,113.7705)(239,170.7025)(240,56.9295)(241,113.91275)(242,113.939)(243,227.933)(244,113.997)(245,228.10299999999998)(246,38.02633333333333)(247,152.142)(248,57.077)(249,76.171)(250,228.8095)(251,114.443)(252,57.2355)(253,229.00300000000001)(254,229.063)(255,76.37400000000001)(256,137.5458)(257,76.45100000000001)(258,76.46733333333333)(259,91.8012)(260,45.9326)(261,76.57433333333334)(262,229.773)(263,76.60766666666667)(264,114.939)(265,91.9932)(266,76.714)(267,92.0792)(268,115.126)(269,230.353)(270,76.83766666666666)
}; % plot of the first histogram
\addplot[color=Green] coordinates{(0,6.269947368421053)(1,19.238105263157895)(2,9.993226666666668)(3,12.472402439024389)(4,15.068785714285717)(5,12.649514285714286)(6,15.602934426229504)(7,16.927360544217688)(8,15.636286956521742)(9,22.09299285714286)(10,14.483969230769228)(11,24.013836065573766)(12,25.22273983739837)(13,28.05565322580645)(14,26.04418633540373)(15,23.145680722891566)(16,26.591709459459455)(17,19.71788957055215)(18,22.207423357664236)(19,27.16217322834646)(20,29.886428571428574)(21,28.517231292517003)(22,25.611677165354322)(23,33.73540677966101)(24,38.842352941176465)(25,36.27164179104478)(26,30.182766666666673)(27,27.731357615894037)(28,33.73150000000001)(29,36.6622582781457)(30,36.88842635658915)(31,27.76277702702703)(32,36.23405633802817)(33,34.930475728155336)(34,39.654281250000004)(35,35.22670072992701)(36,43.79735537190082)(37,21.91342608695652)(38,42.82026851851851)(39,32.198388429752065)(40,35.85663095238095)(41,15.536910256410255)(42,29.077845238095236)(43,27.069635294117646)(44,32.529422535211275)(45,39.496957446808516)(46,44.45512987012987)(47,47.713341463414636)(48,33.207802816901406)(49,49.839515789473694)(50,43.51751648351648)(51,39.76614473684211)(52,39.38186301369863)(53,35.449532467532464)(54,33.06864102564102)(55,41.96157142857143)(56,47.16081818181818)(57,49.87901333333333)(58,38.84431746031746)(59,21.141274193548387)(60,38.377700000000004)(61,55.5458695652174)(62,45.76226153846154)(63,43.316246153846144)(64,48.421513888888896)(65,43.81915789473684)(66,35.75139285714286)(67,37.529913793103454)(68,34.66412698412699)(69,46.44994827586206)(70,46.66595744680851)(71,55.03067441860465)(72,59.55551851851854)(73,49.330145454545445)(74,43.386234042553184)(75,52.11879591836735)(76,59.737275)(77,35.854064516129036)(78,53.01214285714285)(79,22.89622222222222)(80,44.08184615384615)(81,20.492714285714285)(82,48.49637500000001)(83,13.99727027027027)(84,35.59579411764705)(85,40.39113333333333)(86,34.65125714285714)(87,46.27786666666667)(88,34.76864)(89,15.821545454545452)(90,34.848466666666674)(91,18.72585714285714)(92,41.15776470588236)(93,43.776)(94,17.566)(95,51.71347058823529)(96,33.539476190476186)(97,44.0598)(98,16.034181818181818)(99,11.775533333333334)(100,39.282555555555554)(101,47.17463333333333)(102,44.2625)(103,50.69871428571429)(104,74.01325000000001)(105,28.078)(106,18.77626315789474)(107,18.794947368421056)(108,64.9979090909091)(109,82.66346153846153)(110,44.81275000000001)(111,107.62449999999998)(112,47.881466666666675)(113,49.94255555555556)(114,34.27433333333333)(115,60.06966666666667)(116,10.613117647058823)(117,68.83904761904762)(118,60.287000000000006)(119,67.92275000000001)(120,67.9744375)(121,48.41226666666667)(122,45.41908333333333)(123,53.47794117647059)(124,51.98664285714286)(125,60.738388888888885)(126,72.98010000000001)(127,60.889199999999995)(128,45.722875)(129,43.113882352941175)(130,52.41966666666667)(131,13.109785714285715)(132,66.766)(133,51.03777777777778)(134,61.271666666666675)(135,81.72355555555555)(136,70.74876923076923)(137,69.040375)(138,78.93478571428571)(139,55.28340000000001)(140,15.361666666666666)(141,24.586866666666666)(142,36.892266666666664)(143,11.534875)(144,30.77808333333333)(145,55.4208)(146,55.44655)(147,39.625214285714286)(148,41.105555555555554)(149,20.552666666666667)(150,46.278375)(151,50.51654545454546)(152,42.759307692307694)(153,67.404)(154,37.0986)(155,53.03285714285715)(156,32.765)(157,33.77381818181818)(158,61.934999999999995)(159,20.656444444444443)(160,53.13457142857143)(161,26.58042857142857)(162,93.106)(163,23.287375)(164,74.61959999999999)(165,37.330799999999996)(166,62.29844444444445)(167,37.4002)(168,62.37733333333333)(169,62.391333333333336)(170,20.81566666666667)(171,53.54885714285714)(172,83.34666666666666)(173,80.40428571428572)(174,26.815285714285714)(175,18.7884)(176,112.76820000000001)(177,62.669000000000004)(178,94.06716666666667)(179,37.6496)(180,125.55466666666666)(181,62.77733333333333)(182,31.413833333333333)(183,37.7352)(184,125.83399999999999)(185,53.947428571428574)(186,37.7776)(187,104.97333333333334)(188,189.124)(189,47.281)(190,126.27266666666667)(191,63.18633333333333)(192,37.9404)(193,63.268)(194,47.46975)(195,94.9785)(196,38.036)(197,38.0544)(198,84.60833333333332)(199,47.6115)(200,31.751333333333335)(201,95.2985)(202,47.666)(203,95.3685)(204,95.41083333333334)(205,76.4352)(206,27.315142857142856)(207,63.84366666666667)(208,31.936166666666665)(209,38.3752)(210,115.1778)(211,96.02575)(212,64.03566666666667)(213,54.91628571428571)(214,48.082875)(215,192.51975)(216,77.1152)(217,48.22775)(218,144.91424999999998)(219,64.43966666666667)(220,64.55333333333333)(221,96.88166666666666)(222,48.48975)(223,97.10675)(224,64.84633333333333)(225,48.673)(226,130.10533333333333)(227,39.051)(228,117.2124)(229,48.85925)(230,65.19766666666666)(231,32.61983333333333)(232,65.289)(233,65.335)(234,98.081)(235,98.146)(236,130.93266666666668)(237,196.659)(238,98.3845)(239,73.838375)(240,32.84366666666667)(241,39.4308)(242,197.377)(243,65.85366666666667)(244,197.669)(245,65.95233333333333)(246,66.00583333333333)(247,66.03633333333333)(248,49.57275)(249,99.2845)(250,66.24566666666666)(251,99.413)(252,99.468)(253,99.512)(254,149.34175)(255,133.114)(256,199.937)(257,100.277)(258,40.2258)(259,50.46675)(260,101.0335)(261,40.4508)(262,33.73)(263,50.61975)(264,101.385)(265,202.99699999999999)(266,204.104)
}; % plot of the second histogram
\addplot[color=LimeGreen] coordinates{(0,12.336324324324325)(1,17.754189655172414)(2,25.20265573770492)(3,19.37559090909091)(4,24.81955072463768)(5,21.86809615384615)(6,19.684626168224302)(7,23.164016528925618)(8,19.619286821705423)(9,18.449016393442626)(10,16.805043103448277)(11,21.936201388888886)(12,22.543167832167835)(13,18.847725352112672)(14,25.121321678321678)(15,21.699317241379315)(16,23.771346153846153)(17,23.213057142857142)(18,25.521496774193544)(19,24.220975903614463)(20,36.11550295857988)(21,39.36123308270677)(22,25.548274509803917)(23,27.2139209039548)(24,26.529821656050952)(25,36.88392)(26,36.324126865671644)(27,37.715157303370795)(28,47.793956521739126)(29,40.34365269461077)(30,35.53759712230215)(31,29.74753913043478)(32,34.516297872340424)(33,31.920186567164176)(34,34.8882421875)(35,53.151047244094464)(36,35.969175925925924)(37,35.29840833333333)(38,49.728366071428574)(39,53.78323913043478)(40,35.632838709677415)(41,36.1041686746988)(42,43.32275294117647)(43,27.659726027397262)(44,38.710132530120475)(45,46.647549450549455)(46,45.764804878048785)(47,45.24411111111111)(48,41.95324358974359)(49,44.60423655913979)(50,51.78498701298701)(51,50.413811594202905)(52,44.269582089552244)(53,50.37580459770115)(54,39.630492957746476)(55,49.37076)(56,62.24873239436619)(57,43.0278)(58,57.5084411764706)(59,48.47675714285715)(60,56.73261666666667)(61,65.6569423076923)(62,49.751763157894736)(63,45.72678666666667)(64,58.273355932203394)(65,54.427780000000006)(66,53.274741379310356)(67,43.98127586206896)(68,38.82895744680851)(69,47.00913513513513)(70,63.64965306122449)(71,54.29727272727273)(72,54.91776119402986)(73,50.546117647058814)(74,39.482571428571426)(75,59.80091891891893)(76,48.13326086956522)(77,48.58039473684211)(78,56.84361538461539)(79,17.88751612903226)(80,37.83543181818182)(81,59.7014193548387)(82,26.455342857142853)(83,44.734)(84,37.080475)(85,92.77122727272726)(86,53.03911428571429)(87,39.79960714285714)(88,39.81557142857143)(89,28.6025)(90,51.303482758620696)(91,32.07765517241379)(92,35.796)(93,93.10745)(94,42.33745454545454)(95,26.633928571428573)(96,43.934176470588234)(97,40.03857142857142)(98,74.79235000000001)(99,35.0848125)(100,70.2780625)(101,35.1683125)(102,50.52476923076922)(103,66.27182352941178)(104,40.25607142857143)(105,67.13914285714286)(106,62.68683333333334)(107,35.83914285714286)(108,70.587625)(109,41.85005555555555)(110,75.36973333333334)(111,40.39328571428572)(112,47.15015)(113,58.06976923076923)(114,24.63317391304348)(115,94.47818749999999)(116,51.564727272727275)(117,35.4658125)(118,36.04514285714286)(119,56.7958)(120,37.889466666666664)(121,37.9052)(122,47.40116666666666)(123,40.647285714285715)(124,34.49581818181818)(125,60.39813636363636)(126,47.47733333333333)(127,27.139499999999998)(128,34.558)(129,47.545500000000004)(130,71.3463125)(131,50.7568)(132,19.0422)(133,63.49600000000001)(134,40.84157142857142)(135,95.34)(136,20.08221052631579)(137,86.75600000000001)(138,29.380769230769236)(139,50.9516)(140,81.92242857142857)(141,27.319499999999998)(142,38.268)(143,23.929)(144,44.19084615384615)(145,45.07129411764706)(146,68.4325)(147,73.73576923076922)(148,47.95216666666666)(149,15.992083333333333)(150,17.464181818181817)(151,73.95007692307692)(152,38.476)(153,17.498272727272727)(154,35.01290909090909)(155,48.166)(156,96.446)(157,27.576857142857143)(158,32.21816666666667)(159,86.01622222222221)(160,27.66857142857143)(161,24.240125)(162,43.111999999999995)(163,35.290818181818175)(164,48.588)(165,111.11957142857143)(166,129.701)(167,44.92061538461538)(168,24.343125)(169,48.7285)(170,65.00099999999999)(171,65.03377777777779)(172,55.77314285714285)(173,48.824)(174,48.893)(175,83.86057142857143)(176,48.9655)(177,78.55319999999999)(178,39.312799999999996)(179,49.172)(180,73.923)(181,157.8788)(182,21.939222222222224)(183,65.84733333333334)(184,65.92566666666666)(185,56.53742857142857)(186,79.19000000000001)(187,99.148)(188,66.228)(189,79.5162)(190,66.35833333333333)(191,44.26022222222222)(192,79.7124)(193,49.874875)(194,39.9802)(195,66.76166666666667)(196,28.67085714285714)(197,100.427)(198,133.99)(199,67.057)(200,201.40333333333334)(201,50.41)(202,101.0575)(203,28.873428571428573)(204,101.212)(205,40.550599999999996)(206,67.69333333333334)(207,40.6432)(208,50.8605)(209,122.23440000000001)(210,136.085)(211,102.172)(212,68.16083333333334)(213,82.0578)(214,205.46866666666668)(215,68.74900000000001)(216,51.662)(217,41.3584)(218,137.936)(219,69.03099999999999)(220,51.8135)(221,103.743)(222,124.68980000000002)(223,103.9925)(224,104.0425)(225,41.6716)(226,125.11419999999998)(227,29.816285714285716)(228,83.6292)(229,52.3075)(230,104.7435)(231,26.1975)(232,125.8158)(233,209.865)(234,52.66825)(235,210.904)(236,158.34375)(237,211.124)(238,140.93266666666668)(239,84.68039999999999)(240,52.9495)(241,105.95474999999999)(242,70.661)(243,90.97114285714285)(244,212.35299999999998)(245,106.215)(246,70.94083333333333)(247,106.45349999999999)(248,71.03)(249,142.19066666666666)(250,213.40666666666667)(251,71.25733333333334)(252,107.124)(253,107.1635)(254,107.224)(255,71.5335)(256,214.748)(257,128.88899999999998)(258,215.134)(259,86.10839999999999)(260,107.783)(261,53.90575)(262,53.937)(263,71.96366666666667)(264,216.444)(265,108.4275)(266,54.2675)(267,217.152)(268,72.58266666666667)(269,109.195)
}; % plot of the third histogram
\addplot[color=RedOrange] coordinates{(0,10.7019375)(1,18.482236842105266)(2,16.898564705882354)(3,20.80188709677419)(4,22.33293421052631)(5,13.586298245614037)(6,18.597135338345865)(7,22.873608695652166)(8,24.111883928571434)(9,14.251795275590549)(10,15.565396825396823)(11,21.35778145695364)(12,25.154979310344824)(13,23.94607092198582)(14,20.904993710691826)(15,24.114499999999996)(16,25.146580246913576)(17,23.343536423841055)(18,25.377053892215567)(19,32.56318243243244)(20,36.284835937500006)(21,27.390019736842106)(22,29.18396226415095)(23,30.984857142857145)(24,36.38582010582011)(25,38.14634899328859)(26,42.17005555555555)(27,43.59585119047618)(28,47.666527272727286)(29,47.5991726618705)(30,39.334844827586195)(31,36.42077966101696)(32,36.57064227642276)(33,34.981328000000005)(34,43.75146721311475)(35,34.76763414634147)(36,37.1270776699029)(37,27.872886956521743)(38,30.67895238095238)(39,51.54540000000001)(40,33.00740449438203)(41,41.45039759036144)(42,29.039282352941175)(43,49.01011267605633)(44,32.912938271604936)(45,47.41920895522388)(46,48.29232876712328)(47,44.16646428571429)(48,51.24968604651163)(49,29.624855072463767)(50,49.82477777777778)(51,48.27542253521126)(52,52.556027777777786)(53,35.73996825396826)(54,45.02776543209876)(55,46.78708955223881)(56,47.435345679012336)(57,49.446376470588234)(58,44.56633333333334)(59,50.960000000000015)(60,49.811211267605636)(61,51.93365517241379)(62,45.85288709677419)(63,32.56601219512194)(64,62.071760869565225)(65,49.835016393442615)(66,52.1364)(67,65.86479999999999)(68,53.32170370370371)(69,74.48263043478262)(70,39.68253658536585)(71,61.49832075471699)(72,32.60225641025641)(73,58.44908928571429)(74,48.5238)(75,52.98610909090909)(76,49.35670833333333)(77,30.3941875)(78,43.64110869565217)(79,30.431299999999997)(80,36.541799999999995)(81,32.64339285714286)(82,38.80330303030303)(83,45.75725)(84,52.32632142857142)(85,48.23668421052632)(86,20.38374074074074)(87,28.98805263157895)(88,52.47895238095238)(89,66.82463636363637)(90,75.7169411764706)(91,54.107000000000006)(92,32.475)(93,69.0443125)(94,32.502529411764705)(95,40.94266666666667)(96,28.36523076923077)(97,36.90645000000001)(98,46.166625)(99,46.19319999999999)(100,37.02573333333333)(101,66.17985714285714)(102,69.5679375)(103,10.91764705882353)(104,42.853)(105,61.93255555555556)(106,54.68735294117647)(107,76.59623529411765)(108,37.224599999999995)(109,69.8315)(110,118.57527272727272)(111,55.9307)(112,116.56924999999998)(113,46.664249999999996)(114,15.563666666666668)(115,20.765666666666664)(116,74.7941)(117,11.005764705882354)(118,49.259842105263154)(119,51.07190909090909)(120,53.53321428571429)(121,82.0230625)(122,40.188)(123,72.16553846153846)(124,62.58166666666667)(125,78.27125000000001)(126,68.33427272727273)(127,67.14421428571428)(128,59.39257894736843)(129,62.72666666666667)(130,80.69114285714285)(131,35.316562499999996)(132,86.98007692307692)(133,37.71346666666667)(134,47.1635625)(135,17.157818181818183)(136,94.42108333333334)(137,47.241875)(138,63.01558333333333)(139,37.8275)(140,40.54571428571428)(141,34.42227272727273)(142,43.711769230769235)(143,89.1684705882353)(144,63.186666666666675)(145,63.215333333333334)(146,43.788384615384615)(147,37.964099999999995)(148,47.47391666666667)(149,40.71635714285714)(150,54.326928571428574)(151,69.19845454545454)(152,121.15327272727274)(153,69.26763636363636)(154,54.454571428571434)(155,76.33253333333332)(156,63.67416666666667)(157,21.23922222222222)(158,34.783272727272724)(159,63.838)(160,82.15414285714284)(161,82.20957142857142)(162,192.561)(163,48.22825)(164,115.9224)(165,64.52133333333333)(166,193.765)(167,32.309000000000005)(168,48.6)(169,65.00666666666667)(170,48.87025)(171,65.246)(172,27.99142857142857)(173,130.70666666666668)(174,98.134)(175,196.459)(176,196.59)(177,65.622)(178,98.524)(179,65.73)(180,39.4944)(181,65.85466666666666)(182,149.25275)(183,199.103)(184,99.734)(185,199.63)(186,66.63466666666666)(187,40.0144)(188,100.113)(189,50.2925)(190,201.358)(191,201.626)(192,134.55766666666668)(193,101.035)(194,50.5395)(195,101.17066666666666)(196,50.60625)(197,28.928714285714285)(198,202.772)(199,81.2852)(200,162.75439999999998)(201,102.01925)(202,102.1415)(203,51.17575)(204,82.0376)(205,68.39783333333334)(206,68.48133333333332)(207,205.57966666666667)(208,82.3136)(209,51.5115)(210,155.44125)(211,69.12433333333333)(212,69.14866666666667)(213,207.99349999999998)(214,41.6592)(215,69.51133333333333)(216,69.642)(217,125.6872)(218,139.70133333333334)(219,209.642)(220,52.47575)(221,210.119)(222,52.60575)(223,42.1018)(224,105.2925)(225,105.4405)(226,211.18)(227,70.45733333333334)(228,211.52249999999998)(229,211.765)(230,106.0225)(231,106.101)(232,70.76233333333333)(233,106.22175)(234,212.522)(235,70.94433333333333)(236,71.00166666666667)
}; % plot of the fourth histogram
\addplot[color=Orange] coordinates{(0,13.370972222222221)(1,20.617375)(2,27.29604878048781)(3,6.791766233766234)(4,17.95771428571428)(5,20.344983050847457)(6,19.23830303030303)(7,19.145393333333338)(8,19.844079207920792)(9,30.650852713178303)(10,23.99923188405797)(11,27.212699300699306)(12,29.79961971830985)(13,28.24623880597015)(14,21.477946236559138)(15,30.761333333333337)(16,18.063348684210524)(17,16.333132653061227)(18,25.020846153846154)(19,40.57404605263159)(20,35.0015987654321)(21,41.05929861111112)(22,40.37859119496855)(23,36.77924277456648)(24,48.595736842105254)(25,43.859869281045754)(26,36.82990184049079)(27,46.074418079096056)(28,59.43903921568626)(29,44.65153086419753)(30,39.49182051282051)(31,31.218532258064517)(32,33.67151886792452)(33,59.81677777777777)(34,40.6584)(35,38.30319512195123)(36,37.841629310344835)(37,40.50426271186441)(38,44.49015476190476)(39,55.87920833333333)(40,41.11768674698795)(41,40.133430555555556)(42,43.95410526315789)(43,47.7962375)(44,58.14736231884058)(45,55.809013888888884)(46,30.475208333333335)(47,61.08647826086956)(48,50.831215384615376)(49,31.51162857142857)(50,51.11263888888889)(51,35.818)(52,38.34883116883116)(53,52.011500000000005)(54,52.66581481481482)(55,68.7332602739726)(56,62.04644444444446)(57,39.538242424242426)(58,61.20101639344262)(59,57.91815492957747)(60,52.651281250000004)(61,38.44678205128205)(62,51.843578947368414)(63,39.85239393939394)(64,51.29576363636364)(65,37.665)(66,50.26978333333333)(67,42.37408163265306)(68,74.33060655737705)(69,56.32176595744681)(70,39.42520833333334)(71,33.831374999999994)(72,60.66848)(73,46.16137837837837)(74,60.62521276595745)(75,55.94609803921569)(76,53.8904716981132)(77,46.58991111111111)(78,42.42933333333333)(79,57.93296969696969)(80,16.88858823529412)(81,31.952366666666663)(82,56.87140740740741)(83,7.68856)(84,38.47968)(85,33.980117647058826)(86,22.668823529411764)(87,14.836461538461538)(88,84.45481249999999)(89,35.12854545454545)(90,74.37853846153847)(91,73.74266666666668)(92,21.52433333333333)(93,30.608263157894736)(94,64.70966666666668)(95,38.8692)(96,35.392272727272726)(97,68.83476470588235)(98,53.253)(99,75.14269230769231)(100,48.876583333333336)(101,83.9152142857143)(102,48.98425)(103,75.42253846153845)(104,78.49640000000001)(105,28.068214285714287)(106,78.63820000000001)(107,73.7815)(108,84.41278571428572)(109,30.336692307692303)(110,19.7396)(111,43.89533333333333)(112,61.8314375)(113,45.702076923076916)(114,54.12318181818182)(115,46.75929411764706)(116,62.170187500000004)(117,36.20127272727273)(118,19.9247)(119,70.42823529411764)(120,26.644333333333332)(121,140.0235)(122,42.910357142857144)(123,91.09254545454546)(124,40.1222)(125,40.147000000000006)(126,89.27588888888889)(127,40.218599999999995)(128,28.74857142857143)(129,73.2500909090909)(130,20.1646)(131,31.04869230769231)(132,36.73436363636363)(133,47.591941176470584)(134,40.4794)(135,50.625750000000004)(136,60.810500000000005)(137,81.128)(138,22.56566666666667)(139,50.80087499999999)(140,58.098)(141,58.144214285714284)(142,62.68353846153847)(143,45.298944444444444)(144,50.993)(145,95.27739999999999)(146,18.572)(147,68.1711111111111)(148,54.58293333333334)(149,20.482599999999998)(150,68.343)(151,102.5713)(152,17.105166666666666)(153,73.39314285714286)(154,77.109375)(155,34.37416666666667)(156,123.8126)(157,20.6518)(158,17.219583333333333)(159,51.735375)(160,41.4828)(161,69.17366666666666)(162,103.798)(163,69.40133333333334)(164,104.5355)(165,104.76050000000001)(166,140.00333333333333)(167,42.0318)(168,35.051)(169,105.2065)(170,210.51466666666667)(171,105.3585)(172,105.516)(173,35.19433333333333)(174,70.43833333333333)(175,52.864)(176,70.53099999999999)(177,53.0285)(178,70.7445)(179,106.167)(180,106.311)(181,35.464)(182,85.1522)(183,213.206)(184,53.3515)(185,213.60500000000002)(186,213.76)(187,53.5035)(188,107.097)(189,71.62766666666667)(190,107.521)(191,43.0272)(192,53.81875)(193,215.393)(194,143.74300000000002)(195,80.909375)(196,107.9575)(197,43.275400000000005)(198,72.353)(199,108.66749999999999)(200,72.46266666666666)(201,54.40675)(202,108.935)(203,54.5205)(204,109.066)(205,72.75066666666667)(206,109.156)(207,72.789)(208,87.3924)(209,109.4735)(210,109.67775)(211,109.7135)(212,109.748)(213,109.831)(214,87.8928)(215,87.9604)(216,73.32066666666667)(217,73.34400000000001)(218,73.38066666666667)(219,73.42399999999999)(220,44.0694)(221,73.46733333333333)(222,220.532)(223,73.52733333333333)(224,110.3265)(225,220.874)(226,73.66366666666667)(227,110.531)(228,221.2375)(229,73.79833333333333)(230,147.65)(231,55.388)(232,221.64)(233,110.9075)(234,221.972)(235,222.032)(236,111.12933333333335)(237,222.322)(238,111.463)(239,63.77628571428572)(240,223.508)(241,224.324)
}; % plot of the fifth histogram
\addplot[color=Melon] coordinates{(0,12.066225806451612)(1,15.584325581395348)(2,13.351530303030305)(3,14.369126984126982)(4,14.761272727272724)(5,20.675901960784312)(6,17.152598484848486)(7,22.72184821428571)(8,20.303270491803282)(9,22.594763513513517)(10,14.680286764705883)(11,21.307066666666667)(12,28.62722151898734)(13,19.901410256410255)(14,14.837098765432097)(15,24.669576923076914)(16,23.94859863945578)(17,32.58796478873239)(18,26.929194117647064)(19,23.177959677419356)(20,28.300985915492955)(21,33.63326356589147)(22,38.71163636363636)(23,43.16252500000001)(24,35.82096129032258)(25,42.55500671140939)(26,32.52592753623189)(27,43.34418902439025)(28,46.28307194244602)(29,41.1470253164557)(30,28.462129870129875)(31,40.06815671641791)(32,29.307228070175437)(33,36.22077419354838)(34,49.66233846153847)(35,41.262194915254234)(36,44.366300970873795)(37,30.134816513761464)(38,39.59837)(39,41.439225)(40,34.89)(41,40.92644186046512)(42,30.15374626865671)(43,33.81208235294118)(44,26.446064935064932)(45,22.737639999999995)(46,42.86483333333334)(47,32.1570989010989)(48,48.57796875)(49,31.928460526315792)(50,49.04215384615385)(51,47.385914285714286)(52,38.93319444444444)(53,48.15602739726027)(54,36.71302777777778)(55,43.492969230769226)(56,58.16070149253733)(57,55.91598148148149)(58,65.09164864864866)(59,39.762063492063504)(60,44.8524375)(61,70.234671875)(62,69.90697959183673)(63,46.74236206896552)(64,37.383476190476195)(65,36.303672727272726)(66,41.96986153846155)(67,55.154674418604635)(68,46.53707272727273)(69,43.81976086956522)(70,44.16638888888889)(71,63.30470689655173)(72,49.765874999999994)(73,36.77131111111112)(74,41.65996226415094)(75,64.43067499999998)(76,57.011333333333326)(77,48.08447826086957)(78,42.26266666666667)(79,38.86010526315789)(80,13.515243902439025)(81,18.4848)(82,40.45903125)(83,75.72045454545454)(84,44.89863636363636)(85,17.646619047619048)(86,39.72278571428571)(87,75.56592592592591)(88,37.1126)(89,61.88226666666666)(90,84.42836363636364)(91,43.72164705882353)(92,15.496916666666666)(93,62.034416666666665)(94,10.953411764705882)(95,46.57966666666667)(96,55.9481)(97,8.484)(98,37.35973333333333)(99,54.97329411764706)(100,41.574888888888886)(101,58.4980625)(102,33.047941176470594)(103,41.638444444444445)(104,62.50277777777777)(105,57.723769230769236)(106,72.9995)(107,51.221363636363634)(108,17.07981818181818)(109,37.59373333333334)(110,77.4369411764706)(111,43.43215384615385)(112,50.214800000000004)(113,22.167117647058824)(114,39.68878947368421)(115,23.579875)(116,23.592937499999998)(117,94.4079)(118,70.834)(119,34.364363636363635)(120,27.020642857142857)(121,81.09407142857143)(122,31.550666666666668)(123,50.50893333333333)(124,37.900099999999995)(125,21.067111111111114)(126,47.42225)(127,37.9601)(128,31.644833333333334)(129,88.6522)(130,63.348200000000006)(131,47.53666666666667)(132,67.94185714285715)(133,15.862250000000001)(134,51.937000000000005)(135,71.45362499999999)(136,47.65525)(137,63.56686666666667)(138,34.68745454545455)(139,109.0767857142857)(140,40.92507142857143)(141,44.08407692307693)(142,38.2192)(143,95.5734)(144,67.49188235294116)(145,13.664285714285715)(146,82.02285714285715)(147,27.348142857142857)(148,23.94225)(149,21.290888888888887)(150,44.259692307692305)(151,88.56476923076924)(152,23.99625)(153,27.43885714285714)(154,48.041)(155,96.1685)(156,27.536)(157,77.14959999999999)(158,32.199)(159,96.64966666666668)(160,38.6834)(161,96.76966666666668)(162,48.483)(163,64.76633333333332)(164,48.615)(165,77.8516)(166,64.90533333333333)(167,97.40674999999999)(168,73.153875)(169,65.149)(170,195.69299999999998)(171,97.9125)(172,98.0495)(173,196.567)(174,118.15279999999998)(175,98.5025)(176,39.469)(177,65.85133333333333)(178,32.940333333333335)(179,66.00433333333334)(180,99.06325)(181,66.07433333333334)(182,66.11)(183,39.6864)(184,49.93325)(185,100.08725)(186,50.373)(187,134.465)(188,67.28433333333334)(189,67.32033333333334)(190,50.584)(191,67.56233333333334)(192,50.69825)(193,101.505)(194,33.949999999999996)(195,68.015)(196,81.7612)(197,68.234)(198,102.42575)(199,102.641)(200,68.47133333333333)(201,68.52233333333334)(202,206.181)(203,68.761)(204,41.318200000000004)(205,138.08033333333333)(206,34.58566666666667)(207,69.26333333333334)(208,207.876)(209,52.03)(210,208.553)(211,125.1778)(212,69.66666666666667)(213,41.833800000000004)(214,52.35275)(215,41.912400000000005)(216,30.003142857142855)(217,52.553)(218,70.26766666666667)(219,84.4138)(220,211.212)(221,211.361)(222,211.445)(223,84.7212)(224,105.981)(225,70.68366666666667)(226,106.0675)(227,53.09325)(228,70.815)(229,53.15725)(230,141.856)(231,212.976)(232,213.307)(233,213.841)(234,91.67771428571427)(235,106.9995)(236,107.165)(237,214.717)(238,107.3915)(239,107.4615)(240,215.105)(241,72.14566666666667)(242,216.873)(243,216.97)
}; % plot of the first histogram
\legend{glass \#1,  glass \#2, glass \#3, no glass \#1,  no glass \#2, no glass \#3,} % legend of the plot

\end{axis}
\end{tikzpicture}

%% file: plots/panic_fraction/evac_speed.tex
\begin{tikzpicture}[scale=0.75]

% Set the axis labels and title
\begin{axis}[    xlabel={Time [s]},    ylabel={ppl/s},    title={Evacuation Speed - Panic fraction},   legend style={at={(1,1)}, anchor=north east},]

\addplot[color=Turquoise] coordinates{(0,36)(1,48)(2,60)(3,79)(4,70)(5,114)(6,102)(7,127)(8,148)(9,134)(10,140)(11,130)(12,160)(13,150)(14,159)(15,132)(16,139)(17,153)(18,150)(19,139)(20,141)(21,185)(22,137)(23,149)(24,158)(25,132)(26,164)(27,167)(28,147)(29,137)(30,150)(31,139)(32,117)(33,114)(34,121)(35,155)(36,100)(37,93)(38,103)(39,82)(40,96)(41,74)(42,70)(43,66)(44,69)(45,68)(46,77)(47,72)(48,80)(49,82)(50,69)(51,69)(52,85)(53,58)(54,84)(55,71)(56,76)(57,76)(58,63)(59,79)(60,79)(61,76)(62,67)(63,76)(64,67)(65,53)(66,59)(67,51)(68,41)(69,52)(70,48)(71,48)(72,45)(73,41)(74,57)(75,47)(76,39)(77,45)(78,48)(79,34)(80,36)(81,35)(82,31)(83,26)(84,20)(85,17)(86,14)(87,13)(88,12)(89,15)(90,14)(91,19)(92,13)(93,19)(94,24)(95,16)(96,16)(97,13)(98,19)(99,14)(100,22)(101,12)(102,16)(103,19)(104,13)(105,21)(106,13)(107,16)(108,15)(109,15)(110,9)(111,12)(112,12)(113,10)(114,15)(115,5)(116,15)(117,11)(118,16)(119,13)(120,13)(121,21)(122,8)(123,15)(124,11)(125,14)(126,7)(127,15)(128,15)(129,15)(130,9)(131,13)(132,11)(133,11)(134,11)(135,13)(136,12)(137,8)(138,16)(139,22)(140,9)(141,10)(142,14)(143,8)(144,14)(145,11)(146,20)(147,9)(148,9)(149,6)(150,5)(151,14)(152,10)(153,16)(154,10)(155,9)(156,17)(157,7)(158,4)(159,5)(160,6)(161,4)(162,6)(163,13)(164,5)(165,8)(166,2)(167,1)(168,2)(169,4)(170,3)(171,4)(172,4)(173,3)(174,5)(175,6)(176,3)(177,2)(178,3)(179,3)(180,2)(181,0)(182,6)(183,1)(184,7)(185,1)(186,2)(187,2)(188,2)(189,4)(190,1)(191,7)(192,0)(193,2)(194,5)(195,1)(196,0)(197,5)(198,3)(199,0)(200,0)(201,0)(202,4)(203,5)(204,3)(205,3)(206,6)(207,1)(208,3)(209,3)(210,3)(211,3)(212,2)(213,1)(214,4)(215,6)(216,2)(217,3)(218,5)(219,2)(220,4)(221,7)(222,5)(223,3)(224,2)(225,3)(226,3)(227,3)(228,1)(229,1)(230,2)(231,3)(232,6)(233,1)(234,1)(235,2)(236,2)(237,4)(238,3)(239,4)(240,1)(241,2)(242,2)(243,5)(244,2)(245,3)(246,2)(247,4)(248,1)(249,2)(250,2)(251,3)(252,6)(253,2)(254,4)(255,0)(256,2)(257,4)(258,2)(259,3)(260,0)(261,5)(262,2)(263,0)(264,3)(265,5)(266,8)(267,3)(268,4)(269,1)(270,1)(271,1)(272,2)(273,2)(274,3)(275,4)(276,3)(277,2)(278,4)(279,1)(280,3)(281,0)(282,2)(283,0)(284,2)(285,1)(286,1)(287,1)(288,2)(289,3)(290,4)(291,2)(292,3)(293,1)(294,0)(295,3)(296,1)(297,1)(298,4)(299,6)(300,1)(301,0)(302,3)(303,2)(304,4)(305,0)(306,5)(307,3)(308,1)(309,2)(310,2)(311,2)(312,0)(313,1)(314,1)(315,4)(316,3)(317,3)(318,0)(319,1)(320,2)(321,2)(322,6)(323,0)(324,1)(325,2)(326,0)(327,0)(328,3)(329,2)(330,0)(331,0)(332,0)(333,0)(334,1)(335,0)
}; % plot of the first histogram
\addplot[color=Green] coordinates{(0,18)(1,47)(2,59)(3,91)(4,77)(5,101)(6,111)(7,100)(8,126)(9,134)(10,124)(11,145)(12,129)(13,168)(14,149)(15,132)(16,173)(17,163)(18,131)(19,132)(20,140)(21,159)(22,147)(23,160)(24,137)(25,144)(26,128)(27,153)(28,172)(29,150)(30,123)(31,129)(32,143)(33,118)(34,135)(35,120)(36,113)(37,100)(38,85)(39,126)(40,97)(41,80)(42,69)(43,95)(44,87)(45,83)(46,82)(47,67)(48,82)(49,88)(50,54)(51,78)(52,82)(53,62)(54,66)(55,70)(56,60)(57,61)(58,71)(59,66)(60,72)(61,73)(62,62)(63,68)(64,56)(65,48)(66,50)(67,50)(68,56)(69,49)(70,46)(71,42)(72,46)(73,54)(74,50)(75,55)(76,47)(77,43)(78,49)(79,45)(80,22)(81,28)(82,27)(83,28)(84,28)(85,10)(86,13)(87,19)(88,21)(89,24)(90,11)(91,20)(92,24)(93,18)(94,17)(95,18)(96,20)(97,16)(98,20)(99,12)(100,17)(101,10)(102,19)(103,19)(104,18)(105,20)(106,14)(107,13)(108,8)(109,12)(110,12)(111,18)(112,20)(113,13)(114,13)(115,15)(116,9)(117,12)(118,9)(119,19)(120,13)(121,13)(122,11)(123,10)(124,10)(125,14)(126,9)(127,11)(128,14)(129,12)(130,12)(131,13)(132,13)(133,14)(134,18)(135,16)(136,20)(137,13)(138,12)(139,17)(140,13)(141,16)(142,14)(143,14)(144,21)(145,10)(146,10)(147,14)(148,13)(149,12)(150,5)(151,11)(152,15)(153,11)(154,10)(155,10)(156,9)(157,6)(158,11)(159,8)(160,8)(161,7)(162,8)(163,9)(164,11)(165,3)(166,6)(167,3)(168,2)(169,8)(170,10)(171,6)(172,5)(173,1)(174,5)(175,2)(176,2)(177,1)(178,2)(179,2)(180,2)(181,4)(182,8)(183,4)(184,1)(185,2)(186,0)(187,5)(188,0)(189,1)(190,1)(191,2)(192,2)(193,3)(194,1)(195,4)(196,5)(197,5)(198,5)(199,5)(200,1)(201,4)(202,4)(203,5)(204,3)(205,2)(206,6)(207,3)(208,1)(209,3)(210,4)(211,3)(212,2)(213,4)(214,2)(215,3)(216,4)(217,4)(218,3)(219,4)(220,3)(221,2)(222,3)(223,4)(224,1)(225,2)(226,5)(227,3)(228,1)(229,3)(230,3)(231,1)(232,1)(233,2)(234,3)(235,3)(236,3)(237,4)(238,0)(239,4)(240,5)(241,2)(242,1)(243,1)(244,6)(245,3)(246,5)(247,2)(248,1)(249,2)(250,2)(251,4)(252,3)(253,1)(254,2)(255,0)(256,1)(257,3)(258,3)(259,2)(260,1)(261,4)(262,0)(263,2)(264,3)(265,2)(266,2)(267,2)(268,4)(269,2)(270,0)(271,3)(272,5)(273,5)(274,5)(275,5)(276,1)(277,2)(278,1)(279,2)(280,2)(281,1)(282,2)(283,2)(284,3)(285,2)(286,3)(287,1)(288,3)(289,2)(290,6)(291,8)(292,5)(293,2)(294,2)(295,3)(296,4)(297,1)(298,5)(299,2)(300,3)(301,4)(302,6)(303,5)(304,4)(305,3)(306,1)(307,7)(308,1)(309,1)(310,5)(311,2)(312,2)(313,3)(314,5)(315,2)(316,2)(317,2)(318,2)(319,0)(320,1)(321,1)(322,2)(323,3)(324,1)(325,1)(326,0)(327,0)(328,0)(329,0)(330,0)(331,0)(332,0)(333,1)(334,0)(335,2)(336,1)(337,1)(338,1)(339,1)(340,0)(341,1)(342,2)(343,0)(344,0)(345,0)(346,1)(347,0)(348,0)(349,1)(350,0)(351,1)(352,0)(353,0)(354,0)(355,0)(356,0)(357,0)(358,0)(359,0)(360,0)(361,0)(362,0)(363,0)(364,0)(365,0)(366,0)(367,0)(368,0)(369,1)(370,0)(371,0)(372,0)(373,1)(374,0)(375,0)(376,0)(377,0)(378,0)(379,0)(380,0)(381,0)(382,0)(383,0)(384,0)(385,2)(386,0)(387,0)(388,0)(389,1)(390,0)(391,0)(392,0)(393,1)(394,0)(395,0)(396,0)(397,0)(398,0)(399,0)(400,0)(401,0)(402,0)(403,0)(404,0)(405,0)(406,0)(407,0)(408,0)(409,0)(410,0)(411,0)(412,0)(413,0)(414,0)(415,0)(416,0)(417,0)(418,0)(419,0)(420,0)(421,0)(422,1)(423,0)(424,0)(425,0)(426,0)(427,0)(428,0)(429,0)(430,0)(431,0)(432,1)(433,0)(434,0)(435,1)(436,0)(437,0)(438,0)(439,0)(440,0)(441,0)(442,0)(443,0)(444,0)(445,0)(446,0)(447,1)(448,0)(449,0)(450,0)(451,0)(452,0)(453,0)(454,0)(455,0)(456,0)(457,0)(458,0)(459,0)(460,0)(461,0)(462,1)(463,0)(464,0)(465,0)(466,0)(467,0)(468,0)(469,0)(470,0)(471,0)(472,0)(473,0)(474,0)(475,0)(476,0)(477,0)(478,0)(479,0)(480,0)(481,0)(482,0)(483,0)(484,0)(485,0)(486,0)(487,0)(488,0)(489,0)(490,0)(491,0)(492,0)(493,0)(494,0)(495,0)(496,0)(497,0)(498,0)(499,0)(500,0)(501,1)(502,0)(503,1)(504,0)(505,0)(506,0)(507,0)(508,0)(509,0)(510,0)(511,0)(512,0)(513,0)(514,0)(515,0)(516,0)(517,0)(518,0)(519,0)(520,0)(521,0)(522,0)(523,0)(524,0)(525,0)(526,0)(527,0)(528,0)(529,0)(530,0)(531,0)(532,0)(533,0)(534,0)(535,0)(536,0)(537,0)(538,0)(539,0)(540,0)(541,0)(542,1)(543,0)(544,0)(545,0)(546,0)(547,0)(548,0)(549,0)(550,0)(551,0)(552,0)(553,0)(554,0)(555,0)(556,0)(557,0)(558,0)(559,0)(560,0)(561,0)(562,0)(563,0)(564,0)(565,0)(566,0)(567,0)(568,0)(569,0)(570,0)(571,0)(572,0)(573,0)(574,0)(575,0)(576,1)(577,0)(578,0)(579,0)(580,0)(581,0)(582,0)(583,0)(584,0)(585,0)(586,0)(587,0)(588,0)(589,0)(590,0)(591,1)(592,0)(593,0)(594,0)(595,0)(596,0)(597,0)(598,0)(599,0)(600,0)(601,0)(602,0)(603,0)(604,0)(605,0)(606,0)(607,0)(608,0)(609,0)(610,0)(611,0)(612,0)(613,0)(614,0)(615,0)(616,0)(617,1)(618,0)(619,0)(620,0)(621,0)(622,0)(623,0)(624,0)(625,0)(626,0)(627,0)(628,0)(629,0)(630,0)(631,0)(632,0)(633,0)(634,0)(635,0)(636,1)(637,0)(638,0)(639,0)(640,0)(641,0)(642,0)(643,0)(644,0)(645,0)(646,0)(647,0)(648,0)(649,0)(650,0)(651,0)(652,0)(653,0)(654,0)(655,0)(656,0)(657,0)(658,0)(659,0)(660,0)(661,0)(662,0)(663,0)(664,1)(665,0)
}; % plot of the second histogram
\addplot[color=Orange] coordinates{(0,32)(1,30)(2,66)(3,74)(4,93)(5,95)(6,101)(7,123)(8,129)(9,142)(10,140)(11,120)(12,154)(13,122)(14,149)(15,164)(16,159)(17,119)(18,153)(19,141)(20,149)(21,147)(22,130)(23,126)(24,138)(25,138)(26,150)(27,133)(28,167)(29,140)(30,143)(31,125)(32,137)(33,122)(34,112)(35,135)(36,123)(37,92)(38,98)(39,107)(40,92)(41,85)(42,66)(43,96)(44,82)(45,76)(46,78)(47,81)(48,77)(49,69)(50,76)(51,75)(52,59)(53,82)(54,71)(55,90)(56,91)(57,78)(58,76)(59,61)(60,69)(61,76)(62,64)(63,51)(64,51)(65,58)(66,54)(67,51)(68,69)(69,56)(70,61)(71,54)(72,55)(73,49)(74,50)(75,57)(76,50)(77,62)(78,45)(79,35)(80,39)(81,34)(82,35)(83,23)(84,29)(85,25)(86,28)(87,39)(88,21)(89,26)(90,28)(91,24)(92,27)(93,25)(94,27)(95,17)(96,27)(97,26)(98,27)(99,18)(100,24)(101,21)(102,16)(103,22)(104,21)(105,24)(106,19)(107,19)(108,22)(109,20)(110,25)(111,17)(112,22)(113,18)(114,17)(115,21)(116,20)(117,18)(118,16)(119,15)(120,17)(121,21)(122,18)(123,20)(124,22)(125,17)(126,15)(127,16)(128,17)(129,21)(130,19)(131,23)(132,18)(133,19)(134,24)(135,19)(136,14)(137,17)(138,16)(139,15)(140,15)(141,16)(142,17)(143,20)(144,14)(145,16)(146,20)(147,16)(148,20)(149,19)(150,20)(151,19)(152,13)(153,20)(154,12)(155,12)(156,12)(157,14)(158,11)(159,13)(160,12)(161,11)(162,7)(163,15)(164,12)(165,13)(166,6)(167,11)(168,8)(169,6)(170,8)(171,10)(172,11)(173,6)(174,9)(175,8)(176,9)(177,8)(178,8)(179,8)(180,8)(181,7)(182,12)(183,9)(184,9)(185,8)(186,8)(187,8)(188,8)(189,6)(190,7)(191,8)(192,7)(193,10)(194,7)(195,7)(196,10)(197,8)(198,10)(199,8)(200,7)(201,10)(202,8)(203,11)(204,10)(205,9)(206,13)(207,8)(208,8)(209,8)(210,8)(211,6)(212,7)(213,7)(214,8)(215,6)(216,6)(217,6)(218,9)(219,7)(220,9)(221,8)(222,10)(223,7)(224,7)(225,7)(226,11)(227,8)(228,12)(229,9)(230,9)(231,7)(232,8)(233,7)(234,8)(235,7)(236,5)(237,7)(238,7)(239,7)(240,9)(241,9)(242,10)(243,5)(244,8)(245,8)(246,9)(247,8)(248,8)(249,5)(250,5)(251,11)(252,5)(253,9)(254,9)(255,6)(256,8)(257,6)(258,9)(259,6)(260,7)(261,10)(262,8)(263,6)(264,8)(265,10)(266,8)(267,6)(268,5)(269,10)(270,9)(271,9)(272,8)(273,8)(274,9)(275,9)(276,9)(277,8)(278,8)(279,8)(280,6)(281,8)(282,7)(283,7)(284,7)(285,8)(286,10)(287,6)(288,5)(289,7)(290,8)(291,7)(292,9)(293,7)(294,6)(295,6)(296,5)(297,8)(298,11)(299,7)(300,8)(301,6)(302,9)(303,6)(304,6)(305,7)(306,7)(307,8)(308,5)(309,7)(310,9)(311,7)(312,6)(313,7)(314,5)(315,5)(316,6)(317,6)(318,7)(319,8)(320,7)(321,9)(322,7)(323,7)(324,6)(325,6)(326,8)(327,6)(328,5)(329,6)(330,6)(331,7)(332,5)(333,7)(334,6)(335,5)(336,6)(337,5)(338,5)(339,7)(340,5)(341,5)(342,5)(343,5)(344,5)(345,5)(346,6)(347,5)(348,5)(349,5)(350,6)(351,5)(352,5)(353,5)(354,5)(355,6)(356,5)(357,5)(358,6)(359,5)(360,5)(361,5)(362,5)(363,5)(364,6)(365,5)(366,5)(367,5)(368,5)(369,5)(370,5)(371,5)(372,5)(373,5)(374,5)(375,5)(376,5)(377,5)(378,5)(379,5)(380,5)(381,6)(382,5)(383,5)(384,5)(385,5)(386,6)(387,5)(388,5)(389,5)(390,5)(391,6)(392,5)(393,5)(394,5)(395,5)(396,5)(397,5)(398,5)(399,5)(400,5)(401,5)(402,5)(403,5)(404,6)(405,5)(406,5)(407,5)(408,5)(409,5)(410,5)(411,5)(412,5)(413,5)(414,5)(415,5)(416,5)(417,5)(418,6)(419,5)(420,5)(421,5)(422,5)(423,5)(424,5)(425,5)(426,5)(427,5)(428,5)(429,5)(430,5)(431,5)(432,5)(433,5)(434,5)(435,6)(436,6)(437,5)(438,5)(439,5)(440,5)(441,5)(442,5)(443,5)(444,5)(445,5)(446,5)(447,5)(448,5)(449,5)(450,5)(451,5)(452,5)(453,5)(454,5)(455,5)(456,5)(457,6)(458,5)(459,5)(460,5)(461,5)(462,5)(463,5)(464,5)(465,5)(466,5)(467,5)(468,5)(469,5)(470,5)(471,5)(472,5)(473,5)(474,5)(475,5)(476,5)(477,5)(478,5)(479,5)(480,5)(481,5)(482,5)(483,5)(484,5)(485,5)(486,5)(487,5)(488,5)(489,5)(490,5)(491,5)(492,5)(493,6)(494,5)(495,5)(496,5)(497,5)(498,5)(499,6)(500,5)(501,5)(502,5)(503,5)(504,5)(505,5)(506,5)(507,5)(508,5)(509,5)(510,5)(511,5)(512,5)(513,5)(514,5)(515,5)(516,5)(517,5)(518,5)(519,5)(520,5)(521,5)(522,5)(523,5)(524,5)(525,5)(526,5)(527,5)(528,5)(529,5)(530,5)(531,5)(532,5)(533,5)(534,6)(535,5)(536,5)(537,5)(538,5)(539,5)(540,6)(541,5)(542,5)(543,5)(544,5)(545,5)(546,5)(547,5)(548,6)(549,5)(550,5)(551,5)(552,5)(553,5)(554,5)(555,5)(556,5)(557,5)(558,5)(559,5)(560,5)(561,5)(562,5)(563,5)(564,5)(565,5)(566,5)(567,5)(568,5)(569,5)(570,5)(571,5)(572,5)(573,5)(574,5)(575,5)(576,5)(577,5)(578,5)(579,5)(580,5)(581,5)(582,5)(583,5)(584,5)(585,5)(586,5)(587,5)(588,5)(589,5)(590,5)(591,5)(592,5)(593,5)(594,5)(595,5)(596,5)(597,5)(598,5)(599,5)(600,5)(601,5)(602,5)(603,6)(604,5)(605,5)(606,5)(607,5)(608,5)(609,5)(610,5)(611,5)(612,5)(613,5)(614,5)(615,5)(616,5)(617,5)(618,5)(619,5)(620,5)(621,5)(622,5)(623,5)(624,5)(625,5)(626,5)(627,5)(628,5)(629,5)(630,5)(631,5)(632,5)(633,5)(634,5)(635,5)(636,5)(637,5)(638,5)(639,6)(640,5)(641,5)(642,5)(643,5)(644,5)(645,5)(646,5)(647,6)(648,5)
}; % plot of the first histogram
\addplot[color=Red] coordinates{(0,33)(1,43)(2,73)(3,76)(4,73)(5,91)(6,113)(7,105)(8,132)(9,109)(10,129)(11,140)(12,117)(13,161)(14,153)(15,145)(16,140)(17,143)(18,136)(19,123)(20,152)(21,133)(22,137)(23,149)(24,125)(25,139)(26,149)(27,144)(28,156)(29,137)(30,125)(31,121)(32,127)(33,101)(34,130)(35,136)(36,118)(37,91)(38,94)(39,90)(40,88)(41,74)(42,81)(43,85)(44,77)(45,72)(46,87)(47,84)(48,75)(49,74)(50,84)(51,68)(52,83)(53,73)(54,69)(55,72)(56,87)(57,54)(58,79)(59,51)(60,63)(61,40)(62,51)(63,71)(64,58)(65,61)(66,60)(67,44)(68,48)(69,54)(70,49)(71,43)(72,45)(73,49)(74,54)(75,33)(76,42)(77,47)(78,41)(79,48)(80,26)(81,32)(82,27)(83,33)(84,12)(85,20)(86,19)(87,27)(88,20)(89,22)(90,18)(91,20)(92,22)(93,22)(94,14)(95,14)(96,20)(97,17)(98,20)(99,18)(100,19)(101,19)(102,20)(103,16)(104,17)(105,8)(106,14)(107,12)(108,17)(109,28)(110,18)(111,13)(112,14)(113,16)(114,17)(115,22)(116,18)(117,16)(118,19)(119,12)(120,19)(121,18)(122,13)(123,21)(124,10)(125,13)(126,8)(127,15)(128,16)(129,20)(130,21)(131,14)(132,17)(133,19)(134,13)(135,16)(136,10)(137,11)(138,14)(139,12)(140,11)(141,8)(142,12)(143,19)(144,12)(145,15)(146,9)(147,11)(148,12)(149,8)(150,12)(151,16)(152,11)(153,8)(154,8)(155,12)(156,6)(157,6)(158,7)(159,12)(160,10)(161,7)(162,4)(163,6)(164,9)(165,5)(166,7)(167,4)(168,1)(169,4)(170,7)(171,4)(172,3)(173,3)(174,5)(175,4)(176,2)(177,2)(178,5)(179,2)(180,7)(181,4)(182,5)(183,5)(184,7)(185,5)(186,1)(187,3)(188,4)(189,1)(190,5)(191,4)(192,2)(193,6)(194,3)(195,6)(196,6)(197,3)(198,1)(199,4)(200,3)(201,0)(202,3)(203,3)(204,4)(205,2)(206,2)(207,3)(208,4)(209,6)(210,1)(211,2)(212,5)(213,4)(214,1)(215,2)(216,6)(217,2)(218,4)(219,1)(220,4)(221,2)(222,1)(223,6)(224,4)(225,5)(226,7)(227,3)(228,1)(229,2)(230,2)(231,7)(232,1)(233,5)(234,2)(235,2)(236,6)(237,0)(238,8)(239,4)(240,1)(241,3)(242,3)(243,1)(244,4)(245,5)(246,5)(247,2)(248,3)(249,4)(250,4)(251,0)(252,2)(253,2)(254,5)(255,4)(256,4)(257,3)(258,4)(259,1)(260,4)(261,2)(262,8)(263,2)(264,6)(265,4)(266,6)(267,4)(268,2)(269,3)(270,1)(271,6)(272,0)(273,4)(274,6)(275,0)(276,3)(277,1)(278,1)(279,6)(280,2)(281,1)(282,2)(283,5)(284,1)(285,5)(286,4)(287,3)(288,2)(289,3)(290,3)(291,3)(292,1)(293,3)(294,5)(295,3)(296,3)(297,5)(298,0)(299,0)(300,0)(301,4)(302,2)(303,1)(304,7)(305,4)(306,5)(307,2)(308,3)(309,1)(310,1)(311,1)(312,5)(313,6)(314,2)(315,1)(316,3)(317,2)(318,3)(319,0)(320,1)(321,2)(322,1)(323,2)(324,3)(325,0)(326,2)(327,1)(328,1)(329,2)(330,0)(331,1)(332,1)(333,0)(334,2)(335,1)(336,0)(337,1)(338,0)(339,0)(340,0)(341,1)(342,0)(343,0)(344,0)(345,0)(346,1)(347,0)(348,0)(349,1)(350,1)(351,0)(352,0)(353,0)(354,1)(355,1)(356,1)(357,0)(358,0)(359,1)(360,0)(361,0)(362,0)(363,0)(364,0)(365,0)(366,2)(367,0)(368,0)(369,0)(370,0)(371,0)(372,0)(373,0)(374,0)(375,0)(376,1)(377,0)(378,1)(379,0)(380,1)(381,0)(382,0)(383,1)(384,0)(385,0)(386,1)(387,0)(388,0)(389,0)(390,0)(391,0)(392,0)(393,1)(394,0)(395,0)(396,0)(397,0)(398,0)(399,0)(400,0)(401,0)(402,0)(403,1)(404,0)(405,1)(406,1)(407,0)(408,0)(409,0)(410,0)(411,0)(412,0)(413,0)(414,0)(415,0)(416,0)(417,1)(418,0)(419,0)(420,0)(421,0)(422,0)(423,0)(424,0)(425,0)(426,0)(427,0)(428,0)(429,1)(430,0)(431,0)(432,0)(433,0)(434,0)(435,0)(436,0)(437,1)(438,0)(439,1)(440,0)(441,0)(442,0)(443,0)(444,0)(445,0)(446,0)(447,0)(448,0)(449,0)(450,0)(451,0)(452,0)(453,0)(454,0)(455,0)(456,0)(457,0)(458,0)(459,0)(460,0)(461,0)(462,0)(463,0)(464,0)(465,0)(466,0)(467,0)(468,0)(469,0)(470,0)(471,0)(472,0)(473,0)(474,0)(475,0)(476,0)(477,0)(478,0)(479,0)(480,0)(481,0)(482,0)(483,0)(484,0)(485,0)(486,1)(487,1)(488,0)(489,0)(490,0)(491,0)(492,0)(493,0)(494,0)(495,0)(496,0)(497,0)(498,0)(499,0)(500,0)(501,0)(502,1)(503,0)(504,0)(505,0)(506,0)(507,0)(508,1)(509,0)(510,0)(511,0)(512,0)(513,1)(514,1)(515,0)(516,0)(517,0)(518,1)(519,0)(520,0)(521,0)(522,0)(523,1)(524,0)(525,0)(526,0)(527,0)(528,0)(529,0)(530,0)(531,1)(532,0)(533,0)(534,0)(535,0)(536,0)(537,0)(538,0)(539,0)(540,0)(541,0)(542,0)(543,0)(544,0)(545,0)(546,0)(547,0)(548,0)(549,0)(550,0)(551,0)(552,0)(553,0)(554,0)(555,0)(556,0)(557,0)(558,0)(559,0)(560,0)(561,0)(562,0)(563,0)(564,0)(565,0)(566,0)(567,0)(568,2)(569,0)(570,0)(571,0)(572,0)(573,0)(574,0)(575,0)(576,0)(577,0)(578,0)(579,0)(580,0)(581,0)(582,0)(583,0)(584,0)(585,0)(586,1)(587,0)(588,0)(589,0)(590,0)(591,0)(592,0)(593,0)(594,0)(595,0)(596,0)(597,0)(598,0)(599,0)(600,0)(601,0)(602,0)(603,0)(604,0)(605,0)(606,0)(607,0)(608,0)(609,0)(610,0)(611,0)(612,0)(613,0)(614,0)(615,1)(616,0)(617,0)(618,1)(619,0)(620,1)(621,0)(622,0)(623,1)(624,0)(625,0)(626,0)(627,0)(628,0)(629,0)(630,0)(631,0)(632,0)(633,0)(634,0)(635,0)(636,0)(637,0)(638,0)(639,0)(640,0)(641,0)(642,0)(643,0)(644,0)(645,0)(646,0)(647,0)(648,0)(649,0)(650,0)(651,0)(652,0)(653,0)(654,0)(655,0)(656,0)(657,0)(658,0)(659,0)(660,0)(661,2)(662,0)(663,0)(664,0)(665,0)(666,0)(667,0)(668,0)(669,1)(670,0)
}; % plot of the third histogram
\addplot[color=JungleGreen] coordinates{(0,33)(1,53)(2,54)(3,70)(4,93)(5,90)(6,145)(7,112)(8,113)(9,122)(10,122)(11,112)(12,139)(13,125)(14,133)(15,146)(16,141)(17,125)(18,140)(19,137)(20,127)(21,135)(22,125)(23,128)(24,117)(25,169)(26,159)(27,152)(28,131)(29,141)(30,141)(31,111)(32,131)(33,89)(34,113)(35,121)(36,102)(37,111)(38,86)(39,92)(40,88)(41,92)(42,87)(43,88)(44,67)(45,77)(46,80)(47,84)(48,81)(49,79)(50,67)(51,65)(52,70)(53,84)(54,72)(55,82)(56,64)(57,77)(58,86)(59,68)(60,78)(61,66)(62,65)(63,60)(64,62)(65,63)(66,47)(67,59)(68,46)(69,53)(70,51)(71,61)(72,57)(73,69)(74,42)(75,51)(76,55)(77,40)(78,40)(79,39)(80,38)(81,43)(82,35)(83,31)(84,31)(85,29)(86,26)(87,25)(88,20)(89,19)(90,30)(91,30)(92,27)(93,27)(94,24)(95,24)(96,21)(97,34)(98,22)(99,26)(100,20)(101,30)(102,21)(103,32)(104,22)(105,25)(106,17)(107,21)(108,26)(109,17)(110,16)(111,21)(112,24)(113,30)(114,13)(115,24)(116,21)(117,20)(118,19)(119,15)(120,17)(121,18)(122,16)(123,18)(124,22)(125,23)(126,14)(127,19)(128,12)(129,18)(130,21)(131,21)(132,22)(133,19)(134,12)(135,12)(136,22)(137,18)(138,17)(139,15)(140,24)(141,15)(142,17)(143,22)(144,25)(145,15)(146,15)(147,15)(148,17)(149,18)(150,14)(151,14)(152,13)(153,14)(154,16)(155,17)(156,16)(157,12)(158,13)(159,14)(160,15)(161,14)(162,14)(163,14)(164,11)(165,12)(166,12)(167,13)(168,9)(169,12)(170,9)(171,7)(172,9)(173,8)(174,6)(175,6)(176,8)(177,12)(178,8)(179,11)(180,7)(181,9)(182,6)(183,4)(184,8)(185,12)(186,6)(187,6)(188,7)(189,8)(190,6)(191,7)(192,8)(193,8)(194,11)(195,9)(196,9)(197,8)(198,7)(199,11)(200,10)(201,5)(202,8)(203,9)(204,8)(205,6)(206,5)(207,6)(208,7)(209,9)(210,8)(211,10)(212,12)(213,9)(214,8)(215,6)(216,7)(217,8)(218,9)(219,9)(220,9)(221,13)(222,9)(223,11)(224,7)(225,9)(226,9)(227,10)(228,6)(229,8)(230,4)(231,7)(232,9)(233,5)(234,6)(235,9)(236,5)(237,9)(238,6)(239,6)(240,7)(241,5)(242,7)(243,8)(244,5)(245,5)(246,6)(247,9)(248,6)(249,13)(250,8)(251,8)(252,7)(253,7)(254,9)(255,6)(256,6)(257,10)(258,8)(259,6)(260,7)(261,4)(262,5)(263,7)(264,6)(265,7)(266,8)(267,9)(268,5)(269,5)(270,7)(271,12)(272,10)(273,8)(274,7)(275,5)(276,7)(277,6)(278,5)(279,6)(280,7)(281,7)(282,7)(283,9)(284,5)(285,9)(286,7)(287,8)(288,7)(289,6)(290,5)(291,5)(292,5)(293,8)(294,6)(295,7)(296,7)(297,5)(298,8)(299,7)(300,4)(301,7)(302,6)(303,5)(304,8)(305,6)(306,8)(307,7)(308,5)(309,5)(310,5)(311,6)(312,7)(313,6)(314,9)(315,7)(316,6)(317,6)(318,4)(319,7)(320,4)(321,10)(322,8)(323,6)(324,5)(325,8)(326,4)(327,4)(328,7)(329,5)(330,6)(331,6)(332,5)(333,7)(334,5)(335,4)(336,5)(337,4)(338,6)(339,4)(340,5)(341,7)(342,4)(343,5)(344,4)(345,4)(346,5)(347,4)(348,4)(349,6)(350,4)(351,5)(352,5)(353,4)(354,4)(355,5)(356,4)(357,5)(358,4)(359,4)(360,4)(361,6)(362,4)(363,5)(364,6)(365,4)(366,5)(367,4)(368,4)(369,4)(370,4)(371,4)(372,4)(373,5)(374,4)(375,6)(376,5)(377,5)(378,4)(379,4)(380,4)(381,4)(382,4)(383,4)(384,4)(385,4)(386,4)(387,4)(388,4)(389,4)(390,4)(391,4)(392,5)(393,5)(394,4)(395,4)(396,4)(397,4)(398,4)(399,4)(400,4)(401,4)(402,4)(403,4)(404,5)(405,4)(406,5)(407,4)(408,4)(409,4)(410,6)(411,4)(412,4)(413,5)(414,4)(415,4)(416,4)(417,4)(418,4)(419,4)(420,4)(421,4)(422,4)(423,4)(424,4)(425,4)(426,4)(427,4)(428,4)(429,4)(430,4)(431,4)(432,5)(433,4)(434,4)(435,5)(436,4)(437,5)(438,4)(439,4)(440,4)(441,4)(442,4)(443,4)(444,4)(445,4)(446,4)(447,4)(448,5)(449,4)(450,4)(451,4)(452,4)(453,4)(454,4)(455,4)(456,4)(457,4)(458,4)(459,4)(460,5)(461,4)(462,4)(463,4)(464,4)(465,4)(466,4)(467,5)(468,5)(469,4)(470,4)(471,4)(472,5)(473,4)(474,4)(475,4)(476,4)(477,4)(478,4)(479,4)(480,4)(481,4)(482,4)(483,5)(484,4)(485,4)(486,4)(487,4)(488,5)(489,4)(490,4)(491,4)(492,4)(493,4)(494,4)(495,4)(496,4)(497,4)(498,4)(499,5)(500,4)(501,4)(502,4)(503,4)(504,4)(505,4)(506,4)(507,5)(508,4)(509,4)(510,4)(511,4)(512,4)(513,4)(514,4)(515,4)(516,4)(517,4)(518,4)(519,4)(520,5)(521,4)(522,4)(523,4)(524,5)(525,4)(526,4)(527,4)(528,4)(529,4)(530,4)(531,4)(532,4)(533,4)(534,4)(535,5)(536,4)(537,5)(538,4)(539,4)(540,4)(541,4)(542,4)(543,4)(544,4)(545,4)(546,4)(547,5)(548,4)(549,4)(550,4)(551,5)(552,4)(553,4)(554,4)(555,4)(556,4)(557,5)(558,4)(559,4)(560,4)(561,4)(562,4)(563,4)(564,4)(565,4)(566,4)(567,4)(568,4)(569,4)(570,4)(571,4)(572,4)(573,4)(574,4)(575,4)(576,4)(577,4)(578,4)(579,4)(580,4)(581,4)(582,4)(583,4)(584,4)(585,4)(586,4)(587,4)(588,4)(589,4)(590,4)(591,4)(592,4)(593,4)(594,4)(595,4)(596,5)(597,4)(598,4)(599,4)(600,4)(601,4)(602,4)(603,4)(604,4)(605,4)(606,4)(607,4)(608,4)(609,4)(610,4)(611,4)(612,4)(613,4)(614,4)(615,4)(616,4)(617,4)(618,4)(619,4)(620,4)(621,4)(622,4)(623,5)(624,4)(625,4)(626,4)(627,4)(628,5)(629,4)(630,4)(631,4)(632,4)(633,4)(634,4)(635,4)(636,4)(637,5)(638,4)(639,4)(640,4)(641,4)(642,4)(643,4)(644,4)(645,5)(646,4)(647,4)(648,4)(649,4)(650,4)(651,4)(652,4)(653,4)(654,4)(655,4)(656,4)(657,4)(658,4)(659,4)(660,4)(661,4)(662,4)(663,4)(664,4)(665,4)(666,4)(667,4)(668,4)(669,4)(670,4)(671,4)(672,4)(673,4)(674,4)(675,4)(676,4)(677,4)(678,4)(679,4)(680,5)(681,4)
}; % plot of the fifth histogram
\addplot[color=Magenta] coordinates{(0,36)(1,46)(2,57)(3,71)(4,85)(5,99)(6,97)(7,116)(8,106)(9,103)(10,130)(11,125)(12,124)(13,133)(14,124)(15,139)(16,141)(17,135)(18,134)(19,119)(20,136)(21,148)(22,141)(23,126)(24,151)(25,129)(26,152)(27,123)(28,133)(29,162)(30,135)(31,115)(32,130)(33,107)(34,105)(35,121)(36,119)(37,111)(38,102)(39,98)(40,99)(41,81)(42,88)(43,68)(44,84)(45,97)(46,76)(47,68)(48,70)(49,87)(50,65)(51,74)(52,60)(53,62)(54,75)(55,68)(56,78)(57,65)(58,56)(59,79)(60,86)(61,75)(62,65)(63,66)(64,59)(65,62)(66,71)(67,59)(68,63)(69,55)(70,49)(71,68)(72,61)(73,60)(74,62)(75,47)(76,58)(77,52)(78,54)(79,38)(80,26)(81,33)(82,37)(83,37)(84,37)(85,24)(86,33)(87,33)(88,35)(89,24)(90,28)(91,26)(92,19)(93,27)(94,30)(95,29)(96,24)(97,27)(98,27)(99,33)(100,26)(101,37)(102,24)(103,25)(104,30)(105,27)(106,24)(107,22)(108,18)(109,28)(110,27)(111,18)(112,18)(113,22)(114,16)(115,22)(116,22)(117,17)(118,24)(119,23)(120,18)(121,24)(122,16)(123,28)(124,21)(125,26)(126,20)(127,20)(128,21)(129,17)(130,26)(131,16)(132,20)(133,24)(134,25)(135,21)(136,21)(137,23)(138,23)(139,18)(140,23)(141,20)(142,19)(143,18)(144,27)(145,11)(146,18)(147,21)(148,20)(149,15)(150,15)(151,21)(152,17)(153,24)(154,17)(155,14)(156,20)(157,13)(158,14)(159,14)(160,15)(161,17)(162,17)(163,19)(164,16)(165,8)(166,12)(167,13)(168,13)(169,15)(170,12)(171,10)(172,10)(173,9)(174,11)(175,12)(176,9)(177,10)(178,12)(179,13)(180,12)(181,12)(182,9)(183,7)(184,11)(185,12)(186,13)(187,10)(188,10)(189,9)(190,8)(191,9)(192,11)(193,14)(194,8)(195,14)(196,9)(197,15)(198,12)(199,11)(200,12)(201,11)(202,10)(203,8)(204,11)(205,9)(206,10)(207,10)(208,13)(209,10)(210,9)(211,8)(212,8)(213,7)(214,9)(215,7)(216,9)(217,7)(218,10)(219,7)(220,9)(221,9)(222,9)(223,8)(224,8)(225,11)(226,9)(227,16)(228,9)(229,11)(230,9)(231,7)(232,8)(233,9)(234,9)(235,8)(236,11)(237,8)(238,10)(239,7)(240,7)(241,9)(242,11)(243,9)(244,11)(245,9)(246,8)(247,9)(248,9)(249,9)(250,8)(251,9)(252,9)(253,8)(254,9)(255,8)(256,11)(257,8)(258,14)(259,9)(260,9)(261,10)(262,9)(263,8)(264,9)(265,11)(266,9)(267,9)(268,9)(269,10)(270,11)(271,8)(272,6)(273,13)(274,12)(275,7)(276,8)(277,14)(278,7)(279,10)(280,8)(281,8)(282,8)(283,9)(284,11)(285,8)(286,8)(287,11)(288,9)(289,7)(290,9)(291,7)(292,8)(293,10)(294,10)(295,12)(296,7)(297,6)(298,11)(299,8)(300,8)(301,10)(302,13)(303,8)(304,11)(305,10)(306,8)(307,8)(308,9)(309,7)(310,9)(311,9)(312,8)(313,6)(314,10)(315,12)(316,8)(317,7)(318,9)(319,10)(320,9)(321,7)(322,8)(323,7)(324,9)(325,7)(326,8)(327,8)(328,9)(329,7)(330,6)(331,8)(332,10)(333,9)(334,11)(335,6)(336,7)(337,8)(338,8)(339,7)(340,7)(341,6)(342,6)(343,6)(344,7)(345,6)(346,7)(347,7)(348,6)(349,6)(350,6)(351,6)(352,6)(353,6)(354,6)(355,7)(356,6)(357,6)(358,6)(359,6)(360,6)(361,8)(362,6)(363,6)(364,8)(365,6)(366,6)(367,9)(368,7)(369,6)(370,6)(371,7)(372,6)(373,7)(374,7)(375,7)(376,6)(377,6)(378,6)(379,6)(380,6)(381,6)(382,6)(383,6)(384,6)(385,7)(386,6)(387,6)(388,7)(389,7)(390,8)(391,6)(392,6)(393,6)(394,6)(395,6)(396,6)(397,6)(398,8)(399,6)(400,6)(401,8)(402,6)(403,6)(404,6)(405,6)(406,6)(407,8)(408,7)(409,6)(410,6)(411,6)(412,6)(413,6)(414,6)(415,6)(416,6)(417,7)(418,6)(419,6)(420,6)(421,6)(422,7)(423,7)(424,7)(425,6)(426,6)(427,6)(428,7)(429,6)(430,6)(431,6)(432,7)(433,6)(434,6)(435,6)(436,6)(437,6)(438,7)(439,6)(440,6)(441,6)(442,7)(443,6)(444,6)(445,6)(446,6)(447,6)(448,7)(449,7)(450,6)(451,6)(452,6)(453,6)(454,6)(455,6)(456,6)(457,6)(458,6)(459,7)(460,6)(461,7)(462,6)(463,6)(464,7)(465,7)(466,6)(467,6)(468,7)(469,6)(470,6)(471,6)(472,6)(473,6)(474,6)(475,6)(476,6)(477,6)(478,6)(479,6)(480,6)(481,6)(482,6)(483,6)(484,6)(485,6)(486,6)(487,6)(488,6)(489,6)(490,6)(491,6)(492,7)(493,6)(494,6)(495,7)(496,6)(497,6)(498,6)(499,6)(500,6)(501,6)(502,6)(503,6)(504,6)(505,6)(506,6)(507,6)(508,7)(509,7)(510,6)(511,6)(512,6)(513,6)(514,6)(515,6)(516,6)(517,6)(518,6)(519,6)(520,6)(521,6)(522,6)(523,6)(524,6)(525,6)(526,6)(527,6)(528,6)(529,6)(530,6)(531,6)(532,6)(533,6)(534,6)(535,6)(536,6)(537,6)(538,6)(539,6)(540,7)(541,6)(542,6)(543,6)(544,6)(545,7)(546,6)(547,7)(548,6)(549,6)(550,6)(551,6)(552,6)(553,6)(554,7)(555,6)(556,6)(557,6)(558,6)(559,6)(560,6)(561,6)(562,6)(563,6)(564,6)(565,6)(566,6)(567,6)(568,6)(569,6)(570,6)(571,6)(572,6)(573,6)(574,6)(575,6)(576,6)(577,7)(578,7)(579,6)(580,6)(581,6)(582,7)(583,6)(584,6)(585,6)(586,6)(587,6)(588,6)(589,6)(590,6)(591,6)(592,6)(593,6)(594,6)(595,6)(596,6)(597,6)(598,6)(599,6)(600,6)(601,6)(602,6)(603,6)(604,7)(605,6)(606,6)(607,6)(608,6)(609,6)(610,6)(611,6)(612,6)(613,6)(614,6)(615,6)(616,6)(617,6)(618,6)(619,6)(620,6)(621,6)(622,6)(623,7)(624,6)(625,6)(626,6)(627,6)(628,6)(629,6)(630,6)(631,6)(632,6)(633,7)(634,6)(635,6)(636,6)(637,7)(638,6)
}; % plot of the fourth histogram
\legend{0\%, 10\%, 20\%, 30\%, 40\%, 50\%} % legend of the plot

\end{axis}
\end{tikzpicture}

%% file: plots/panic_fraction/injury_levels.tex
\begin{tikzpicture}[scale=0.75]
\begin{axis}[
ybar, % style of the histogram: clustered columns
width=\columnwidth, % width of the plot
height=8cm, % height of the plot
title={Injury Levels - Panic fraction},
ylabel={Amount}, % label for the y-axis
symbolic x coords={HEALTHY,MINOR,MODERATE,SERIOUS}, % labels for the x-ticks
xtick=data, % position the x-ticks at the data points
ymin=0, % minimum value of the y-axis
legend style={at={(1,1)}, anchor=north east}, % position of the legend
legend cell align=left, % alignment of the legend cells
ymajorgrids=true, % display major grids
bar width=0.15cm % width of the bars
]
\addplot[color=Turquoise,fill] coordinates{(HEALTHY,23944)(MINOR,5036)(MODERATE,805)(SERIOUS,184)
}; % plot of the first histogram
\addplot[color=Green,fill] coordinates{(HEALTHY,23932)(MINOR,4961)(MODERATE,858)(SERIOUS,209)
}; % plot of the second histogram
\addplot[color=Orange,fill] coordinates{(HEALTHY,23617)(MINOR,5278)(MODERATE,811)(SERIOUS,204)
}; % plot of the first histogram
\addplot[color=Red,fill,fill] coordinates{(HEALTHY,23607)(MINOR,5329)(MODERATE,786)(SERIOUS,186)
}; % plot of the third histogram
\addplot[color=JungleGreen,fill] coordinates{(HEALTHY,23295)(MINOR,5620)(MODERATE,728)(SERIOUS,193)
}; % plot of the fifth histogram
\addplot[color=Magenta,fill] coordinates{(HEALTHY,23120)(MINOR,5743)(MODERATE,747)(SERIOUS,194)
}; % plot of the fourth histogram
\legend{0\%, 10\%, 20\%, 30\%, 40\%, 50\%} % legend of the plot
\end{axis}
\end{tikzpicture}

\begin{tikzpicture}[scale=0.75]
\begin{axis}[
ybar, % style of the histogram: clustered columns
width=\columnwidth, % width of the plot
height=8cm, % height of the plot
ylabel={Amount}, % label for the y-axis
symbolic x coords={SEVERE,CRITICAL,FATAL}, % labels for the x-ticks
xtick=data, % position the x-ticks at the data points
ymin=0, % minimum value of the y-axis
legend style={at={(0.75,1)}, anchor=north east}, % position of the legend
legend cell align=left, % alignment of the legend cells
ymajorgrids=true, % display major grids
bar width=0.15cm % width of the bars
]
\addplot[color=Turquoise,fill] coordinates{(SEVERE,28)(CRITICAL,2)(FATAL,1)
}; % plot of the first histogram
\addplot[color=Green,fill] coordinates{(SEVERE,37)(CRITICAL,3)(FATAL,0)
}; % plot of the second histogra
\addplot[color=Orange,fill] coordinates{(SEVERE,58)(CRITICAL,20)(FATAL,12)
}; % plot of the first histogram
\addplot[color=Red,fill,fill] coordinates{(SEVERE,55)(CRITICAL,24)(FATAL,13)
}; % plot of the third histogram
\addplot[color=JungleGreen,fill] coordinates{(SEVERE,92)(CRITICAL,47)(FATAL,25)
}; % plot of the fifth histogram
\addplot[color=Magenta,fill] coordinates{(SEVERE,80)(CRITICAL,51)(FATAL,65)
}; % plot of the fourth histogram
\legend{0\%, 10\%, 20\%, 30\%, 40\%, 50\%} % legend of the plot
\end{axis}
\end{tikzpicture}

%% file: sections/conclusion.tex
In this paper, we created a model of a specific evacuation scenario in order to have a simulation model that is quite realistic and to experiment with the impact that different safety measures would have made if not overlooked. \par
The simulation model was able the fairly replicate the real event in terms of the number of fatalities, which was our main focus for the descriptive scenario. The experiments performed in the speculative scenario showed that reducing the number of people admitted to the event and not allowing glass bottles would have considerably lowered the number of injuries and also avoided victims. The mobile application has revealed itself as a double-edged sword: on the one hand, it resulted in a reduction in the evacuation time, but on the other hand, more fatalities were registered. \par
Future work could focus on using the same model in different scenarios since it would be enough to redefine the map, or to understand the impact of the various combinations of factors (for example reducing the number of people and not allowing glass bottles at the same time), or to introduce a social force model to regulate how people move inside the simulation (an example of a Netlogo implementation is available at \cite{b9}).
In conclusion, this work demonstrated how when planning an event that involves a high number of people all the necessary safety procedures must be observed thoroughly because even the slightest carelessness can result in injuries and victims.

%% file: main.bbl
\begin{thebibliography}{00}
\bibitem{b1} Almeida, J. E., Kokkinogenis, Z., \& Rossetti, R. J. F. (2012). NetLogo implementation of an evacuation scenario. Iberian Conference on Information Systems and Technologies, CISTI, 1–4.
\bibitem{b4} Abbreviated Injury Scale. (2022, July 1). In Wikipedia. https://en.wikipedia.org/wiki/Abbreviated\_Injury\_Scale
\bibitem{b5} Guardian News and Media. (2017, June 4). More than 1,500 Juventus fans in Turin injured after stampede. The Guardian. Retrieved January 1, 2023, from https://www.theguardian.com/football/2017/jun/03/juventus-fans-injured-turin-square-panic-firecrackers 
\bibitem{b6} Jamaluddin, M., Chen, H., Razek, R., Kwon, J., \&amp; Damanhoury, K. E. (2022, October 3). At least 125 killed in Indonesia Soccer Stadium Crush. CNN. Retrieved January 1, 2023, from https://edition.cnn.com/2022/10/01/asia/indonesia-soccer-stadium-stampede-persebaya-surabaya-arema-fc-intl-hnk/index.html 
\bibitem{b7} Guardian News and Media. (2018, December 8). Six dead and dozens hurt in nightclub stampede in Italy. The Guardian. Retrieved January 1, 2023, from https://www.theguardian.com/world/2018/dec/08/italy-nightclub-stampede-deaths-corinaldo
\bibitem{b8}Cultor - Atlante di Torino - Piazza San Carlo. Retrieved January 1, 2023, from http://www.atlanteditorino.it/monografie/pSanCarlo.html 
\bibitem{b9}Antoine Tordeux - Social Force Model, from https://github.com/chraibi/SocialForceModel
\bibitem{b10}Foini, D. Murino, S., Baschmakov, K. \& Rzyska, M. - CrowdLogo, from https://github.com/DavideFoini/CrowdLogo
\bibitem{b11} Danial Muhammed, Soran Saeed, Tarik A. Rashid (2019). A Comprehensive Study on
Pedestrians’ Evacuation, International Journal of Recent Contributions from Engineering,
Science \& IT ( iJES), Vol. 7, No. 4 : DOi: 10.3991/ijes.v7i4.11767
\bibitem{b12}Keith Still, G. (2007), Review of Pedestrian and Evacuation Simulations, International Journal of Critical Infrastructures, Inderscience Enterprises Ltd, vol. 3(3/4), pages 376-388.
\bibitem{b13}Knetemann, K.(2021), Understanding Group Behaviour During Evacuations Inside Buildings - An Exploratory Agent-Based Modelling Approach, Master Thesis, TU Delft.
\bibitem{b14}Li, Y. (2019), Utilizing Dynamic Context Semantics in Smart Behaviour of Informing Cyber-Physical Systems, Dissertation, TU Delft.
\end{thebibliography}
